\documentclass[footinbib,a4paper,aps,prx,superscriptaddress,reprint,twocolumn,preprintnumbers,amsmath,amssymb,nobalancelastpage,10pt,accepted=2024-07-03]{quantumarticle}

\pdfoutput=1

\usepackage[T1]{fontenc}
\usepackage{tikz}
\usepackage{lipsum}
\usepackage{relsize}
\usepackage{bigints}
\usepackage{graphicx}
\usepackage{dcolumn}
\usepackage{bm}
\usepackage{graphicx} 
\usepackage{float}
\usepackage[spanish, english]{babel}  
\usepackage[utf8]{inputenc}
\usepackage{amsmath}
\usepackage{physics}
\usepackage{bm}
\usepackage{amsfonts}
\usepackage{tablefootnote}
\usepackage[compat=1.1.0]{tikz-feynman}
\usepackage[subnum]{cases}
\usepackage{hyperref}
\hypersetup{
    colorlinks=true,
    linkcolor=blue,
    citecolor=blue,
    filecolor=blue,      
    urlcolor=blue,
    }

\usepackage{subfigure}
\usepackage{float}
\usepackage{slashed}
\usepackage{amssymb}
\usepackage{amsmath}
\usepackage{amsfonts}
\usepackage{array}

\usepackage{bm}
\usepackage{xcolor}
\graphicspath{{figures/}}

\usepackage{braket}

\definecolor{mygrey}{gray}{0.35}
\definecolor{myblue}{rgb}{0.2,0.2,0.8}
\definecolor{mygreen}{rgb}{0.2,0.8,0.5}
\definecolor{myzard}{cmyk}{0,0,0.05,0}
\definecolor{mywhite}{rgb}{1,1,1}
\definecolor{myred}{rgb}{1,0.,0.3}

 \def\ee{\mathord{\rm e}}
 
 \def\ii{\mathord{\rm i}}

\def\fourth{\textstyle\frac{1}{4}}

\renewcommand{\ii}{{\rm i}}
\renewcommand{\ee}{{\rm e}}
\def\beq{\begin{equation}}
\def\eeq{\end{equation}}
\def\barray{\begin{eqnarray}}
\def\earray{\end{eqnarray}}

\begin{document}


\title{Thermal masses and trapped-ion quantum  spin models: a self-consistent    approach to  Yukawa-type interactions in the $\lambda\!\phi^4$ model}

\author{P. Vi\~{n}as}
\author{E. L\'{o}pez}%
\author{A. Bermudez}%
\affiliation{Instituto de F\'isica Te\'orica, UAM-CSIC, Universidad Aut\'onoma de Madrid, Cantoblanco, 28049 Madrid, Spain.}%

\maketitle

\begin{abstract}
The quantum simulation of magnetism in trapped-ion systems makes use of the   crystal  vibrations to mediate pairwise  interactions between spins, which are encoded in the internal electronic states of the ions, and  measured  in experiments that probe the real-time dynamics. These interactions   can be  accounted for by a long-wavelength relativistic theory, where  the phonons are described by a coarse-grained   Klein-Gordon field $\phi(x)$ locally  coupled to the spins that acts as a carrier,  leading to an analogue of pion-mediated Yukawa interactions.  In the vicinity of a structural  transition of the ion crystal, one must go beyond the Klein-Gordon fields,  and include additional $\lambda\phi^4$ terms responsible for   phonon-phonon scattering. This leads to quantum effects that can be expressed by Feynman loop integrals that  modify the range of the Yukawa-type spin interactions; an effect that could be used to probe the  underlying fixed point of this  quantum field theory (QFT).
Unfortunately, the rigidity of the trapped-ion crystal makes it  challenging to  observe genuine quantum effects, such as the flow of the critical point with the quartic coupling $\lambda$. We hereby show that thermal effects, which can be controlled by laser cooling, can unveil this flow through the appearance of thermal masses in interacting QFTs. We perform self-consistent calculations that resum certain Feynman diagrams and, additionally, go beyond mean-field theory 
to predict  how measurements on the  trapped-ion spin system can  probe  key  properties of  the  $\lambda\phi^4$ QFT. 
\end{abstract}

\setcounter{tocdepth}{2}
\begingroup
\hypersetup{linkcolor=black}
\tableofcontents
\endgroup

\section{{\bf Introduction} \label{introduction}}
The field of quantum technologies, which aims at developing quantum devices that  provide novel functionalities with a quantifiable advantage with respect  their classical counterparts, has become a promising area of research in both the academic sector and the technological industry, e.g.~\cite{Acin_2018,Mohseni2017}. Regarding the application of these technologies  to quantum computation~\cite{nielsen00},  there has been a  remarkable  recent progress~\cite{Postler2022,Krinner2022,PhysRevLett.129.030501,https://doi.org/10.48550/arxiv.2208.01863,https://doi.org/10.48550/arxiv.2207.06431} towards the long-term goal of  a  large-scale fault-tolerant error-corrected device~\cite{FTQEC} that can outperform classical computers in  relevant tasks~\cite{Montanaro2016}. However, before these large-scale devices become available,    one is restricted to operate with small to mid-scale  prototypes in  the so-called {noisy intermediate-scale quantum} (NISQ) era~\cite{NISQ}. Here, one aims at realizing specific circuits, or even prototype quantum  algorithms~\cite{RevModPhys.94.015004}, on the largest possible number of qubits and gates, evading  the overhead of active quantum error correction. Remarkably, even in the presence of noise, these NISQ  devices have already enabled the demonstration of {quantum advantage}~\cite{google,doi:10.1126/science.abe8770}, that is, using a quantum device to solve  a problem  that  would require an unfeasible amount of time   with a classical machine. A current research goal is to extend these  demonstrations of quantum advantage to problems that can be of practical relevance in various areas of science.

In this respect, the  simulation of  quantum many-body models  is a problem of considerable interest  in different disciplines, ranging from quantum chemistry, to condensed matter and high-energy physics. As originally emphasised by Richard Feynman~\cite{feynman}, the characteristic exponential scaling of the size of the Hilbert space of a quantum many-body system 
hints at the inherent complexity of this type of problems, and the inefficiency of a brute-force numerical approach based on  classical computers. Although various numerical methods have been developed over the years to overcome  these difficulties, there are still many open questions regarding real-time dynamics, finite-fermion densities and, generally, strongly-correlated phenomena.  The  idea of quantum simulations~\cite{Cirac2012,PRXQuantum.2.017003,potential3} is to use a quantum device instead of a classical one, which can be controlled so as to reproduce the properties and dynamics of the  model of interest. This has  already found several applications in the aforementioned  areas~\cite{RevModPhys.92.015003,Bloch2012,Blatt2012,RevModPhys.93.025001,Banuls2020}. 

A quantum simulation proceeds by, first, encoding the degrees of freedom of the target model into  those of the quantum device, preparing a specific initial state, and then letting the system evolve in time before a measurement stage. This can be achieved in two ways. One can  use the same building blocks as in  quantum computers, i.e. qubits, which are then acted upon by a sequence of quantum logic gates. This sequence reproduces approximately the real-time dynamics of the model under a  Suzuki-Trotter expansion~\cite{cmp/1103900351}, and leads to the so-called {digital quantum simulations}~\cite{doi:10.1126/science.273.5278.1073}. Note that, in spite of working with qubits, one can  simulate fermionic and bosonic degrees of freedom at the expense of an overhead in the number of gates and/or qubit-register size, e.g.~\cite{Somma_2003}.   Alternatively, one can use special-purpose quantum simulators that already have spins, fermions, bosons, or combinations thereof, as the relevant degrees of freedom. This advantage comes at the price of certain limitations in the range of models that can be simulated  since, in general, one cannot  realize an arbitrary  unitary on  the exponentially-large Hilbert space. In fact, these quantum simulators are not acted upon by concatenating gates drawn from a universal gate set, but rather by letting the system evolve  continuously in time under  approximate effective Hamiltonians with a restricted set of terms. By tuning the strength of these terms, one can  mimic  approximately  the target model in a specific parameter  regime. These  devices are known as analog quantum simulators~\cite{PhysRevLett.81.3108}.

In this manuscript, we are interested in the use of long trapped-ion chains as analog  quantum simulators~\cite{Blatt2012,RevModPhys.93.025001}, targeting  models from high-energy physics. The potential advantage of working with a large number of laser-cooled ions for frequency standards and clocks led to the development of linear Paul traps~\cite{PhysRevA.45.6493}, which can store  long ion chains along the  trap symmetry axis while reducing the micromotion.  In the digital approach, these chains are operated by laser/microwave radiation to implement a high-fidelity gate set, which  has been exploited for  quantum simulations of  small-scale spin models in condensed matter, either following a Hamiltonian~\cite{doi:10.1126/science.1208001,https://doi.org/10.48550/arxiv.2209.05691} or a Lindbladian~\cite{Barreiro2011,Schindler2013} time evolution. Additionally, digital quantum simulations of a lattice gauge theory~\cite{PhysRevD.10.2445,PhysRevD.11.395,Gattringer:2010zz} have also been performed~\cite{Martinez2016,PRXQuantum.3.020324}.

For further scaling, we consider in this work the analog quantum simulation approach. Here, instead of using sequences of single and  two-qubit gates~\cite{PhysRevLett.82.1971,PhysRevA.62.022311}, one can obtain an effective spin model for the whole ion chain by acting with  always-on  far-detuned lasers, which typically lead to long-range spin-spin interactions mediated by the phonons of the chain~\cite{PhysRevLett.92.207901}. This idea has turned out to be extremely fruitful for the analog quantum simulation of  magnetism~\cite{Friedenauer2008,Kim2010,Britton2012,doi:10.1126/science.1232296,doi:10.1126/science.1232296,doi:10.1126/science.1251422,Jurcevic2014,Richerme2014,Smith2016,Zhang2017,Tan2021,Morong2021}. In all of these experiments, the focus lies on the interacting spins, while the phonons are mere auxiliary degrees of freedom. In contrast, as advocated in  recent works~\cite{a1, a2}, including these bosonic degrees of freedom in the quantum simulation offers a more complete picture,  and provides a  neat understanding of some characteristics of the spin models. 
In particular, the decay of the spin-spin couplings with the inter-ion distance can be described by  a dipolar decay with an additional exponential tail that is controlled by the  value of the laser detuning~\cite{PhysRevA.93.013625,PhysRevB.95.024431}. The origin of this particular distance dependence is  clarified by a long-wavelength description of the model~\cite{a2}, in which the phonons are described as quantised sound waves in terms of a relativistic  Klein-Gordon field $\phi(x)$~\cite{Klein1926,Gordon1926}. This scalar  field  has a Yukawa-type coupling to the spins, and can be used to mediate the spin-spin interactions and explore the  entanglement dynamics~\cite{a2,https://doi.org/10.48550/arxiv.2211.09441}.

An interesting question from the perspective of QFTs is to push this analogy further, and move to situations  beyond the free Klein-Gordon QFT. Inspired by the fermion-Higgs sector of the electroweak theory~\cite{PhysRevLett.19.1264}, 
our trapped-ion quantum simulation would become richer in the presence of  $\lambda\phi^4$ interactions~\cite{WILSON197475}. As discussed in~\cite{a2}, this type of self-interactions become the relevant ones  in the vicinity of a structural phase transition~\cite{PhysRevB.77.064111},  and lead to a variety of scattering  events for the bosonic excitations of the scalar field. These quantum effects  can be expressed in terms of Feynman loop diagrams leading to changes in the physical mass of the bosons. 
Interestingly, in the trapped-ion context,  this renormalization effect  can affect the effective spin-spin interactions of the  quantum simulator as one approaches the structural phase transition,  opening an original route for probing the underlying effective QFT via the real-time dynamics of the spins. Unfortunately, for typical realizations, the rigidity of the ion chain tends to mask these quantum effects~\cite{a2,PhysRevB.89.214408}, such that 
the change of the physical mass 
with the quartic coupling $\lambda$ is predicted to be very small. In this work, we argue that one can overcome this limitation 
by working at finite temperatures   which, in QFTS,  lead to the so-called thermal masses and the phenomenon of symmetry restoration~\cite{PhysRevD.9.3320,PhysRevD.9.3357}. We present a non-perturbative self-consistent approach to predict the renormalization  of the spin-spin  interactions at finite temperatures, and discuss how these can be used to probe the  thermal QFT that controls the structural phase transition of the ion crystal at long wavelengths.
Our presentation is organised as follows.

In Sec.~\ref{experimental}, we  discuss  how to  describe a  structural phase transition in a trapped-ion chain   by an effective  $\lambda\phi^4$ QFT, and how the trapped-ion quantum simulators of spin models can be understood in light of this QFT as a Yukawa-type problem of interactions mediated by a scalar field. We finish by discussing qualitatively the effects  of self-interactions of the scalar  field and non-zero temperatures in these Yukawa-type spin-spin interactions. In Sec.~\ref{sec:self_cons}, we derive a set of equations to deal quantitatively with self-interactions and non-zero temperatures in this field theory, which requires resuming certain types of Feynman diagrams in a non-perturbative approach. We emphasise how to deal with ultraviolet and infrared divergences in this approach, both of which arise for the low spacetime dimensions relevant to the trapped-ion problem. In Sec.~\ref{sec:self_cons_ions}, we discuss in detail how to adapt these techniques  to the specific details of trapped ions, and describe our numerical approach to solve the aforementioned equations. Finally, we discuss how these results predict a temperature-dependent contribution to the physical mass of the scalar particles, which changes the range of the  Yukawa-type spin-spin interactions,  giving concrete predictions for a realistic trapped-ion experiment. Finally, in Sec.~\ref{sec:conclusion}, we present our conclusions and outlook.  

\section{\bf Quantum simulation of Yukawa-type models \label{experimental}}

For the shake of completeness, and to fix our notation, we  review in this section  some concepts about trapped ions,  as well as the connection of the phonon-mediated spin-spin interactions in the vicinity of a structural  transition to the quantum simulation of a Yukawa-type QFT at  non-zero temperatures.

\subsection{From trapped ions to $\lambda\phi^4$ quantum fields\label{ions_to_fields}}

As advanced in the introduction, we consider a system of $N_1$ atomic ions of charge $e$ confined in a linear Paul trap~\cite{ions1,doi:10.1063/1.5088164},  and assume that the symmetry axis of this trap lies along the $x$ direction. In this kind of devices, ions are trapped using a combination of  {AC} and {DC} potentials, and can reach a stable crystalline distribution for hours or even days, depending on the trap design, and the cooling  and vacuum conditions~\cite{ghosh1995ion}. In the pseudo-potential approximation~\cite{PhysRevA.45.6493,RevModPhys.75.281}, the secular motion of the ions can be described by an effective quadratic potential with constant trap frequency $\{\omega_{\alpha}\}$ that aims at confining the ions along each axis $\alpha\in\{x,y,z\}$. Since the ions are charged, there is a   competition between this overall trapping potential and the inter-particle Coulomb repulsion, leading to
\beq
\label{eq:motional_ham}
H=\sum_{i,\alpha}\left(\frac{1}{2m_a}p_{i,\alpha}^2+\frac{1}{2}m_a\omega_{\alpha}^2r_{i,\alpha}^2\right)+\frac{1}{2}\sum_{i\neq j}\frac{e^2}{4\pi\epsilon_0}\frac{1}{|\boldsymbol{r}_i-\boldsymbol{r}_{j}|},
\eeq
where $\boldsymbol{r}_{i},\boldsymbol{p}_{i}$ are the canonical position-momentum operators of the ions with mass $m_a$ and charge $e$, and $\epsilon_0$ is the vacuum permittivity. This competition 
leads to a set of equilibrium positions $\{\boldsymbol{r}_i^0=l\tilde{\boldsymbol{r}}_i^0\}_{i=1}^N$, where we have introduced a constant with the units of length $l$ that fulfills  
$l^3={e^2}/{4\pi\epsilon_0m_a\omega_x^2}$~\cite{James1998,Marquet2003}, which are obtained by solving the non-linear equations
\beq
\label{eq:positions}
\tilde{r}_{i,\alpha}^0-\kappa_\alpha\sum_{j\neq i}\frac{\tilde{r}^0_{i,\alpha}-\tilde{r}^0_{j,\alpha}}{|\tilde{\boldsymbol{r}}^0_{i}-\tilde{\boldsymbol{r}}^0_{j}|^3}
= 0,
\eeq
where $\kappa_\alpha=(\omega_x/\omega_\alpha)^2$.
For $\omega_x\ll\omega_y,\omega_z$, these equilibrium positions lie along the symmetry axis $\boldsymbol{r}_i^0={\rm x}_i{\bf e}_x$, and present a typical lattice spacing  that is rather homogeneous in the bulk of the ion chain $|\boldsymbol{r}_i^0-\boldsymbol{r}_{i+1}^0|\approx d={\rm min}_{i\neq j}\{{\boldsymbol{r}}^0_{i}-{\boldsymbol{r}}^0_{j}|\}$. 

After expanding the Coulomb potential in Eq.~\eqref{eq:motional_ham} to quadratic order in the small displacements  $\boldsymbol{r}_i(t)=\boldsymbol{r}_i^0+\boldsymbol{u}_i(t)$, the secular ion motion  is accurately described by  $H\approx H_2$ with  
\beq
H_2=\sum_{i,\alpha}\left(\frac{\pi_{i,\alpha}^2}{2m_a}+\frac{1}{2}k_{\alpha}u^2_{i,\alpha}+\frac{1}{4}\sum_{i\neq j}k^{\alpha}_{i,j}(u_{i,\alpha}-u_{j,\alpha})^2\right),
\label{22}
\eeq
Here, $\pi_{i,\alpha}=m_a\partial_t u_{i,\alpha}$ are the  momenta conjugate to the displacement operators  $[u_{i,\alpha}(t),\pi_{j,\beta}(t)]=\ii\hbar\delta_{i,j}\delta_{\alpha,\beta}$, and we have introduced the effective spring constants
\beq
\label{eq:spring_constants}
k_{i,j}^z=k_{i,j}^y=-\frac{1}{2}k_{i,j}^x=-\frac{e^2}{4\pi\epsilon_0}\frac{1}{|{\rm x}_i-{\rm x}_j|^3}.
\eeq

\begin{figure}[t]
	\centering
	\includegraphics[width=0.95\columnwidth]{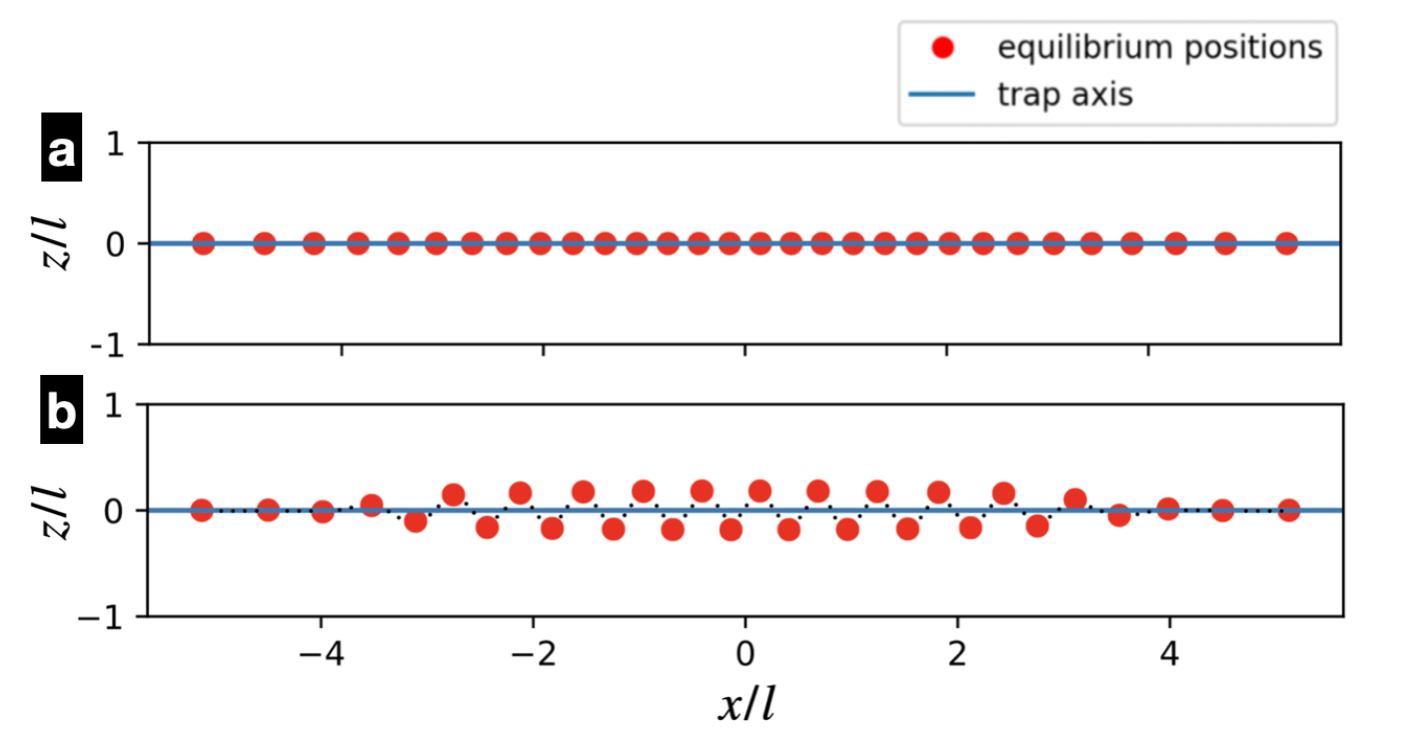}
	\caption{\label{fig:ladder}  \textbf{Trapped-ion chain and zigzag ladder:} The equilibrium positions  have been calculated numerically solving the non-linear system of equations~\eqref{eq:positions}  for a chain with $N=30$ ions, to illustrate the linear and zigzag configurations. \textbf{(a)} Linear-chain configuration for  a trap-frequency ratio  $\kappa_z=(\omega_x/\omega_z)^2=10^{-4}$. \textbf{(b)} Zigzag-ladder distribution for a trap frequency ratio of $\kappa_z=0.02$.   }

\end{figure}

According to this approximation,   the small displacements of the ion crystal are governed by a model that  resembles a simple elastic/harmonic chain~\cite{altland_simons_2010}. In comparison to this textbook example, we note that the effective  spring constants do not restrict to nearest neighbors and, moreover, one also finds  an additional local elastic term $k_{\alpha}=m_a\omega_{\alpha}^2$. For the longitudinal displacements  $u_{i,x}$, these differences do not introduce qualitative changes in the physics: the ions vibrate in the form of collective excitations in analogy to the acoustic phonons in a solid, which are associated to quantised compressional sound waves. 
In contrast, the different sign of   the spring constants~\eqref{eq:spring_constants} that couple to  the transverse displacements     $u_{i,z}$ can actually lead to a   situation that differs markedly from the quantised shear sound waves of an elastic solid. In fact, as one  rises the ratio of the Paul-trap frequencies $\kappa_z=(\omega_x/\omega_z)^2$ above a critical value $\kappa_z>\kappa_{z,{\rm c}}(N_1)$~\cite{PhysRevLett.70.818,PhysRevLett.71.2753},  there is a structural  transition from the linear chain into what is known as a zigzag ladder, as first observed experimentally in~\cite{PhysRevA.45.6493}.  For  a specific ion number and frequency ratio, one can numerically solve  the system of non-linear equations for the equilibrium positions~\eqref{eq:positions},  leading to the configurations displayed in Fig.~\ref{fig:ladder}. The critical value  decreases with  the number of ions   as a power law 
$
\kappa_{z,{\rm c}}(N_1)=aN_1^b
\label{alpha}
$
for certain constants $a>0$ and $b<0$~\cite{PhysRevLett.70.818}; a scaling  verified in experiments~\cite{PhysRevLett.85.2466}.  

These structural changes are the finite-size precursors of a quantum phase transition, which can be  characterised by the spontaneous breaking of  a discrete inversion symmetry with respect to the trap axis. Even working far away from the thermodynamic limit $N_1\to\infty$, these mesoscopic structural  transitions have a well-defined soft mode~\cite{PhysRevB.77.064111}, which allows one to derive a low-energy/long-wavelength approximation that goes beyond the elastic limit. This requires going beyond the quadratic Hamiltonian in Eq.~\eqref{22} by considering higher orders in the expansion of the Coulomb interaction~\eqref{eq:motional_ham}. This long-wavelength theory is the counterpart of the  $\lambda\phi^4$ model QFTs in which the aforementioned inversion symmetry is $\phi(x)\to-\phi(x)$ and gets spontaneously broken at a certain critical point. In the Hamiltonian formulation, this QFT can be written as
\beq
\label{eq:lambda_phi_4_nu}
H\!=\!\!\!\bigintsss\!\!\!{\rm d}{\rm x}\left(\frac{1}{2}\pi^2(x)+\frac{1}{2}(\partial_{\rm x}\phi(x))^2+\frac{m_0^2}{2}\phi^2(x)+\frac{\lambda_0}{4!}\phi^4(x)\right)\!,
\eeq
where $x=(t,{\rm x})$ are the Minkowski spacetime coordinates, and  $m_0,\lambda_0$ are the bare mass and bare coupling constants, respectively. In this expression,  we have used natural units $\hbar=c=1$, as customary in high-energy physics. For the connection to  trapped ions, it is more appropriate to work in SI units~\cite{Greiner:1996zu}, such that  
\beq
\label{eq:lambda_phi_4}
H\!=\!\!\!\bigintsss\!\!\!{\rm d}{\rm x}\!\left(\!\frac{c^2\pi^2(x)}{2\hbar^2}+\frac{\hbar^2}{2}(\partial_{\rm x}\phi(x))^2+\frac{m_0^2c^2}{2}\phi^2(x)+\frac{\lambda_0}{4!}\phi^4(x)\!\!\right).
\eeq
In comparison to  natural units, where the field operator and its conjugate momentum have scaling dimensions $[\phi]=\mathsf{ L}^0$, $[\pi]=\mathsf{ L}^{-1}$, the dimensional analysis in SI units leads to  the following scaling $[\phi]=1/\mathsf{M}^{1/2}\mathsf{L}^{1/2}$, whereas $[\pi]=\mathsf{ M}^{3/2}\mathsf{ L}^{3/2}/\mathsf{T}$.

Let us discuss first how the quadratic Klein-Gordon part of the  QFT~\eqref{eq:lambda_phi_4} 
arises through a coarse-graining procedure of Eq.~\eqref{eq:motional_ham}, when focusing on the quadratic approximation~\eqref{22}. In this way, we will highlight the important differences with respect to the effective QFT for  compressional sound waves in an elastic chain~\cite{altland_simons_2010}. By focusing on the bulk of the ion crystal, we can use periodic boundary conditions, and 
move to a Fourier representation for the small displacements
\begin{align}
\label{eq:Fourier}
u_{i,z}=\frac{1}{\sqrt{N_1}}\mathlarger{\sum}_{{\rm k}\in {\rm BZ}}\ee^{\ii{\rm k}di}u_{z}({\rm k}).
\end{align}
Here, the quasi-momentum is ${\rm k}=\frac{2\pi}{dN_1}n_1$ for $n_1\in\{1,\cdots,N_1\}$, and thus lies within the first Brillouin zone ${\rm BZ}=(0,2\pi/d]$, where $d$ is the lattice spacing in the bulk of the ion chain, which is approximately constant (see Fig.~\ref{fig:ladder} {\bf (a)}).  Using this transformation, and  focusing first on the quadratic part~\eqref{22}, one can derive the dispersion relation
\begin{align}
\omega({\rm k})=\omega_z\sqrt{1-\kappa_z\!\left(\frac{l}{d}\right)^{\!\!\!\!3}\,\mathlarger{\sum}_{r=1}^{\frac{1}{2}N_1}\frac{4}{r^3}\sin^{2}\!\left(\frac{{\rm k}dr}{2}\right)}.
\label{29}
\end{align}

This dispersion relation resembles that of a standard lattice regularization of the Klein-Gordon QFT in Eq.~\eqref{eq:lambda_phi_4}. In a Hamiltonian lattice field theory~\cite{PhysRevD.11.395}, the spatial components are discretised as ${\rm x}\mapsto {\rm x}_i=i a$, where $i\in\{1,...,N_1\}$, which require introducing an artificial lattice spacing $a$ that regularises the QFT by an ultra-violet cutoff $\Lambda_{\rm c}=\pi/a$. Following the lattice approach~\cite{lattice}, one performs the following substitutions   
\beq 
\int \!{\rm d}{\rm x} \mapsto a\sum_{i},\quad
\partial_{\rm x}\phi(x)\mapsto \frac{1}{a}\bigg(\phi(t,(i+1)a)-\phi(t,ia)\bigg).
\label{9}
\eeq
By applying a similar Fourier transformation~\eqref{eq:Fourier} to the discretised  QFT~\eqref{eq:lambda_phi_4}, one finds  the dispersion relation
\begin{align}
\label{eq:lattice_dispersion}
\omega({\rm k})=\sqrt{\frac{m_0^2c^4}{\hbar^2}+\frac{4c^2}{a^2}\sin^{2}\left(\frac{{\rm k}a}{2}\right)},
\end{align}
where ${\rm k}=\frac{2\pi}{aN_1}n_1$ for $n_1\in\{1,\cdots,N_1\}$. Although this expression clearly resembles the trapped-ion case~\eqref{29}, there are important differences. The most apparent one is that the dispersion relation in Eq.~\eqref{29} contains a dipolar tail due to the long-range nature of the effective spring constants~\eqref{eq:spring_constants}. In addition, there  is  a sign difference with respect to~\eqref{eq:lattice_dispersion} that will be crucial for  the long-wavelength coarse-graining. 

 The  dispersion relation of the  Klein-Gordon QFT $\omega^2({\rm k})={m_0^2c^4/\hbar^2+c^2 {\rm  k}^2}$ is recovered from Eq.~\eqref{eq:lattice_dispersion} by considering $|{\rm k}|\ll\Lambda_{\rm c}$.
This particular low-energy limit corresponds to a coarse-graining approximation, where the relevant scale  is much larger than the lattice spacing $\xi_0\propto 1/m_0\gg a$, and one  says that the continuum QFT is recovered by sending $a\to 0$. The  coarse graining in the trapped-ion case~\eqref{29} is slightly different. Due to the  sign difference in Eq.~\eqref{29}, the lowest-energy mode corresponds to the so-called zigzag mode, and one has to expand around ${\rm k}=\pi/d+\delta{\rm k}$. From this expansion, one can readily recover  the same Klein-Gordon dispersion, identifying  the speed of the  transversal sound waves  as
\beq
c_{\rm t}^2=d^2\omega_x^2\left(\frac{l}{d}\right)^{\!\!\!3}\!\!\eta_{N_1}\!(1),
\label{30b}
\eeq
which plays the role of an effective speed of light in the quantum simulator of the QFT~\eqref{eq:lambda_phi_4}.
In this expression,  we have  used a truncated version of the Dirichlet eta function, namely
\beq
\eta_{N_1}\!(s)=\mathlarger{\sum}_{r=1}^{\frac{1}{2} N_1}\frac{(-1)^{r+1}}{r^{s}},
\eeq
such that $\eta_{N_1}(1)\to \log 2$ in the thermodynamic limit $N_1\to\infty$.
In addition, one obtains the effective bare mass 
\beq
m_0^2=:\frac{\hbar^2\omega_{\rm zz}^2}{c_{\rm t}^4}=\frac{\hbar^2}{c_{\rm t}^4}\!\left(\omega_z^2-\omega_x^2\,\frac{7}{2}\!\left(\frac{l}{d}\right)^{\!\!\!3}\!\!\zeta_{N_1}\!(3)\right),
\label{30a}
\eeq
 where  $\omega_{\rm zz}$ is the frequency of the zigzag mode, and we have introduced a truncated version of the Riemann zeta function 
\beq
\label{eq:riemann_zeta}
\zeta_{N_1}\!(s)=\mathlarger{\sum}_{r=1}^{\frac{1}{2} N_1}\frac{1}{r^{s}},
\eeq
such that $\zeta_{N_1}(3)\to 1.202$ in the thermodynamic limit $N_1\to\infty$, corresponding to Ap\'ery's constant.

At this point, we emphasise that the coarse-graining procedure  in a physical trapped-ion lattice of spacing $d$ does not require sending the spacing to $d\to 0$. Alternatively, one can tune the parameters close to a critical point where the bare mass~\eqref{30a} would vanish $m_0\to 0$, such that the effective Compton wavelength fulfills  $\xi_0=\hbar/m_0c_{\rm t}\gg d$, and the low-energy physics does not depend on microscopic lattice details, but rather on the universal properties captured by  a  QFT. As it can be checked from Eq.~\eqref{30a}, this critical point $m_0^2|^{\phantom{2}}_{\rm c}=0$ coincides precisely with the linear-to-zigzag   transition at
\beq
\label{eq:critical_point_classical}
1-\kappa_{z,{\rm c}}\,\frac{7}{2}\!\left(\frac{l}{d}\right)^{\!\!\!3}\!\!\zeta_{N_1}\!(3)=0.
\eeq
This critical point $\kappa_{z,{\rm c}}(N_1)$ has a scaling with the number of ions that  agrees with the previously-mentioned  power laws~\cite{PhysRevLett.70.818,PhysRevLett.71.2753}, already for moderate  ion numbers~\cite{PhysRevLett.109.010501}.

So far, our discussion has focused on the elastic/quadratic terms and the dispersion relation, but we have not identified the fields yet. In order to find their trapped-ion analogue,  we need to separate the rapid oscillations around the zigzag mode~\cite{PhysRevLett.106.010401,a2} from a slowly-varying envelope that will play the role of the coarse-grained field~\cite{altland_simons_2010,Affleck1990}. In addition, one has to rescale the  position and momentum operators in Eq.~\eqref{22} to achieve the correct scaling dimensions  below Eq.~\eqref{eq:lambda_phi_4}
\beq
\label{eq:trapped_ion_coarse_garining}
\phi(x)=\frac{(-1)^i}{\sqrt{m_ad^3}}{u_{i,z}(t)},\hspace{2ex}\pi(x)={(-1)^i}\sqrt{m_ad}\,\,\pi_{i,z}(t),
\eeq
such that one  recovers  the  canonical algebra $[\phi(t,{\rm x}),\pi(t,{\rm y})]=\ii\hbar{\delta_{i,j}}/{d}\to\ii\hbar\delta({\rm x}-{\rm y})$. Since the coarse-grained fields vary slowly, one can perform a gradient expansion $ \phi(t,{\rm y})\approx \phi(t,{\rm x})+({\rm y}-{\rm x})\partial_{\rm x}\phi(t,{\rm x})$, and obtain  the Klein-Gordon part of Eq.~\eqref{eq:lambda_phi_4} from the original microscopic  theory~\eqref{22}. 

Let us note that, at the level of the effective QFT~\eqref{eq:lambda_phi_4}, the critical point $m_0^2=0$ is stable thanks to the additional quartic potential with $\lambda_0>0$. In the trapped-ion case,  one needs to  extend the microscopic theory~\eqref{22} to  quartic order~\cite{Marquet2003}, and apply  again the    gradient expansion to identify the analogue of the bare quartic coupling. In this procedure,  the rigidity of the trapped-ion chain becomes relevant.
As occurs with other long-wavelength  descriptions in condensed-matter and high-energy physics~\cite{doi:10.1142/1882,Gonzalez1995,Son:2002zn}, one can use thermodynamic arguments  when constructing the effective field theory. In the present case, where the coarse-grained QFT can be directly obtained from the trapped-ion model via the gradient expansion, one finds that, in addition to the effective speed of light~\eqref{30b} and bare mass~\eqref{30a}, the rigidity modulus of the trapped-ion chain~\cite{1986iv,2017mcp..book.....T} gives rise to an additional dimensionless Luttinger paramater~\cite{a2}, namely
\beq
\label{eq:lutt_param}
K_0=\frac{m_adc_{\rm t}}{\hbar}.
\eeq
This parameter quantifies the rigidity of the trapped-ion chain under a shear strain that aims  at deforming it transversely, and becomes important when identifying the coupling constant of the $\lambda\phi^4$ model. In fact, after expanding Eq.~\eqref{eq:motional_ham} to  fourth order and identifying the terms that are more important for the structural phase transition, one finds $H\approx H_2+H_4$ where 
\beq
H_4=\frac{1}{2}\sum_{i\neq j}\frac{\beta^{z}_{i,j}}{4!}(u_{i,z}-u_{j,z})^4,
\eeq
and we have introduced the quartic coupling matrix
\beq
\beta_{i,j}^z=\frac{e^2}{4\pi\epsilon_0}\frac{9}{|{\rm x}_i-{\rm x}_j|^5}.
\eeq
At this point, one can perform again  the gradient expansion below Eq.~\eqref{eq:trapped_ion_coarse_garining}, which allows one to find the final microscopic expression for the $\lambda\phi^4$ coupling  
\begin{align}
\lambda_0=\frac{729\zeta_{N_1}\!(5)}{2K_0^4}m_a^3\omega_x^2l^3.
\label{lambda_iones}
\end{align}

We have thus discussed how  a trapped-ion chain in the vicinity of a structural phase transition serves as a quantum simulator of a regularised self-interacting QFT~\eqref{eq:lambda_phi_4} with the effective speed of light in Eq.~\eqref{30b}, and  the bare parameters  in Eqs.~\eqref{30a},~\eqref{eq:lutt_param} and~\eqref{lambda_iones}. In this context, note that the critical point~\eqref{eq:critical_point_classical} is obtained by setting the bare mass~\eqref{30a} to zero, and thus corresponds to the classical field-theory calculation in which the minimum of the  quartic potential  underlying Eq.~\eqref{eq:lambda_phi_4} changes from a single to a double well. At this point,  the
$\mathbb{Z}_2$ inversion symmetry of the real scalar field $\phi\to-\phi$  gets spontaneously broken. From the perspective of QFTs, one knows that  the classical quartic potential gets quantum contributions in terms of Feynman loop diagrams~\cite{PhysRevD.7.1888,PhysRevD.9.1686}, such that the single to double-well will no longer be characterised by the classical critical point. Instead, the excitations are dressed leading to a physical mass $m_0^2\to m_{\rm P}^2$, and one finds  that the phase transition $m_{\rm P}^2=0$ yields a critical line in parameter space $(m_0^2,\lambda_0)$ that separates the symmetry-broken and symmetry-preserved regions. Going back to the  trapped-ion problem,   the critical point~\eqref{eq:critical_point_classical} will flow with the coupling strength $\kappa_{z,{\rm c}}\to\kappa_{z,{\rm c}}(\lambda_0)$, defining a line that separates the linear chain $\kappa_z<\kappa_{z,{\rm c}}(\lambda_0)$ from the  zigzag ladder $\kappa_z>\kappa_{z,{\rm c}}(\lambda_0)$. 

Note that, if one takes the continuum limit $a\to 0$ in the lattice field approach~\eqref{9},  the UV divergences of the loop integrals must be subtracted from the bare mass in order to get finite parameters and draw a meaningful phase diagram. In the trapped-ion case, on the contrary,  the lattice spacing remains constant as one approaches the critical point, and it is the  physical Compton wavelength which becomes very large  $\xi_0\mapsto\xi_{ \rm P}\gg d$, justifying the long-wavelength description. Accordingly,  one can stick to the bare parameters without any additional subtraction, and still find a meaningful phase diagram. One must  bear in mind that  the critical line will depend on the lattice spacing and other non-universal microscopic properties. In contrast,  as one approaches this critical line,  the universal properties of the phase transition, i.e. scaling critical  exponents, should be controlled by the fixed point of the continuum QFT~\eqref{eq:lambda_phi_4}. This corresponds to the so-called Wilson-Fisher fixed point, which can be characterised by an $\epsilon$-expansion in higher dimensions~\cite{PhysRevLett.28.240}.  In $D=1+1$ dimensions, however,  the  perturbative renormalization-group techniques~\cite{WILSON197475} underlying this $\epsilon$-expansion break down, as all perturbations are relevant in the renormalization-group sense. Localising the  critical line of the lattice model, as well as the critical exponents of the corresponding fixed point,  requires using non-perturbative techniques, such as Monte Carlo or tensor-network methods~\cite{PhysRevD.58.076003,PhysRevD.79.056008,Takanori_Sugihara_2004,montecarlo,PhysRevD.88.085030,Kadoh2019,PhysRevD.99.034508,PhysRevResearch.2.033278,PhysRevD.106.L071501}. The continuum limit of these studies is consistent with a QFT presenting a second-order  phase transition where the $\mathbb{Z}_2$  symmetry gets broken~\cite{PhysRevD.13.2778}, and where the fixed point lies in the  universality class of the two-dimensional Ising model.

\subsection{Yukawa-type  interactions and real-time spin dynamics \label{experimental_sub}}

As advanced in the introduction, we are interested in the use of trapped-ions as quantum simulators of Yukawa-type spin models, and how the real-time dynamics of the spins can serve to probe the underlying interacting QFT. So far, however, we have only discussed the motional degrees of freedom of the ions, leading to the effective QFT~\eqref{eq:lambda_phi_4}  in the vicinity of the linear-to-zigzag transition~\eqref{eq:critical_point_classical}. As noted in the introduction, the  ions also have an atomic  structure with many electronic levels, among which one can select a pair of long-lived states to encode the degrees of freedom of a chain of   spins  $\{\ket{\uparrow_i},\ket{\downarrow_i}\}_{i=1}^{N_1}$. These two states can correspond to the so-called optical, hyperfine or Zeeman qubits in trapped-ion quantum computing~\cite{doi:10.1063/1.5088164}. Building on this seminal proposal, the subsequent experimental and theoretical efforts have turned trapped ions  into one of the leading platforms in the quest of building a large-scale fault-tolerant quantum computer~\cite{doi:10.1063/1.5088164}. In addition to the   experiments that we have already mentioned~\cite{Postler2022,https://doi.org/10.48550/arxiv.2208.01863}, which contribute to  the continuous effort~\cite{Chiaverini2004,Schindler1059,Nigg302,Linkee1701074,Negnevitsky2018,Fluhmann2019,Stricker2020,https://doi.org/10.48550/arxiv.2010.09681,Erhard2021,Egan2021,PhysRevLett.127.240501,PhysRevX.11.041058,PhysRevX.12.011032} towards trapped-ion quantum error correction~\cite{PhysRevX.7.041061},  a variety of NISQ algorithms have also  been realized over the years~\cite{Gulde2003,Barrett2004,Riebe2004,doi:10.1126/science.1110335,doi:10.1126/science.aad9480,Figgatt2017,doi:10.1126/science.aaw9415}. The success of these implementations relies on the very accurate performance of the universal gate set~\cite{fidelities1,PhysRevLett.117.060504,PhysRevLett.117.060505,PhysRevLett.117.140501,Erhard2019,PhysRevLett.123.260503}. This also includes  high-accuracy two-qubit  gates which, typically,  are the most challenging part of the gate set in any  platform. Moreover, either by exploiting ion shuttling~\cite{Kielpinski2002,doi:10.1116/1.5126186,doi:10.1126/science.1177077} or  individually-addressed laser beams ~\cite{Debnath2016,Figgatt2019,PRXQuantum.2.020343}, one can implement these entangling  gates between arbitrary qubit pairs in the  crystal. This  allows for  a programmable connectivity that has allowed to demonstrate  complex protocols with current NISQ error rates,  e.g.~\cite{Postler2022,https://doi.org/10.48550/arxiv.2208.01863},   which would not be possible  if the connectivity was local. We note that the digital methods can also be combined with  classical variational methods in a hybrid approach, which has found applications in  quantum chemistry~\cite{PhysRevX.8.031022,Nam2020} and also lattice gauge theories~\cite{Kokail2019}. Particularly in the context of lattice field theories, where one  eventually aims at recovering the continuum limit, increasing the size of these quantum simulators will be important in the near future. In addition, including both the matter particles and interaction carriers in the quantum simulation will also be important, allowing one to get closer to the higher-dimensional non-Abelian gauge theories of the  standard model, which is one of the longer-term goals~\cite{https://doi.org/10.1002/andp.201300104,Zohar_2016,doi:10.1080/00107514.2016.1151199,Bannuls2020,Carmen_Ba_uls_2020,doi:10.1098/rsta.2021.0064,Klco_2022,https://doi.org/10.48550/arxiv.2204.03381}.

In the analog approach to  quantum simulation~\cite{Friedenauer2008,Kim2010,Britton2012,doi:10.1126/science.1232296,doi:10.1126/science.1232296,doi:10.1126/science.1251422,Jurcevic2014,Richerme2014,Smith2016,Zhang2017,Tan2021,Morong2021}, these spins are coupled to the transverse phonons of the ion crystal by applying an off-resonant spin-dependent dipole force. This force can be obtained by a bichromatic laser beam with a pair of tones $n=1,2$ of frequency (wave-vector) $\omega_{{\rm L},n} (\boldsymbol{k}_{{\rm L},n})$, which are either {\it (i)} symmetrically detuned with respect to the red and blue motional sidebands~\cite{PhysRevLett.82.1971,PhysRevA.62.022311}, or {\it (ii)} far-detuned from the atomic transition~\cite{Leibfried2003,doi:10.1098/rsta.2003.1205}. In the following, we consider the second scheme although, as  discussed   later, the formalism also applies  to the former, albeit in a different spin basis. 

In contrast to the accumulation of errors in digital quantum simulators, which arise  from both the  imperfect operations in a gate sequence and the approximations inherent to the Suzuki-Trotter expansion, it is  more difficult to account for the growth of errors in  analog quantum simulators. Nonetheless, one expects that their accumulation in time will not be as detrimental as  in a generic digital approach, especially when one is interested in recovering intensive observables~\cite{Cirac2012}. Accordingly, the common expectation is that one will be able to demonstrate quantum advantage using near-term  experiments with  these analog quantum simulators~\cite{Bloch2012,Blatt2012}. In fact, some experiments have already been able to track  real-time dynamics of a many-body model, going beyond the capabilities of current classical computers with state-of-the-art numerical algorithms~\cite{Trotzky2012}. Even if it is   difficult to provide mathematical proofs of  quantum advantage, as one is departing from the quantum-computing framework in which the scaling of required resources for a target accuracy is routinely  estimated,  there has been recent  progress   in this direction~\cite{https://doi.org/10.48550/arxiv.2204.13644,https://doi.org/10.48550/arxiv.2212.04924}.

Working in the  Lamb-Dicke regime~\cite{doi:10.1098/rsta.2003.1205}, the light-matter interaction of the ions with the bichromatic laser beam leads to a local interaction between the spins and the ion displacements.  If the laser  illuminates the entire ion chain,  this reads 
\beq
\label{eq:spin_depn_force}
V(t)=\sum_ig\sin(\Delta\omega_{\rm L}t-\Delta{k}_{{\rm L},x}{\rm x}_i)\sigma_i^z u_{i,z}^{\phantom{z}}(t),
\eeq
where we have introduced the beatnote frequency (wave-vector) $\Delta\omega_{\rm L}=\omega_{{\rm L},1}-\omega_{{\rm L},2}$ ($\Delta\boldsymbol{k}_{\rm L}=\boldsymbol{k}_{{\rm L},1}-\boldsymbol{k}_{{\rm L},2}$), and the Pauli operator $\sigma_i^z=\ket{\uparrow_i}\bra{\uparrow_i}-\ket{\downarrow_i}\bra{\downarrow_i}$. In the above expression, the force strength reads $g=\hbar\Omega_{\rm L}\Delta{k}_{{\rm L},z}$, where $\Omega_{\rm L}$ is the differential ac-Stark shift between the two electronic states~\cite{doi:10.1098/rsta.2003.1205}. For simplicity, we  will assume that $\Delta\boldsymbol{k}_{{\rm L}}||{\bf e}_z$, such that $\Delta{k}_{{\rm L},x}=0$ from now on. Working in the weak-force regime 
\beq
\label{eq:far_detuned_condition}
\frac{|g|}{\sqrt{2m_a\omega({\rm k})}}\ll |\omega({\rm k})-\Delta\omega_{\rm L}|,
\eeq
it is possible to   obtain an effective spin model  with long-range interactions mediated by the transverse phonons~\cite{PhysRevLett.92.207901}, which governs the slower dynamics of the spins~\cite{Friedenauer2008,Kim2010,Britton2012,doi:10.1126/science.1232296,doi:10.1126/science.1232296,doi:10.1126/science.1251422,Jurcevic2014,Richerme2014,Smith2016,Zhang2017,Tan2021,Morong2021}.  
Using the coarse-graining in Eq.~\eqref{eq:trapped_ion_coarse_garining}, this coupling can be expressed as 
\begin{gather}
{V}(t)=-\sum_iJ(t,{\rm x}_i)\phi(t,{\rm x}_i)\sigma_i^z,
\label{32}
\end{gather}
which can be understood as a Yukawa-type coupling if one writes $\sigma_i^z=2\psi^{\dagger}(t,{\rm x}_i)\psi(t,{\rm x}_i)-1$ in terms of a local fermionic field. Here,  we have introduced the harmonic source terms
\beq
\label{eq:source_term}
J(t,{\rm x}_i)=\frac{(-1)^{i+1}}{\sqrt{m_ad^3}}\frac{g}{K}\sin(\Delta\omega_{\rm L}t-\Delta{k}_{{\rm L},x}{\rm x}_i).
\eeq

Let us  now address an important point by discussing when the coarse-grained description is expected to capture the properties of the long-range spin-spin interactions. The idea is that, whenever the harmonic sources~\eqref{eq:source_term}  oscillate at a frequency that is close to the frequency $\omega({\rm k})$~\eqref{29} of the lowest  zigzag mode at ${\rm k}=\pi/d$, namely $\Delta\omega_{\rm L}\approx\omega(\pi/d)$, then the long-wavelength approximation will provide reliable results. This is actually more general than  working at the vicinity of the structural phase transition, which is a low-energy approximation since the zigzag mode becomes the soft mode of the transition $\omega(\pi/d)\approx 0$. Accordingly, the long-wavelength approximation  is also valid at other parameter regimes far from the structural  transition, in which the additional non-linearities are unimportant. It is in these regimes,  in which the elastic terms~\eqref{22} suffice to describe the problem, where most of the experimental trapped-ion quantum simulators work~\cite{Friedenauer2008,Kim2010,Britton2012,doi:10.1126/science.1232296,doi:10.1126/science.1232296,doi:10.1126/science.1251422,Jurcevic2014,Richerme2014,Smith2016,Zhang2017,Tan2021,Morong2021}.

In this case, the coarse-grained theory corresponds to a  Klein-Gordon field with the dispersion relation of Eq.~\eqref{29}, and the real-time dynamics of the spins is governed   by a unitary evolution operator  with an effective Ising Hamiltonian
\begin{gather}
U_{\rm eff}(t)\approx\ee^{-\ii(t_{\rm f}-t_0)H_{\rm eff}/\hbar}, \hspace{2ex} H_{\rm eff}=\frac{1}{2}\mathlarger{\sum}_{i,j} J_{ij}\sigma_i^z\sigma_j^z,
\label{34}
\end{gather}
where we have introduced the spin-spin coupling strengths
\beq
\label{eq:couplings_propagator}
J_{ij}=-\hbar \Omega_{\rm L}^2\eta_{x}^22\omega_x\,G^{\rm E}_{m_{\rm eff}}({\rm x}_i-{\rm x}_i).
\eeq
As first noted in the experimental works~\cite{Britton2012,doi:10.1126/science.1232296}, by controlling the detuning of the laser beams used to generate the long-range  interactions in a trapped-ion crystal, the decay of the spin-spin couplings with the inter-ion distance can be  approximately fitted to a power law with a tunable exponent.
In the expression above, the distance decay of the spin-spin couplings is  controlled by  a dimensionally-reduced Euclidean propagator of the Klein-Gordon field, which is obtained after integrating the temporal  components in $x_1-x_2=(\tau,{\rm x}_i-{\rm x}_j)$, and reads as follows
\beq
G^{\rm E}_{ m_{\rm eff}}({\rm x}_i-{\rm x}_i)=d\bigintss_0^{\!\frac{2\pi}{d}}\!\!\!\!\!\!\!{\rm d}{\rm k}\,\frac{\ee^{\ii {\rm k}({\rm x}_i-{\rm x}_j)}}{\omega^2({\rm k})-\Delta\omega_{\rm L}^2}.
\label{eq:Euclidean_propagator}
\eeq
This propagator can be interpreted in terms of excitations of the scalar field with an effective mass/Compton wavelength
\beq
\label{eq:eff_compton1}
m_{\rm eff}=\sqrt{m_0^2-\left(\frac{\hbar\Delta\omega_{\rm L}}{c_{\rm t}^2}\right)^{\!\!2}}=\frac{\hbar}{\xi_{\rm eff}c_{\rm t}},
\eeq
where we recall that the trapped-ion bare mass has been defined in Eq.~\eqref{30a}. The interpretation in terms of the coarse-grained picture is really transparent, the spin-spin interactions are mediated by the Klein-Gordon field, and the spin-spin couplings are controlled by the distance decay of the corresponding propagator and, thus, by the effective Compton wavelength.  It is important to emphasize that, as will be discussed further below,  a classical numerical simulation of the real-time dynamics~\eqref{34} of the full Yukawa-coupled QFT of spins and scalar fields is a very complicated problem especially in the presence of further non-linearities, finite temperatures, and certain microscopic details of the trapped-ion system. This connects to the prospects  of using experimental quantum simulations in the context of quantum advantage, as mentioned in the introduction. 

In the continuum QFT, the propagator~\eqref{eq:Euclidean_propagator} can be expressed in terms of modified Bessel functions~\cite{fradkin} which,  for the case of one spatial dimension,  lead to exponentially-decaying spin-spin couplings with a typical decay length controlled by the Compton wavelength~\cite{a2}. For a standard lattice regularization of the Klein-Gordon QFT, which  only has nearest-neighbor couplings leading to Eq.~\eqref{eq:lattice_dispersion}, the previous integral can be   evaluated by extending the momentum to the   complex plane ${\rm k}d\mapsto z\in\mathbb{C}$, and by noticing that the integrand contains a simple pole that contributes with an exponential distance decay~\cite{fradkin}. However, for the trapped-ion regularization, the dispersion relation~\eqref{29} also presents a branch cut, which contributes with an additional  term with a dipolar distance decay. Altogether, the spin-spin couplings read  
 \beq
\label{eq:spin_spin_couplings_ions}
J_{ij}=	J_{\rm eff}\!\left(\frac{\omega_{x}^4\eta_{N_1}\!(1)}{(\omega_{{z}}^2-\Delta\omega_{\rm L}^2)^2}\frac{l^3}{|{\rm x}_i-{\rm x}_j|^3}-(-1)^{i-j}\frac{\xi_{\rm eff}d^2}{l^3}\ee^{-\frac{|{\rm x}_i-{\rm x}_j|}{\xi_{\rm eff}}}\right)\!,
\eeq
where we have introduced an effective strength
\beq
J_{\rm eff}=\frac{\hbar\Omega_{\rm L}^2\eta_{{x}}^2}{\omega_{{x}}}\frac{2}{\eta_{N_1}\!(1)},
\eeq
and $\eta_{{x}}=k_{{\rm L},x}\sqrt{\hbar/2m_a\omega_x}$ is the  Lamb-Dicke parameter. In the final expression~\eqref{eq:spin_spin_couplings_ions}, one can substitute the inhomogeneous  equilibrium positions ${\rm x}_i$ that stem from the solution of Eq.~\eqref{eq:positions}.  Given the relation of the bare mass with    the zigzag mode frequency in Eq.~\eqref{30a}, we see that the effective Compton wavelength~\eqref{eq:eff_compton1} that sets the range of the interactions  in turn depends on by how close the laser beatnote is with respect to this vibrational mode.

As discussed in detail in~\cite{a2}, the exponential part of the spin-spin interactions is the typical contribution  of a Yukawa interaction in $D=1+1$ dimensions~\cite{fradkin,peskin}. On the other hand, the physical lattice regularization  of this QFT  stemming from the trapped-ion microscopic model also includes long-range couplings that modify the   dispersion relation. This leads to a branch-cut discontinuity  that is responsible for an additional dipolar-decaying part in the spin-spin couplings. As discussed in~\cite{a2}, the effective coarse-grained description  of these Yukawa-type interactions is  very accurate already for moderate-size chains  with  tens of   ions, considering realistic parameters and inhomogeneous ion crystals. 

Since the distance dependence of the spin-spin couplings can be inferred from various  experimental techniques that measure the real-time dynamics of the effective spin model~\cite{Britton2012,Jurcevic2014,doi:10.1126/science.1251422}, it follows that one could use the spins  as  probes of the underlying effective relativistic QFT. One can thus use experiments in which the crystal sizes are sufficiently large to admit a continuum limit as  quantum simulators of a relativistic Yukawa-type problem. In analogy to the simulations of lattice gauge theories~\cite{PhysRevD.10.2445,PhysRevD.11.395,Gattringer:2010zz} mentioned above,  the trapped-ion spins would mimic the matter degrees of freedom since, under a Jordan-Wigner transformation~\cite{Jordan1928},  they can be interpreted as  fermionic matter. In the present case, this matter would be quenched, and sit on  the sites of a real physical lattice  corresponding to the ion crystal. Instead of having a gauge field to mediate the interactions, which must be defined on the links of the lattice to allow for a local  symmetry, here one has a global inversion symmetry of a real scalar field $\phi(x)$  defined on the lattice sites. Note that the continuum limit is not recovered by sending the physical lattice spacing to zero, but rather by working in parameter regimes where only the long-wavelength properties are  of relevance. These regimes correspond to the vicinity of a second-order phase transition.

To test the validity of this expression~\eqref{eq:spin_spin_couplings_ions} in a specific realistic setup, one may consider crystals of up to $N=51$ atomic $^{40}{\rm Ca}^+$  ions  in a  linear Paul trap, forming a chain with an  overall length of $L\approx 250\,\mu$m~\cite{insbruck}. Reference~\cite{insbruck} reviews several aspects of the coherent manipulation of such  long strings. It is also shown that stability of the chain is characterised by a lifetime of approximately $27\,$s, a period after which the collisions of the ions with the background gas become important and can even melt the crystal~\cite{Ob_il_2019,PhysRevA.100.033419}. This sets an upper bound for the possible time of the experimental runs, and demands  fast and efficient   laser cooling in order to reach a low vibrational state of all axial and transverse  modes,  as required for high-fidelity quantum control. We note that these constraints  can be met  by using resolved-sideband cooling for the transverse modes~\cite{doi:10.1063/1.1354345},  and {polarization-gradient} cooling for the axial ones~\cite{polarization}. For the  transverse modes relevant for our work, we will assume that the mean phonon number ranges between $1$-$30$, which will become important below.

The spins are encoded as optical qubits in the long-lived electronic  states $\downarrow\,=4{\rm S}_{1/2}(m = +1/2)$, and $\uparrow\,=4{\rm D}_{5/2
}(m = +5/2)$, which have long coherence  $T_2=64\,$ms and decay  $T_1=1\,$s times~\cite{insbruck}. The motional degrees of freedom depend on the average lattice spacing $d\approx 5\,\mu$m, and the  transverse and axial trap frequencies, which have typical values of $\omega_y/2\pi=2.93\,$MHz, $\omega_z/2\pi=2.89\,$MHz and  $\omega_x/2\pi=127\,$kHz in most of the experiments discussed in~\cite{insbruck}. In the following, we assume  a  $N_1=30$ ion chain, in which the classical estimate of the linear-to-zigzag transition~\eqref{eq:critical_point_classical}  for a fixed axial trap  frequency of $\omega_x/2\pi=127\,$kHz corresponds to the critical transverse  frequency $\omega_{z,{\rm c}}/2\pi\approx2.7\,$MHz. We thus consider modifying this trap frequency  in the range $\omega_z/2\pi\in[2.75,2.89]\,$MHz to cover the regimes in the symmetry-preserved phase, i.e. a stable chain configuration, which  are well described by  either the  Klein-Gordon effective QFT  for $\omega_z\gg \omega_{z,{\rm c}}$, or otherwise by the full  $\lambda\phi^4$  model for $\omega_z\gtrsim \omega_{z,{\rm c}}$. 

\begin{figure}[t]
	\centering
	\includegraphics[width=0.95\columnwidth]{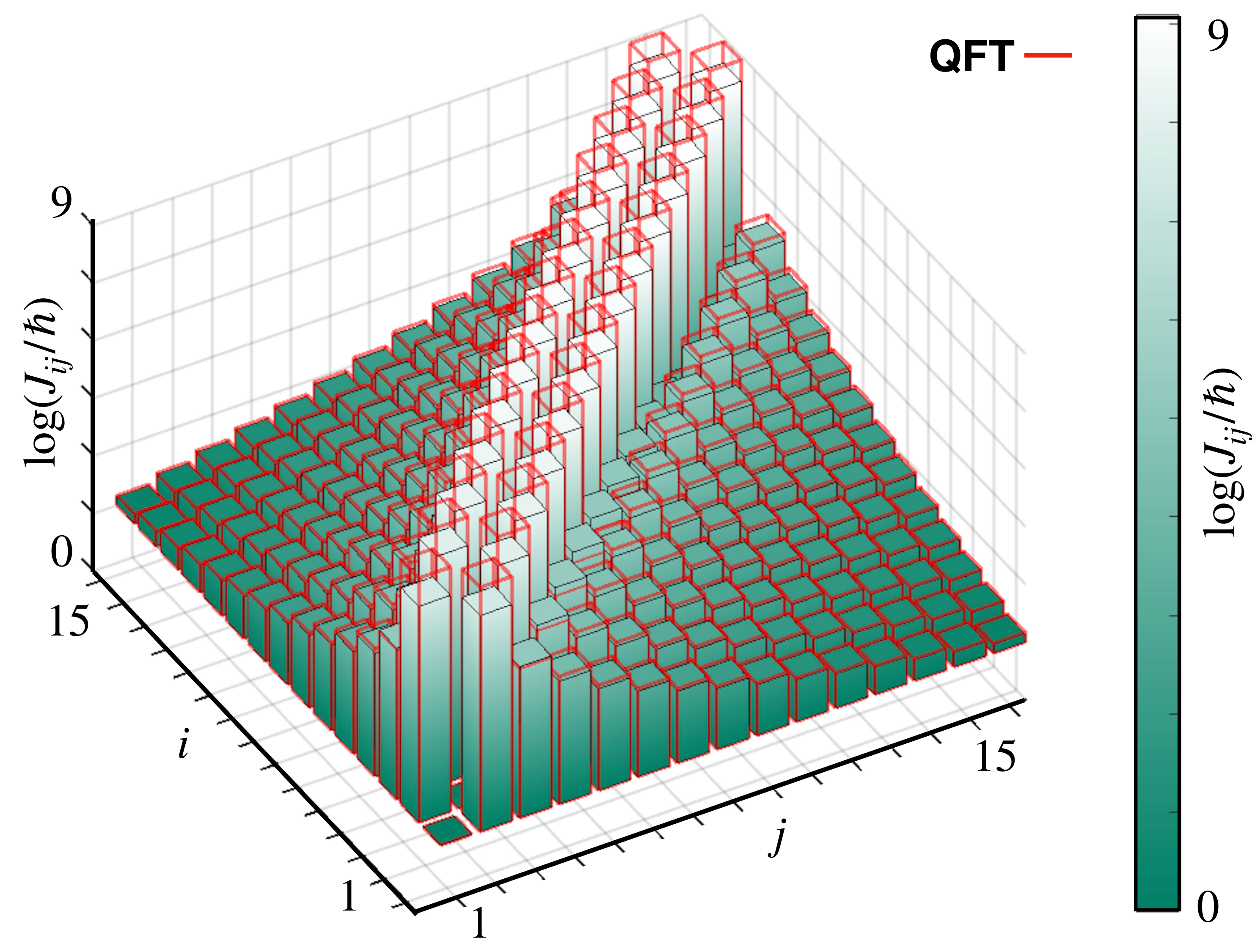}
	\caption{\label{fig:spin_cpuplings}  \textbf{Spin-spin interactions and Yukawa-type QFT:} We represent the spin-spin interactions $J_{ij}$ in log-scale for half of a chain of $N_1=30$ ions (the remaining half is related by inversion symmetry). The green bars represent the exact expression~\eqref{eq:exact_J_ij}, whereas the red lines stand for the coarse-grained approximation of Yukawa-type interactions mediated by an effective Klein-Gordon field~\eqref{eq:spin_spin_couplings_ions}.  }
\end{figure}

In the experiments discussed in~\cite{insbruck}, they use a bichromatic laser scheme with a pair of beams  symmetrically detuned with respect to the red and blue motional sidebands~\cite{PhysRevLett.82.1971,PhysRevA.62.022311}. This leads to a state-dependent dipole force that acts in a different spin-basis with respect to Eq.~\eqref{eq:spin_depn_force}. As discussed in~\cite{PhysRevLett.92.207901,insbruck,Monroe_varenna}, the spin-spin interactions mediated by the transverse phonons are given by the following expression
\begin{gather}
\label{eq:exact_J_ij}
H_{\rm eff}=\frac{1}{2}\mathlarger{\sum}_{i,j}\, J_{ij}\sigma_i^x\sigma_j^x,\hspace{2ex} J_{ij}=|\Omega_{\rm L}|^2E_{\rm R}\mathlarger{\sum}_{n_1=1}^{N_1}\frac{\mathcal{M}_{in_1}\mathcal{M}_{jn_1}}{\Delta\omega_{\rm L}^2-\omega_{z,n_1}^2},
\end{gather}
where the recoil energy is $E_{\rm R}=(\hbar\Delta k_{\rm L})^2/2m_a$, and the laser beatnote  is now referenced to the qubit transition $\omega_{{\rm L},1}-\omega_{{\rm L},2}=\omega_0+\Delta\omega_L$, with $\omega_0/2\pi=411.5\,$THz. Also, in contrast to the previous case~\eqref{eq:spin_depn_force}, $\Omega_{\rm L}$ is now the quadrupole Rabi frequency~\cite{Benhelm_thesis} instead of the differential ac-Stark shift.
In this expression~\eqref{eq:exact_J_ij}, we have introduced the normal-mode frequencies $\omega_{z,n_{1}}$ and wavevectors $\mathcal{M}_{in_1}$ of the transverse phonons~\cite{Marquet2003}. We note that Eq.~\eqref{eq:exact_J_ij} does not rely on   the  approximations used to obtain the coarse-grained QFT prediction~\eqref{eq:spin_spin_couplings_ions}. The only common implicit assumption is that one neglects the higher-order quartic terms, as well as other off-resonant contributions beyond the  dipole force~\eqref{eq:spin_depn_force}.  

Considering the experimental trap frequencies $\omega_z/2\pi=2.89\,$MHz and $\omega_x/2\pi=127\,$kHz, and setting $\omega_{{\rm L},1}-\omega_{{\rm L},2}=\omega_0+\Delta\omega_{\rm L}$ where $\Delta\omega_{\rm L}/2\pi=2.43\,$MHz is red-detuned with respect to the zigzag mode at $\omega_{z,N_1}/2\pi=2.7\,$MHz, we compare in Fig.~\ref{fig:spin_cpuplings} the spin interactions in a trapped-ion chain of $N_1=30$ ions obtained from the exact expressions of the inhomogeneous  crystal~\eqref{eq:exact_J_ij} with those of the coarse-grained Yukawa-mediated expression~\eqref{eq:spin_spin_couplings_ions}. In this figure, we have set   $\Omega_{\rm L}/2\pi=1\,$MHz, such that $J_{i_0,i_0+1}/h\approx1.4\,$kHz for the ion at the center of the chain $i_0=15$. The agreement of the coarse-grained QFT prediction, which has no fitting parameter,  is rather remarkable for such moderate-size chains. As expected, there are some deviations at distances around the UV lattice cutoff, but the accuracy becomes very good as the ion distance increases.

In summary, we can conclude that trapped-ion quantum simulators of spin models, like those of the experiments in~\cite{Friedenauer2008,Kim2010,doi:10.1126/science.1232296,doi:10.1126/science.1232296,doi:10.1126/science.1251422,Jurcevic2014,Richerme2014,Smith2016,Zhang2017,Tan2021,Morong2021} but performed on chains with a few tens of ions and a spin-dependent force that is red-detuned with respect to the transverse zigzag mode,  can be used to probe the physics of a relativistic QFT when interpreted in the light of a Yukawa-type interaction. As seen from the glasses of QFT, the problem becomes more interesting in the presence of quartic interactions and non-zero temperatures, as discussed below.

\subsection{Effect of $\lambda\phi^4$ term and non-zero temperatures \label{thermal_sub}}
\label{sec:thermal_masses_discussion}

Let us now discuss  how to take into account the  non-linearities~\eqref{eq:lambda_phi_4} that go beyond the harmonic approximation~\eqref{22}, as well as a non-zero temperature,  both of which become relevant in a trapped-ion experiment. In the parameter regime underlying Fig.~\ref{fig:spin_cpuplings}, the trapping conditions are far from the linear-to-zigzag critical point $\omega_{z}\gg\omega_{z,{\rm c}}$, such that the harmonic-crystal approximation~\eqref{22} is  an accurate description of the collective phonons. In terms of the coarse-grained  QFT~\eqref{eq:lambda_phi_4}, the bare quartic term~\eqref{lambda_iones} is negligible in comparison to the bare mass~\eqref{30a}, and the effective QFT can be reduced to that of a real Klein-Gordon field. Temperature  enters in the conditions that must be imposed on the strength of the Yukawa-type 
coupling~\eqref{32}, which we recall  had to fulfil the weak-coupling constraint~\eqref{eq:far_detuned_condition}  in the zero-temperature limit. In the presence of thermal fluctuations, there can be some bosonic enhancement of the Yukawa-type  coupling, and the condition~\eqref{eq:far_detuned_condition} must be upgraded to
\beq
\label{eq:weak_constraint_couplings}
\frac{|g|\sqrt{1+2\overline{n}({\rm k}})}{\sqrt{2m_a\omega({\rm k})}}\ll |\omega({\rm k})-\Delta\omega_{\rm L}|,
\eeq
 where $\overline{n}({\rm k})$ is the mean excitation number of the scalar field. In this regime,  the real-time time-evolution of the spins is still described by a long-range Ising model~\eqref{34} in complete analogy to the zero-temperature case.  Once again, the spin-spin couplings~\eqref{eq:couplings_propagator}  are proportional to a dimensionally-reduced Euclidean propagator~\eqref{eq:Euclidean_propagator}.  The important difference is that this is not the free propagator of a Klein-Gordon QFT, and can get a temperature-dependent contribution as  one goes beyond the harmonic non-interacting limit.   

As mentioned previously, one can control $\overline{n}({\rm k})$ by means of   laser cooling. For a single trapped ion~\cite{PhysRevA.46.2668}, one can show that the resulting state corresponds to a thermal Gibbs state, and that the mean-phonon number can be controlled by adjusting the ratio of the cooling and heating rates, which depend themselves on the  detuning and intensity  of the cooling laser. For a 
trapped-ion chain, the steady state that results from the cooling  will depend on the laser-cooling scheme. For instance, in the resolved-sideband limit, one can individually cool each of the transverse modes to a target mean excitation number $\overline{n}({\rm k})$. 
In this work, we assume that, after such cooling stage, the trapped-ion crystal is allowed to thermalize, and one can then define a single effective temperature  according to the Bose-Einstein distribution of a thermal Gibbs state
\beq
\label{eq:BE_dist}
T=\frac{\hbar\omega({\rm k})}{k_{\rm B}\log(1+1/\overline{n}({\rm k}))}.
\eeq
Given the nature of the coarse-grained approximation underlying~\eqref{eq:lambda_phi_4}, we believe that, even if there are deviations to this idealised thermalization, and the effective temperature varies for the different modes, the only relevant quantity is the effective temperature around the zigzag mode ${\rm k}=\pi/d$. 

Let us now discuss why we need to go beyond the harmonic limit.   In this limit, increasing  the  temperature only results in more stringent conditions for the Yukawa-coupling strength~\eqref{eq:weak_constraint_couplings}, which in turn result in a weaker magnitude for the spin-spin interactions~\eqref{eq:spin_spin_couplings_ions}. The situation becomes  more interesting in the presence of non-linearities, such as the quartic coupling~\eqref{lambda_iones}, which become more important  as   one approaches the linear-to-zigzag transition $\omega_{z}\to\omega_{z,{\rm c}}$. At zero temperature~\cite{a2}, a  path integral can be used to show that the spin-spin couplings~\eqref{eq:couplings_propagator} are still described by Eq.~\eqref{eq:couplings_propagator}, but this time 
 controlled by a dressed propagator and renormalised sources. This propagator depends on all of the possible scattering processes of the self-interacting scalar field, as its excitations propagate between a pair of distant ions. These interaction effects will shift the physical mass of the carrier from $m^2_0\mapsto m_{\rm P}^2$, changing  the effective Compton wavelength~\eqref{eq:eff_compton}, which will in turn change the  range of the Yukawa-type interactions~\eqref{eq:spin_spin_couplings_ions}. This  can  be inferred  experimentally from  changes of Fig.~\ref{fig:spin_cpuplings} as one approaches the structural transition. 

Perturbatively, one  expects a zero-temperature shift of the bare mass that scales with the quartic coupling and stems from the so-called tadpole Feynman diagram $\delta m_0^2|_{T=0}\propto\lambda_0$~\cite{fradkin, peskin}, which will be discussed at length below. As estimated in~\cite{a2}, these type of effects are inhibited by the very large rigidity of the ion chain, as the quartic coupling scales with the inverse fourth power of the rigidity modulus~\eqref{lambda_iones}. Moreover, given the fact that the effective coarse-grained parameters in Eqs.~\eqref{30b},~\eqref{30a},~\eqref{eq:lutt_param} and~\eqref{lambda_iones} depend on the microscopic experimental parameters in a convoluted manner, it is not straightforward to modify $\lambda_0$ independently of the others to see its effect on the range of the interactions. 

In addition to the above zero-temperature shift of the bare mass, the tadpole diagram  also contributes to the so-called thermal mass~\cite{Laine:2016hma,kapusta_gale_2006}. Perturbatively, this reads as follows
\beq
\delta m_0^2|_{T}\propto\frac{\lambda_0}{2}\int\frac{{\rm d}{\rm k}}{2\pi}\frac{\overline{n}({\rm k})}{\omega({\rm k})},
\label{eq:thermal_mass_pert}
\eeq
where the proportionality  stems from changes that would be required to convert from natural units into SI units. Regardless of the proportionality factor, the important result of thermal field theory is that, in spite of having a small quartic coupling, the above integral can lead to a shift proportional to some power of the ratio $\delta m_0^2|_{T}\propto\lambda_0(T/m_0)^\alpha$~\cite{Laine:2016hma}. For $D=1+1$ dimensions, we find $\delta m_0^2|_{T}\propto\lambda_0(T/|m_0|)$ in the high-temperature regime   $T\gg m_0$, which  can thus amount    to a  large shift even for a perturbative $\lambda_0$. In the present context, this thermal shift will  change the  Compton wavelength~\eqref{eq:eff_compton1}, and the range of the spin-spin interactions~\eqref{eq:spin_spin_couplings_ions}; an effect that will be characterised non-perturbatively below.

Coming back to the trapped-ion regularization of the $\lambda\phi^4$ QFT~\eqref{eq:lambda_phi_4}, the classical critical point~\eqref{eq:critical_point_classical} of the $\mathbb{Z}_2$-breaking phase transition, obtained by setting the bare mass~\eqref{30a} to zero $m_0^2=0$, will  get contributions from the thermal masses. Hence,  the physical mass will be dressed with   both temperature and quartic-coupling contributions, such that the phase transition at  $m_{\rm P}^2(\lambda_0,T)|_{\rm c}=0$ now corresponds  to a critical surface in parameter space $(m_0^2,\lambda_0,T)$. Determining how the critical point flows with  temperature and coupling strength in the lattice model is a non-perturbative problem that requires going beyond the previous discussion and will be addressed in the following sections.  In general, if one starts in a symmetry-broken phase $m_{\rm P}^2(\lambda_0,T)<0$, by solely increasing the temperature, symmetry restoration can take place $m_{\rm P}^2(\lambda_0,T+\delta T)>0$~\cite{PhysRevD.9.3320,PhysRevD.9.3357}, such that one ends in a symmetry-preserved phase. In terms of a linear-to-zigzag phase transition at finite temperatures~\cite{PhysRevLett.105.265703},  an analogue of this restoration of symmetry has actually been  observed already in  experiments with small trapped-ion chains~\cite{PhysRevA.99.063402,PhysRevA.102.033116,PhysRevB.103.104106}. To the best of our knowledge,   the connection to relavistic QFTs and thermal masses has not been previously noticed in the trapped-ion literature. In these experiments, the cooling lasers are not only used to prepare an initial thermal state, but are actually applied continuously during the experiment, such that  one is exploring  the steady state of a driven-dissipative system. In spite of these differences, the observations resemble the phenomenon of thermal masses and the restoration of symmetry. As discussed in~\cite{PhysRevA.99.063402}, a linear shift of the critical ratio $\kappa_{z,{\rm c}}$~\eqref{eq:critical_point_classical} with temperature has been reported, which is somewhat reminiscent of the previous thermal mass shift. In the following section, we will present a self-consistent non-perturbative method that can be used to derive quantitative predictions of how the critical point flows, and how the range of the spin-spin interactions changes with temperature.

\section{\bf Self-consistency beyond mean field theory}
\label{sec:self_cons}

In this section, we present a detailed account  of our self-consistent  prediction for the range of the spin-spin interactions, and how it changes with temperature as one approaches the linear-to-zigzag transition. We start by reviewing the functional approach to the self-interacting scalar field theory, and then move on to discuss our approach to get a set of finite self-consistent equations in spite of UV and IR divergences.

\subsection{Perturbative generating functional of $\lambda\phi^4$ fields}

In this subsection, we review the functional approach to the diagrammatic perturbation theory of the self-interacting scalar field~\cite{peskin,fradkin,ryder}.
The central object in this approach is the generating functional, obtained by adding a source term to the path integral. In  its Euclidean version, where the  $\boldsymbol{x}=(\tau,{\rm x})$ is obtained from $x=(t,{\rm x})$ after a Wick rotation   $\tau=\ii t$, the generating functional is given by 
\begin{equation}
\label{eq:gen_fucntional}
    Z[J]=\int \!{\cal D} \phi \ee^{- \!\!\bigintssss\!\!\! {\rm d}^2 x \big( {\cal L} - J(\boldsymbol{x}) \phi(\boldsymbol{x}) \big)}, 
\end{equation}
where ${\cal L}$ the Lagrangian density associated to the  Hamiltonian field theory in Eq.~\eqref{eq:lambda_phi_4_nu}, provided one works in imaginary time. 
The generating functional  gives the $n$-point Green's functions upon derivation with respect to the sources
\begin{equation}
\langle \phi(\boldsymbol{x}_1)...\phi(\boldsymbol{x}_n)\rangle
=\frac{\delta^n {\cal Z}[J]}{\delta J(\boldsymbol{x}_1)...\delta J(\boldsymbol{x}_n)}\eval_{J=0}, 
\end{equation}
where the expectation value is taken on the vacuum, and ${\cal Z}[J]=Z[J]/Z[0]$ is the normalized generating functional. 

In the absence of interactions, the Lagrangian ${\cal L}={\cal L}_0$ reduces to a real Klein-Gordon theory quadratic in the fields,
and the normalised generating functional finds a simple analytical expression 
\begin{equation}
\label{eq:free_Z}
    {\cal Z}_0[J]= \ee^{{1 \over 2}\!\!\bigintssss\!\!\! {\rm d}^2 x \!\!\bigintssss\!\!\! {\rm d}^2 y \, J(\boldsymbol{x}) \Delta_0 (\boldsymbol{x}-\boldsymbol{y}) J(\boldsymbol{y}) } \ ,
\end{equation}
where $\Delta_0 (\boldsymbol{x}-\boldsymbol{y})=\langle \phi(\boldsymbol{x})\phi(\boldsymbol{y})\rangle$ is the Euclidean propagator of a free scalar field. The propagator is most conveniently written in terms of its Fourier decomposition
\begin{equation} 
\Delta_0(\boldsymbol{x})=\!\!\bigintsss\!\!\! \frac{{\rm d}^2k}{(2\pi)^2}\,\tilde{\Delta}_0(\boldsymbol{k})\ee^{\ii \boldsymbol{k}\cdot\boldsymbol{x}},\quad \tilde{\Delta}_0(\boldsymbol{k})=\frac{1}{\boldsymbol{k}^2+m_0^2},
\label{3} 
\end{equation}
where the Euclidean momentum $\boldsymbol{k}=(k_0,{\rm k})$ is related to the 2-momentum $k=(\omega,{\rm k})$ in Minkowski spacetime by $k_0=-\ii\omega$.
In an interacting theory ${\cal L}={\cal L}_0 + {\cal L}_{\rm int}$, such as ${\cal L}_{\rm int}={\lambda_0 \over 4!} \phi^4$ in our case, the generating functional can be obtained using the following identity in functional analysis
\begin{equation}
\mathcal{Z}[J]=\frac{\hspace{2ex}\exp\left[-\bigintsss\!\!{\rm d}^2z\mathcal{L}_{\rm int}\left(\frac{\delta}{\delta J(\boldsymbol{z})}\right)\right]\!\!\mathcal{Z}_0[J]\hspace{3.5ex}}{\hspace{2ex}\exp\left[-\bigintsss\!\!{\rm d}^2z\,\mathcal{L}_{\rm int}\left(\frac{\delta}{\delta J(\boldsymbol{z})}\right)\right]\!\!\mathcal{Z}_0[J]\,\bigg|_{J=0}} \ .
\label{6}
\end{equation}

This expression can be expanded in a power series in $\lambda_0$ to the desired perturbative order. The expansion can be graphically represented in terms of Feynman diagrams. Taking into account that the denominator in Eq.~(\ref{6}) cancels out the so-called vacuum diagrams, i.e. those diagrams that do not contain any external source terms, the final expression reads

\begin{widetext}
\begin{eqnarray*}
\mathcal{Z}[J]=\bigg[1-\frac{1}{4}
\begin{tikzpicture}
\begin{feynman}
\node (a);
\node[right=0.5cm of a, dot] (b);
\node[right=0.5cm of b] (c);
\diagram* {
(a) -- [insertion=0] (b) -- [insertion=1] (c) 
};
\path (b)--++(90:0.3) coordinate (A);
\draw (A) circle(0.3);
\end{feynman}
\end{tikzpicture}
+\frac{1}{8}
\begin{tikzpicture}
\begin{feynman}
\node (a);
\node[right=0.5cm of a, dot] (b);
\node[above=0.6cm of b, dot] (e);
\node[right=0.5cm of b] (c);
\diagram* {
(a) -- [insertion=0] (b) -- [insertion=1] (c) 
};
\path (b)--++(90:0.3) coordinate (A);
\draw (A) circle(0.3);
\path (b)--++(90:0.9) coordinate (B);
\draw (B) [dot];
\draw (B) circle(0.3);
\end{feynman}
\end{tikzpicture}
+\frac{1}{8}
\begin{tikzpicture}
\begin{feynman}
\node (a);
\node[right=0.5cm of a, dot] (b);
\node[right=0.7cm of b, dot] (c);
\node[right=0.5cm of c] (d);
\diagram* {
  (a) -- [insertion=0] (b) -- (c) -- [insertion=1] (d)
};
\path (b)--++(90:0.3) coordinate (A);
\draw (A) circle(0.3);
\path (c)--++(90:0.3) coordinate (B);
\draw (B) circle(0.3);
\end{feynman}
\end{tikzpicture}
+&\frac{1}{12}
\begin{tikzpicture}[baseline={(0,-0.1)}]
\begin{feynman};
\node (a);
\node[right=0.5cm of a, dot] (b);
\vertex[right=0.3cm of b] (i);
\node[right=0.23cm of i, dot] (c);
\node[right=0.5cm of c] (d);
\diagram* {
   (a) -- [insertion=0] (b) -- (i) -- (c) -- [insertion=1] (d)
};
\path (i)--++(90:0) coordinate (A);
\draw (A) circle(0.3);
\end{feynman}
\end{tikzpicture}
-\frac{1}{24}
\begin{tikzpicture}[baseline={(0,-0.1)}]
\begin{feynman}
\node (a);
\node[right=0.5cm of a, dot] (b);
\node[right=0.5cm of b] (c);
\node[above=0.5cm of b] (d);
\node[below=0.5cm of b] (e);
\diagram* {
  (a) -- [insertion=0] (b) -- [insertion=1] (c),
  (d) -- [insertion=0] (b) -- [insertion=1](e)
};
\end{feynman}
\end{tikzpicture}
+\frac{1}{12}
\begin{tikzpicture}[baseline={(0,-0.1)}]
\begin{feynman}
\node (a);
\node[right=0.5cm of a, dot] (i);
\node[right=0.5cm of i, dot] (b);
\node[right=0.5cm of b] (c);
\node[above=0.5cm of b] (d);
\node[below=0.5cm of b] (e);
\diagram* {
  (a) -- [insertion=0] (i) -- (b) -- [insertion=1] (c),
  (d) -- [insertion=0] (b) -- [insertion=1] (e)
};
\path (i)--++(90:0.3) coordinate (A);
\draw (A) circle(0.3);
\end{feynman}
\end{tikzpicture} \hspace{24ex}\phantom{+}\nonumber\\
&
+\frac{1}{32}
\begin{tikzpicture}[baseline={(0,0.55)}]
\begin{feynman}
\node (a);
\node[right=0.5cm of a, dot] (b);
\node[right=0.5cm of b] (c);

\node[above=1.3cm of a] (a2);
\node[above=1.3cm of b, dot] (b2);
\node[above=1.3cm of c] (c2);
\diagram*{
  (a) -- [insertion=0] (b) -- [insertion=1] (c),
  (a2) -- [insertion=0] (b2) -- [insertion=1] (c2)
};
\path (b)--++(90:0.3) coordinate (A);
\draw (A) circle(0.3);
\path (b2)--++(270:0.3) coordinate (B);
\draw (B) circle(0.3);
\end{feynman}
\end{tikzpicture}
+\frac{1}{16}
\begin{tikzpicture}[baseline={(0,0.25)}]
\begin{feynman}
\node (a);
\node[right=0.5cm of a, dot] (b);
\node[right=0.5cm of b] (c);

\node[above=0.6cm of a] (a2);
\node[above=0.6cm of b, dot] (b2);
\node[above=0.6cm of c] (c2);
\diagram*{
  (a) -- [insertion=0] (b) -- [insertion=1] (c),
  (a2) -- [insertion=0] (b2) -- [insertion=1] (c2)
};
\path (b)--++(90:0.3) coordinate (A);
\draw (A) circle(0.3);
\end{feynman}
\end{tikzpicture}
+\cdots
\bigg]\mathcal{Z}_0[J].
\end{eqnarray*}
\begin{equation}
\label{8}
\end{equation}
\end{widetext}
The crosses $\times=J(\boldsymbol{x})$ represent sources, the blobs $\begin{tikzpicture}
\begin{feynman}
\node[dot] (b);
\diagram* {
(b)
};
\end{feynman}
\end{tikzpicture}=\lambda_0$ interaction vertices, and the different free propagators obey $\begin{tikzpicture}
\begin{feynman}
\node (a);
\node[right=0.5cm of a, dot] (b);
\diagram* {
(a) -- [insertion=0] (b)
};
\path (b)--++(90:0.3) coordinate (A);
\end{feynman}
\end{tikzpicture}=\Delta_0(\boldsymbol{x}-\boldsymbol{z}_1)$,$\begin{tikzpicture}
\begin{feynman}
\node[dot] (b);
\diagram* {
(b)
};
\path (b)--++(90:0.15) coordinate (A);
\draw (A) circle(0.15);
\end{feynman}
\end{tikzpicture}=\Delta_0(0)$ and $\begin{tikzpicture}
\begin{feynman}
\node[dot] (a);
\node[right=0.5cm of a, dot] (b);
\diagram* {
(a) -- (b)
};
\end{feynman}
\end{tikzpicture}=\Delta_0(\boldsymbol{z}_1-\boldsymbol{z}_2)$.
As usual, we recall that one has to integrate over the location of the interaction vertices, here denoted by $\{\boldsymbol{z}_i\}$.

\subsection{Feynman diagrams and self-consistent equations}
\label{sec:SC_eqs}

The Feynman diagrams in Eq.~(\ref{8}) provide quantum corrections to the $n$-point correlation functions of the QFT. For instance,  the propagator  of the interacting theory, which corresponds to  the 2-point function,  can be written as
\beq
 \label{eq:self_energy}
 \tilde{\Delta}(\boldsymbol{k})={1 \over \boldsymbol{k}^2+m_0^2+\mathsf{\Sigma}(\boldsymbol{k})} \ ,
 \eeq
where $\mathsf{\Sigma}(\boldsymbol{k})$, the self-energy~\cite{PhysRev.75.1736,sch1,sch2}, contains the contributions of all 1-particle irreducible diagrams with two external legs. We recall that these diagrams are those that cannot be separated into  disconnected
pieces by cutting an internal propagator~\cite{peskin}. At order $\lambda_0^2$, the self-energy is given by the first, second and fourth diagrams in Eq.~(\ref{8}). The third diagram is not 1-particle irreducible and, thus, does not contribute  to $\mathsf{\Sigma}(\boldsymbol{k})$. The first two diagrams belong to the tadpole type, namely, each loop integral depends on a single internal momentum that is  independent of the external momentum of the propagator. As a consequence, the tadpole  amounts to a renormalization of the mass and, in the $(1+1)$-dimensional case, is responsible for the only ultra-violet divergence of the QFT. The fourth diagram is the so-called sunrise or sunset diagram and, in contrast, also depends also on external momenta. In this way, this diagram  contributes to both a mass and a wavefunction renormalization.

Evaluating the self-energy non-perturbatively is of course out of reach. It is however possible to resum the tadpole family to all orders in the quartic coupling. The result is encoded in the self-consistent equation
\beq
\mathsf{\Sigma}_{\rm td}= \frac{\lambda_0}{2}\!\!\bigintsss\!\!\!\frac{{\rm d}^2k}{(2\pi)^2} \frac{1}{\boldsymbol{k}^2+m_0^2+\mathsf{\Sigma}_{\rm td}}.
\label{tad}
\eeq
This approximation to the self-energy becomes exact for an $O(N)$ vector model in the limit  $N\to\infty$~\cite{coleman}. In condensed matter, it corresponds to the Hartree method of mean-field theory~\cite{PhysRevD.10.2042}. At this point,   we must discuss  the occurrence of divergences in the QFT~\eqref{eq:lambda_phi_4}. In fact, the integral in Eq.~\eqref{tad}   contains a UV logarithmic divergence in $D=1+1$ dimensions, which  must be regularized by introducing a   cutoff. Since
we ultimately want  to use this self-consistent resumation of Feynman diagrams to predict  changes of the Yukawa-type spin-spin interactions in the trapped-ion chain~\eqref{eq:spin_spin_couplings_ions},  the regularisation scheme should correspond to a lattice. In the standard approach of lattice field theories, where the spatial derivatives are exchanged for discrete differences~\eqref{9} that only lead to nearest-neighbor couplings,  the continuum  propagator appearing in Eq.~\eqref{eq:self_energy} with the tadpole-resummed self-energy ~\eqref{tad}, must be substituted by 
\beq
\label{eq:lattice_prop}
\tilde{\Delta}_{\rm td}({\boldsymbol k})=\frac{1}{k_0^2+\hat{\rm k}^2+\mu^2 },
\eeq
where the analogue of the spatial momentum is
\beq
\label{eq:lattice_mom}
\hat{\rm k}=\frac{2}{a}\sin\!\left(\frac{{\rm k}a}{2}\right).
\eeq
 As already mentiond above, the quasi-momentum lies within the First Brillouin zone ${\rm k}={2 \pi  \over N_1 a}n_1$ for $n_1\in\{1,\cdots,N_1\}$. In the propagator,  we have also defined the tadpole-renormalised mass through the following self-consistent equation, sometimes  refereed to as the gap equation,
\beq
\label{eq:mu}
\mu^2 =m_0^2+\mathsf{\Sigma}_{\rm td}=m_0^2+\frac{\lambda_0}{2}\!\!\!\bigintsss\!\!\!\frac{{\rm d}^2 k}{(2\pi)^2}\tilde{\Delta}_{\rm td}(\boldsymbol{k}).
\eeq
We note that the integrals over  quasi-momenta are to be understood as mode sums  $\int\!\!\frac{{\rm dk}}{2\pi}\to\frac{1}{N_1a}\sum_{n_1}$. In  the thermodynamic limit, one sends $N_1 \rightarrow \infty$, such that ${\rm k}\in[0,{2\pi \over a})$, and the corresponding integral can be evaluated analytically
\beq
\label{eq:self_cons_lattice}
\mu^2=m_0^2+\frac{\lambda_0}{4\pi}\frac{1}{\sqrt{1+\fourth \mu^2 a^2}}\,\mathsf{K}\!\left(\frac{1}{1+\fourth \mu^2 a^2}\right).
\eeq
Here, $\mathsf{K}(x)=\int_0^{\pi/2}\!\!{\rm d}\theta(1-x\sin^2\theta)^{-1/2}$ is the complete elliptic integral of the first kind~\cite{elliptic1}, which is finite,  showing that the UV divergence is regularised by the non-zero lattice spacing. For our scalar field theory, this is the only UV divergence. In Sec.~\ref{sec:self_cons_ions}, we will discuss how this self-consistent equation, as well as the expression that follow, can be adapted to the trapped-ion case where the dispersion relation~\eqref{29} includes the not only nearest-neighbors but the full dipolar tail. For the moment, however, we continue with the standard Hamiltonian lattice regularization, and the corresponding propagator~\eqref{eq:lattice_prop}.

Let us now discuss the connection of this renormalised mass with the aforementioned phase transition. 
The physical mass of the interacting QFT is determined by the pole of the quantum corrected propagator~\eqref{eq:self_energy} in Minkowski spacetime. The analytical prolongation   is simply achieved by replacing $k_0 \to -\ii \omega$, such that the propagator  in Minkowski spacetime  $\tilde{\Delta}(k)$ is obtained from $-\ii\tilde{\Delta}(-\ii\omega,{\rm k})\to \tilde{\Delta}(k)$. 
In the lattice version~\eqref{eq:lattice_prop}, the Lorenz invariance between  energy  and  spatial momentum  is broken. The physical mass is then  defined as the on-shell energy at vanishing spatial momentum  $m_{\rm P}^2={\bar \omega}^2$, and   determined by the pole of the propagator
\beq
{\bar \omega}^2-m_0^2-\mathsf{\Sigma}(-\ii{\bar \omega},0)=0.
\label{pole}
\eeq
 In the mean-field tadpole approximation, the physical mass would thus be  $m_{\rm P}^2=\mu^2$, which implies that the classical critical point  for the $\mathbb{Z}_2$-breaking phase transition $m_0^2=0$ will flow to a different value that can be obtained by solving for $\mu^2=0$. When trying to solve this equation, one faces a different type of divergence in the QFT, namely an infra-red (IR) divergence.
Note that the elliptic integral in Eq.~\eqref{eq:self_cons_lattice} inherits the logarithmic IR divergence of the tadpole~\eqref{tad} when $\mu^2=0$, since $\lim_{x\to 1} \mathsf{K}(x) = \infty$. This prevents criticality 
to be achieved for any finite negative value of the bare mass $m_0^2$, independently of the value of the quartic coupling  $\lambda_0$ and the non-zero lattice spacing. Accordingly, the vacuum of the 1+1 $\lambda\phi^4$ theory would always remain in the unbroken phase if one sticks to this mean-field tadpole approximation, which is clearly wrong in light of other studies~\cite{PhysRevD.58.076003,PhysRevD.79.056008,Takanori_Sugihara_2004,montecarlo,PhysRevD.88.085030,Kadoh2019,PhysRevD.99.034508,PhysRevResearch.2.033278,PhysRevD.106.L071501}. We note that this caveat  only appears in $(1+1)$ dimensions, and forces us to go beyond the mean-field approximation.

In order to circumvent this problem, and obtain a line of critical points in parameter space $(m_0^2,\lambda_0)$, further quantum corrections need to be included in the self-energy. We will make the self-energy exact at second order in the quartic coupling  by adding the sunrise contribution, which has a  diagrammatic representation  given by the fourth Feynman diagram in  Eq.~\eqref{8}. Contrary to the previous tadpole terms, this contribution depends on the external momenta of the propagator, making a full self-consistent treatment that includes tadpole- and sunrise-like diagrams to all orders of the quartic coupling impractical. We can, however, include all tadpole decorations in the internal propagators of the sunrise diagram, leading in this way to an improved self-energy $\mathsf{\Sigma}(-\ii\omega,{\rm k})$. This can be achieved by introducing again the tadpole-resummed propagator in Eq.~(\ref{eq:lattice_prop}) in the mentioned self-energy,  including thus the tadpole-like decorations to every order of the quartic coupling. Paralleling our discussion around Eq.~\eqref{pole}, 
the critical line determined by $m_{\rm P}^2=0$ will thus be defined by the new condition
\beq
\label{crit1}
\mu^2+ \mathsf{\Sigma}_{\rm sr}({\boldsymbol{0}})=0 \ ,
\eeq
where tadpole corrections come from Eq.~\eqref{eq:mu}, and 
\beq
\label{eq:sunrise_self_energy}
\mathsf{\Sigma}_{\rm sr}(\boldsymbol{0})=-\frac{\lambda_0^2}{6}\!\!\bigintsss\!\!\!\frac{{\rm d}^2k{\rm d}^2q}{(2\pi)^4}\tilde{\Delta}_{\rm td}(\boldsymbol{k})\tilde{\Delta}_{\rm td}(\boldsymbol{q})\tilde{\Delta}_{\rm td}(\boldsymbol{k}+\boldsymbol{q}).
\eeq

Eventually, we will also be interested in moving out of criticality, since it is the non-zero value of the  physical mass $m_{\rm P}^2$ via its associated effective Compton wavelength $\xi_{\rm eff,P}$, which controls the range of the spin-spin interactions~\eqref{eq:spin_spin_couplings_ions}. Considering that the effective $\lambda\phi^4$ model is only relevant close to the structural phase transition of the ion chain,  we can assume that in the region of interest for the experiment, the physical mass  $m_{\rm P}^2$ will be small.
In order to solve for the pole of the improved propagator in Eq.~\eqref{pole}, it will  then suffice to consider
\beq
\mathsf{\Sigma}_{\rm sr}(-\ii \omega,0) \approx \mathsf{\Sigma}_{\rm sr}(\boldsymbol{0})- \left.{\partial \mathsf{\Sigma}_{\rm sr}\over \partial k_0^2}\right|_{\boldsymbol{0}} \, \omega^2 \ ,
\eeq
where the Taylor expansion of the sunrise diagram yields
\begin{eqnarray}
\label{eq:wf_sr}
\left.{\partial \mathsf{\Sigma}_{\rm sr}\over \partial k_0^2}\right|_{\boldsymbol{0}} \!\!\!\!\!\!\! && = 
\frac{\lambda_0^2}{6}\!\!\bigintsss\!\!\!\frac{{\rm d}^2p{\rm d}^2q}{(2\pi)^4}\tilde{\Delta}_{\rm td}(\boldsymbol{p})\tilde{\Delta}_{\rm td}(\boldsymbol{q}) \\
&&\times \left( \tilde{\Delta}_{\rm td}^2(\boldsymbol{p+q})-(p_0+q_0)^2 \, \tilde{\Delta}_{\rm td}^3(\boldsymbol{p+q}) \right) \nonumber \ . 
\end{eqnarray}
Close to this pole, the  propagator with contributions from all tadpoles and the sunrise diagram has the expression
\beq
-\ii\tilde{\Delta}(-\ii \omega, 0 )\to\tilde{\Delta}(\omega, 0 ) \approx {\ii\mathsf{z} \over \omega^2 - m_{\rm P}^2},
\eeq
  where we have rotated back to  Minkowski spacetime. Here, one can readily identify the physical mass to be 
\beq
m_{\rm P}^2 =\mathsf{z}\big( \mu^2 +\left.\mathsf{\Sigma}_{\rm sr}\right|_{{\boldsymbol 0}}\big). \;\;\; 
\label{mass}
\eeq
In addition to the additive contribution of these Feynman diagrams to the physical mass, one  also observes the appearance of a multiplicative contribution from  the residue at the pole
\beq
\label{eq:wv_ren}
\mathsf{z}=\left( 1+{\partial \mathsf{\Sigma}_{\rm sr}\over \partial k_0^2}(\boldsymbol{0})\right)^{-1} .
\eeq
We thus see that quantum corrections do not only modify the mass of the theory, but also the normalization of the field itself. This is the physical meaning of $\mathsf{z}$, the so-called wave-function renormalization. In order to have the field canonically normalized, it is necessary to rescale $\phi(x)\to \sqrt{\mathsf{z}} \phi(x)$. This explains the appearance of $\mathsf{z}$ in  the physical mass~\eqref{mass}. 

Let us now discuss the improved prediction  of the critical line, which is obtained by  solving  $m_{\rm P}^2=0$. This  is no longer impeded by the IR divergence of the tadpole integrals,  as the resummed  contributions to the mass $\mu^2$ are no longer required to be  zero. In some sense, the  lattice regularization provides a  cutoff at large momenta that allows us to get finite results in the UV limit, while the addition of the sunrise diagram  provides an effective  ``mass cutoff'' at low energies that allows us to get finite results in the IR limit. In summary, the equation that must be solved is given by 
\begin{gather}
{m}_{\rm P}^2=\mathsf{z}\left(\mu^2 -\frac{\lambda_0^2}{6}\!\!\bigintsss\!\!\!\frac{{\rm d}^2k{\rm d}^2q}{(2\pi)^4}\tilde{\Delta}_{\rm td}(\boldsymbol{k})\tilde{\Delta}_{\rm td}(\boldsymbol{q})\tilde{\Delta}_{\rm td}(\boldsymbol{k}+\boldsymbol{q})\right),
\label{17a}
\end{gather}
where the wavefunction renormalization $\mathsf{z}$ is given by Eqs.~\eqref{eq:wv_ren} and~\eqref{eq:wf_sr}, and the tadpole contribution $\mu^2$ is that of Eqs.~\eqref{eq:mu} and~\eqref{tad}. Even though Eq.~\eqref{17a} is not a proper self-consistency equation when written in this form, we opt for simplicity and refer to the whole set of equations~\eqref{eq:mu} and~\eqref{17a} as the self-consistency equations.
 In the following sections, we will present numerical solutions of this set of equations applied to the Yukawa-type interactions in  trapped ions. Let us, however, first discuss how the formalism of thermal field theories can account for non-zero temperatures.

\subsection{Non-zero temperature and thermal field theories}

As stated in Subsec.~\ref{thermal_sub}, the main goal of this work is to explore the effect of non-zero temperatures in the Yukawa-mediated spin-spin interactions of a trapped-ion quantum simulator~\eqref{34}. We argued that, in the vicinity of a structural phase transition, the non-linearities will lead to a shift of the physical mass of    the effective $\lambda\phi^4$ QFT, which will change the distance decay of the spin-spin couplings~\eqref{eq:spin_spin_couplings_ions}. Moreover, there can be additional contributions at non-zero temperatures related to the perturbative thermal mass of Eq.~\eqref{eq:thermal_mass_pert}. In this subsection, we discuss how  to generalise the previous  self-consistency equations~\eqref{17a} to a non-zero temperature  using the formalism of thermal field theories~\cite{Laine:2016hma,kapusta_gale_2006}. 

According to a path integral approach~\cite{fradkin} in  Euclidean time for a system at thermal equilibrium~\cite{10.1143/PTP.14.351}, the  generating functional of Eq.~\eqref{eq:gen_fucntional} 
 corresponds
to the partition function of the  $\lambda\phi^4$ model in the limit of a vanishing temperature. For non-zero temperatures $T>0$, the  integrals in Eq.~\eqref{eq:free_Z} must be modified, as  the  spacetime $x=(\tau,{\rm x})$ is no longer the Euclidean plane.  The  field  in the path integral must fulfill  periodic boundary conditions 
\begin{align}
\phi(\tau,{\rm x})=\phi(\tau+\beta,{\rm x}),
\label{eq:per_field}
\end{align}
where $\beta={1}/{T}$ is the inverse temperature in natural units, such that the Euclidean plane is effectively compactified   into a cylinder $\mathbb{R}^2\mapsto S_r\times\mathbb{R}$ of radius $r=\beta/2\pi$. Given the periodicity of the scalar field~\eqref{eq:per_field},  the  propagator obeys the Kubo-Martin-Schwinger condition~\cite{doi:10.1143/JPSJ.12.570,PhysRev.115.1342}, namely $\Delta_0(\tau,{\rm x})=\Delta_0(\tau+\beta,{\rm x})$. Accordingly,  the transformation to momentum space~\eqref{3} requires using  a Fourier series~\cite{Laine:2016hma,kapusta_gale_2006,thermal} in the temporal coordinate instead of a Fourier transform
\beq
\label{eq:matsubara}
\Delta_0(\tau,{\rm x})=\frac{1}{\beta}\mathlarger{\sum}_{n_0\in\mathbb{Z}}\bigintsss\!\!\! \frac{{\rm d}{\rm k}}{2\pi}\,\,\tilde{\Delta}_0(\omega_{n_0},{\rm k})\ee^{\ii(\omega_{ n_0}\!\tau+{\rm k}{\rm x})},
\eeq
where $\omega_{n_0}={2\pi n_0}/{\beta}$ are the bosonic {Matsubara frequencies}.
The important aspect of the Matsubara formalism is that the generating functional of  the equilibrium $n$-point functions of the thermal field theory has the  same functional form as the $T=0$  case~\eqref{6},  provided one substitutes  the frequency integrals by a series in the Matsubara frequencies. Accordingly, one can use the previous diagrammatic results~\eqref{8}, as well as the self-consistency equations~\eqref{17a} with the lattice propagator~\eqref{eq:lattice_prop}, by substituting: {\it(i)} the 2-momentum by $k\to(\frac{2\pi}{\beta}n_0,\frac{2\pi}{aN_1}n_1)$, where $n_0\in\mathbb{Z}$ and $n_1\in\{1,\cdots,N_1\}$; {\it (ii)} the Euclidean lattice propagator by the Matsubara propagator $\Delta _0(k)\to\Delta _0(\omega_n,{\rm k})$; and {\it (iii)} the momentum integrals by mode sums $\bigintsss\!\!\!\frac{{\rm d}^2k}{(2\pi)^2}\to\frac{1}{\beta}\sum_{n_0\in\mathbb{Z}}\frac{1}{aN_1}\sum_{n_1=1}^{N_1}$.

 \begin{figure}
\resizebox{0.5\textwidth}{!}{
\includegraphics{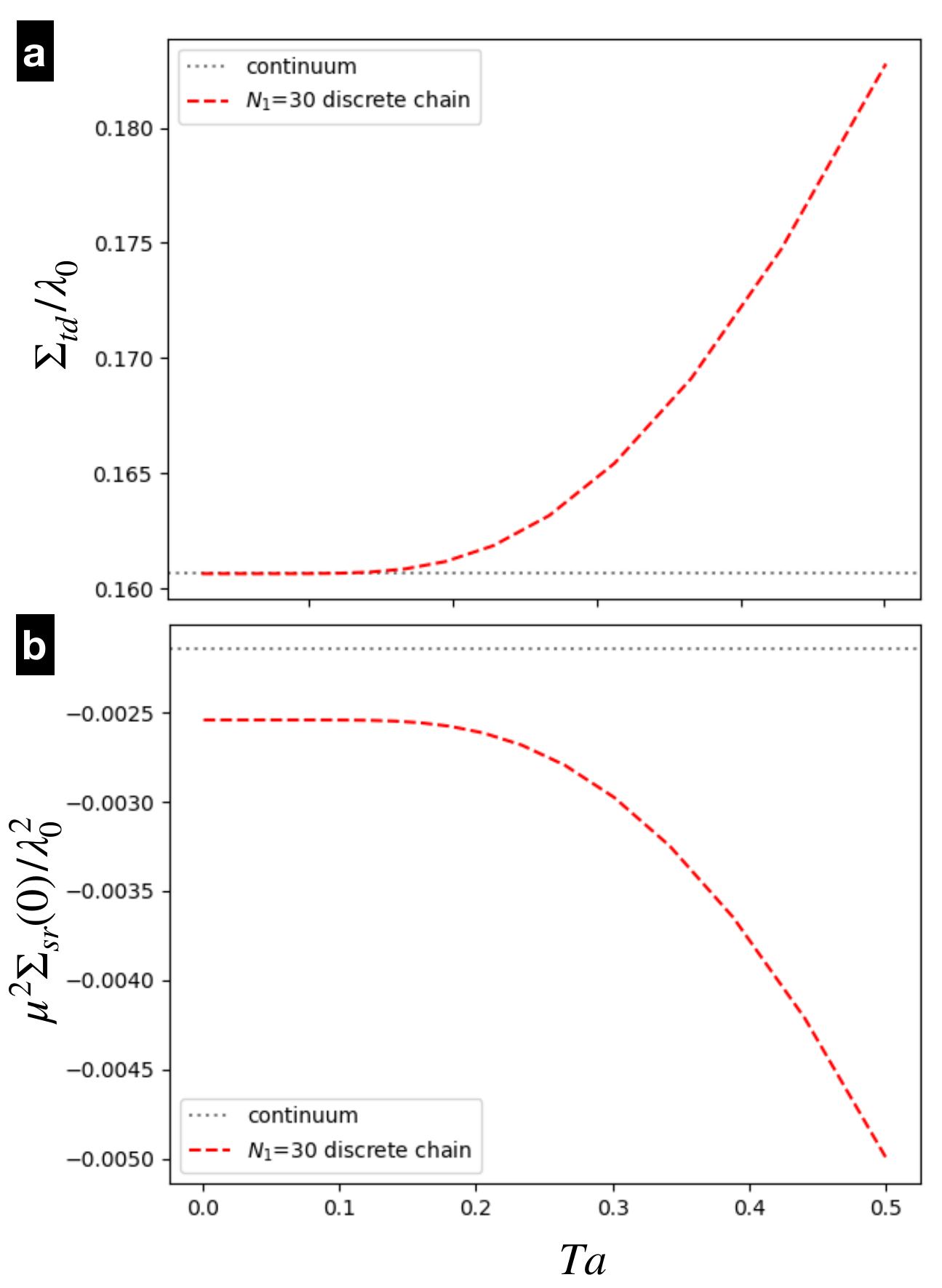}
}
\caption{\label{fig:sums}  \textbf{Tadpole and sunrise  mass shifts:} \textbf{(a)} The tadpole renormalised mass  $m_0^2\to \mu^2=m_0^2+\mathsf{\Sigma}_{\rm td}$  can be expressed in terms of a dimensionless mass shift $\mathsf{\Sigma}_{\rm td}/\lambda_0$ independent of the coupling. 
We set  $\mu a=1$, and depict both the  analytical expression  for the  $T=0$  mass shift~\eqref{eq:self_cons_lattice} (black dotted line),  obtained  in the thermodynamic limit $N_1\to\infty$, and the finite-temperature expression (red dashed line),  obtained by performing the Matsubara sum explicitly~\eqref{eq:ratio_pert} for a  lattice with $N_1=30$ sites.  \textbf{(b)} The sunrise renormalised mass  $\mu^2\to {m_P}^2=\mu^2+\mathsf{\Sigma}_{\rm sr}(\boldsymbol{0})$ is proportional to a dimensionless ratio $\mu^2\mathsf{\Sigma}_{\rm sr}(\boldsymbol{0})/\lambda_0^2$ that is, again, independent of the quartic coupling. We compare  the exact $T=0$ result in the thermodynamic limit (black dotted line), discussed in Appendix~\ref{sec: continuum}, with the finite-temperature expression obtained by performing the Matsubara sums explicitly~\eqref{eq:sr_mass} (see Appendix~\ref{sec:analytical_sums}), for a lattice with $N_1=30$ sites (red dashed line).}
\end{figure}

Let us illustrate this procedure by  considering the lattice-regularised tadpole  contribution~\eqref{tad}.
At finite temperatures, we have
\begin{gather}
 \mathsf{\Sigma}_{\rm td}= \frac{\lambda_0}{2}\frac{T}{a N_1}\!\mathlarger{\sum}_{n_0,n_1}\!\frac{1}{(2\pi T n_0)^2+\hat{\rm k}^2+\mu^2}\ .
 \label{eq:dimensionless_self_cons}
\end{gather}
The Matsubara sum can be performed explicitly, yielding 
 \beq
\label{eq:ratio_pert} 
 \mathsf{\Sigma}_{\rm td}=\frac{\lambda_0}{4N_1}\!\mathlarger{\sum}_{n_1=1}^{N_1}
 {{\rm coth} \big( {\sqrt{{\hat {\rm k}}^2+\mu^2} \over 2T }\big) \over a\sqrt{{\hat {\rm k}}^2+\mu^2}} .
 \eeq
This contribution is the mean field counterpart of Eq.~\eqref{eq:thermal_mass_pert}, as detailed in Appendix~\ref{sec:analytical_sums}. We plot in Fig.~\ref{fig:sums}{\bf (a)} the dimensionless quotient $\mathsf{\Sigma}_{\rm td}/\lambda_0$ for $N_1=30$ as a function of the temperature, measured in lattice units, and for the value of the tadpole renormalized mass $\mu a=1$. The dotted line in that figure corresponds to the zero temperature thermodynamic limit, evaluated in Eq.~\eqref{eq:self_cons_lattice}. We observe a quite small deviation from the thermodynamic limit already for the moderate number of sites we have chosen. 

Remarkably, the Matsubara mode sums involved in the sunrise contribution $\mathsf{\Sigma}_{\rm sr} (\boldsymbol{0})$ \eqref{mass} can also be performed analytically. The resulting expression can be found in Eq.~\eqref{eq:sr_mass} of Appendix~\ref{sec:analytical_sums}. The dimensionless combination $\mu^2 \mathsf{\Sigma}_{\rm sr} (\boldsymbol{0})/\lambda_0^2$ is shown in Fig.~\ref{fig:sums}{\bf (b)}, where again a small deviation from the zero temperature thermodynamic limit is found for $N_1=30$.
Conversely, the wavefunction renormalization~\eqref{eq:wv_ren} at finite temperature needs to be computed numerically, as no analytical expression for the Matsubara mode sums was found. 
This requires the truncation of the Matsubara sums to a finite number of modes $n_0\in\{-N_0,\cdots, N_0\}$, and thus raises the issue of the convergence with $N_0$. This is studied in Appendix~\ref{sec:analytical_sums}, obtaining a fast convergence  which will be important for the efficiency of our numerical approach, even in the typically more demanding low temperature limit.

\section{\bf Critical line and  trapped-ion  spin-spin couplings }
\label{sec:self_cons_ions}

In this section, we will numerically solve the self-consistent equations~\eqref{eq:mu} and~\eqref{17a} at non-zero temperatures. We will obtain results for the standard Hamiltonian lattice discretization of the scalar field in Eqs~\eqref{eq:lattice_prop}-\eqref{eq:lattice_mom}, but also  introduce the dipolar tail of the dispersion~\eqref{29} to be able to make explicit predictions for the trapped-ion case.

\subsection{Numerical estimate of the critical line}

As discussed above,  the $\mathbb{Z}_2$-breaking phase transition in the $\lambda\phi^4$ model is characterised by a 
classical critical point at $m_0^2=0$, which will flow with temperature and quartic coupling  due to thermal and quantum effects that shift the pole of the propagator to the physical mass $m_0^2\to m_{\rm P}^2$. As a consequence, the classical critical point will become a critical surface in parameter space $(m_0^2a^2,\lambda_0a^2,Ta)$ determined by solving the equation ${m}_{\rm P}^2(\lambda_0a^2,Ta)=0$. If one fixes the temperature to a specific value, we want to obtain the corresponding critical line for the bare mass $m_{0}^2|^{\phantom{2}}_{\rm c}$ as a function of the bare coupling strength $\lambda_0$. This line will separate  the symmetry-broken $m_0^2<m_{0}^2|^{\phantom{2}}_{\rm c}$ from the symmetry-preserved $m_0^2>m_{0}^2|^{\phantom{2}}_{\rm c}$ phase at the given temperature $Ta$.

In order to numerically obtain these critical lines for various non-zero temperatures, one needs to impose the condition ${m}^2_{{\rm P}}=0$ in Eq.~\eqref{17a}, and solve the self-consistency equations to extract the bare critical parameters $(m_0^2a^2,\lambda_0a^2)|^{\phantom{2}}_{\rm c}$. We note that the wavefunction renormalization, contributing multiplicatively to the physical mass~\eqref{mass},  does not play any role in the determination of the critical points.
Then, the routine followed to compute the critical line reads as follows 
\begin{figure}[!h]
\begin{flushleft}
\texttt{ Routine {R1}}\\
\vspace{0.5ex}
\texttt{1 Select an interval $\left(\mu_1^2a^2,\mu_2^2a^2\right)$ in the $\mu^2 a^2$ axis}\\
\texttt{2 For each $\mu^2a^2$ in $\left(\mu_1^2a^2,\mu_2^2a^2\right)$ and  a given  $Ta$:}\\
\texttt{3 \quad Impose ${m}_{\rm P}^2=0$  in the lattice counterpart of \\
\quad \quad Eq.~(\ref{17a}) using Matsubara  sums to obtain \\ \quad \quad $\lambda_0a^2$ as a function of $\mu a$, $T a$ and $N_1$}\\
\texttt{4 \quad Substitute $\mu^2a^2$ in Eq.~\eqref{eq:mu} to obtain $m_0^2a^2|_{\rm c}$.}\\
\texttt{5 Return the list of critical points  $\{\left(m_{0}^2a^2,\lambda_{0}a^2\right)|_{\rm c}\}$}
\end{flushleft}
\end{figure}

\begin{figure}
\resizebox{.5\textwidth}{!}{
\includegraphics{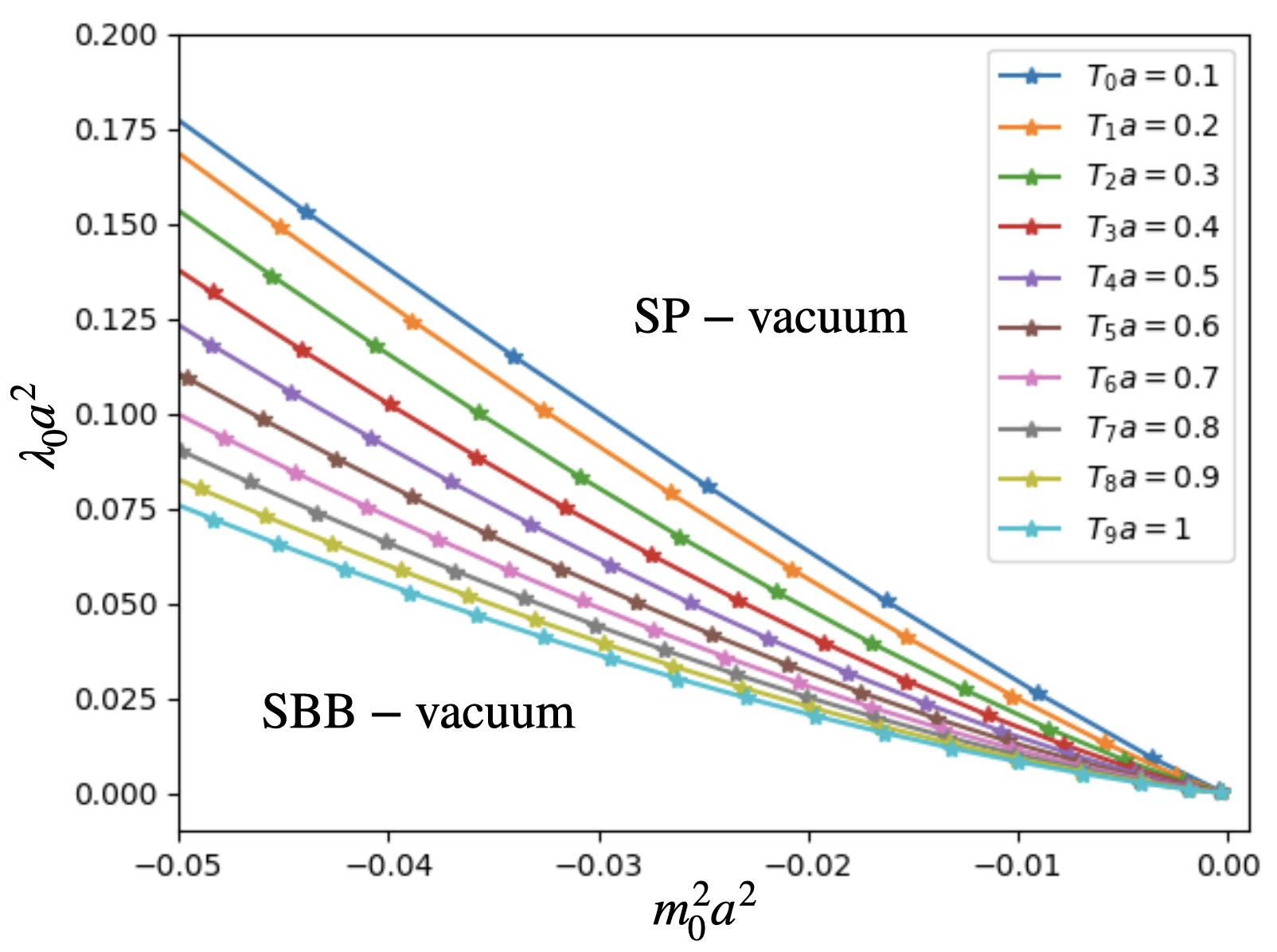}
}
\caption{\label{fig:critical_field}{\bf Critical lines for the parity-breaking phase transition:} We solve numerically the self-consistency equations~\eqref{eq:mu} and~\eqref{17a} for the critical point ${m}^2_{\rm P}|_{\rm c}=0$. We  consider a lattice  regularization~\eqref{eq:lattice_prop} with $N_1=30$ sites and use the analytical Matsubara mode sums discussed in Appendix~\ref{sec:analytical_sums} along with the \texttt{Routine R1} described in the text. When the temperature increases, the effect of a thermal mass makes the critical line take lower values for the quartic coupling. For each colored line, the upper region corresponds to the symmetry-preserved ground state (SP-vacuum), whereas the region below corresponds to the ground state with  spontaneous breaking of the parity symmetry (SSB-vacuum). }
\end{figure}

In Fig.~\ref{fig:critical_field}, we represent the critical line for different temperatures.  For a fixed value of the temperature ${T}a$, the region above the critical line represents a symmetric  phase, adiabatically connected to the Klein-Gordon thermal state, whereas the region below is the symmetry-broken phase in which the scalar field acquires a non-zero  expectation value and can no longer be adiabatically connected to the Klein-Gordon vacuum.
As  can be observed in this figure, the critical lines take lower values of the quartic coupling as the temperature increases. This behaviour is consistent with the appearance of a {thermal mass}~\eqref{eq:thermal_mass_pert} and the phenomenon of restoration of symmetry~\cite{PhysRevD.9.3320,PhysRevD.9.3357}. For a fixed value of the coupling strength $\lambda_0a^2$, if one starts at a point $(m_0^2a^2,T_1a)$ in which the symmetry is broken (i.e. a point below the corresponding critical line of Fig.~\ref{fig:critical_field}), one may increase the temperature towards $T_2>T_1$, such that the corresponding equilibrium state lies now above the  critical line, and thus belongs to  the symmetry-preserved phase.  Let us note that the dependence of the critical line with the quartic coupling is similar to the zero-temperature results obtained using other numerical methods, such as Monte Carlo~\cite{PhysRevD.58.076003,montecarlo}. If one aims at exploring regimes beyond those of relevance in the trapped-ion realization (see our discussion in Sec.~\ref{sec:thermal_masses_discussion}), there are certain limitations of the current approach that are discussed in detail in Appendix~\ref{sec: crossings}.

Once the critical lines have been derived, a more informative figure for the connection to the trapped-ion case and the Yukawa-mediated interactions discussed in the following section  would be a  contour plot for the  non-zero values of the physical  mass ${m}_{\rm P}^2a^2$ in the plane $({m}_0^2a^2,{T}a)$ at  a fixed quartic coupling ${\lambda_0}a^2$. This will be the first step  to make a connection with the distance  decay of the spin-spin interactions in future sections. Achieving this goal requires incorporating the wavefunction renormalization~\eqref{eq:wv_ren}, which contributes multiplicatively to the physical mass~\eqref{17a}, into a different numerical routine
\begin{figure}[H]
\begin{flushleft}
\texttt{ Routine {R2}}\\
\vspace{0.5ex}
\texttt{1 Select the interval of dimensionless temperatures\\ \quad $({T}_1a,{T}_2a)$ and a fixed $\lambda_0 a^2$ value}\\
\texttt{2 For each ${Ta}$ in $({T}_1a,{T}_2a)$:}\\
\texttt{3 \quad Initialize $\mu={\mu}_{\rm 0}^2a^2>0$ and ${m}_{\rm P}^2a^2=1$}\\
\texttt{4 \quad Set $\epsilon$}\\
\texttt{5 \quad While ${m}_{\rm P}^2a^2\mathsf{z}^{-1}\geq 0$:}\\
\texttt{6 \quad \quad  Update ${m}_{\rm P}^2a^2\mathsf{z}^{-1}$ and compute ${m}_{\rm 0}^2a^2$ from Eq.~(\ref{17a})\\ \quad \quad \quad and Eq.~(\ref{eq:mu}) respectively}\\
\texttt{7 \quad\quad Compute $\mathsf{z}$ using \eqref{eq:wv_ren}}\\
\texttt{8 \quad\quad $\mu \leftarrow \mu-\epsilon$}\\
\texttt{9 Using $\mathsf{z}$ return the ${m}_{\rm P}^2a^2$ values in $({\mu}_{0}^2a^2,{\mu}_{f}^2a^2)$;\\ \quad $({m_0}_{0}^2a^2,{m_0}_{f}^2a^2)$;$({T}_1a,{T}_2a)$ }
\end{flushleft}
\end{figure}

\begin{figure}
\resizebox{.5\textwidth}{!}{
\includegraphics{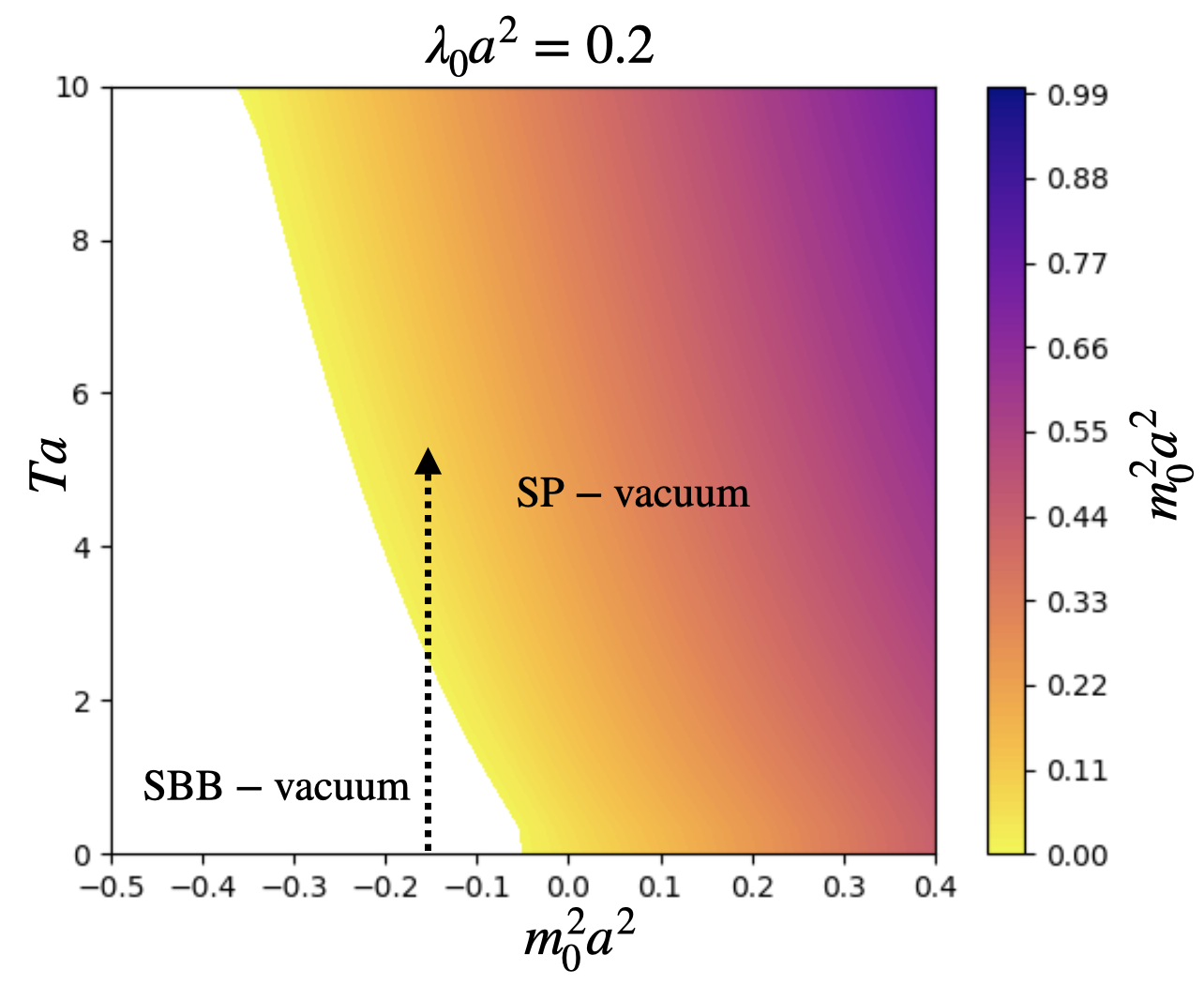}
}
\caption{\label{fig:critical_field_labels}{\bf Physical mass  for the $\lambda\phi^4$ model on a lattice:} Contour plot for non-zero values of the tadpole and sunrise contributions to the physical mass ${m}_{\rm P}^2a^2$ within the symmetric ground state (coloured region). The physical mass is obtained by solving  the self-consistent equations~\eqref{eq:mu} and~\eqref{17a}. The numerical solution uses  the   Matsubara mode sums discussed in Appendix~\ref{sec:analytical_sums} and the \texttt{Routine R2}, described in the text, considering $N_1=30$ lattice sites. The dashed arrow indicates the phenomenon of symmetry restoration  by increasing the temperature.}
\end{figure}

In practice, the same $\epsilon$ variation of $\mu$ leads to different size increments on ${m}_{\rm P}^2a^2$ and ${m}_{\rm 0}^2a^2$, depending on $Ta$ and the previous ${m}_{\rm P}^2a^2$ value. To obtain uniform increments independently of position in parameter space, we heuristically define an adapted step $\epsilon_i \propto {m}_{\rm P,i-1}^2a^2/(1+Ta)^{(1/3)}$. Now, we can run a simulation based on this numerical routine to obtain a contour plot of the physical mass. 
The numerical results are shown in Fig.~\ref{fig:critical_field_labels}, where the region in parameter space with a non-zero physical mass is coloured according to the scale specified in the rightmost inset. The white region corresponds to the symmetry-broken phase, where  the physical mass would become negative, leading to a transition from a single to a double well in the effective potential~\cite{PhysRevD.7.1888,PhysRevD.9.1686}. This potential, via the self-consistency equations,  has quantum corrections stemming from all the tadpole  diagrams, as well as perturbative corrections of the sunrise  diagram. We can see how the symmetry-broken region changes with temperature for the various quartic couplings. In this contour plots, the restoration of symmetry by increasing temperature becomes very clear, as it corresponds to any vertical line connecting the symmetry-broken and symmetry-preserved phases.

 As explained before, our self-consistent equations are exact to order $\lambda_0^2$, but miss many higher-order contributions. It is therefore important to benchmark the performance of our approach. We will consider a specific universal quantity, namely the ratio  $f_{\rm c}=\lambda_0/\mu^2|_{\rm c}$ at the $T=0$ critical point of the $\lambda\phi^4$ QFT in the continuum, which has also been computed with other numerical approaches. In the continuum, the UV divergence of the tadpole contribution~\eqref{tad}  would make the dimensionless ratio $\lambda_0/m_0^2|_{\rm c}$ vanish. It is then customary to replace the bare mass by the UV-finite tadpole renormalised mass $\mu^2$~\eqref{eq:mu}.
Using our lattice discretization, we can access the critical ratio as a function of the lattice spacing $f_{\rm c }(a)$, or, alternatively, as a function of $f_{\rm c }(\lambda_0a^2)$. Its value in the continuum QFT can be thus obtained by estimating $f_{\rm c }(\lambda_0a^2)$ as the dimensionless coupling strength $\lambda_0a^2\to 0$.

\begin{figure}
\resizebox{.5\textwidth}{!}{
\includegraphics{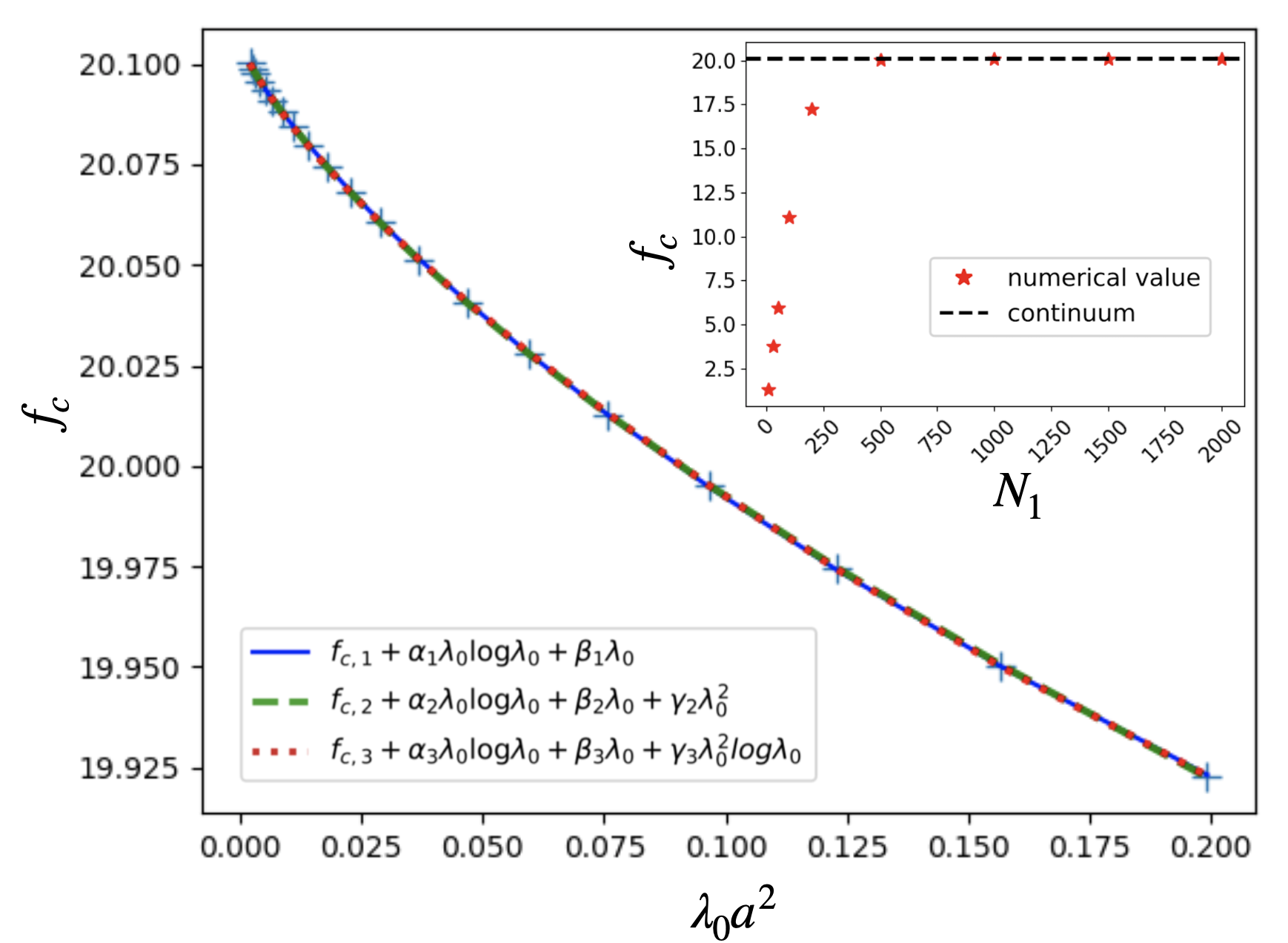}
}
\caption{\label{fig:critical_ratio}{\bf Critical ratio as a function of $\lambda_0$:} We use \texttt{Routine R1} and $N_1=2000$ to compute $\lambda_0a^2$ and $f_{\rm c}$ for each $\mu^2a^2$ value. Three different logarithmic models are fitted (blue line, green dashed line and red dotted line respectively), obtaining $f_{1,c}=20.10(4),\quad f_{2,c}\simeq f_{3,c} = 20.10(5).$ The inset shows how the the critical ratio $f_{3,c}$ converges to the continuum for different $N_1$ values to check the validity of the discretization used.}
\end{figure}

To compute $f_{\rm c}$, one can make use of \texttt{Routine R1} with some minor modifications. For a fixed number of spatial lattice sites $N_1$, and using a small-enough interval of values for $\mu^2a^2$, one can naively  extract the critical ratio  by plotting $\lambda_0a^2|_c$ vs $\mu^2a^2|_c$, and  fitting the graph to a linear function in order to extract the slope. We find $f_{\rm c}\simeq 20.105$ using $N_1=2000$.  As discussed in detail in~\cite{Delcamp_2020}, logarithmic contributions must be taken into account in order to be get a more precise estimate. Guided by this work, we employ three logarithmic models to fit our results, as depicted in Fig.~\ref{fig:critical_ratio}. As shown in this figure, the three models give the same critical ratio value up to the second decimal place, $f_{\rm c} = 20.10$. We again use $N_1=2000$, as we appreciate a clear convergence to the continuum (see the inset in Fig.~\ref{fig:critical_ratio}).
Within our approach,  the universal ratio can actually be computed directly in the continuum. From equations~\eqref{crit1} and \eqref{eq:sunrise_self_energy}, we have $f_{\rm c} =\sqrt{6 (2\pi)^4/ I}$
with
\begin{equation}
    I=  \!\!\bigintsss\!\!\! {\rm d}^2k{\rm d}^2q{ 1\over (k^2 +1)(q^2+1)((k+q)^2+1)} \ .
\end{equation}
In Appendix~\ref{sec: continuum} we evaluate analytically this integral with the help of Feynman parameters obtaining $f_{\rm c}=20.1055$, which certifies the good behaviour of the continuum limit of the lattice discretization.

The quantum critical point of the $\lambda \phi^4$ theory in 1+1 dimensions has been studied with a variety of non-perturbative methods, such as Tensor Networks ~\cite{Delcamp_2020, Milsted_2013, Kadoh_2019} or Monte Carlo ~\cite{PhysRevD.99.034508}, among others ~\cite{Serone_2018, Elias_Mir__2017}. All these works find values near $f_{\rm c}\simeq 66.0$ with increasing levels of precision. 
The deviation of our self-consistent approach from these more-accurate estimates  is common to  in quantum many-body models with critical points, as illustrated for instance by the mean-field underestimation $J/h|_{\rm c}=1/2$ of the critical point $h/J|_{\rm c}=1$ of the Ising model in a transverse field~\cite{PFEUTY197079}. Common to this type of mean-field treatments, our treatment  overestimates the stability of the ordered symmetry-broken phase under an increase in the mass. We can say that the simplicity of our method, which reduces drastically the computational complexity to a solution of self-consistency equations, comes at a price. On the other hand, the interest of our approach  lies in being the most economic one capable of coping with the divergences and still detecting the phase transition and study the phase diagram. Indeed, this simplicity will be crucial in the next section, where we will apply it to the trapped-ion quantum simulators of spin models~\cite{Friedenauer2008,Kim2010,doi:10.1126/science.1232296,doi:10.1126/science.1232296,doi:10.1126/science.1251422,Jurcevic2014,Richerme2014,Smith2016,Zhang2017,Tan2021,Morong2021}.
Indeed, the non-standard dispersion relation \eqref{29}, responsible for the dipolar tail of the spin-spin interactions in Eq.~\eqref{eq:spin_spin_couplings_ions},
might be difficult to treat accurately with more precise methods, such as the aforementioned tensor networks which can approximate long-range couplings by using the so-called matrix-product operators.
Moreover, since we are ultimately interested in dealing with non-zero temperatures, this would further increase the computational complexity of tensor-network methods, requiring again to work with matrix-product operators. Finally, in the context of the trapped-ion implementation, addressing the complete problem of the Yukawa-type spin-spin interactions would require solving the real-time dynamics of the spins under locally coupled to the phonons, which evolve on a much slower timescale in comparison to the effective $\lambda\phi^4$ model.  This would require simulation long-time dynamics, which  is a notably difficult for Tensor Networks, and out of reach for Monte Carlo methods due to the sign problem. Our self-consistent treatment should  thus be seen as a proof of concept for the proposal to use thermal effects to study interacting QFTs with trapped-ion simulators.

\subsection{Estimates for the trapped-ion quantum simulator \label{critical2}}

Let us now discuss how  the previous results can be connected to the trapped-ion quantum simulators of spin models~\cite{Friedenauer2008,Kim2010,doi:10.1126/science.1232296,doi:10.1126/science.1232296,doi:10.1126/science.1251422,Jurcevic2014,Richerme2014,Smith2016,Zhang2017,Tan2021,Morong2021}. In Sec.~\ref{experimental}, we  described in detail how, far from the linear-to-zigzag structural phase transition~\eqref{eq:critical_point_classical}, the spin-spin couplings mediated by the transverse phonons~\eqref{eq:exact_J_ij} of a trapped-ion chain  can be described accurately by the Yukawa-type  interactions mediated by a Klein-Gordon field~\eqref{eq:spin_spin_couplings_ions} (see the comparison in Fig.~\ref{fig:spin_cpuplings}). As the trap frequencies are modified and one gets closer to the structural phase transition, the Coulomb non-linearities start to play a bigger role, and transform this QFT into the $\lambda\phi^4$ model~\eqref{eq:lambda_phi_4}  with  an effective speed of light in Eq.~\eqref{30b}, and bare parameters  in Eqs.~\eqref{30a},~\eqref{eq:lutt_param} and-\eqref{lambda_iones}. We argued that the specific dispersion relation $\omega({\rm k})$ for the transverse modes of the trapped-ion chain~\eqref{29} is very similar to that of a discretized scalar field~\eqref{eq:lattice_dispersion}, which underlies the Feynman propagator~\eqref{eq:lattice_prop} we used in the  solution of the self-consistent equations~\eqref{17a} of the previous subsection. Note that this propagator, as well as the   results in Figs.~\ref{fig:sums}-\ref{fig:critical_ratio}, have all been obtained using natural units $\hbar=c=k_{\rm B}=1$. Moreover, in the lattice discretization at finite temperature, these self-consistency equations~\eqref{eq:mu} and~\eqref{17a} are rewritten in terms of finite mode sums, and depend on dimensionless parameters obtained through a specific power of the lattice constant $m_0^2a^2,\lambda_0a^2,Ta$,   

In the trapped-ion case, the effective speed of light $c_{\rm t}$~\eqref{30b} only appears after the gradient expansion leading to Eq.~\eqref{eq:lambda_phi_4}. This gradient expansion cannot account for the branch-cut discontinuity of the dispersion relation~\eqref{29} when extended to the complex plane, and would thus miss the dipolar part~\eqref{eq:spin_spin_couplings_ions} of the Yukawa-mediated interactions. Therefore, rather than setting $\hbar=c_{\rm t}=k_{\rm B}=1$ in the coarse-grained trapped-ion case, it would be better to work with the full phonon propagator prior to the long-wavelength approximation. This requires reformulating  the self-consistency equations~\eqref{eq:mu} and \eqref{17a} in terms of dimensionless trapped-ion parameters, which can no longer  be obtained by multiplying the microscopic parameters with a power of the ion  lattice spacing $d$, as we need to use SI units.

In order to find such a formulation, we start by revisiting 
the tadpole-resummed propagator on the lattice~\eqref{eq:lattice_prop}. 
For trapped ions, the   tadpole-resummed propagator is analogous to Eq.~\eqref{eq:lattice_prop}, but has  inverse squared-energy dimension
\beq
\label{eq:lattice_prop_ions}
\tilde{\Delta}_{\rm td}(k_0,{\rm k})=\frac{1}{(\hbar k_0)^2+(\hbar\hat{\rm k})^2+(\mu c_{\rm t}^2)^2},
\eeq
where  $k_0$ has dimensions of inverse time, and will be substituted by the Matsubara frequencies for a non-zero temperature. Therefore, the trapped-ion analogue of the spatial lattice momentum $\hat{\rm k}$ in Eq.~\eqref{eq:lattice_mom} must also have units of inverse time. Moreover, in the absence of quartic interactions,  the effective bare mass in Eq.~\eqref{30a} should appear as a  pole in this propagator $\mu^2=m_0^2$ upon the substitution of $k_0\to-\ii\omega$. Taking into account these two conditions, we find that the analogue of the lattice spatial momentum~\eqref{eq:lattice_mom} that appears for the nearest-neighbor discretization of the scalar field~\eqref{9} is
\beq
\label{eq:spatial_part_prop}
 \hat{\rm k}^2=\frac{7}{2}\omega_x^2\!\left(\frac{l}{d}\right)^{\!\!\!3}\!\zeta_{N_1}\!(3)-\omega_x^2\left(\frac{l}{d}\right)^{\!\!\!3}\mathlarger{\sum}_{r=1}^{\frac{1}{2}N_1}\frac{4}{r^3}\sin^2\!\!\left(\frac{{\rm k}dr}{2}\right),
\eeq
where we have used the truncated  Riemann zeta in Eq.~\eqref{eq:riemann_zeta}. 
Accordingly, $\hat{\rm k}$  has the desired dimension of inverse time, and one sees how the dipolar tail of the phonons dispersion relation~\eqref{29} enters in the propagator.

\begin{figure*}[t]
\centering
\includegraphics[width=2.1\columnwidth]{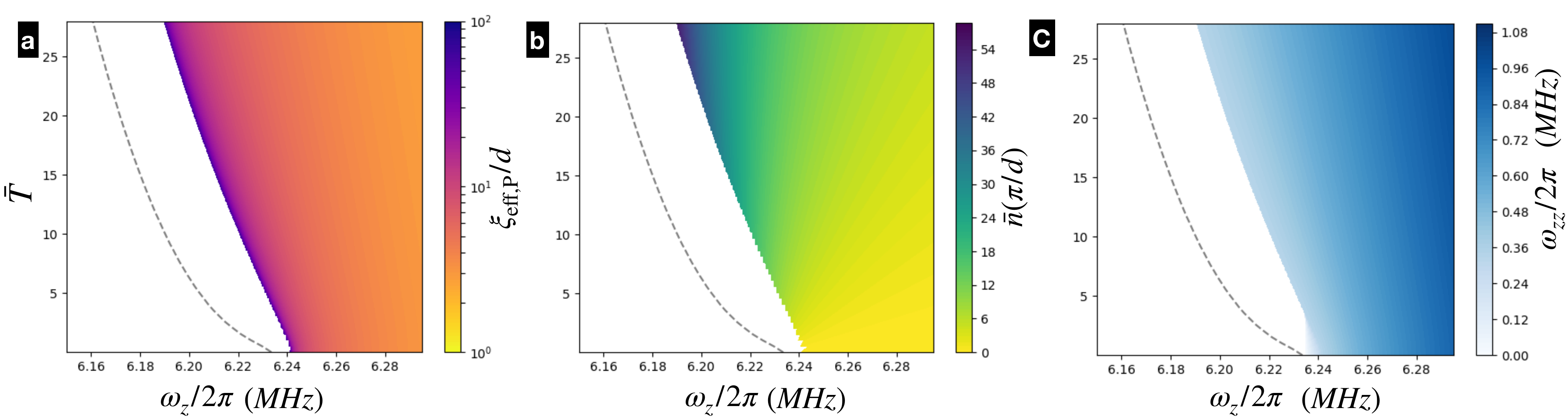}
\caption{ {\bf Renormalization of the range of the spin-spin interactions in a trapped-ion chain:} {\bf (a) } Contour plots of the effective Compton wavelength ${\xi}_{\rm eff,P}/d$ that controls the range of the Yukawa-type spin-spin interactions. In the expression of the spin-spin couplings $J_{ij}$~\eqref{eq:spin_spin_couplings_ions}, one substitutes $J_{\rm eff}\to J_{\rm eff,P}$ and $\xi_{\rm eff}\to \xi_{\rm eff,P}$, due to  the quantum and thermal contributions in Eq.~\eqref{eq:eff_compton}. Thermal effects are encoded in the dependence of the effective Compton wavelength with the temperature.
In the $x$ axis, we represent the radial frequency, while the $y$ axis is reserved for the dimensionless temperature~\eqref{44a}. We use analytical Matsubara mode sums and $N_1=30$ spatial points. Experimental parameters for these simulations have been considered according to a string of  $^{40}$Ca$^{+}$ ions~\cite{insbruck}, and fixing the axial trap frequency to $\omega_x/2\pi=0.45\,$MHz, and the laser detuning with respect to the qubit transition to  $\Delta\omega_L/2\pi=0.318$MHz. The explored  region in $(\omega_z,\overline{T})$ parameter space  avoids getting extremely close to the structrural  phase transition, here  represented by the blue dotted line, since the constraint~\eqref{eq:weak_constraint_couplings} for red detunings $\Delta\omega_{\rm L}<\omega({\rm k})$ would imply very slow spin dynamics (see our discussion in the main text). {\bf (b) } Mean number of phonons $\overline{n}(\pi/d) =(\ee^{(\hbar\omega_{\rm zz,P}/{k_{\rm B}T})}-1)^{-1}$, where $\omega_{\rm zz,P}$ refers to the zigzag-mode frequency. {\bf (c) } Renormalised zigzag mode frequency $\omega_{\rm zz,P}=\overline{m}_P\omega_x$.}
\label{fig:ions}
\end{figure*}

Once the propagator has been identified, we can re-scale  the tadpole self-consistency equation~\eqref{eq:mu} with a certain power of the effective speed of light and Planck's constant, such that the equation has the right dimension of  energy squared. In SI units, the Matsubara frequencies in Eq.~\eqref{eq:matsubara} are expressed  in terms of $\omega_{n_0}={2\pi n_0}/{\beta\hbar}$, where $\beta=1/k_{\rm B}T$, such that the trapped-ion tadpole self-consistency equation reads
\beq
\label{eq:tadpole_ions}
\mu^2c_{\rm t}^4=m_0^2c_{\rm t}^4+  c_{\rm t}^4 \frac{\lambda_0}{2}\frac{k_{\rm B}T}{dN_1}\!\mathlarger{\sum}_{n_0,n_1}\tilde{\Delta}_{\rm td}\!\left(\omega_{n_0},\frac{2\pi}{dN_1}n_1\right).
\eeq
Considering that the quartic coupling~\eqref{lambda_iones} in SI units has dimensions $[\lambda_0]=\mathsf{M}^3\mathsf{L}^3\mathsf{T}^{-2}$, one can check that the above equation~\eqref{eq:tadpole_ions} has the desired squared-energy dimension. In order to get 
an equation involving only dimensionless parameters,
one can simply divide by the squared energy associated to the motional quanta $(\hbar\omega_x)^2$, which allows us to identify the following dimensionless renormalized parameters
\beq
\label{44a}
\begin{split}
\overline{\mu}^2&=\left(\frac{\mu c_{\rm t}^2}{\hbar\omega_x}\right)^{\!\!\!2}, \quad
\overline{T}=\frac{k_BT}{\hbar\omega_x}
\end{split}
\eeq
as well as the following dimensionless bare couplings
\beq
\begin{split}
\overline{m}_0^2&=\left(\frac{m_0^{\phantom{2}}c_{\rm t}^2}{\hbar\omega_x}\right)^{\!\!\!2}=\left(\frac{\omega_z}{\omega_x}\right)^{\!\!\!2}-\frac{7}{2}\!\left(\frac{l}{d}\right)^{\!\!\!3}\!\!
\zeta_{N_1}\!(3),\\
\overline{\lambda}_0&=\left(\frac{c_{\rm t}}{\hbar}\right)^{\!\!\!3}\!
\frac{\lambda_0}{\omega_x^2}=\frac{729\zeta_{N_1}\!(5)}{2}\frac{\hbar}{m_a\omega_x d^2}{\left(\frac{l}{d}\right)^{\!\!\!\frac{3}{2}}\!({\eta_{N_1}\!(1))^{-\frac{1}{2}}}}.\\
\end{split}
\eeq
Finally, one can rewrite the trapped-ion tadpole self-consistency Eq.~(\ref{eq:tadpole_ions}) in terms of  dimensionless quantities as 
\begin{gather}
\overline{\mu}^2=\overline{m}_0^2+\frac{\overline{\lambda}_0}{2}\frac{\overline{T}}{N_1
}\!\mathlarger{\sum}_{n_0,n_1}\frac{\left({l}/{d}\right)^{\!\frac{3}{2}}({\eta_{N_1}\!(1))^{\frac{1}{2}}}}{(2\pi\overline{T}n_0)^2+(\hat{\rm k}/\omega_x)^2+\overline{\mu}^2},\end{gather}
which has the same mathematical structure as the lattice self-consistent equation previously found~\eqref{eq:dimensionless_self_cons}. Again, Matsubara mode sums are to be performed analytically attending to~ \eqref{eq:ratio_pert}. The only difference, apart from the pre-factor  $\left({l}/{d}\right)^{{3/}{2}}({\eta_{N_1}\!(1))^{{1}/{2}}}$, is that the part that depends on the spatial momentum $(\hat{\rm k}/\omega_x)$ in the propagator~\eqref{eq:lattice_prop_ions}  contains now the  dipolar terms~\eqref{eq:spatial_part_prop}. Following this  procedure, we can find the trapped-ion analogue of all the self-consistency equations that include the sunrise diagram~\eqref{17a}, which now depend on   dimensionless parameters that can be numerically adjusted.

The numerical routines introduced in the previous subsection can   be applied to the   trapped-ion case directly. 
We can now use this numerical simulation  to approximate the value of the physical mass~\eqref{17a} as the trap frequencies are modified, and the trapped-ion chain gets closer to the linear-to-zigzag phase transition. With this value, we can estimate the effect of the interactions on the  effective Compton wavelength~\eqref{eq:eff_compton} $\xi_{\rm eff}\to\xi_{{\rm eff},P}$, as well as on the spin-spin coupling strength $J_{\rm eff}\to J_{{\rm eff},P}$. The former is due to the quantum and thermal contributions to the physical mass~\eqref{mass}, whereas the later comes from the field rescaling with  the wavefunction renormalization~\eqref{eq:wv_ren}. Both expressions  would enter in a renormalised version of the spin-spin couplings of Eq.~\eqref{eq:spin_spin_couplings_ions}, as we recall that these are mediated by the excitations of the self-interacting scalar field, which are controlled by the physical pole and the full propagator. In particular, we find
\beq
\begin{split}
\label{eq:eff_compton}
&J_{{\rm eff},P}=J_{{\rm eff}}\sqrt{\mathsf{z}_R},\hspace{2ex} \\
\xi_{{\rm eff},P}=&\frac{(l/d)^{3/2}(\eta_{N_1}(1))^{1/2}}{\sqrt{({{m}}_{\rm P}^{\phantom{2}}c_{\rm t}^2/\hbar\omega_x)^2-(\Delta\omega_{\rm L}/\omega_x)^2}}d. 
\end{split}
\eeq

 In this way, one sees that quantum and thermal effects in the $\lambda\phi^4$ model will change the spin-spin interactions of the trapped-ion quantum simulator. This opens a very interesting perspective, allowing future experiments to probe the nature of the fixed point  of this effective QFT by  measuring  the dynamics of the spins. In fact, the distance dependence of the spin-spin couplings has been inferred using various  experimental techniques in recent years~\cite{Britton2012,Jurcevic2014,doi:10.1126/science.1251422}. Using these techniques while gradually approaching the linear-to-zigzag phase transition would allow one to infer the flow of the critical point, and address universal properties of the QFT in a quantum simulator. For the sake of completeness, we also note that the renormalization of the quartic coupling~\eqref{lambda_iones}, and the associated four-point functions,  would give rise to four- and higher-spin interactions~\cite{a2}. These are, however, much weaker and negligible in a trapped-ion experiment given the constraints considered in this work~\eqref{eq:weak_constraint_couplings}.

To make the numerical results closer to the trapped-ion language,   one can also make use of the Bose-Einstein distribution~\eqref{eq:BE_dist} to obtain a contour plot of the  mean number of phonons $\overline{ n}(\pi/d)$ for the relevant zigzag mode. The final result is that of Fig.~\ref{fig:ions} where, rather than plotting the Compton wavelength as a function of the dimensionless bare mass, we  plot it as a function of the radial trap frequency, which is the standard experimental parameter used to control the shape of the ion crystal. We can see in Fig.~\ref{fig:ions}~{\bf (a)}  how the classical critical point~\eqref{eq:critical_point_classical} flows with the temperature and with the radial trap frequency (blue dotted line). In the coloured region, which lies well within the symmetry-preserved phase (i.e. linear ion chain), we see how the effective Compton wavelength~\eqref{eq:eff_compton}  entering the spin-spin couplings~\eqref{eq:spin_spin_couplings_ions} changes as one approaches the critical line. Therefore, using some of the  experimental techniques based on probing the real-time dynamics of the spins~\cite{Britton2012,Jurcevic2014,doi:10.1126/science.1251422}, one can extract the distance decay and test how the effective Compton wavelength gets renormalised by quantum and thermal effects. In Fig.~\ref{fig:ions}~{\bf (b)}, we also represent the average number of phonons in the zigzag mode as a function of the temperature and the transverse trap frequency. This sets the target detuning of the laser  cooling on the ion crystal used to control the contribution of the thermal masses, and the renormalization of the spin-spin couplings in an experiment.

Note that, in order to comply with the constraint~\eqref{eq:weak_constraint_couplings} and still have spin-spin coupling strengths that are not too slow in comparison to additional experimental sources of noise such as dephasing or motional heating/decoherence, one should avoid getting extremely close to the structural phase transition. There, the renormalised version of the zigzag mode~\eqref{30a}, which is proportional to the physical mass $m_{\rm P}$ of the QFT,  softens $\omega_{\rm zz,P}\to 0$. In light of the constraint in Eq.~\eqref{eq:weak_constraint_couplings}, the laser beatnote $\Delta\omega_{\rm L}<\omega_{\rm zz,P}$ must be very small, leading to very slow spin dynamics. For this reason, our plot in Fig.~\ref{fig:ions} restricts to the colored regions, and does not consider all the parameters down to the critical line.  In fact, the renormalised zigzag frequency, which changes according to Fig.~\ref{fig:ions}{\bf (b)}  is always higher than $\omega_{\rm zz,P}/2\pi\geq 318\,$kHz, leaving enough frequency space for the spin-dependent dipole force to fulfill Eq.~\eqref{eq:weak_constraint_couplings} and still lead to sufficiently-fast spin dynamics in the $0.5$-$10$ms scale.  Altogether, Fig.~\ref{fig:ions}  provides a quantitative prediction of how the range of the spin-spin interactions changes as a function of the temperature, and recall that this is a direct consequence of the underlying interactions and Feynman diagrams of the coarse-grained $\lambda\phi^4$ QFT, and cannot be accounted for if one truncates the description of the system at the quadratic order for the phonons.

\section{\bf Conclusions and outlook}
\label{sec:conclusion}

In this manuscript, we have presented a self-consistent approach to estimate non-zero temperature effects in the trapped-ion quantum simulators of spin models. We have argued that the range of the spin-spin interactions mediated by the transverse phonons of the ion chain can be accurately captured by an effective QFT of a real scalar field that has a Yukawa-type coupling to the spins. In the vicinity of a linear-to-zigzag transition of the ion chain, $\lambda\phi^4$ interactions must be included in this QFT, which can modify the nature of the Yukawa interactions through phonon-phonon scattering. In light of the renormalizability of this QFT, this interaction effects can be recast in a renormalization of the bare quartic coupling, bare mass,  and wavefunction renormalization. We have argued that the later two effects yield a renormalization of the range and magnitude of the Yukawa-type spin-spin couplings $J_{ij}$, which could be inferred from trapped-ion experiments that reconstruct $J_{ij}$ from the real-time dynamics of the spins. Accordingly, the trapped-ion quantum simulator could be used to probe renormalization of this paradigmatic QFT. 

To find a quantitative prediction of these effects, we have presented a self-consistent approach that resums tadpole-like Feynman diagrams to all orders of the quartic coupling~\eqref{tad}. This so-called self-consistent Hartree approximation is afflicted by an infra-red divergence when trying to use it to determine thermal and quantum effects  of the critical point of the linear-to-zigzag transition. We have thus extended our approach beyond mean-field theory by also considering the sunrise diagram, and its additive and multiplicative contributions to the renormalized mass of the QFT~\eqref{17a}. We have discussed how these self-consistent approach can be applied to the trapped-ion case at non-zero temperatures, which requires using   a specific propagator~\eqref{eq:lattice_prop_ions} that includes a dipolar regularization~\eqref{eq:spatial_part_prop} of the QFT stemming from a multipole expansion of the Coulomb interactions among the trapped ions. Moreover, we have also discussed how to apply the Matsubara formalism in Euclidean time to account for non-zero temperatures in the experiment, i.e. laser cooling to a non-zero mean phonon number. Using realistic parameters for recent experiments with long ion chains~\cite{insbruck}, we have been able to derive specific quantitative prediction of thermal effects in the spin-spin interactions, which have been depicted in Fig.~\ref{fig:ions}. 
As an outlook, we note that these predictions  can serve as a guide to future trapped-ion experiments that aim at exploring the present connection between spin-model quantum simulators and this relativistic Yukawa-type problem. We note that the experimental quantum simulation, when working in the long-wavelength regime discussed in this work, will actually go beyond our approximations, and effectively compute the contributions to the Compton wavelength stemming from all possible Feynman diagrams. Moreover, the scaling of this quantity with the experimentally-tunable parameters will unveil the critical exponents of the phase transition of the $\lambda\phi^4$ model, which should correspond to those of the Ising universality class. Finally, we would like to mention that an interesting problem for future study would be to go beyond the schemes for the spin models of the Ising type, and consider further terms that can lead to an effective relativistic QFT of Dirac fermions Yukawa-coupled to the self-interacting scalar field. This trapped-ion quantum simulator would be closer to a lower-dimensional version of the fermion-Higgs sector of the electroweak interactions, and can provide a way to go beyond semi-classical calculations for the fractionalization of charge in fermion-scalar QFTs~\cite{PhysRevD.13.3398}.

\acknowledgements

We acknowledge support from PID2021-127726NB-I00 (MCIU/AEI/FEDER, UE), from the Grant IFT Centro de Excelencia Severo Ochoa CEX2020-001007-S, funded by MCIN/AEI/10.13039/501100011033, and from the CSIC Research Platform on Quantum Technologies PTI-001.  P. V. and A. B. acknowledge support from the EU Quantum Technology Flagship grant AQTION under grant  number 820495. The project leading to this application/publication has received funding from the European Union’s Horizon Europe research and innovation programme under grant agreement No 101114305 (“MILLENION-SGA1” EU Project).

\begin{appendix}

\section{\bf Thermal effects and Matsubara mode sums}
\label{sec:analytical_sums}

In this Appendix, we evaluate the thermal corrections resulting from the tadpole and sunrise diagrams in the lattice regularized $\lambda \phi^4$ QFT.
The contribution of the resumed tadpole family to
the self-energy is
\begin{gather}
  \mathsf{\Sigma}_{\rm td}= \frac{\lambda_0}{2}\frac{T}{ a N_1}\!\mathlarger{\sum}_{n_0,n_1}\!\frac{1}{(2\pi T n_0)^2+\hat{\rm k}^2+\mu^2}.
\end{gather}
The sum over the Matsubara frequencies runs on $n_0 \in {\mathbb Z}$ and $n_1=1,..,N_1$, with $N_1$ the number of sites of the lattice. The lattice analogue of the spatial momentum  $\hat{\rm k}$ is given in \eqref{eq:lattice_mom}. The Matsubara sum can be explicitly performed with the help of Cauchy's theorem \cite{kapusta_gale_2006}, obtaining
 \beq
\label{eq:ratio_pert_app} 
 \mathsf{\Sigma}_{\rm td}=\frac{\lambda_0}{4  N_1}\sum_{n_1=1}^{N_1}{1 \over a \omega(n_1)}\,{\rm coth} 
 \,{\omega(n_1)\over 2T}
 \eeq
where 
\begin{equation}
 \omega(n_1)=
 \sqrt{{4 \over a^2} \sin^2 {\pi n_1 \over N_1}+\mu^2 } \ .
\end{equation}
This can be rewritten as
 \beq
 \mathsf{\Sigma}_{\rm td}=\frac{\lambda_0}{2 N_1}\sum_{n_1=1}^{N_1}{1  \over a \omega(n_1)} \left( {1 \over 2} + {1 \over \ee^{\omega(n_1)/T} -1} \right) .
 \eeq\\\\
 

Being independent of the external momenta, the tadpole diagrams represent a shift on the bare mass.
The first term in the parenthesis is the zero temperature contribution. The second term, which is the mean field analogue of \eqref{34}, contains the thermal effects.
As explained in Sec.~\ref{sec:SC_eqs}, the sunrise diagram contributes both to the mass and wave function renormalization. The finite temperature lattice version of the mass shift \eqref{eq:sunrise_self_energy} is given by

\begin{widetext}
\beq
\mathsf{\Sigma}_{\rm sr}(\boldsymbol{0})=-{\lambda_0^2  \over 6 } {T^2 \over a^2 N_1^2}\;\;\mathlarger{\sum}_{n_0, n_1}\;\mathlarger{\sum}_{l_0,l_1}\hspace{1mm}\frac{1}{(2\pi T n_0)^2+\omega(n_1)^2} 
 \frac{1}{(2\pi T (n_0+l_0))^2+\omega(n_1+l_1)^2}\frac{1}{(2\pi T l_0)^2+\omega(l_1)^2} \ .
\eeq
\end{widetext}
The Matsubara sums can again be performed explicitly, with the result
\begin{widetext}
\beq
\label{eq:sr_mass}
\mathsf{\Sigma}_{\rm sr} (\boldsymbol{0})=-{\lambda_0^2 T^2 \over 24 N_1^2} \;\;\mathlarger{\sum}_{n_1,l_1} {1 \over \omega_1 \omega_2 \omega_3 \big( \sum_i\omega_i\big)}\left(1+ \mathlarger{\sum}_{ i<j}\; {(\omega_i^2+\omega_j^2 -\omega_{ij}^2)\omega_{ij} \over \prod_{k<l}(\omega_k+\omega_l-\omega_{kl})}\Big( \coth { \omega_i \over 2T} \coth {\omega_j \over 2T} -1 \Big)\right) \ ,
\eeq
\end{widetext}
where we have defined $\omega_1=\omega(n_1)$, $\omega_2=\omega(l_1)$ and $\omega_3=\omega(n_1+l_1)$ and $\omega_{ij}=\omega_{k}$ with $k\neq i,j$ and $i,j=1,2,3$. The first term in parenthesis in the zero temperature contribution, and the second one the thermal correction. 

\begin{figure}
	\centering
	\includegraphics[width=0.9\columnwidth]{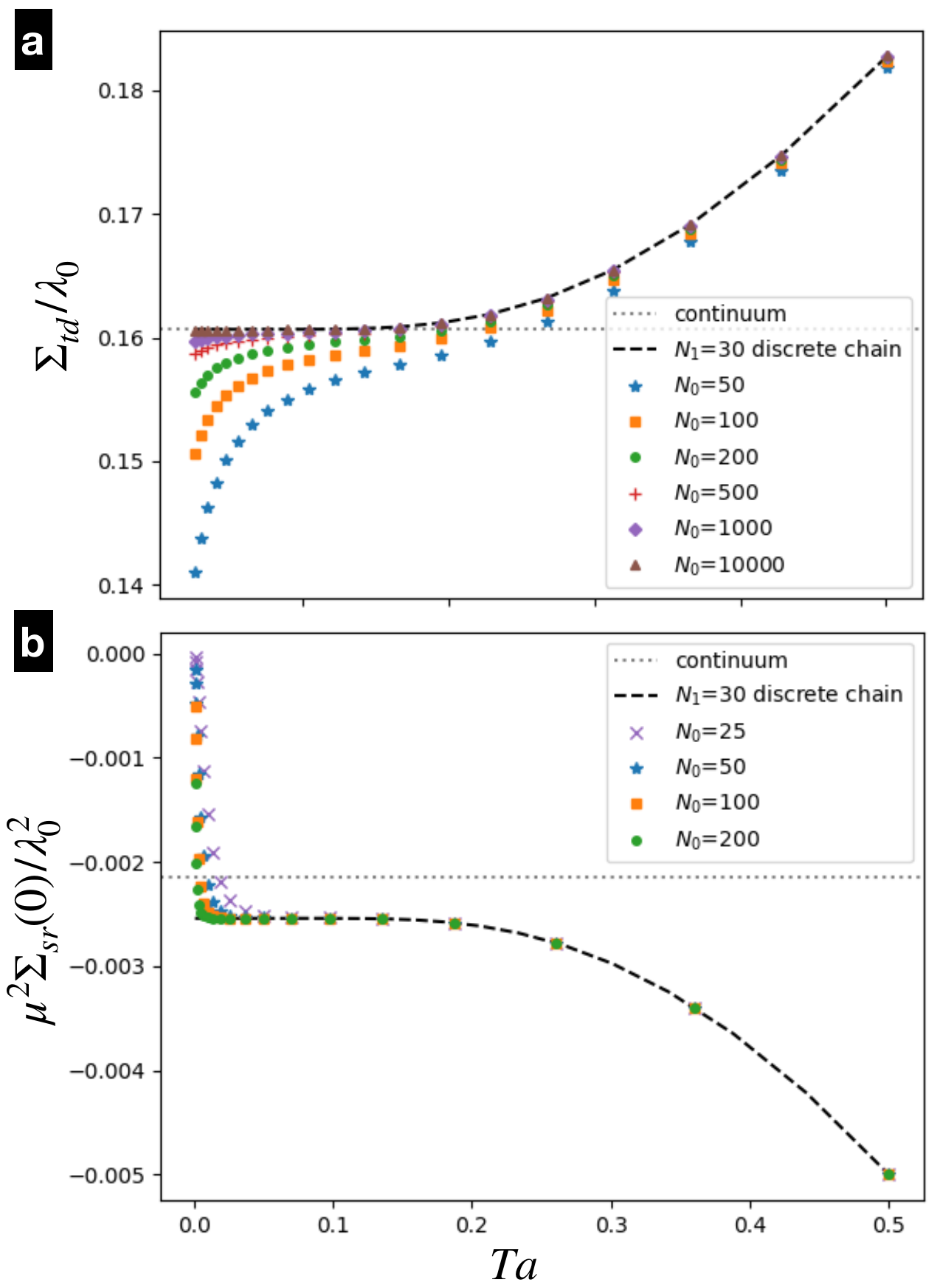}
	\caption{\label{fig:sums_thermal}  \textbf{  Convergence of tadpole and sunrise  Matsubara mode sums for the mass shifts:} Convergence of the Matsubara  sums displayed in Fig.~\ref{fig:sums} as a function the number of modes $N_0$, both  for the tadpole \textbf{(a)} and the sunrise contributions \textbf{(b)}.}
\end{figure}

\begin{figure}
\resizebox{.5\textwidth}{!}{
\includegraphics{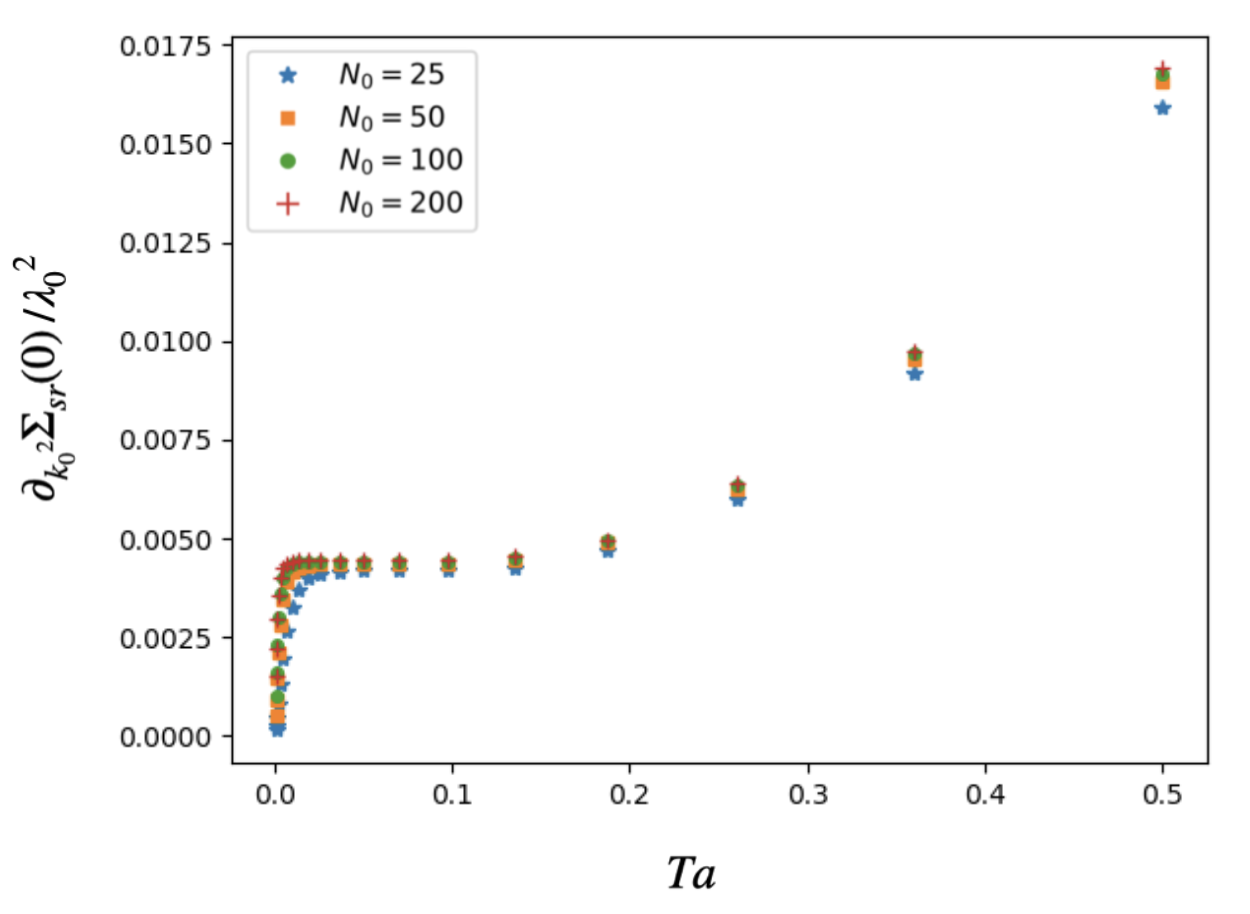}
}
\caption{\label{fig:matsubara_wf}{\bf Convergence of the Matsubara mode sums for the wavefunction renormalization:} Convergence of the Matsubara sums as a function of the number of modes $N_0$ for the quantity $\partial_{{k_0}^2}\Sigma_{sr}(0)/{{\lambda_0}^2}$ appearing in the wavefunction renormalization contribution.}
\end{figure}

Finally, let us address the contribution to the wave function renormalization of the sunrise diagram, Eq.~\eqref{eq:wf_sr}. At finite temperature, we have
\begin{widetext}
\beq
{\partial \mathsf{\Sigma}_{\rm sr}\over \partial k_0^2}(\boldsymbol{0}) ={\lambda_0^2  \over 6 } {T^2 \over a^2 N_1^2}\;\;\mathlarger{\sum}_{n_0, n_1}\;\mathlarger{\sum}_{l_0, l_1}\hspace{1mm}\frac{1}{(2\pi T  n_0)^2+\omega(n_1)^2} \frac{1}{(2\pi T l_0)^2+\omega(m_1)^2} \frac{\omega(n_1+l_1)^2}{\big((2\pi T (n_0+l_0))^2+\omega(n_1+l_1)^2\big)^3}  \ .
\label{wave}
\eeq
\end{widetext}
Unfortunately, we have not found an analytical expression for the Matsubara sums. They have to be evaluated numerically, which implies truncating the sums to a finite domain $n_0,m_0=-N_0,\cdots, N_0$.  It is thus important to 
analyze the convergence of the truncated sums with $N_0$. For the sake of completeness, we compare in Fig.~\ref{fig:sums_thermal} the numerical evaluation of the tadpole and sunrise mass shifts with the analytical results Eqs.~\eqref{eq:ratio_pert_app} and~\eqref{eq:sr_mass}. Whereas the sunrise contribution converges very fast for moderate $N_0$ at the temperatures of interest, an accurate estimate of the tadpole shift requires much larger values. Therefore, incorporating these expressions in the numerical routines that solve the self-consistent equations makes them much more efficient. 
The result of the numerical evaluation of the wavefunction renormalization contribution \eqref{wave} is shown in Fig.~\ref{fig:matsubara_wf}. As it was the case for the sunrise mass shift, the convergence is very good even in the low-temperature limit for moderate values of $N_0$.



\section{\bf  Analytical estimate of the critical ratio}
\label{sec: continuum}

We evaluate here the $T=0$ critical point of the $\lambda \phi^4$ theory in the continuum.
As explained in text main text, it is customary to describe the critical point in terms of the dimensionless ratio $\lambda_0/\mu^2$, with $\mu^2$ UV-finite tadpole renormalized mass~\eqref{eq:mu}. Within our self-consistent approach, the critical point is determined by Eq.~\eqref{crit1} and \eqref{eq:sunrise_self_energy}
\beq
\mu^2=\frac{\lambda_0^2}{6}\!\!\bigintsss\!\!\!\frac{{\rm d}^2k{\rm d}^2q}{(2\pi)^4} {1 \over k^2+\mu^2}{1 \over q^2+\mu^2}{1 \over (k+q)^2+\mu^2} \ .
\eeq
Rescaling the momenta in the integral $p,q \to \mu p,\mu q$, we obtain
\beq
\label{eq: continuum_critical_ratio}
f_{\rm c}=\left. {\lambda_0 \over \mu^2}\right|_{\rm c}=\sqrt{\frac{6(2\pi)^4}{I}} \ ,
\eeq
with 
\beq
I=\bigintsss\!\!\!\frac{{\rm d}^2k{\rm d}^2q}{(2\pi)^4} {1 \over k^2+1}{1 \over q^2+1}{1 \over (k+q)^2+1} \ .
\eeq


We evaluate this integral making use of Feynman parameters \cite{peskin}, namely
\beq
I=\bigintsss{\rm d}^2k{\rm d}^2q\frac{1}{k^2+1}\bigintsss_{0}^{1}{\rm d}x{\rm d}y \frac{\delta(1+x+y)}{(q^2+k^2xy+1)^2} \ .
\eeq
Integrating over $q$ we obtain
\beq
I=\pi \bigintsss {\rm d}^2k \bigintsss_{0}^{1}{\rm d}x  \frac{1}{k^2+1} \frac{1}{k^2x(1-x)+1}.
\eeq
Finally, we use Feynman parameters again to integrate over $p$, with the result
\beq
\begin{split}
I&=\pi^2\bigintsss_{0}^{1}{\rm d}x{\rm d}z \frac{1}{1-z(1-x(1-x))}= \nonumber \\
&=\frac{\pi^2}{18}\left(\psi^{1}\!\!\left(\frac{1}{6}\right)+\psi^{1}\!\!\left(\frac{1}{3}\right)-\psi^{1}\!\!\left(\frac{2}{3}\right)-\psi^{1}\!\!\left(\frac{5}{6}\right)\right),
\end{split}
\eeq
where we have introduced the PolyGamma functions defined in terms of derivatives of Euler's gamma function $\Gamma(z)=\int_0^{\infty}{\rm d}t t^z\ee^{-t}$, namely $\psi^n(z)={\rm d}^{n+1}\log\Gamma(z)/{\rm d}z^{n+1}$. 

\section{\bf  Critical line crossings}
\label{sec: crossings}

\begin{figure}
\resizebox{.5\textwidth}{!}{
\includegraphics{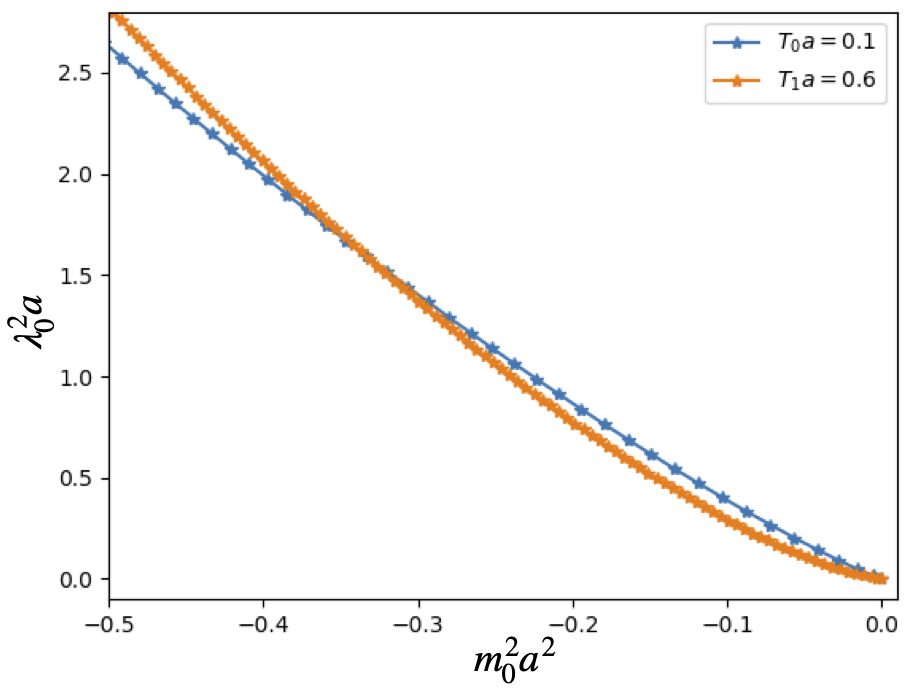}
}
\caption{\label{fig:crossing}{\bf Critical lines with crossings:} Two critical lines for different temperatures show crossings in the $(m_0^2a^2,\lambda_0 a^2)$ plane as a consequence of the approximations in our procedure.}
\end{figure}

\begin{figure}[!t]
\resizebox{.49\textwidth}{!}{
\includegraphics{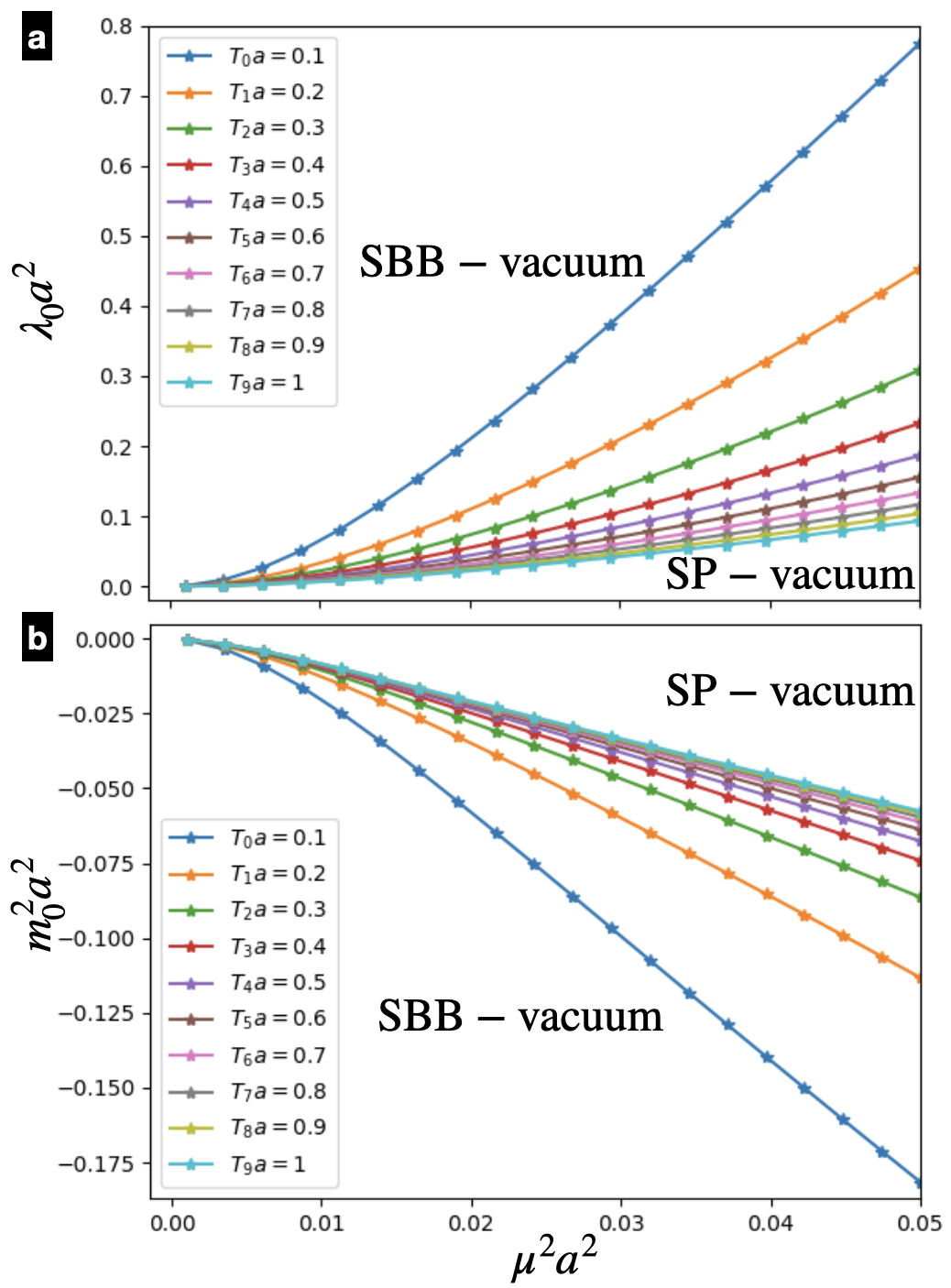}
}
\caption{\label{fig:critical_field_non_crossing}{\bf Critical lines with no crossings:} {\bf (a)} We solve the self-consistent equations for the critical point ${m}^2_{\rm P}|_{\rm c}=0$ and plot $\lambda_0a^2$ with respect to $\mu^2a^2$, which leads to non-crossing critical lines. {\bf (b)} We also plot the bare mass $m_0^2a^2$ as a function of $\mu a^2$ to confirm that no crossings occur.}
\end{figure}

\begin{figure*}[!t]
\resizebox{\textwidth}{!}{
\includegraphics{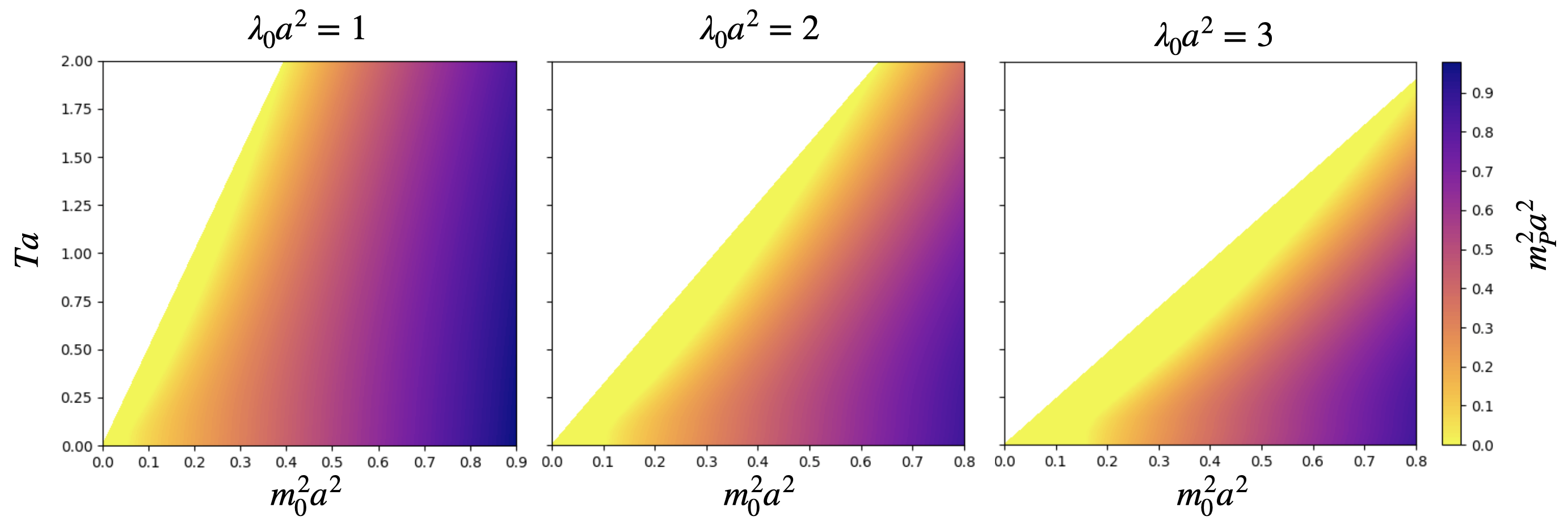}
}
\caption{\label{fig:critical_field1_non_crossing} {\bf Finite-physical mass  of the $\lambda\phi^4$ model on a lattice with respect to $\mu^2 a^2$:} Contour plot for finite values of the tadpole and sunrise contributions to the physical mass ${m}_{\rm P}^2a^2$, which are obtained by solving  the self-consistent equation~\eqref{17a}. The numerical solution uses analytical Matsubara mode sums and $N_1=30$ spatial points to calculate the tadpole and sunrise diagrams. The white region corresponds to the symmetry-broken phase, and the coloured region to the symmetric one. As $\lambda_0a^2$ increases, the critical line ${m}_{\rm P}^2=0$ separating both phases is folded to the right. As shown, no crossings occur when plotting with respect to $\mu^2a^2$.}
\end{figure*}

\begin{figure*}[!t]
\resizebox{\textwidth}{!}{
\includegraphics{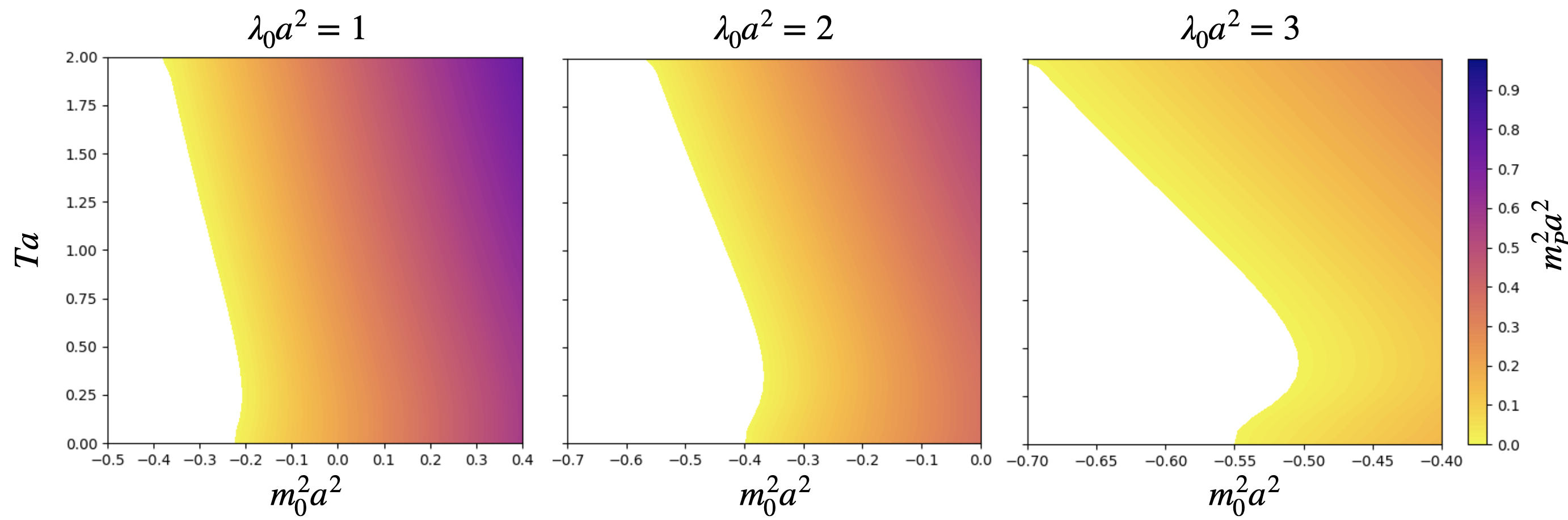}
}
\caption{\label{fig:critical_field1} {\bf Finite-physical mass  of the $\lambda\phi^4$ model on a lattice with respect to $m_0^2a^2$:} Contour plot for finite values of the tadpole and sunrise contributions to the physical mass ${m}_{\rm P}^2a^2$, which are obtained by solving  the self-consistent equation ~\eqref{17a}. The numerical solution uses analytical Matsubara mode sums and $N_1=30$ spatial points to calculate the tadpole and sunrise diagrams. The white region corresponds to the symmetry-broken phase, and the coloured region to the symmetric one. As $\lambda_0a^2$ increases, the critical line ${m}_{\rm P}^2=0$ separating both phases is shifted to the left. Crossings can be appreciated for low temperatures.}
\end{figure*}

In this Appendix, we describe certain limitations of the current approach that can become important away for the regime of interest discussed in the main text, namely that of large couplings and masses.
As it can be seen in Fig.~\ref{fig:crossing}, the critical lines separating the broken and unbroken phases 
have crossing points in the $(m_0^2a^2,\lambda_0 a^2)$ plane. Actually, the critical lines of 
any two temperatures always cross for sufficiently negative $m_0^2$ and large $\lambda_0$. This behavior is an artefact that stems from the approximations underlying   our procedure, as not only many Feynman diagrams are being discarded in the loop expansion, but also the self-consistency resummations make tadpole-like diagrams prevail upon the rest.
Fig.~\ref{fig:critical_field_non_crossing} shows that these crossing are however not manifest when the critical lines are plotted as a function of the tadpole renormalized mass $\mu^2$.
In Fig.~\ref{fig:critical_field1_non_crossing}, we also present the contour plots of the physical mass as a function of $(\mu^2a^2,Ta)$, which show a very regular behavior for large values of the couplings and masses. On the other hand, when displaying these contour plots in terms of the bare mass, we encounter an unphysical re-entrance of the symmetry-broken phase at low temperatures, that becomes more evident as one increases the quartic coupling (see Fig.~\ref{fig:critical_field1}). This reentrance is again a consequence of the aforementioned crossings of critical lines. 

We now show that indeed equations
\eqref{eq:mu} and \eqref{eq:sunrise_self_energy}
allow for pairs of parameters $\mu_1$, $\mu_2$ and $T_1$, $T_2$ such that  $m_{0,1}^2=m_{0,2}^2$ and $\lambda_{0,1}=\lambda_{0,2}$.
Let us defined the dimensioneless combinations $\Gamma_{\rm td}=\Sigma_{\rm td}/\lambda_0$
and $\Gamma_{\rm sr}=-\Sigma_{\rm sr}(\boldsymbol{0})/\lambda_0^2 a^2$, which are functions of $\mu a$ and $T a$. From \eqref{eq:mu} and \eqref{eq:sunrise_self_energy}, at the crossing we have
\beq
\label{eq:corssing_conditions1}
{\left(\frac{\mu_1}{\mu_2}\right)}^2=\frac{\Gamma_{\rm sr,1}}{\Gamma_{\rm sr,2}} \ ,
\eeq
together with
\beq
\mu_1^2-\lambda_0\Gamma_{\rm td,1}=\mu_2^2-\lambda_0\Gamma_{\rm td,2},
\eeq
where $\Gamma_{\rm sr,1}$ and $\Gamma_{\rm sr,2}$ refer to $\Gamma_{\rm sr}(\mu_1a,T_1a)$ and $\Gamma_{\rm sr}(\mu_2a,T_2a)$, and equivalently for $\Gamma_{\rm td}$.
Once $T_1$ and $T_2$ are chosen, the first equation determines $\mu_2=\mu_2(\mu_1)$, while the second singles out a value for $\mu_1$.
On the contrary, it is immediate to see that  crossings are absent when
the critical values of the coupling are plotted as a function of $\mu a$. 
From Eq.~\eqref{eq:corssing_conditions1}, such a crossing
would imply $\Gamma_{\rm sr,1} =\Gamma_{\rm sr,2}$.
Since the sunrise mass shift is a monotonic function of the temperature (see Fig.~\ref{fig:sums}), this condition is never satisfied.

\newpage



\end{appendix}
\clearpage

\bibliographystyle{apsrev4-1}
\bibliography{refs}

\begin{thebibliography}{181}%
\makeatletter
\providecommand \@ifxundefined [1]{%
 \@ifx{#1\undefined}
}%
\providecommand \@ifnum [1]{%
 \ifnum #1\expandafter \@firstoftwo
 \else \expandafter \@secondoftwo
 \fi
}%
\providecommand \@ifx [1]{%
 \ifx #1\expandafter \@firstoftwo
 \else \expandafter \@secondoftwo
 \fi
}%
\providecommand \natexlab [1]{#1}%
\providecommand \enquote  [1]{``#1''}%
\providecommand \bibnamefont  [1]{#1}%
\providecommand \bibfnamefont [1]{#1}%
\providecommand \citenamefont [1]{#1}%
\providecommand \href@noop [0]{\@secondoftwo}%
\providecommand \href [0]{\begingroup \@sanitize@url \@href}%
\providecommand \@href[1]{\@@startlink{#1}\@@href}%
\providecommand \@@href[1]{\endgroup#1\@@endlink}%
\providecommand \@sanitize@url [0]{\catcode `\\12\catcode `\$12\catcode `\&12\catcode `\#12\catcode `\^12\catcode `\_12\catcode `\%12\relax}%
\providecommand \@@startlink[1]{}%
\providecommand \@@endlink[0]{}%
\providecommand \url  [0]{\begingroup\@sanitize@url \@url }%
\providecommand \@url [1]{\endgroup\@href {#1}{\urlprefix }}%
\providecommand \urlprefix  [0]{URL }%
\providecommand \Eprint [0]{\href }%
\providecommand \doibase [0]{http://dx.doi.org/}%
\providecommand \selectlanguage [0]{\@gobble}%
\providecommand \bibinfo  [0]{\@secondoftwo}%
\providecommand \bibfield  [0]{\@secondoftwo}%
\providecommand \translation [1]{[#1]}%
\providecommand \BibitemOpen [0]{}%
\providecommand \bibitemStop [0]{}%
\providecommand \bibitemNoStop [0]{.\EOS\space}%
\providecommand \EOS [0]{\spacefactor3000\relax}%
\providecommand \BibitemShut  [1]{\csname bibitem#1\endcsname}%
\let\auto@bib@innerbib\@empty
\bibitem [{\citenamefont {Acín}\ \emph {et~al.}(2018)\citenamefont {Acín}, \citenamefont {Bloch}, \citenamefont {Buhrman}, \citenamefont {Calarco}, \citenamefont {Eichler}, \citenamefont {Eisert}, \citenamefont {Esteve}, \citenamefont {Gisin}, \citenamefont {Glaser}, \citenamefont {Jelezko}, \citenamefont {Kuhr}, \citenamefont {Lewenstein}, \citenamefont {Riedel}, \citenamefont {Schmidt}, \citenamefont {Thew}, \citenamefont {Wallraff}, \citenamefont {Walmsley},\ and\ \citenamefont {Wilhelm}}]{Acin_2018}%
  \BibitemOpen
  \bibfield  {author} {\bibinfo {author} {\bibfnamefont {A.}~\bibnamefont {Acín}}, \bibinfo {author} {\bibfnamefont {I.}~\bibnamefont {Bloch}}, \bibinfo {author} {\bibfnamefont {H.}~\bibnamefont {Buhrman}}, \bibinfo {author} {\bibfnamefont {T.}~\bibnamefont {Calarco}}, \bibinfo {author} {\bibfnamefont {C.}~\bibnamefont {Eichler}}, \bibinfo {author} {\bibfnamefont {J.}~\bibnamefont {Eisert}}, \bibinfo {author} {\bibfnamefont {D.}~\bibnamefont {Esteve}}, \bibinfo {author} {\bibfnamefont {N.}~\bibnamefont {Gisin}}, \bibinfo {author} {\bibfnamefont {S.~J.}\ \bibnamefont {Glaser}}, \bibinfo {author} {\bibfnamefont {F.}~\bibnamefont {Jelezko}}, \bibinfo {author} {\bibfnamefont {S.}~\bibnamefont {Kuhr}}, \bibinfo {author} {\bibfnamefont {M.}~\bibnamefont {Lewenstein}}, \bibinfo {author} {\bibfnamefont {M.~F.}\ \bibnamefont {Riedel}}, \bibinfo {author} {\bibfnamefont {P.~O.}\ \bibnamefont {Schmidt}}, \bibinfo {author} {\bibfnamefont {R.}~\bibnamefont {Thew}}, \bibinfo {author} {\bibfnamefont {A.}~\bibnamefont
  {Wallraff}}, \bibinfo {author} {\bibfnamefont {I.}~\bibnamefont {Walmsley}}, \ and\ \bibinfo {author} {\bibfnamefont {F.~K.}\ \bibnamefont {Wilhelm}},\ }\href {\doibase 10.1088/1367-2630/aad1ea} {\bibfield  {journal} {\bibinfo  {journal} {New Journal of Physics}\ }\textbf {\bibinfo {volume} {20}},\ \bibinfo {pages} {080201} (\bibinfo {year} {2018})}\BibitemShut {NoStop}%
\bibitem [{\citenamefont {Mohseni}\ \emph {et~al.}(2017)\citenamefont {Mohseni}, \citenamefont {Read}, \citenamefont {Neven}, \citenamefont {Boixo}, \citenamefont {Denchev}, \citenamefont {Babbush}, \citenamefont {Fowler}, \citenamefont {Smelyanskiy},\ and\ \citenamefont {Martinis}}]{Mohseni2017}%
  \BibitemOpen
  \bibfield  {author} {\bibinfo {author} {\bibfnamefont {M.}~\bibnamefont {Mohseni}}, \bibinfo {author} {\bibfnamefont {P.}~\bibnamefont {Read}}, \bibinfo {author} {\bibfnamefont {H.}~\bibnamefont {Neven}}, \bibinfo {author} {\bibfnamefont {S.}~\bibnamefont {Boixo}}, \bibinfo {author} {\bibfnamefont {V.}~\bibnamefont {Denchev}}, \bibinfo {author} {\bibfnamefont {R.}~\bibnamefont {Babbush}}, \bibinfo {author} {\bibfnamefont {A.}~\bibnamefont {Fowler}}, \bibinfo {author} {\bibfnamefont {V.}~\bibnamefont {Smelyanskiy}}, \ and\ \bibinfo {author} {\bibfnamefont {J.}~\bibnamefont {Martinis}},\ }\href {\doibase 10.1038/543171a} {\bibfield  {journal} {\bibinfo  {journal} {Nature}\ }\textbf {\bibinfo {volume} {543}},\ \bibinfo {pages} {171} (\bibinfo {year} {2017})}\BibitemShut {NoStop}%
\bibitem [{\citenamefont {Nielsen}\ and\ \citenamefont {Chuang}(2000)}]{nielsen00}%
  \BibitemOpen
  \bibfield  {author} {\bibinfo {author} {\bibfnamefont {M.~A.}\ \bibnamefont {Nielsen}}\ and\ \bibinfo {author} {\bibfnamefont {I.~L.}\ \bibnamefont {Chuang}},\ }\href@noop {} {\emph {\bibinfo {title} {Quantum Computation and Quantum Information}}}\ (\bibinfo  {publisher} {Cambridge University Press},\ \bibinfo {year} {2000})\BibitemShut {NoStop}%
\bibitem [{\citenamefont {Postler}\ \emph {et~al.}(2022)\citenamefont {Postler}, \citenamefont {Heu$\beta$en}, \citenamefont {Pogorelov}, \citenamefont {Rispler}, \citenamefont {Feldker}, \citenamefont {Meth}, \citenamefont {Marciniak}, \citenamefont {Stricker}, \citenamefont {Ringbauer}, \citenamefont {Blatt}, \citenamefont {Schindler}, \citenamefont {M{\"u}ller},\ and\ \citenamefont {Monz}}]{Postler2022}%
  \BibitemOpen
  \bibfield  {author} {\bibinfo {author} {\bibfnamefont {L.}~\bibnamefont {Postler}}, \bibinfo {author} {\bibfnamefont {S.}~\bibnamefont {Heu$\beta$en}}, \bibinfo {author} {\bibfnamefont {I.}~\bibnamefont {Pogorelov}}, \bibinfo {author} {\bibfnamefont {M.}~\bibnamefont {Rispler}}, \bibinfo {author} {\bibfnamefont {T.}~\bibnamefont {Feldker}}, \bibinfo {author} {\bibfnamefont {M.}~\bibnamefont {Meth}}, \bibinfo {author} {\bibfnamefont {C.~D.}\ \bibnamefont {Marciniak}}, \bibinfo {author} {\bibfnamefont {R.}~\bibnamefont {Stricker}}, \bibinfo {author} {\bibfnamefont {M.}~\bibnamefont {Ringbauer}}, \bibinfo {author} {\bibfnamefont {R.}~\bibnamefont {Blatt}}, \bibinfo {author} {\bibfnamefont {P.}~\bibnamefont {Schindler}}, \bibinfo {author} {\bibfnamefont {M.}~\bibnamefont {M{\"u}ller}}, \ and\ \bibinfo {author} {\bibfnamefont {T.}~\bibnamefont {Monz}},\ }\href {\doibase 10.1038/s41586-022-04721-1} {\bibfield  {journal} {\bibinfo  {journal} {Nature}\ }\textbf {\bibinfo {volume} {605}},\ \bibinfo {pages} {675}
  (\bibinfo {year} {2022})}\BibitemShut {NoStop}%
\bibitem [{\citenamefont {Krinner}\ \emph {et~al.}(2022)\citenamefont {Krinner}, \citenamefont {Lacroix}, \citenamefont {Remm}, \citenamefont {Di~Paolo}, \citenamefont {Genois}, \citenamefont {Leroux}, \citenamefont {Hellings}, \citenamefont {Lazar}, \citenamefont {Swiadek}, \citenamefont {Herrmann}, \citenamefont {Norris}, \citenamefont {Andersen}, \citenamefont {M{\"u}ller}, \citenamefont {Blais}, \citenamefont {Eichler},\ and\ \citenamefont {Wallraff}}]{Krinner2022}%
  \BibitemOpen
  \bibfield  {author} {\bibinfo {author} {\bibfnamefont {S.}~\bibnamefont {Krinner}}, \bibinfo {author} {\bibfnamefont {N.}~\bibnamefont {Lacroix}}, \bibinfo {author} {\bibfnamefont {A.}~\bibnamefont {Remm}}, \bibinfo {author} {\bibfnamefont {A.}~\bibnamefont {Di~Paolo}}, \bibinfo {author} {\bibfnamefont {E.}~\bibnamefont {Genois}}, \bibinfo {author} {\bibfnamefont {C.}~\bibnamefont {Leroux}}, \bibinfo {author} {\bibfnamefont {C.}~\bibnamefont {Hellings}}, \bibinfo {author} {\bibfnamefont {S.}~\bibnamefont {Lazar}}, \bibinfo {author} {\bibfnamefont {F.}~\bibnamefont {Swiadek}}, \bibinfo {author} {\bibfnamefont {J.}~\bibnamefont {Herrmann}}, \bibinfo {author} {\bibfnamefont {G.~J.}\ \bibnamefont {Norris}}, \bibinfo {author} {\bibfnamefont {C.~K.}\ \bibnamefont {Andersen}}, \bibinfo {author} {\bibfnamefont {M.}~\bibnamefont {M{\"u}ller}}, \bibinfo {author} {\bibfnamefont {A.}~\bibnamefont {Blais}}, \bibinfo {author} {\bibfnamefont {C.}~\bibnamefont {Eichler}}, \ and\ \bibinfo {author} {\bibfnamefont
  {A.}~\bibnamefont {Wallraff}},\ }\href {\doibase 10.1038/s41586-022-04566-8} {\bibfield  {journal} {\bibinfo  {journal} {Nature}\ }\textbf {\bibinfo {volume} {605}},\ \bibinfo {pages} {669} (\bibinfo {year} {2022})}\BibitemShut {NoStop}%
\bibitem [{\citenamefont {Zhao}\ \emph {et~al.}(2022)\citenamefont {Zhao}, \citenamefont {Ye}, \citenamefont {Huang}, \citenamefont {Zhang}, \citenamefont {Wu}, \citenamefont {Guan}, \citenamefont {Zhu}, \citenamefont {Wei}, \citenamefont {He}, \citenamefont {Cao}, \citenamefont {Chen}, \citenamefont {Chung}, \citenamefont {Deng}, \citenamefont {Fan}, \citenamefont {Gong}, \citenamefont {Guo}, \citenamefont {Guo}, \citenamefont {Han}, \citenamefont {Li}, \citenamefont {Li}, \citenamefont {Li}, \citenamefont {Liang}, \citenamefont {Lin}, \citenamefont {Qian}, \citenamefont {Rong}, \citenamefont {Su}, \citenamefont {Sun}, \citenamefont {Wang}, \citenamefont {Wu}, \citenamefont {Xu}, \citenamefont {Ying}, \citenamefont {Yu}, \citenamefont {Zha}, \citenamefont {Zhang}, \citenamefont {Huo}, \citenamefont {Lu}, \citenamefont {Peng}, \citenamefont {Zhu},\ and\ \citenamefont {Pan}}]{PhysRevLett.129.030501}%
  \BibitemOpen
  \bibfield  {author} {\bibinfo {author} {\bibfnamefont {Y.}~\bibnamefont {Zhao}}, \bibinfo {author} {\bibfnamefont {Y.}~\bibnamefont {Ye}}, \bibinfo {author} {\bibfnamefont {H.-L.}\ \bibnamefont {Huang}}, \bibinfo {author} {\bibfnamefont {Y.}~\bibnamefont {Zhang}}, \bibinfo {author} {\bibfnamefont {D.}~\bibnamefont {Wu}}, \bibinfo {author} {\bibfnamefont {H.}~\bibnamefont {Guan}}, \bibinfo {author} {\bibfnamefont {Q.}~\bibnamefont {Zhu}}, \bibinfo {author} {\bibfnamefont {Z.}~\bibnamefont {Wei}}, \bibinfo {author} {\bibfnamefont {T.}~\bibnamefont {He}}, \bibinfo {author} {\bibfnamefont {S.}~\bibnamefont {Cao}}, \bibinfo {author} {\bibfnamefont {F.}~\bibnamefont {Chen}}, \bibinfo {author} {\bibfnamefont {T.-H.}\ \bibnamefont {Chung}}, \bibinfo {author} {\bibfnamefont {H.}~\bibnamefont {Deng}}, \bibinfo {author} {\bibfnamefont {D.}~\bibnamefont {Fan}}, \bibinfo {author} {\bibfnamefont {M.}~\bibnamefont {Gong}}, \bibinfo {author} {\bibfnamefont {C.}~\bibnamefont {Guo}}, \bibinfo {author} {\bibfnamefont
  {S.}~\bibnamefont {Guo}}, \bibinfo {author} {\bibfnamefont {L.}~\bibnamefont {Han}}, \bibinfo {author} {\bibfnamefont {N.}~\bibnamefont {Li}}, \bibinfo {author} {\bibfnamefont {S.}~\bibnamefont {Li}}, \bibinfo {author} {\bibfnamefont {Y.}~\bibnamefont {Li}}, \bibinfo {author} {\bibfnamefont {F.}~\bibnamefont {Liang}}, \bibinfo {author} {\bibfnamefont {J.}~\bibnamefont {Lin}}, \bibinfo {author} {\bibfnamefont {H.}~\bibnamefont {Qian}}, \bibinfo {author} {\bibfnamefont {H.}~\bibnamefont {Rong}}, \bibinfo {author} {\bibfnamefont {H.}~\bibnamefont {Su}}, \bibinfo {author} {\bibfnamefont {L.}~\bibnamefont {Sun}}, \bibinfo {author} {\bibfnamefont {S.}~\bibnamefont {Wang}}, \bibinfo {author} {\bibfnamefont {Y.}~\bibnamefont {Wu}}, \bibinfo {author} {\bibfnamefont {Y.}~\bibnamefont {Xu}}, \bibinfo {author} {\bibfnamefont {C.}~\bibnamefont {Ying}}, \bibinfo {author} {\bibfnamefont {J.}~\bibnamefont {Yu}}, \bibinfo {author} {\bibfnamefont {C.}~\bibnamefont {Zha}}, \bibinfo {author} {\bibfnamefont {K.}~\bibnamefont
  {Zhang}}, \bibinfo {author} {\bibfnamefont {Y.-H.}\ \bibnamefont {Huo}}, \bibinfo {author} {\bibfnamefont {C.-Y.}\ \bibnamefont {Lu}}, \bibinfo {author} {\bibfnamefont {C.-Z.}\ \bibnamefont {Peng}}, \bibinfo {author} {\bibfnamefont {X.}~\bibnamefont {Zhu}}, \ and\ \bibinfo {author} {\bibfnamefont {J.-W.}\ \bibnamefont {Pan}},\ }\href {\doibase 10.1103/PhysRevLett.129.030501} {\bibfield  {journal} {\bibinfo  {journal} {Phys. Rev. Lett.}\ }\textbf {\bibinfo {volume} {129}},\ \bibinfo {pages} {030501} (\bibinfo {year} {2022})}\BibitemShut {NoStop}%
\bibitem [{\citenamefont {Ryan-Anderson}\ \emph {et~al.}(2022)\citenamefont {Ryan-Anderson}, \citenamefont {Brown}, \citenamefont {Allman}, \citenamefont {Arkin}, \citenamefont {Asa-Attuah}, \citenamefont {Baldwin}, \citenamefont {Berg}, \citenamefont {Bohnet}, \citenamefont {Braxton}, \citenamefont {Burdick}, \citenamefont {Campora}, \citenamefont {Chernoguzov}, \citenamefont {Esposito}, \citenamefont {Evans}, \citenamefont {Francois}, \citenamefont {Gaebler}, \citenamefont {Gatterman}, \citenamefont {Gerber}, \citenamefont {Gilmore}, \citenamefont {Gresh}, \citenamefont {Hall}, \citenamefont {Hankin}, \citenamefont {Hostetter}, \citenamefont {Lucchetti}, \citenamefont {Mayer}, \citenamefont {Myers}, \citenamefont {Neyenhuis}, \citenamefont {Santiago}, \citenamefont {Sedlacek}, \citenamefont {Skripka}, \citenamefont {Slattery}, \citenamefont {Stutz}, \citenamefont {Tait}, \citenamefont {Tobey}, \citenamefont {Vittorini}, \citenamefont {Walker},\ and\ \citenamefont
  {Hayes}}]{https://doi.org/10.48550/arxiv.2208.01863}%
  \BibitemOpen
  \bibfield  {author} {\bibinfo {author} {\bibfnamefont {C.}~\bibnamefont {Ryan-Anderson}}, \bibinfo {author} {\bibfnamefont {N.~C.}\ \bibnamefont {Brown}}, \bibinfo {author} {\bibfnamefont {M.~S.}\ \bibnamefont {Allman}}, \bibinfo {author} {\bibfnamefont {B.}~\bibnamefont {Arkin}}, \bibinfo {author} {\bibfnamefont {G.}~\bibnamefont {Asa-Attuah}}, \bibinfo {author} {\bibfnamefont {C.}~\bibnamefont {Baldwin}}, \bibinfo {author} {\bibfnamefont {J.}~\bibnamefont {Berg}}, \bibinfo {author} {\bibfnamefont {J.~G.}\ \bibnamefont {Bohnet}}, \bibinfo {author} {\bibfnamefont {S.}~\bibnamefont {Braxton}}, \bibinfo {author} {\bibfnamefont {N.}~\bibnamefont {Burdick}}, \bibinfo {author} {\bibfnamefont {J.~P.}\ \bibnamefont {Campora}}, \bibinfo {author} {\bibfnamefont {A.}~\bibnamefont {Chernoguzov}}, \bibinfo {author} {\bibfnamefont {J.}~\bibnamefont {Esposito}}, \bibinfo {author} {\bibfnamefont {B.}~\bibnamefont {Evans}}, \bibinfo {author} {\bibfnamefont {D.}~\bibnamefont {Francois}}, \bibinfo {author} {\bibfnamefont
  {J.~P.}\ \bibnamefont {Gaebler}}, \bibinfo {author} {\bibfnamefont {T.~M.}\ \bibnamefont {Gatterman}}, \bibinfo {author} {\bibfnamefont {J.}~\bibnamefont {Gerber}}, \bibinfo {author} {\bibfnamefont {K.}~\bibnamefont {Gilmore}}, \bibinfo {author} {\bibfnamefont {D.}~\bibnamefont {Gresh}}, \bibinfo {author} {\bibfnamefont {A.}~\bibnamefont {Hall}}, \bibinfo {author} {\bibfnamefont {A.}~\bibnamefont {Hankin}}, \bibinfo {author} {\bibfnamefont {J.}~\bibnamefont {Hostetter}}, \bibinfo {author} {\bibfnamefont {D.}~\bibnamefont {Lucchetti}}, \bibinfo {author} {\bibfnamefont {K.}~\bibnamefont {Mayer}}, \bibinfo {author} {\bibfnamefont {J.}~\bibnamefont {Myers}}, \bibinfo {author} {\bibfnamefont {B.}~\bibnamefont {Neyenhuis}}, \bibinfo {author} {\bibfnamefont {J.}~\bibnamefont {Santiago}}, \bibinfo {author} {\bibfnamefont {J.}~\bibnamefont {Sedlacek}}, \bibinfo {author} {\bibfnamefont {T.}~\bibnamefont {Skripka}}, \bibinfo {author} {\bibfnamefont {A.}~\bibnamefont {Slattery}}, \bibinfo {author} {\bibfnamefont
  {R.~P.}\ \bibnamefont {Stutz}}, \bibinfo {author} {\bibfnamefont {J.}~\bibnamefont {Tait}}, \bibinfo {author} {\bibfnamefont {R.}~\bibnamefont {Tobey}}, \bibinfo {author} {\bibfnamefont {G.}~\bibnamefont {Vittorini}}, \bibinfo {author} {\bibfnamefont {J.}~\bibnamefont {Walker}}, \ and\ \bibinfo {author} {\bibfnamefont {D.}~\bibnamefont {Hayes}},\ }\href {\doibase 10.48550/ARXIV.2208.01863} {\  (\bibinfo {year} {2022}),\ 10.48550/ARXIV.2208.01863}\BibitemShut {NoStop}%
\bibitem [{\citenamefont {Acharya}\ \emph {et~al.}(2022)\citenamefont {Acharya}, \citenamefont {Aleiner},\ and\ \citenamefont {Allen}}]{https://doi.org/10.48550/arxiv.2207.06431}%
  \BibitemOpen
  \bibfield  {author} {\bibinfo {author} {\bibfnamefont {R.}~\bibnamefont {Acharya}}, \bibinfo {author} {\bibfnamefont {I.}~\bibnamefont {Aleiner}}, \ and\ \bibinfo {author} {\bibfnamefont {R.~e.~{\it al.}.}\ \bibnamefont {Allen}},\ }\href {\doibase 10.1038/s41586-022-05434-1} {\  (\bibinfo {year} {2022}),\ 10.1038/s41586-022-05434-1}\BibitemShut {NoStop}%
\bibitem [{\citenamefont {Aharonov}\ and\ \citenamefont {Ben-Or}(1998)}]{FTQEC}%
  \BibitemOpen
  \bibfield  {author} {\bibinfo {author} {\bibfnamefont {D.}~\bibnamefont {Aharonov}}\ and\ \bibinfo {author} {\bibfnamefont {M.}~\bibnamefont {Ben-Or}},\ }\href {\doibase 10.1137/S0097539799359385} {\bibfield  {journal} {\bibinfo  {journal} {SIAM J. Comput.}\ }\textbf {\bibinfo {volume} {38}},\ \bibinfo {pages} {1207} (\bibinfo {year} {1998})}\BibitemShut {NoStop}%
\bibitem [{\citenamefont {Montanaro}(2016)}]{Montanaro2016}%
  \BibitemOpen
  \bibfield  {author} {\bibinfo {author} {\bibfnamefont {A.}~\bibnamefont {Montanaro}},\ }\href {\doibase 10.1038/npjqi.2015.23} {\bibfield  {journal} {\bibinfo  {journal} {npj Quantum Information}\ }\textbf {\bibinfo {volume} {2}},\ \bibinfo {pages} {15023} (\bibinfo {year} {2016})}\BibitemShut {NoStop}%
\bibitem [{\citenamefont {Preskill}(2018)}]{NISQ}%
  \BibitemOpen
  \bibfield  {author} {\bibinfo {author} {\bibfnamefont {J.}~\bibnamefont {Preskill}},\ }\href {\doibase 10.22331/q-2018-08-06-79} {\bibfield  {journal} {\bibinfo  {journal} {Quantum}\ }\textbf {\bibinfo {volume} {2}},\ \bibinfo {pages} {79} (\bibinfo {year} {2018})}\BibitemShut {NoStop}%
\bibitem [{\citenamefont {Bharti}\ \emph {et~al.}(2022)\citenamefont {Bharti}, \citenamefont {Cervera-Lierta}, \citenamefont {Kyaw}, \citenamefont {Haug}, \citenamefont {Alperin-Lea}, \citenamefont {Anand}, \citenamefont {Degroote}, \citenamefont {Heimonen}, \citenamefont {Kottmann}, \citenamefont {Menke}, \citenamefont {Mok}, \citenamefont {Sim}, \citenamefont {Kwek},\ and\ \citenamefont {Aspuru-Guzik}}]{RevModPhys.94.015004}%
  \BibitemOpen
  \bibfield  {author} {\bibinfo {author} {\bibfnamefont {K.}~\bibnamefont {Bharti}}, \bibinfo {author} {\bibfnamefont {A.}~\bibnamefont {Cervera-Lierta}}, \bibinfo {author} {\bibfnamefont {T.~H.}\ \bibnamefont {Kyaw}}, \bibinfo {author} {\bibfnamefont {T.}~\bibnamefont {Haug}}, \bibinfo {author} {\bibfnamefont {S.}~\bibnamefont {Alperin-Lea}}, \bibinfo {author} {\bibfnamefont {A.}~\bibnamefont {Anand}}, \bibinfo {author} {\bibfnamefont {M.}~\bibnamefont {Degroote}}, \bibinfo {author} {\bibfnamefont {H.}~\bibnamefont {Heimonen}}, \bibinfo {author} {\bibfnamefont {J.~S.}\ \bibnamefont {Kottmann}}, \bibinfo {author} {\bibfnamefont {T.}~\bibnamefont {Menke}}, \bibinfo {author} {\bibfnamefont {W.-K.}\ \bibnamefont {Mok}}, \bibinfo {author} {\bibfnamefont {S.}~\bibnamefont {Sim}}, \bibinfo {author} {\bibfnamefont {L.-C.}\ \bibnamefont {Kwek}}, \ and\ \bibinfo {author} {\bibfnamefont {A.}~\bibnamefont {Aspuru-Guzik}},\ }\href {\doibase 10.1103/RevModPhys.94.015004} {\bibfield  {journal} {\bibinfo  {journal} {Rev.
  Mod. Phys.}\ }\textbf {\bibinfo {volume} {94}},\ \bibinfo {pages} {015004} (\bibinfo {year} {2022})}\BibitemShut {NoStop}%
\bibitem [{\citenamefont {et~al.}(2019)}]{google}%
  \BibitemOpen
  \bibfield  {author} {\bibinfo {author} {\bibfnamefont {F.~A.}\ \bibnamefont {et~al.}},\ }\href {\doibase 10.1038/s41586-019-1666-5} {\bibfield  {journal} {\bibinfo  {journal} {Nature}\ }\textbf {\bibinfo {volume} {574}},\ \bibinfo {pages} {505–510} (\bibinfo {year} {2019})}\BibitemShut {NoStop}%
\bibitem [{\citenamefont {Zhong}\ \emph {et~al.}(2020)\citenamefont {Zhong}, \citenamefont {Wang}, \citenamefont {Deng}, \citenamefont {Chen}, \citenamefont {Peng}, \citenamefont {Luo}, \citenamefont {Qin}, \citenamefont {Wu}, \citenamefont {Ding}, \citenamefont {Hu}, \citenamefont {Hu}, \citenamefont {Yang}, \citenamefont {Zhang}, \citenamefont {Li}, \citenamefont {Li}, \citenamefont {Jiang}, \citenamefont {Gan}, \citenamefont {Yang}, \citenamefont {You}, \citenamefont {Wang}, \citenamefont {Li}, \citenamefont {Liu}, \citenamefont {Lu},\ and\ \citenamefont {Pan}}]{doi:10.1126/science.abe8770}%
  \BibitemOpen
  \bibfield  {author} {\bibinfo {author} {\bibfnamefont {H.-S.}\ \bibnamefont {Zhong}}, \bibinfo {author} {\bibfnamefont {H.}~\bibnamefont {Wang}}, \bibinfo {author} {\bibfnamefont {Y.-H.}\ \bibnamefont {Deng}}, \bibinfo {author} {\bibfnamefont {M.-C.}\ \bibnamefont {Chen}}, \bibinfo {author} {\bibfnamefont {L.-C.}\ \bibnamefont {Peng}}, \bibinfo {author} {\bibfnamefont {Y.-H.}\ \bibnamefont {Luo}}, \bibinfo {author} {\bibfnamefont {J.}~\bibnamefont {Qin}}, \bibinfo {author} {\bibfnamefont {D.}~\bibnamefont {Wu}}, \bibinfo {author} {\bibfnamefont {X.}~\bibnamefont {Ding}}, \bibinfo {author} {\bibfnamefont {Y.}~\bibnamefont {Hu}}, \bibinfo {author} {\bibfnamefont {P.}~\bibnamefont {Hu}}, \bibinfo {author} {\bibfnamefont {X.-Y.}\ \bibnamefont {Yang}}, \bibinfo {author} {\bibfnamefont {W.-J.}\ \bibnamefont {Zhang}}, \bibinfo {author} {\bibfnamefont {H.}~\bibnamefont {Li}}, \bibinfo {author} {\bibfnamefont {Y.}~\bibnamefont {Li}}, \bibinfo {author} {\bibfnamefont {X.}~\bibnamefont {Jiang}}, \bibinfo {author}
  {\bibfnamefont {L.}~\bibnamefont {Gan}}, \bibinfo {author} {\bibfnamefont {G.}~\bibnamefont {Yang}}, \bibinfo {author} {\bibfnamefont {L.}~\bibnamefont {You}}, \bibinfo {author} {\bibfnamefont {Z.}~\bibnamefont {Wang}}, \bibinfo {author} {\bibfnamefont {L.}~\bibnamefont {Li}}, \bibinfo {author} {\bibfnamefont {N.-L.}\ \bibnamefont {Liu}}, \bibinfo {author} {\bibfnamefont {C.-Y.}\ \bibnamefont {Lu}}, \ and\ \bibinfo {author} {\bibfnamefont {J.-W.}\ \bibnamefont {Pan}},\ }\href {\doibase 10.1126/science.abe8770} {\bibfield  {journal} {\bibinfo  {journal} {Science}\ }\textbf {\bibinfo {volume} {370}},\ \bibinfo {pages} {1460} (\bibinfo {year} {2020})}\BibitemShut {NoStop}%
\bibitem [{\citenamefont {Feynman}(1982)}]{feynman}%
  \BibitemOpen
  \bibfield  {author} {\bibinfo {author} {\bibfnamefont {R.~P.}\ \bibnamefont {Feynman}},\ }\href {\doibase 10.1007/BF02650179} {\bibfield  {journal} {\bibinfo  {journal} {International journal of theoretical physics}\ }\textbf {\bibinfo {volume} {21}},\ \bibinfo {pages} {467} (\bibinfo {year} {1982})}\BibitemShut {NoStop}%
\bibitem [{\citenamefont {Cirac}\ and\ \citenamefont {Zoller}(2012)}]{Cirac2012}%
  \BibitemOpen
  \bibfield  {author} {\bibinfo {author} {\bibfnamefont {J.~I.}\ \bibnamefont {Cirac}}\ and\ \bibinfo {author} {\bibfnamefont {P.}~\bibnamefont {Zoller}},\ }\href {\doibase 10.1038/nphys2275} {\bibfield  {journal} {\bibinfo  {journal} {Nature Physics}\ }\textbf {\bibinfo {volume} {8}},\ \bibinfo {pages} {264} (\bibinfo {year} {2012})}\BibitemShut {NoStop}%
\bibitem [{\citenamefont {Altman}\ \emph {et~al.}(2021)\citenamefont {Altman}, \citenamefont {Brown}, \citenamefont {Carleo}, \citenamefont {Carr}, \citenamefont {Demler}, \citenamefont {Chin}, \citenamefont {DeMarco}, \citenamefont {Economou}, \citenamefont {Eriksson}, \citenamefont {Fu}, \citenamefont {Greiner}, \citenamefont {Hazzard}, \citenamefont {Hulet}, \citenamefont {Koll\'ar}, \citenamefont {Lev}, \citenamefont {Lukin}, \citenamefont {Ma}, \citenamefont {Mi}, \citenamefont {Misra}, \citenamefont {Monroe}, \citenamefont {Murch}, \citenamefont {Nazario}, \citenamefont {Ni}, \citenamefont {Potter}, \citenamefont {Roushan}, \citenamefont {Saffman}, \citenamefont {Schleier-Smith}, \citenamefont {Siddiqi}, \citenamefont {Simmonds}, \citenamefont {Singh}, \citenamefont {Spielman}, \citenamefont {Temme}, \citenamefont {Weiss}, \citenamefont {Vu\ifmmode \check{c}\else \v{c}\fi{}kovi\ifmmode~\acute{c}\else \'{c}\fi{}}, \citenamefont {Vuleti\ifmmode~\acute{c}\else \'{c}\fi{}}, \citenamefont {Ye},\ and\
  \citenamefont {Zwierlein}}]{PRXQuantum.2.017003}%
  \BibitemOpen
  \bibfield  {author} {\bibinfo {author} {\bibfnamefont {E.}~\bibnamefont {Altman}}, \bibinfo {author} {\bibfnamefont {K.~R.}\ \bibnamefont {Brown}}, \bibinfo {author} {\bibfnamefont {G.}~\bibnamefont {Carleo}}, \bibinfo {author} {\bibfnamefont {L.~D.}\ \bibnamefont {Carr}}, \bibinfo {author} {\bibfnamefont {E.}~\bibnamefont {Demler}}, \bibinfo {author} {\bibfnamefont {C.}~\bibnamefont {Chin}}, \bibinfo {author} {\bibfnamefont {B.}~\bibnamefont {DeMarco}}, \bibinfo {author} {\bibfnamefont {S.~E.}\ \bibnamefont {Economou}}, \bibinfo {author} {\bibfnamefont {M.~A.}\ \bibnamefont {Eriksson}}, \bibinfo {author} {\bibfnamefont {K.-M.~C.}\ \bibnamefont {Fu}}, \bibinfo {author} {\bibfnamefont {M.}~\bibnamefont {Greiner}}, \bibinfo {author} {\bibfnamefont {K.~R.}\ \bibnamefont {Hazzard}}, \bibinfo {author} {\bibfnamefont {R.~G.}\ \bibnamefont {Hulet}}, \bibinfo {author} {\bibfnamefont {A.~J.}\ \bibnamefont {Koll\'ar}}, \bibinfo {author} {\bibfnamefont {B.~L.}\ \bibnamefont {Lev}}, \bibinfo {author} {\bibfnamefont
  {M.~D.}\ \bibnamefont {Lukin}}, \bibinfo {author} {\bibfnamefont {R.}~\bibnamefont {Ma}}, \bibinfo {author} {\bibfnamefont {X.}~\bibnamefont {Mi}}, \bibinfo {author} {\bibfnamefont {S.}~\bibnamefont {Misra}}, \bibinfo {author} {\bibfnamefont {C.}~\bibnamefont {Monroe}}, \bibinfo {author} {\bibfnamefont {K.}~\bibnamefont {Murch}}, \bibinfo {author} {\bibfnamefont {Z.}~\bibnamefont {Nazario}}, \bibinfo {author} {\bibfnamefont {K.-K.}\ \bibnamefont {Ni}}, \bibinfo {author} {\bibfnamefont {A.~C.}\ \bibnamefont {Potter}}, \bibinfo {author} {\bibfnamefont {P.}~\bibnamefont {Roushan}}, \bibinfo {author} {\bibfnamefont {M.}~\bibnamefont {Saffman}}, \bibinfo {author} {\bibfnamefont {M.}~\bibnamefont {Schleier-Smith}}, \bibinfo {author} {\bibfnamefont {I.}~\bibnamefont {Siddiqi}}, \bibinfo {author} {\bibfnamefont {R.}~\bibnamefont {Simmonds}}, \bibinfo {author} {\bibfnamefont {M.}~\bibnamefont {Singh}}, \bibinfo {author} {\bibfnamefont {I.}~\bibnamefont {Spielman}}, \bibinfo {author} {\bibfnamefont {K.}~\bibnamefont
  {Temme}}, \bibinfo {author} {\bibfnamefont {D.~S.}\ \bibnamefont {Weiss}}, \bibinfo {author} {\bibfnamefont {J.}~\bibnamefont {Vu\ifmmode \check{c}\else \v{c}\fi{}kovi\ifmmode~\acute{c}\else \'{c}\fi{}}}, \bibinfo {author} {\bibfnamefont {V.}~\bibnamefont {Vuleti\ifmmode~\acute{c}\else \'{c}\fi{}}}, \bibinfo {author} {\bibfnamefont {J.}~\bibnamefont {Ye}}, \ and\ \bibinfo {author} {\bibfnamefont {M.}~\bibnamefont {Zwierlein}},\ }\href {\doibase 10.1103/PRXQuantum.2.017003} {\bibfield  {journal} {\bibinfo  {journal} {PRX Quantum}\ }\textbf {\bibinfo {volume} {2}},\ \bibinfo {pages} {017003} (\bibinfo {year} {2021})}\BibitemShut {NoStop}%
\bibitem [{\citenamefont {Daley}\ \emph {et~al.}(2022)\citenamefont {Daley}, \citenamefont {Bloch}, \citenamefont {Kokail}, \citenamefont {Flannigan}, \citenamefont {Pearson}, \citenamefont {Troyer},\ and\ \citenamefont {Zoller}}]{potential3}%
  \BibitemOpen
  \bibfield  {author} {\bibinfo {author} {\bibfnamefont {A.}~\bibnamefont {Daley}}, \bibinfo {author} {\bibfnamefont {I.}~\bibnamefont {Bloch}}, \bibinfo {author} {\bibfnamefont {C.}~\bibnamefont {Kokail}}, \bibinfo {author} {\bibfnamefont {S.}~\bibnamefont {Flannigan}}, \bibinfo {author} {\bibfnamefont {N.}~\bibnamefont {Pearson}}, \bibinfo {author} {\bibfnamefont {M.}~\bibnamefont {Troyer}}, \ and\ \bibinfo {author} {\bibfnamefont {P.}~\bibnamefont {Zoller}},\ }\href {\doibase 10.1038/s41586-022-04940-6} {\bibfield  {journal} {\bibinfo  {journal} {Nature}\ }\textbf {\bibinfo {volume} {607}},\ \bibinfo {pages} {667} (\bibinfo {year} {2022})}\BibitemShut {NoStop}%
\bibitem [{\citenamefont {McArdle}\ \emph {et~al.}(2020)\citenamefont {McArdle}, \citenamefont {Endo}, \citenamefont {Aspuru-Guzik}, \citenamefont {Benjamin},\ and\ \citenamefont {Yuan}}]{RevModPhys.92.015003}%
  \BibitemOpen
  \bibfield  {author} {\bibinfo {author} {\bibfnamefont {S.}~\bibnamefont {McArdle}}, \bibinfo {author} {\bibfnamefont {S.}~\bibnamefont {Endo}}, \bibinfo {author} {\bibfnamefont {A.}~\bibnamefont {Aspuru-Guzik}}, \bibinfo {author} {\bibfnamefont {S.~C.}\ \bibnamefont {Benjamin}}, \ and\ \bibinfo {author} {\bibfnamefont {X.}~\bibnamefont {Yuan}},\ }\href {\doibase 10.1103/RevModPhys.92.015003} {\bibfield  {journal} {\bibinfo  {journal} {Rev. Mod. Phys.}\ }\textbf {\bibinfo {volume} {92}},\ \bibinfo {pages} {015003} (\bibinfo {year} {2020})}\BibitemShut {NoStop}%
\bibitem [{\citenamefont {Bloch}\ \emph {et~al.}(2012)\citenamefont {Bloch}, \citenamefont {Dalibard},\ and\ \citenamefont {Nascimb{\`e}ne}}]{Bloch2012}%
  \BibitemOpen
  \bibfield  {author} {\bibinfo {author} {\bibfnamefont {I.}~\bibnamefont {Bloch}}, \bibinfo {author} {\bibfnamefont {J.}~\bibnamefont {Dalibard}}, \ and\ \bibinfo {author} {\bibfnamefont {S.}~\bibnamefont {Nascimb{\`e}ne}},\ }\href {\doibase 10.1038/nphys2259} {\bibfield  {journal} {\bibinfo  {journal} {Nature Physics}\ }\textbf {\bibinfo {volume} {8}},\ \bibinfo {pages} {267} (\bibinfo {year} {2012})}\BibitemShut {NoStop}%
\bibitem [{\citenamefont {Blatt}\ and\ \citenamefont {Roos}(2012)}]{Blatt2012}%
  \BibitemOpen
  \bibfield  {author} {\bibinfo {author} {\bibfnamefont {R.}~\bibnamefont {Blatt}}\ and\ \bibinfo {author} {\bibfnamefont {C.~F.}\ \bibnamefont {Roos}},\ }\href {\doibase 10.1038/nphys2252} {\bibfield  {journal} {\bibinfo  {journal} {Nature Physics}\ }\textbf {\bibinfo {volume} {8}},\ \bibinfo {pages} {277} (\bibinfo {year} {2012})}\BibitemShut {NoStop}%
\bibitem [{\citenamefont {Monroe}\ \emph {et~al.}(2021)\citenamefont {Monroe}, \citenamefont {Campbell}, \citenamefont {Duan}, \citenamefont {Gong}, \citenamefont {Gorshkov}, \citenamefont {Hess}, \citenamefont {Islam}, \citenamefont {Kim}, \citenamefont {Linke}, \citenamefont {Pagano}, \citenamefont {Richerme}, \citenamefont {Senko},\ and\ \citenamefont {Yao}}]{RevModPhys.93.025001}%
  \BibitemOpen
  \bibfield  {author} {\bibinfo {author} {\bibfnamefont {C.}~\bibnamefont {Monroe}}, \bibinfo {author} {\bibfnamefont {W.~C.}\ \bibnamefont {Campbell}}, \bibinfo {author} {\bibfnamefont {L.-M.}\ \bibnamefont {Duan}}, \bibinfo {author} {\bibfnamefont {Z.-X.}\ \bibnamefont {Gong}}, \bibinfo {author} {\bibfnamefont {A.~V.}\ \bibnamefont {Gorshkov}}, \bibinfo {author} {\bibfnamefont {P.~W.}\ \bibnamefont {Hess}}, \bibinfo {author} {\bibfnamefont {R.}~\bibnamefont {Islam}}, \bibinfo {author} {\bibfnamefont {K.}~\bibnamefont {Kim}}, \bibinfo {author} {\bibfnamefont {N.~M.}\ \bibnamefont {Linke}}, \bibinfo {author} {\bibfnamefont {G.}~\bibnamefont {Pagano}}, \bibinfo {author} {\bibfnamefont {P.}~\bibnamefont {Richerme}}, \bibinfo {author} {\bibfnamefont {C.}~\bibnamefont {Senko}}, \ and\ \bibinfo {author} {\bibfnamefont {N.~Y.}\ \bibnamefont {Yao}},\ }\href {\doibase 10.1103/RevModPhys.93.025001} {\bibfield  {journal} {\bibinfo  {journal} {Rev. Mod. Phys.}\ }\textbf {\bibinfo {volume} {93}},\ \bibinfo {pages}
  {025001} (\bibinfo {year} {2021})}\BibitemShut {NoStop}%
\bibitem [{\citenamefont {Ba{\~{n}}uls}\ \emph {et~al.}(2020)\citenamefont {Ba{\~{n}}uls}, \citenamefont {Blatt}, \citenamefont {Catani}, \citenamefont {Celi}, \citenamefont {Cirac}, \citenamefont {Dalmonte}, \citenamefont {Fallani}, \citenamefont {Jansen}, \citenamefont {Lewenstein}, \citenamefont {Montangero}, \citenamefont {Muschik}, \citenamefont {Reznik}, \citenamefont {Rico}, \citenamefont {Tagliacozzo}, \citenamefont {Van~Acoleyen}, \citenamefont {Verstraete}, \citenamefont {Wiese}, \citenamefont {Wingate}, \citenamefont {Zakrzewski},\ and\ \citenamefont {Zoller}}]{Banuls2020}%
  \BibitemOpen
  \bibfield  {author} {\bibinfo {author} {\bibfnamefont {M.~C.}\ \bibnamefont {Ba{\~{n}}uls}}, \bibinfo {author} {\bibfnamefont {R.}~\bibnamefont {Blatt}}, \bibinfo {author} {\bibfnamefont {J.}~\bibnamefont {Catani}}, \bibinfo {author} {\bibfnamefont {A.}~\bibnamefont {Celi}}, \bibinfo {author} {\bibfnamefont {J.~I.}\ \bibnamefont {Cirac}}, \bibinfo {author} {\bibfnamefont {M.}~\bibnamefont {Dalmonte}}, \bibinfo {author} {\bibfnamefont {L.}~\bibnamefont {Fallani}}, \bibinfo {author} {\bibfnamefont {K.}~\bibnamefont {Jansen}}, \bibinfo {author} {\bibfnamefont {M.}~\bibnamefont {Lewenstein}}, \bibinfo {author} {\bibfnamefont {S.}~\bibnamefont {Montangero}}, \bibinfo {author} {\bibfnamefont {C.~A.}\ \bibnamefont {Muschik}}, \bibinfo {author} {\bibfnamefont {B.}~\bibnamefont {Reznik}}, \bibinfo {author} {\bibfnamefont {E.}~\bibnamefont {Rico}}, \bibinfo {author} {\bibfnamefont {L.}~\bibnamefont {Tagliacozzo}}, \bibinfo {author} {\bibfnamefont {K.}~\bibnamefont {Van~Acoleyen}}, \bibinfo {author} {\bibfnamefont
  {F.}~\bibnamefont {Verstraete}}, \bibinfo {author} {\bibfnamefont {U.-J.}\ \bibnamefont {Wiese}}, \bibinfo {author} {\bibfnamefont {M.}~\bibnamefont {Wingate}}, \bibinfo {author} {\bibfnamefont {J.}~\bibnamefont {Zakrzewski}}, \ and\ \bibinfo {author} {\bibfnamefont {P.}~\bibnamefont {Zoller}},\ }\href {\doibase 10.1140/epjd/e2020-100571-8} {\bibfield  {journal} {\bibinfo  {journal} {The European Physical Journal D}\ }\textbf {\bibinfo {volume} {74}},\ \bibinfo {pages} {165} (\bibinfo {year} {2020})}\BibitemShut {NoStop}%
\bibitem [{\citenamefont {Suzuki}(1976)}]{cmp/1103900351}%
  \BibitemOpen
  \bibfield  {author} {\bibinfo {author} {\bibfnamefont {M.}~\bibnamefont {Suzuki}},\ }\href {\doibase 10.1007/BF01609348} {\bibfield  {journal} {\bibinfo  {journal} {Communications in Mathematical Physics}\ }\textbf {\bibinfo {volume} {51}},\ \bibinfo {pages} {183 } (\bibinfo {year} {1976})}\BibitemShut {NoStop}%
\bibitem [{\citenamefont {Lloyd}(1996)}]{doi:10.1126/science.273.5278.1073}%
  \BibitemOpen
  \bibfield  {author} {\bibinfo {author} {\bibfnamefont {S.}~\bibnamefont {Lloyd}},\ }\href {\doibase 10.1126/science.273.5278.1073} {\bibfield  {journal} {\bibinfo  {journal} {Science}\ }\textbf {\bibinfo {volume} {273}},\ \bibinfo {pages} {1073} (\bibinfo {year} {1996})}\BibitemShut {NoStop}%
\bibitem [{\citenamefont {Somma}\ \emph {et~al.}(2003)\citenamefont {Somma}, \citenamefont {Ortiz}, \citenamefont {Knill},\ and\ \citenamefont {Gubernatis}}]{Somma_2003}%
  \BibitemOpen
  \bibfield  {author} {\bibinfo {author} {\bibfnamefont {R.~D.}\ \bibnamefont {Somma}}, \bibinfo {author} {\bibfnamefont {G.}~\bibnamefont {Ortiz}}, \bibinfo {author} {\bibfnamefont {E.~H.}\ \bibnamefont {Knill}}, \ and\ \bibinfo {author} {\bibfnamefont {J.}~\bibnamefont {Gubernatis}},\ }in\ \href {\doibase 10.1117/12.487249} {\emph {\bibinfo {booktitle} {{SPIE} Proceedings}}},\ \bibinfo {editor} {edited by\ \bibinfo {editor} {\bibfnamefont {E.}~\bibnamefont {Donkor}}, \bibinfo {editor} {\bibfnamefont {A.~R.}\ \bibnamefont {Pirich}}, \ and\ \bibinfo {editor} {\bibfnamefont {H.~E.}\ \bibnamefont {Brandt}}}\ (\bibinfo  {publisher} {{SPIE}},\ \bibinfo {year} {2003})\BibitemShut {NoStop}%
\bibitem [{\citenamefont {Jaksch}\ \emph {et~al.}(1998)\citenamefont {Jaksch}, \citenamefont {Bruder}, \citenamefont {Cirac}, \citenamefont {Gardiner},\ and\ \citenamefont {Zoller}}]{PhysRevLett.81.3108}%
  \BibitemOpen
  \bibfield  {author} {\bibinfo {author} {\bibfnamefont {D.}~\bibnamefont {Jaksch}}, \bibinfo {author} {\bibfnamefont {C.}~\bibnamefont {Bruder}}, \bibinfo {author} {\bibfnamefont {J.~I.}\ \bibnamefont {Cirac}}, \bibinfo {author} {\bibfnamefont {C.~W.}\ \bibnamefont {Gardiner}}, \ and\ \bibinfo {author} {\bibfnamefont {P.}~\bibnamefont {Zoller}},\ }\href {\doibase 10.1103/PhysRevLett.81.3108} {\bibfield  {journal} {\bibinfo  {journal} {Phys. Rev. Lett.}\ }\textbf {\bibinfo {volume} {81}},\ \bibinfo {pages} {3108} (\bibinfo {year} {1998})}\BibitemShut {NoStop}%
\bibitem [{\citenamefont {Raizen}\ \emph {et~al.}(1992)\citenamefont {Raizen}, \citenamefont {Gilligan}, \citenamefont {Bergquist}, \citenamefont {Itano},\ and\ \citenamefont {Wineland}}]{PhysRevA.45.6493}%
  \BibitemOpen
  \bibfield  {author} {\bibinfo {author} {\bibfnamefont {M.~G.}\ \bibnamefont {Raizen}}, \bibinfo {author} {\bibfnamefont {J.~M.}\ \bibnamefont {Gilligan}}, \bibinfo {author} {\bibfnamefont {J.~C.}\ \bibnamefont {Bergquist}}, \bibinfo {author} {\bibfnamefont {W.~M.}\ \bibnamefont {Itano}}, \ and\ \bibinfo {author} {\bibfnamefont {D.~J.}\ \bibnamefont {Wineland}},\ }\href {\doibase 10.1103/PhysRevA.45.6493} {\bibfield  {journal} {\bibinfo  {journal} {Phys. Rev. A}\ }\textbf {\bibinfo {volume} {45}},\ \bibinfo {pages} {6493} (\bibinfo {year} {1992})}\BibitemShut {NoStop}%
\bibitem [{\citenamefont {Lanyon}\ \emph {et~al.}(2011)\citenamefont {Lanyon}, \citenamefont {Hempel}, \citenamefont {Nigg}, \citenamefont {Müller}, \citenamefont {Gerritsma}, \citenamefont {Zähringer}, \citenamefont {Schindler}, \citenamefont {Barreiro}, \citenamefont {Rambach}, \citenamefont {Kirchmair}, \citenamefont {Hennrich}, \citenamefont {Zoller}, \citenamefont {Blatt},\ and\ \citenamefont {Roos}}]{doi:10.1126/science.1208001}%
  \BibitemOpen
  \bibfield  {author} {\bibinfo {author} {\bibfnamefont {B.~P.}\ \bibnamefont {Lanyon}}, \bibinfo {author} {\bibfnamefont {C.}~\bibnamefont {Hempel}}, \bibinfo {author} {\bibfnamefont {D.}~\bibnamefont {Nigg}}, \bibinfo {author} {\bibfnamefont {M.}~\bibnamefont {Müller}}, \bibinfo {author} {\bibfnamefont {R.}~\bibnamefont {Gerritsma}}, \bibinfo {author} {\bibfnamefont {F.}~\bibnamefont {Zähringer}}, \bibinfo {author} {\bibfnamefont {P.}~\bibnamefont {Schindler}}, \bibinfo {author} {\bibfnamefont {J.~T.}\ \bibnamefont {Barreiro}}, \bibinfo {author} {\bibfnamefont {M.}~\bibnamefont {Rambach}}, \bibinfo {author} {\bibfnamefont {G.}~\bibnamefont {Kirchmair}}, \bibinfo {author} {\bibfnamefont {M.}~\bibnamefont {Hennrich}}, \bibinfo {author} {\bibfnamefont {P.}~\bibnamefont {Zoller}}, \bibinfo {author} {\bibfnamefont {R.}~\bibnamefont {Blatt}}, \ and\ \bibinfo {author} {\bibfnamefont {C.~F.}\ \bibnamefont {Roos}},\ }\href {\doibase 10.1126/science.1208001} {\bibfield  {journal} {\bibinfo  {journal} {Science}\
  }\textbf {\bibinfo {volume} {334}},\ \bibinfo {pages} {57} (\bibinfo {year} {2011})}\BibitemShut {NoStop}%
\bibitem [{\citenamefont {Katz}\ \emph {et~al.}(2022)\citenamefont {Katz}, \citenamefont {Feng}, \citenamefont {Risinger}, \citenamefont {Monroe},\ and\ \citenamefont {Cetina}}]{https://doi.org/10.48550/arxiv.2209.05691}%
  \BibitemOpen
  \bibfield  {author} {\bibinfo {author} {\bibfnamefont {O.}~\bibnamefont {Katz}}, \bibinfo {author} {\bibfnamefont {L.}~\bibnamefont {Feng}}, \bibinfo {author} {\bibfnamefont {A.}~\bibnamefont {Risinger}}, \bibinfo {author} {\bibfnamefont {C.}~\bibnamefont {Monroe}}, \ and\ \bibinfo {author} {\bibfnamefont {M.}~\bibnamefont {Cetina}},\ }\href {\doibase 10.1038/s41567-023-02102-7} {\  (\bibinfo {year} {2022}),\ 10.1038/s41567-023-02102-7}\BibitemShut {NoStop}%
\bibitem [{\citenamefont {Barreiro}\ \emph {et~al.}(2011)\citenamefont {Barreiro}, \citenamefont {M{\"u}ller}, \citenamefont {Schindler}, \citenamefont {Nigg}, \citenamefont {Monz}, \citenamefont {Chwalla}, \citenamefont {Hennrich}, \citenamefont {Roos}, \citenamefont {Zoller},\ and\ \citenamefont {Blatt}}]{Barreiro2011}%
  \BibitemOpen
  \bibfield  {author} {\bibinfo {author} {\bibfnamefont {J.~T.}\ \bibnamefont {Barreiro}}, \bibinfo {author} {\bibfnamefont {M.}~\bibnamefont {M{\"u}ller}}, \bibinfo {author} {\bibfnamefont {P.}~\bibnamefont {Schindler}}, \bibinfo {author} {\bibfnamefont {D.}~\bibnamefont {Nigg}}, \bibinfo {author} {\bibfnamefont {T.}~\bibnamefont {Monz}}, \bibinfo {author} {\bibfnamefont {M.}~\bibnamefont {Chwalla}}, \bibinfo {author} {\bibfnamefont {M.}~\bibnamefont {Hennrich}}, \bibinfo {author} {\bibfnamefont {C.~F.}\ \bibnamefont {Roos}}, \bibinfo {author} {\bibfnamefont {P.}~\bibnamefont {Zoller}}, \ and\ \bibinfo {author} {\bibfnamefont {R.}~\bibnamefont {Blatt}},\ }\href {\doibase 10.1038/nature09801} {\bibfield  {journal} {\bibinfo  {journal} {Nature}\ }\textbf {\bibinfo {volume} {470}},\ \bibinfo {pages} {486} (\bibinfo {year} {2011})}\BibitemShut {NoStop}%
\bibitem [{\citenamefont {Schindler}\ \emph {et~al.}(2013)\citenamefont {Schindler}, \citenamefont {M{\"u}ller}, \citenamefont {Nigg}, \citenamefont {Barreiro}, \citenamefont {Martinez}, \citenamefont {Hennrich}, \citenamefont {Monz}, \citenamefont {Diehl}, \citenamefont {Zoller},\ and\ \citenamefont {Blatt}}]{Schindler2013}%
  \BibitemOpen
  \bibfield  {author} {\bibinfo {author} {\bibfnamefont {P.}~\bibnamefont {Schindler}}, \bibinfo {author} {\bibfnamefont {M.}~\bibnamefont {M{\"u}ller}}, \bibinfo {author} {\bibfnamefont {D.}~\bibnamefont {Nigg}}, \bibinfo {author} {\bibfnamefont {J.~T.}\ \bibnamefont {Barreiro}}, \bibinfo {author} {\bibfnamefont {E.~A.}\ \bibnamefont {Martinez}}, \bibinfo {author} {\bibfnamefont {M.}~\bibnamefont {Hennrich}}, \bibinfo {author} {\bibfnamefont {T.}~\bibnamefont {Monz}}, \bibinfo {author} {\bibfnamefont {S.}~\bibnamefont {Diehl}}, \bibinfo {author} {\bibfnamefont {P.}~\bibnamefont {Zoller}}, \ and\ \bibinfo {author} {\bibfnamefont {R.}~\bibnamefont {Blatt}},\ }\href {\doibase 10.1038/nphys2630} {\bibfield  {journal} {\bibinfo  {journal} {Nature Physics}\ }\textbf {\bibinfo {volume} {9}},\ \bibinfo {pages} {361} (\bibinfo {year} {2013})}\BibitemShut {NoStop}%
\bibitem [{\citenamefont {Wilson}(1974)}]{PhysRevD.10.2445}%
  \BibitemOpen
  \bibfield  {author} {\bibinfo {author} {\bibfnamefont {K.~G.}\ \bibnamefont {Wilson}},\ }\href {\doibase 10.1103/PhysRevD.10.2445} {\bibfield  {journal} {\bibinfo  {journal} {Phys. Rev. D}\ }\textbf {\bibinfo {volume} {10}},\ \bibinfo {pages} {2445} (\bibinfo {year} {1974})}\BibitemShut {NoStop}%
\bibitem [{\citenamefont {Kogut}\ and\ \citenamefont {Susskind}(1975)}]{PhysRevD.11.395}%
  \BibitemOpen
  \bibfield  {author} {\bibinfo {author} {\bibfnamefont {J.}~\bibnamefont {Kogut}}\ and\ \bibinfo {author} {\bibfnamefont {L.}~\bibnamefont {Susskind}},\ }\href {\doibase 10.1103/PhysRevD.11.395} {\bibfield  {journal} {\bibinfo  {journal} {Phys. Rev. D}\ }\textbf {\bibinfo {volume} {11}},\ \bibinfo {pages} {395} (\bibinfo {year} {1975})}\BibitemShut {NoStop}%
\bibitem [{\citenamefont {Gattringer}\ and\ \citenamefont {Lang}(2010)}]{Gattringer:2010zz}%
  \BibitemOpen
  \bibfield  {author} {\bibinfo {author} {\bibfnamefont {C.}~\bibnamefont {Gattringer}}\ and\ \bibinfo {author} {\bibfnamefont {C.~B.}\ \bibnamefont {Lang}},\ }\href {\doibase 10.1007/978-3-642-01850-3} {\emph {\bibinfo {title} {{Quantum chromodynamics on the lattice}}}},\ Vol.\ \bibinfo {volume} {788}\ (\bibinfo  {publisher} {Springer},\ \bibinfo {address} {Berlin},\ \bibinfo {year} {2010})\BibitemShut {NoStop}%
\bibitem [{\citenamefont {Martinez}\ \emph {et~al.}(2016)\citenamefont {Martinez}, \citenamefont {Muschik}, \citenamefont {Schindler}, \citenamefont {Nigg}, \citenamefont {Erhard}, \citenamefont {Heyl}, \citenamefont {Hauke}, \citenamefont {Dalmonte}, \citenamefont {Monz}, \citenamefont {Zoller},\ and\ \citenamefont {Blatt}}]{Martinez2016}%
  \BibitemOpen
  \bibfield  {author} {\bibinfo {author} {\bibfnamefont {E.~A.}\ \bibnamefont {Martinez}}, \bibinfo {author} {\bibfnamefont {C.~A.}\ \bibnamefont {Muschik}}, \bibinfo {author} {\bibfnamefont {P.}~\bibnamefont {Schindler}}, \bibinfo {author} {\bibfnamefont {D.}~\bibnamefont {Nigg}}, \bibinfo {author} {\bibfnamefont {A.}~\bibnamefont {Erhard}}, \bibinfo {author} {\bibfnamefont {M.}~\bibnamefont {Heyl}}, \bibinfo {author} {\bibfnamefont {P.}~\bibnamefont {Hauke}}, \bibinfo {author} {\bibfnamefont {M.}~\bibnamefont {Dalmonte}}, \bibinfo {author} {\bibfnamefont {T.}~\bibnamefont {Monz}}, \bibinfo {author} {\bibfnamefont {P.}~\bibnamefont {Zoller}}, \ and\ \bibinfo {author} {\bibfnamefont {R.}~\bibnamefont {Blatt}},\ }\href {\doibase 10.1038/nature18318} {\bibfield  {journal} {\bibinfo  {journal} {Nature}\ }\textbf {\bibinfo {volume} {534}},\ \bibinfo {pages} {516} (\bibinfo {year} {2016})}\BibitemShut {NoStop}%
\bibitem [{\citenamefont {Nguyen}\ \emph {et~al.}(2022)\citenamefont {Nguyen}, \citenamefont {Tran}, \citenamefont {Zhu}, \citenamefont {Green}, \citenamefont {Alderete}, \citenamefont {Davoudi},\ and\ \citenamefont {Linke}}]{PRXQuantum.3.020324}%
  \BibitemOpen
  \bibfield  {author} {\bibinfo {author} {\bibfnamefont {N.~H.}\ \bibnamefont {Nguyen}}, \bibinfo {author} {\bibfnamefont {M.~C.}\ \bibnamefont {Tran}}, \bibinfo {author} {\bibfnamefont {Y.}~\bibnamefont {Zhu}}, \bibinfo {author} {\bibfnamefont {A.~M.}\ \bibnamefont {Green}}, \bibinfo {author} {\bibfnamefont {C.~H.}\ \bibnamefont {Alderete}}, \bibinfo {author} {\bibfnamefont {Z.}~\bibnamefont {Davoudi}}, \ and\ \bibinfo {author} {\bibfnamefont {N.~M.}\ \bibnamefont {Linke}},\ }\href {\doibase 10.1103/PRXQuantum.3.020324} {\bibfield  {journal} {\bibinfo  {journal} {PRX Quantum}\ }\textbf {\bibinfo {volume} {3}},\ \bibinfo {pages} {020324} (\bibinfo {year} {2022})}\BibitemShut {NoStop}%
\bibitem [{\citenamefont {S\o{}rensen}\ and\ \citenamefont {M\o{}lmer}(1999)}]{PhysRevLett.82.1971}%
  \BibitemOpen
  \bibfield  {author} {\bibinfo {author} {\bibfnamefont {A.}~\bibnamefont {S\o{}rensen}}\ and\ \bibinfo {author} {\bibfnamefont {K.}~\bibnamefont {M\o{}lmer}},\ }\href {\doibase 10.1103/PhysRevLett.82.1971} {\bibfield  {journal} {\bibinfo  {journal} {Phys. Rev. Lett.}\ }\textbf {\bibinfo {volume} {82}},\ \bibinfo {pages} {1971} (\bibinfo {year} {1999})}\BibitemShut {NoStop}%
\bibitem [{\citenamefont {S\o{}rensen}\ and\ \citenamefont {M\o{}lmer}(2000)}]{PhysRevA.62.022311}%
  \BibitemOpen
  \bibfield  {author} {\bibinfo {author} {\bibfnamefont {A.}~\bibnamefont {S\o{}rensen}}\ and\ \bibinfo {author} {\bibfnamefont {K.}~\bibnamefont {M\o{}lmer}},\ }\href {\doibase 10.1103/PhysRevA.62.022311} {\bibfield  {journal} {\bibinfo  {journal} {Phys. Rev. A}\ }\textbf {\bibinfo {volume} {62}},\ \bibinfo {pages} {022311} (\bibinfo {year} {2000})}\BibitemShut {NoStop}%
\bibitem [{\citenamefont {Porras}\ and\ \citenamefont {Cirac}(2004)}]{PhysRevLett.92.207901}%
  \BibitemOpen
  \bibfield  {author} {\bibinfo {author} {\bibfnamefont {D.}~\bibnamefont {Porras}}\ and\ \bibinfo {author} {\bibfnamefont {J.~I.}\ \bibnamefont {Cirac}},\ }\href {\doibase 10.1103/PhysRevLett.92.207901} {\bibfield  {journal} {\bibinfo  {journal} {Phys. Rev. Lett.}\ }\textbf {\bibinfo {volume} {92}},\ \bibinfo {pages} {207901} (\bibinfo {year} {2004})}\BibitemShut {NoStop}%
\bibitem [{\citenamefont {Friedenauer}\ \emph {et~al.}()\citenamefont {Friedenauer}, \citenamefont {Schmitz}, \citenamefont {Glueckert}, \citenamefont {Porras},\ and\ \citenamefont {Schaetz}}]{Friedenauer2008}%
  \BibitemOpen
  \bibfield  {author} {\bibinfo {author} {\bibfnamefont {A.}~\bibnamefont {Friedenauer}}, \bibinfo {author} {\bibfnamefont {H.}~\bibnamefont {Schmitz}}, \bibinfo {author} {\bibfnamefont {J.~T.}\ \bibnamefont {Glueckert}}, \bibinfo {author} {\bibfnamefont {D.}~\bibnamefont {Porras}}, \ and\ \bibinfo {author} {\bibfnamefont {T.}~\bibnamefont {Schaetz}},\ }\href {\doibase 10.1038/nphys1032} {\bibfield  {journal} {\bibinfo  {journal} {Nature Physics}\ }10.1038/nphys1032}\BibitemShut {NoStop}%
\bibitem [{\citenamefont {Kim}\ \emph {et~al.}(2010)\citenamefont {Kim}, \citenamefont {Chang}, \citenamefont {Korenblit}, \citenamefont {Islam}, \citenamefont {Edwards}, \citenamefont {Freericks}, \citenamefont {Lin}, \citenamefont {Duan},\ and\ \citenamefont {Monroe}}]{Kim2010}%
  \BibitemOpen
  \bibfield  {author} {\bibinfo {author} {\bibfnamefont {K.}~\bibnamefont {Kim}}, \bibinfo {author} {\bibfnamefont {M.-S.}\ \bibnamefont {Chang}}, \bibinfo {author} {\bibfnamefont {S.}~\bibnamefont {Korenblit}}, \bibinfo {author} {\bibfnamefont {R.}~\bibnamefont {Islam}}, \bibinfo {author} {\bibfnamefont {E.~E.}\ \bibnamefont {Edwards}}, \bibinfo {author} {\bibfnamefont {J.~K.}\ \bibnamefont {Freericks}}, \bibinfo {author} {\bibfnamefont {G.-D.}\ \bibnamefont {Lin}}, \bibinfo {author} {\bibfnamefont {L.-M.}\ \bibnamefont {Duan}}, \ and\ \bibinfo {author} {\bibfnamefont {C.}~\bibnamefont {Monroe}},\ }\href {\doibase 10.1038/nature09071} {\bibfield  {journal} {\bibinfo  {journal} {Nature}\ }\textbf {\bibinfo {volume} {465}},\ \bibinfo {pages} {590} (\bibinfo {year} {2010})}\BibitemShut {NoStop}%
\bibitem [{\citenamefont {Britton}\ \emph {et~al.}(2012)\citenamefont {Britton}, \citenamefont {Sawyer}, \citenamefont {Keith}, \citenamefont {Wang}, \citenamefont {Freericks}, \citenamefont {Uys}, \citenamefont {Biercuk},\ and\ \citenamefont {Bollinger}}]{Britton2012}%
  \BibitemOpen
  \bibfield  {author} {\bibinfo {author} {\bibfnamefont {J.~W.}\ \bibnamefont {Britton}}, \bibinfo {author} {\bibfnamefont {B.~C.}\ \bibnamefont {Sawyer}}, \bibinfo {author} {\bibfnamefont {A.~C.}\ \bibnamefont {Keith}}, \bibinfo {author} {\bibfnamefont {C.-C.~J.}\ \bibnamefont {Wang}}, \bibinfo {author} {\bibfnamefont {J.~K.}\ \bibnamefont {Freericks}}, \bibinfo {author} {\bibfnamefont {H.}~\bibnamefont {Uys}}, \bibinfo {author} {\bibfnamefont {M.~J.}\ \bibnamefont {Biercuk}}, \ and\ \bibinfo {author} {\bibfnamefont {J.~J.}\ \bibnamefont {Bollinger}},\ }\href {\doibase 10.1038/nature10981} {\bibfield  {journal} {\bibinfo  {journal} {Nature}\ }\textbf {\bibinfo {volume} {484}},\ \bibinfo {pages} {489} (\bibinfo {year} {2012})}\BibitemShut {NoStop}%
\bibitem [{\citenamefont {Islam}\ \emph {et~al.}(2013)\citenamefont {Islam}, \citenamefont {Senko}, \citenamefont {Campbell}, \citenamefont {Korenblit}, \citenamefont {Smith}, \citenamefont {Lee}, \citenamefont {Edwards}, \citenamefont {Wang}, \citenamefont {Freericks},\ and\ \citenamefont {Monroe}}]{doi:10.1126/science.1232296}%
  \BibitemOpen
  \bibfield  {author} {\bibinfo {author} {\bibfnamefont {R.}~\bibnamefont {Islam}}, \bibinfo {author} {\bibfnamefont {C.}~\bibnamefont {Senko}}, \bibinfo {author} {\bibfnamefont {W.~C.}\ \bibnamefont {Campbell}}, \bibinfo {author} {\bibfnamefont {S.}~\bibnamefont {Korenblit}}, \bibinfo {author} {\bibfnamefont {J.}~\bibnamefont {Smith}}, \bibinfo {author} {\bibfnamefont {A.}~\bibnamefont {Lee}}, \bibinfo {author} {\bibfnamefont {E.~E.}\ \bibnamefont {Edwards}}, \bibinfo {author} {\bibfnamefont {C.-C.~J.}\ \bibnamefont {Wang}}, \bibinfo {author} {\bibfnamefont {J.~K.}\ \bibnamefont {Freericks}}, \ and\ \bibinfo {author} {\bibfnamefont {C.}~\bibnamefont {Monroe}},\ }\href {\doibase 10.1126/science.1232296} {\bibfield  {journal} {\bibinfo  {journal} {Science}\ }\textbf {\bibinfo {volume} {340}},\ \bibinfo {pages} {583} (\bibinfo {year} {2013})}\BibitemShut {NoStop}%
\bibitem [{\citenamefont {Senko}\ \emph {et~al.}(2014)\citenamefont {Senko}, \citenamefont {Smith}, \citenamefont {Richerme}, \citenamefont {Lee}, \citenamefont {Campbell},\ and\ \citenamefont {Monroe}}]{doi:10.1126/science.1251422}%
  \BibitemOpen
  \bibfield  {author} {\bibinfo {author} {\bibfnamefont {C.}~\bibnamefont {Senko}}, \bibinfo {author} {\bibfnamefont {J.}~\bibnamefont {Smith}}, \bibinfo {author} {\bibfnamefont {P.}~\bibnamefont {Richerme}}, \bibinfo {author} {\bibfnamefont {A.}~\bibnamefont {Lee}}, \bibinfo {author} {\bibfnamefont {W.~C.}\ \bibnamefont {Campbell}}, \ and\ \bibinfo {author} {\bibfnamefont {C.}~\bibnamefont {Monroe}},\ }\href {\doibase 10.1126/science.1251422} {\bibfield  {journal} {\bibinfo  {journal} {Science}\ }\textbf {\bibinfo {volume} {345}},\ \bibinfo {pages} {430} (\bibinfo {year} {2014})}\BibitemShut {NoStop}%
\bibitem [{\citenamefont {Jurcevic}\ \emph {et~al.}(2014)\citenamefont {Jurcevic}, \citenamefont {Lanyon}, \citenamefont {Hauke}, \citenamefont {Hempel}, \citenamefont {Zoller}, \citenamefont {Blatt},\ and\ \citenamefont {Roos}}]{Jurcevic2014}%
  \BibitemOpen
  \bibfield  {author} {\bibinfo {author} {\bibfnamefont {P.}~\bibnamefont {Jurcevic}}, \bibinfo {author} {\bibfnamefont {B.~P.}\ \bibnamefont {Lanyon}}, \bibinfo {author} {\bibfnamefont {P.}~\bibnamefont {Hauke}}, \bibinfo {author} {\bibfnamefont {C.}~\bibnamefont {Hempel}}, \bibinfo {author} {\bibfnamefont {P.}~\bibnamefont {Zoller}}, \bibinfo {author} {\bibfnamefont {R.}~\bibnamefont {Blatt}}, \ and\ \bibinfo {author} {\bibfnamefont {C.~F.}\ \bibnamefont {Roos}},\ }\href {\doibase 10.1038/nature13461} {\bibfield  {journal} {\bibinfo  {journal} {Nature}\ }\textbf {\bibinfo {volume} {511}},\ \bibinfo {pages} {202} (\bibinfo {year} {2014})}\BibitemShut {NoStop}%
\bibitem [{\citenamefont {Richerme}\ \emph {et~al.}(2014)\citenamefont {Richerme}, \citenamefont {Gong}, \citenamefont {Lee}, \citenamefont {Senko}, \citenamefont {Smith}, \citenamefont {Foss-Feig}, \citenamefont {Michalakis}, \citenamefont {Gorshkov},\ and\ \citenamefont {Monroe}}]{Richerme2014}%
  \BibitemOpen
  \bibfield  {author} {\bibinfo {author} {\bibfnamefont {P.}~\bibnamefont {Richerme}}, \bibinfo {author} {\bibfnamefont {Z.-X.}\ \bibnamefont {Gong}}, \bibinfo {author} {\bibfnamefont {A.}~\bibnamefont {Lee}}, \bibinfo {author} {\bibfnamefont {C.}~\bibnamefont {Senko}}, \bibinfo {author} {\bibfnamefont {J.}~\bibnamefont {Smith}}, \bibinfo {author} {\bibfnamefont {M.}~\bibnamefont {Foss-Feig}}, \bibinfo {author} {\bibfnamefont {S.}~\bibnamefont {Michalakis}}, \bibinfo {author} {\bibfnamefont {A.~V.}\ \bibnamefont {Gorshkov}}, \ and\ \bibinfo {author} {\bibfnamefont {C.}~\bibnamefont {Monroe}},\ }\href {\doibase 10.1038/nature13450} {\bibfield  {journal} {\bibinfo  {journal} {Nature}\ }\textbf {\bibinfo {volume} {511}},\ \bibinfo {pages} {198} (\bibinfo {year} {2014})}\BibitemShut {NoStop}%
\bibitem [{\citenamefont {Smith}\ \emph {et~al.}(2016)\citenamefont {Smith}, \citenamefont {Lee}, \citenamefont {Richerme}, \citenamefont {Neyenhuis}, \citenamefont {Hess}, \citenamefont {Hauke}, \citenamefont {Heyl}, \citenamefont {Huse},\ and\ \citenamefont {Monroe}}]{Smith2016}%
  \BibitemOpen
  \bibfield  {author} {\bibinfo {author} {\bibfnamefont {J.}~\bibnamefont {Smith}}, \bibinfo {author} {\bibfnamefont {A.}~\bibnamefont {Lee}}, \bibinfo {author} {\bibfnamefont {P.}~\bibnamefont {Richerme}}, \bibinfo {author} {\bibfnamefont {B.}~\bibnamefont {Neyenhuis}}, \bibinfo {author} {\bibfnamefont {P.~W.}\ \bibnamefont {Hess}}, \bibinfo {author} {\bibfnamefont {P.}~\bibnamefont {Hauke}}, \bibinfo {author} {\bibfnamefont {M.}~\bibnamefont {Heyl}}, \bibinfo {author} {\bibfnamefont {D.~A.}\ \bibnamefont {Huse}}, \ and\ \bibinfo {author} {\bibfnamefont {C.}~\bibnamefont {Monroe}},\ }\href {\doibase 10.1038/nphys3783} {\bibfield  {journal} {\bibinfo  {journal} {Nature Physics}\ }\textbf {\bibinfo {volume} {12}},\ \bibinfo {pages} {907} (\bibinfo {year} {2016})}\BibitemShut {NoStop}%
\bibitem [{\citenamefont {Zhang}\ \emph {et~al.}(2017)\citenamefont {Zhang}, \citenamefont {Pagano}, \citenamefont {Hess}, \citenamefont {Kyprianidis}, \citenamefont {Becker}, \citenamefont {Kaplan}, \citenamefont {Gorshkov}, \citenamefont {Gong},\ and\ \citenamefont {Monroe}}]{Zhang2017}%
  \BibitemOpen
  \bibfield  {author} {\bibinfo {author} {\bibfnamefont {J.}~\bibnamefont {Zhang}}, \bibinfo {author} {\bibfnamefont {G.}~\bibnamefont {Pagano}}, \bibinfo {author} {\bibfnamefont {P.~W.}\ \bibnamefont {Hess}}, \bibinfo {author} {\bibfnamefont {A.}~\bibnamefont {Kyprianidis}}, \bibinfo {author} {\bibfnamefont {P.}~\bibnamefont {Becker}}, \bibinfo {author} {\bibfnamefont {H.}~\bibnamefont {Kaplan}}, \bibinfo {author} {\bibfnamefont {A.~V.}\ \bibnamefont {Gorshkov}}, \bibinfo {author} {\bibfnamefont {Z.-X.}\ \bibnamefont {Gong}}, \ and\ \bibinfo {author} {\bibfnamefont {C.}~\bibnamefont {Monroe}},\ }\href {\doibase 10.1038/nature24654} {\bibfield  {journal} {\bibinfo  {journal} {Nature}\ }\textbf {\bibinfo {volume} {551}},\ \bibinfo {pages} {601} (\bibinfo {year} {2017})}\BibitemShut {NoStop}%
\bibitem [{\citenamefont {Tan}\ \emph {et~al.}(2021)\citenamefont {Tan}, \citenamefont {Becker}, \citenamefont {Liu}, \citenamefont {Pagano}, \citenamefont {Collins}, \citenamefont {De}, \citenamefont {Feng}, \citenamefont {Kaplan}, \citenamefont {Kyprianidis}, \citenamefont {Lundgren}, \citenamefont {Morong}, \citenamefont {Whitsitt}, \citenamefont {Gorshkov},\ and\ \citenamefont {Monroe}}]{Tan2021}%
  \BibitemOpen
  \bibfield  {author} {\bibinfo {author} {\bibfnamefont {W.~L.}\ \bibnamefont {Tan}}, \bibinfo {author} {\bibfnamefont {P.}~\bibnamefont {Becker}}, \bibinfo {author} {\bibfnamefont {F.}~\bibnamefont {Liu}}, \bibinfo {author} {\bibfnamefont {G.}~\bibnamefont {Pagano}}, \bibinfo {author} {\bibfnamefont {K.~S.}\ \bibnamefont {Collins}}, \bibinfo {author} {\bibfnamefont {A.}~\bibnamefont {De}}, \bibinfo {author} {\bibfnamefont {L.}~\bibnamefont {Feng}}, \bibinfo {author} {\bibfnamefont {H.~B.}\ \bibnamefont {Kaplan}}, \bibinfo {author} {\bibfnamefont {A.}~\bibnamefont {Kyprianidis}}, \bibinfo {author} {\bibfnamefont {R.}~\bibnamefont {Lundgren}}, \bibinfo {author} {\bibfnamefont {W.}~\bibnamefont {Morong}}, \bibinfo {author} {\bibfnamefont {S.}~\bibnamefont {Whitsitt}}, \bibinfo {author} {\bibfnamefont {A.~V.}\ \bibnamefont {Gorshkov}}, \ and\ \bibinfo {author} {\bibfnamefont {C.}~\bibnamefont {Monroe}},\ }\href {\doibase 10.1038/s41567-021-01194-3} {\bibfield  {journal} {\bibinfo  {journal} {Nature Physics}\
  }\textbf {\bibinfo {volume} {17}},\ \bibinfo {pages} {742} (\bibinfo {year} {2021})}\BibitemShut {NoStop}%
\bibitem [{\citenamefont {Morong}\ \emph {et~al.}(2021)\citenamefont {Morong}, \citenamefont {Liu}, \citenamefont {Becker}, \citenamefont {Collins}, \citenamefont {Feng}, \citenamefont {Kyprianidis}, \citenamefont {Pagano}, \citenamefont {You}, \citenamefont {Gorshkov},\ and\ \citenamefont {Monroe}}]{Morong2021}%
  \BibitemOpen
  \bibfield  {author} {\bibinfo {author} {\bibfnamefont {W.}~\bibnamefont {Morong}}, \bibinfo {author} {\bibfnamefont {F.}~\bibnamefont {Liu}}, \bibinfo {author} {\bibfnamefont {P.}~\bibnamefont {Becker}}, \bibinfo {author} {\bibfnamefont {K.~S.}\ \bibnamefont {Collins}}, \bibinfo {author} {\bibfnamefont {L.}~\bibnamefont {Feng}}, \bibinfo {author} {\bibfnamefont {A.}~\bibnamefont {Kyprianidis}}, \bibinfo {author} {\bibfnamefont {G.}~\bibnamefont {Pagano}}, \bibinfo {author} {\bibfnamefont {T.}~\bibnamefont {You}}, \bibinfo {author} {\bibfnamefont {A.~V.}\ \bibnamefont {Gorshkov}}, \ and\ \bibinfo {author} {\bibfnamefont {C.}~\bibnamefont {Monroe}},\ }\href {\doibase 10.1038/s41586-021-03988-0} {\bibfield  {journal} {\bibinfo  {journal} {Nature}\ }\textbf {\bibinfo {volume} {599}},\ \bibinfo {pages} {393} (\bibinfo {year} {2021})}\BibitemShut {NoStop}%
\bibitem [{\citenamefont {Bermudez}\ \emph {et~al.}(2017{\natexlab{a}})\citenamefont {Bermudez}, \citenamefont {Aarts},\ and\ \citenamefont {Müller}}]{a1}%
  \BibitemOpen
  \bibfield  {author} {\bibinfo {author} {\bibfnamefont {A.}~\bibnamefont {Bermudez}}, \bibinfo {author} {\bibfnamefont {G.}~\bibnamefont {Aarts}}, \ and\ \bibinfo {author} {\bibfnamefont {M.}~\bibnamefont {Müller}},\ }\href {\doibase 10.1103/physrevx.7.041012} {\bibfield  {journal} {\bibinfo  {journal} {Physical Review X}\ }\textbf {\bibinfo {volume} {7}} (\bibinfo {year} {2017}{\natexlab{a}}),\ 10.1103/physrevx.7.041012}\BibitemShut {NoStop}%
\bibitem [{\citenamefont {Martín-Vázquez}\ \emph {et~al.}(2021)\citenamefont {Martín-Vázquez}, \citenamefont {Aarts}, \citenamefont {Müller},\ and\ \citenamefont {Bermudez}}]{a2}%
  \BibitemOpen
  \bibfield  {author} {\bibinfo {author} {\bibfnamefont {G.}~\bibnamefont {Martín-Vázquez}}, \bibinfo {author} {\bibfnamefont {G.}~\bibnamefont {Aarts}}, \bibinfo {author} {\bibfnamefont {M.}~\bibnamefont {Müller}}, \ and\ \bibinfo {author} {\bibfnamefont {A.}~\bibnamefont {Bermudez}},\ }\href {\doibase 10.1103/PRXQuantum.3.020352} {\  (\bibinfo {year} {2021}),\ 10.1103/PRXQuantum.3.020352}\BibitemShut {NoStop}%
\bibitem [{\citenamefont {Nevado}\ and\ \citenamefont {Porras}(2016)}]{PhysRevA.93.013625}%
  \BibitemOpen
  \bibfield  {author} {\bibinfo {author} {\bibfnamefont {P.}~\bibnamefont {Nevado}}\ and\ \bibinfo {author} {\bibfnamefont {D.}~\bibnamefont {Porras}},\ }\href {\doibase 10.1103/PhysRevA.93.013625} {\bibfield  {journal} {\bibinfo  {journal} {Phys. Rev. A}\ }\textbf {\bibinfo {volume} {93}},\ \bibinfo {pages} {013625} (\bibinfo {year} {2016})}\BibitemShut {NoStop}%
\bibitem [{\citenamefont {Bermudez}\ \emph {et~al.}(2017{\natexlab{b}})\citenamefont {Bermudez}, \citenamefont {Tagliacozzo}, \citenamefont {Sierra},\ and\ \citenamefont {Richerme}}]{PhysRevB.95.024431}%
  \BibitemOpen
  \bibfield  {author} {\bibinfo {author} {\bibfnamefont {A.}~\bibnamefont {Bermudez}}, \bibinfo {author} {\bibfnamefont {L.}~\bibnamefont {Tagliacozzo}}, \bibinfo {author} {\bibfnamefont {G.}~\bibnamefont {Sierra}}, \ and\ \bibinfo {author} {\bibfnamefont {P.}~\bibnamefont {Richerme}},\ }\href {\doibase 10.1103/PhysRevB.95.024431} {\bibfield  {journal} {\bibinfo  {journal} {Phys. Rev. B}\ }\textbf {\bibinfo {volume} {95}},\ \bibinfo {pages} {024431} (\bibinfo {year} {2017}{\natexlab{b}})}\BibitemShut {NoStop}%
\bibitem [{\citenamefont {Klein}(1926)}]{Klein1926}%
  \BibitemOpen
  \bibfield  {author} {\bibinfo {author} {\bibfnamefont {O.}~\bibnamefont {Klein}},\ }\href {\doibase 10.1007/BF01397481} {\bibfield  {journal} {\bibinfo  {journal} {Zeitschrift f{\"u}r Physik}\ }\textbf {\bibinfo {volume} {37}},\ \bibinfo {pages} {895} (\bibinfo {year} {1926})}\BibitemShut {NoStop}%
\bibitem [{\citenamefont {Gordon}(1926)}]{Gordon1926}%
  \BibitemOpen
  \bibfield  {author} {\bibinfo {author} {\bibfnamefont {W.}~\bibnamefont {Gordon}},\ }\href {\doibase 10.1007/BF01390840} {\bibfield  {journal} {\bibinfo  {journal} {Zeitschrift f{\"u}r Physik}\ }\textbf {\bibinfo {volume} {40}},\ \bibinfo {pages} {117} (\bibinfo {year} {1926})}\BibitemShut {NoStop}%
\bibitem [{\citenamefont {Hidaka}\ \emph {et~al.}(2022)\citenamefont {Hidaka}, \citenamefont {Iso},\ and\ \citenamefont {Shimada}}]{https://doi.org/10.48550/arxiv.2211.09441}%
  \BibitemOpen
  \bibfield  {author} {\bibinfo {author} {\bibfnamefont {Y.}~\bibnamefont {Hidaka}}, \bibinfo {author} {\bibfnamefont {S.}~\bibnamefont {Iso}}, \ and\ \bibinfo {author} {\bibfnamefont {K.}~\bibnamefont {Shimada}},\ }\href {\doibase 10.1103/PhysRevD.107.085003} {\  (\bibinfo {year} {2022}),\ 10.1103/PhysRevD.107.085003}\BibitemShut {NoStop}%
\bibitem [{\citenamefont {Weinberg}(1967)}]{PhysRevLett.19.1264}%
  \BibitemOpen
  \bibfield  {author} {\bibinfo {author} {\bibfnamefont {S.}~\bibnamefont {Weinberg}},\ }\href {\doibase 10.1103/PhysRevLett.19.1264} {\bibfield  {journal} {\bibinfo  {journal} {Phys. Rev. Lett.}\ }\textbf {\bibinfo {volume} {19}},\ \bibinfo {pages} {1264} (\bibinfo {year} {1967})}\BibitemShut {NoStop}%
\bibitem [{\citenamefont {Wilson}\ and\ \citenamefont {Kogut}(1974)}]{WILSON197475}%
  \BibitemOpen
  \bibfield  {author} {\bibinfo {author} {\bibfnamefont {K.~G.}\ \bibnamefont {Wilson}}\ and\ \bibinfo {author} {\bibfnamefont {J.}~\bibnamefont {Kogut}},\ }\href {\doibase https://doi.org/10.1016/0370-1573(74)90023-4} {\bibfield  {journal} {\bibinfo  {journal} {Physics Reports}\ }\textbf {\bibinfo {volume} {12}},\ \bibinfo {pages} {75} (\bibinfo {year} {1974})}\BibitemShut {NoStop}%
\bibitem [{\citenamefont {Fishman}\ \emph {et~al.}(2008)\citenamefont {Fishman}, \citenamefont {De~Chiara}, \citenamefont {Calarco},\ and\ \citenamefont {Morigi}}]{PhysRevB.77.064111}%
  \BibitemOpen
  \bibfield  {author} {\bibinfo {author} {\bibfnamefont {S.}~\bibnamefont {Fishman}}, \bibinfo {author} {\bibfnamefont {G.}~\bibnamefont {De~Chiara}}, \bibinfo {author} {\bibfnamefont {T.}~\bibnamefont {Calarco}}, \ and\ \bibinfo {author} {\bibfnamefont {G.}~\bibnamefont {Morigi}},\ }\href {\doibase 10.1103/PhysRevB.77.064111} {\bibfield  {journal} {\bibinfo  {journal} {Phys. Rev. B}\ }\textbf {\bibinfo {volume} {77}},\ \bibinfo {pages} {064111} (\bibinfo {year} {2008})}\BibitemShut {NoStop}%
\bibitem [{\citenamefont {Podolsky}\ \emph {et~al.}(2014)\citenamefont {Podolsky}, \citenamefont {Shimshoni}, \citenamefont {Silvi}, \citenamefont {Montangero}, \citenamefont {Calarco}, \citenamefont {Morigi},\ and\ \citenamefont {Fishman}}]{PhysRevB.89.214408}%
  \BibitemOpen
  \bibfield  {author} {\bibinfo {author} {\bibfnamefont {D.}~\bibnamefont {Podolsky}}, \bibinfo {author} {\bibfnamefont {E.}~\bibnamefont {Shimshoni}}, \bibinfo {author} {\bibfnamefont {P.}~\bibnamefont {Silvi}}, \bibinfo {author} {\bibfnamefont {S.}~\bibnamefont {Montangero}}, \bibinfo {author} {\bibfnamefont {T.}~\bibnamefont {Calarco}}, \bibinfo {author} {\bibfnamefont {G.}~\bibnamefont {Morigi}}, \ and\ \bibinfo {author} {\bibfnamefont {S.}~\bibnamefont {Fishman}},\ }\href {\doibase 10.1103/PhysRevB.89.214408} {\bibfield  {journal} {\bibinfo  {journal} {Phys. Rev. B}\ }\textbf {\bibinfo {volume} {89}},\ \bibinfo {pages} {214408} (\bibinfo {year} {2014})}\BibitemShut {NoStop}%
\bibitem [{\citenamefont {Dolan}\ and\ \citenamefont {Jackiw}(1974)}]{PhysRevD.9.3320}%
  \BibitemOpen
  \bibfield  {author} {\bibinfo {author} {\bibfnamefont {L.}~\bibnamefont {Dolan}}\ and\ \bibinfo {author} {\bibfnamefont {R.}~\bibnamefont {Jackiw}},\ }\href {\doibase 10.1103/PhysRevD.9.3320} {\bibfield  {journal} {\bibinfo  {journal} {Phys. Rev. D}\ }\textbf {\bibinfo {volume} {9}},\ \bibinfo {pages} {3320} (\bibinfo {year} {1974})}\BibitemShut {NoStop}%
\bibitem [{\citenamefont {Weinberg}(1974)}]{PhysRevD.9.3357}%
  \BibitemOpen
  \bibfield  {author} {\bibinfo {author} {\bibfnamefont {S.}~\bibnamefont {Weinberg}},\ }\href {\doibase 10.1103/PhysRevD.9.3357} {\bibfield  {journal} {\bibinfo  {journal} {Phys. Rev. D}\ }\textbf {\bibinfo {volume} {9}},\ \bibinfo {pages} {3357} (\bibinfo {year} {1974})}\BibitemShut {NoStop}%
\bibitem [{\citenamefont {Šašura}\ and\ \citenamefont {Buzek}(2002)}]{ions1}%
  \BibitemOpen
  \bibfield  {author} {\bibinfo {author} {\bibfnamefont {M.}~\bibnamefont {Šašura}}\ and\ \bibinfo {author} {\bibfnamefont {V.}~\bibnamefont {Buzek}},\ }\href {\doibase 10.1080/09500340110115497} {\bibfield  {journal} {\bibinfo  {journal} {Journal of Modern Optics}\ }\textbf {\bibinfo {volume} {49}},\ \bibinfo {pages} {1593} (\bibinfo {year} {2002})}\BibitemShut {NoStop}%
\bibitem [{\citenamefont {Bruzewicz}\ \emph {et~al.}(2019)\citenamefont {Bruzewicz}, \citenamefont {Chiaverini}, \citenamefont {McConnell},\ and\ \citenamefont {Sage}}]{doi:10.1063/1.5088164}%
  \BibitemOpen
  \bibfield  {author} {\bibinfo {author} {\bibfnamefont {C.~D.}\ \bibnamefont {Bruzewicz}}, \bibinfo {author} {\bibfnamefont {J.}~\bibnamefont {Chiaverini}}, \bibinfo {author} {\bibfnamefont {R.}~\bibnamefont {McConnell}}, \ and\ \bibinfo {author} {\bibfnamefont {J.~M.}\ \bibnamefont {Sage}},\ }\href {\doibase 10.1063/1.5088164} {\bibfield  {journal} {\bibinfo  {journal} {Applied Physics Reviews}\ }\textbf {\bibinfo {volume} {6}},\ \bibinfo {pages} {021314} (\bibinfo {year} {2019})}\BibitemShut {NoStop}%
\bibitem [{\citenamefont {Ghosh}(1995)}]{ghosh1995ion}%
  \BibitemOpen
  \bibfield  {author} {\bibinfo {author} {\bibfnamefont {P.}~\bibnamefont {Ghosh}},\ }\href@noop {} {\enquote {\bibinfo {title} {Ion traps. clarendon},}\ } (\bibinfo {year} {1995})\BibitemShut {NoStop}%
\bibitem [{\citenamefont {Leibfried}\ \emph {et~al.}(2003{\natexlab{a}})\citenamefont {Leibfried}, \citenamefont {Blatt}, \citenamefont {Monroe},\ and\ \citenamefont {Wineland}}]{RevModPhys.75.281}%
  \BibitemOpen
  \bibfield  {author} {\bibinfo {author} {\bibfnamefont {D.}~\bibnamefont {Leibfried}}, \bibinfo {author} {\bibfnamefont {R.}~\bibnamefont {Blatt}}, \bibinfo {author} {\bibfnamefont {C.}~\bibnamefont {Monroe}}, \ and\ \bibinfo {author} {\bibfnamefont {D.}~\bibnamefont {Wineland}},\ }\href {\doibase 10.1103/RevModPhys.75.281} {\bibfield  {journal} {\bibinfo  {journal} {Rev. Mod. Phys.}\ }\textbf {\bibinfo {volume} {75}},\ \bibinfo {pages} {281} (\bibinfo {year} {2003}{\natexlab{a}})}\BibitemShut {NoStop}%
\bibitem [{\citenamefont {James}(1998)}]{James1998}%
  \BibitemOpen
  \bibfield  {author} {\bibinfo {author} {\bibfnamefont {D.~F.~V.}\ \bibnamefont {James}},\ }\href {\doibase 10.1007/s003400050373} {\bibfield  {journal} {\bibinfo  {journal} {Applied Physics B}\ }\textbf {\bibinfo {volume} {66}},\ \bibinfo {pages} {181} (\bibinfo {year} {1998})}\BibitemShut {NoStop}%
\bibitem [{\citenamefont {Marquet}\ \emph {et~al.}(2003)\citenamefont {Marquet}, \citenamefont {Schmidt-Kaler},\ and\ \citenamefont {James}}]{Marquet2003}%
  \BibitemOpen
  \bibfield  {author} {\bibinfo {author} {\bibfnamefont {C.}~\bibnamefont {Marquet}}, \bibinfo {author} {\bibfnamefont {F.}~\bibnamefont {Schmidt-Kaler}}, \ and\ \bibinfo {author} {\bibfnamefont {D.~F.~V.}\ \bibnamefont {James}},\ }\href {\doibase 10.1007/s00340-003-1097-7} {\bibfield  {journal} {\bibinfo  {journal} {Applied Physics B}\ }\textbf {\bibinfo {volume} {76}},\ \bibinfo {pages} {199} (\bibinfo {year} {2003})}\BibitemShut {NoStop}%
\bibitem [{\citenamefont {Altland}\ and\ \citenamefont {Simons}(2010)}]{altland_simons_2010}%
  \BibitemOpen
  \bibfield  {author} {\bibinfo {author} {\bibfnamefont {A.}~\bibnamefont {Altland}}\ and\ \bibinfo {author} {\bibfnamefont {B.~D.}\ \bibnamefont {Simons}},\ }\href {\doibase 10.1017/CBO9780511789984} {\emph {\bibinfo {title} {Condensed Matter Field Theory}}},\ \bibinfo {edition} {2nd}\ ed.\ (\bibinfo  {publisher} {Cambridge University Press},\ \bibinfo {year} {2010})\BibitemShut {NoStop}%
\bibitem [{\citenamefont {Schiffer}(1993)}]{PhysRevLett.70.818}%
  \BibitemOpen
  \bibfield  {author} {\bibinfo {author} {\bibfnamefont {J.~P.}\ \bibnamefont {Schiffer}},\ }\href {\doibase 10.1103/PhysRevLett.70.818} {\bibfield  {journal} {\bibinfo  {journal} {Phys. Rev. Lett.}\ }\textbf {\bibinfo {volume} {70}},\ \bibinfo {pages} {818} (\bibinfo {year} {1993})}\BibitemShut {NoStop}%
\bibitem [{\citenamefont {Dubin}(1993)}]{PhysRevLett.71.2753}%
  \BibitemOpen
  \bibfield  {author} {\bibinfo {author} {\bibfnamefont {D.~H.~E.}\ \bibnamefont {Dubin}},\ }\href {\doibase 10.1103/PhysRevLett.71.2753} {\bibfield  {journal} {\bibinfo  {journal} {Phys. Rev. Lett.}\ }\textbf {\bibinfo {volume} {71}},\ \bibinfo {pages} {2753} (\bibinfo {year} {1993})}\BibitemShut {NoStop}%
\bibitem [{\citenamefont {Enzer}\ \emph {et~al.}(2000)\citenamefont {Enzer}, \citenamefont {Schauer}, \citenamefont {Gomez}, \citenamefont {Gulley}, \citenamefont {Holzscheiter}, \citenamefont {Kwiat}, \citenamefont {Lamoreaux}, \citenamefont {Peterson}, \citenamefont {Sandberg}, \citenamefont {Tupa}, \citenamefont {White}, \citenamefont {Hughes},\ and\ \citenamefont {James}}]{PhysRevLett.85.2466}%
  \BibitemOpen
  \bibfield  {author} {\bibinfo {author} {\bibfnamefont {D.~G.}\ \bibnamefont {Enzer}}, \bibinfo {author} {\bibfnamefont {M.~M.}\ \bibnamefont {Schauer}}, \bibinfo {author} {\bibfnamefont {J.~J.}\ \bibnamefont {Gomez}}, \bibinfo {author} {\bibfnamefont {M.~S.}\ \bibnamefont {Gulley}}, \bibinfo {author} {\bibfnamefont {M.~H.}\ \bibnamefont {Holzscheiter}}, \bibinfo {author} {\bibfnamefont {P.~G.}\ \bibnamefont {Kwiat}}, \bibinfo {author} {\bibfnamefont {S.~K.}\ \bibnamefont {Lamoreaux}}, \bibinfo {author} {\bibfnamefont {C.~G.}\ \bibnamefont {Peterson}}, \bibinfo {author} {\bibfnamefont {V.~D.}\ \bibnamefont {Sandberg}}, \bibinfo {author} {\bibfnamefont {D.}~\bibnamefont {Tupa}}, \bibinfo {author} {\bibfnamefont {A.~G.}\ \bibnamefont {White}}, \bibinfo {author} {\bibfnamefont {R.~J.}\ \bibnamefont {Hughes}}, \ and\ \bibinfo {author} {\bibfnamefont {D.~F.~V.}\ \bibnamefont {James}},\ }\href {\doibase 10.1103/PhysRevLett.85.2466} {\bibfield  {journal} {\bibinfo  {journal} {Phys. Rev. Lett.}\ }\textbf {\bibinfo
  {volume} {85}},\ \bibinfo {pages} {2466} (\bibinfo {year} {2000})}\BibitemShut {NoStop}%
\bibitem [{\citenamefont {Greiner}\ and\ \citenamefont {Reinhardt}(1996)}]{Greiner:1996zu}%
  \BibitemOpen
  \bibfield  {author} {\bibinfo {author} {\bibfnamefont {W.}~\bibnamefont {Greiner}}\ and\ \bibinfo {author} {\bibfnamefont {J.}~\bibnamefont {Reinhardt}},\ }\href {\doibase 10.1007/978-3-642-61485-9} {\emph {\bibinfo {title} {{Field quantization}}}}\ (\bibinfo {year} {1996})\BibitemShut {NoStop}%
\bibitem [{\citenamefont {Rothe}(2012)}]{lattice}%
  \BibitemOpen
  \bibfield  {author} {\bibinfo {author} {\bibfnamefont {H.~J.}\ \bibnamefont {Rothe}},\ }\href {\doibase 10.1142/8229} {\emph {\bibinfo {title} {{Lattice Gauge Theories : An Introduction (Fourth Edition)}}}},\ Vol.~\bibinfo {volume} {43}\ (\bibinfo  {publisher} {World Scientific Publishing Company},\ \bibinfo {year} {2012})\ Chap.~\bibinfo {chapter} {3}\BibitemShut {NoStop}%
\bibitem [{\citenamefont {Bermudez}\ and\ \citenamefont {Plenio}(2012)}]{PhysRevLett.109.010501}%
  \BibitemOpen
  \bibfield  {author} {\bibinfo {author} {\bibfnamefont {A.}~\bibnamefont {Bermudez}}\ and\ \bibinfo {author} {\bibfnamefont {M.~B.}\ \bibnamefont {Plenio}},\ }\href {\doibase 10.1103/PhysRevLett.109.010501} {\bibfield  {journal} {\bibinfo  {journal} {Phys. Rev. Lett.}\ }\textbf {\bibinfo {volume} {109}},\ \bibinfo {pages} {010501} (\bibinfo {year} {2012})}\BibitemShut {NoStop}%
\bibitem [{\citenamefont {Shimshoni}\ \emph {et~al.}(2011)\citenamefont {Shimshoni}, \citenamefont {Morigi},\ and\ \citenamefont {Fishman}}]{PhysRevLett.106.010401}%
  \BibitemOpen
  \bibfield  {author} {\bibinfo {author} {\bibfnamefont {E.}~\bibnamefont {Shimshoni}}, \bibinfo {author} {\bibfnamefont {G.}~\bibnamefont {Morigi}}, \ and\ \bibinfo {author} {\bibfnamefont {S.}~\bibnamefont {Fishman}},\ }\href {\doibase 10.1103/PhysRevLett.106.010401} {\bibfield  {journal} {\bibinfo  {journal} {Phys. Rev. Lett.}\ }\textbf {\bibinfo {volume} {106}},\ \bibinfo {pages} {010401} (\bibinfo {year} {2011})}\BibitemShut {NoStop}%
\bibitem [{\citenamefont {Affleck}(1990)}]{Affleck1990}%
  \BibitemOpen
  \bibfield  {author} {\bibinfo {author} {\bibfnamefont {I.}~\bibnamefont {Affleck}},\ }\href {http://inis.iaea.org/search/search.aspx?orig_q=RN:21047685} {\emph {\bibinfo {title} {Field theory methods and quantum critical phenomena Course 10}}}\ (\bibinfo  {publisher} {North-Holland},\ \bibinfo {address} {Netherlands},\ \bibinfo {year} {1990})\BibitemShut {NoStop}%
\bibitem [{\citenamefont {Emery}(1993)}]{doi:10.1142/1882}%
  \BibitemOpen
  \bibfield  {author} {\bibinfo {author} {\bibfnamefont {V.~J.}\ \bibnamefont {Emery}},\ }\href {\doibase 10.1142/1882} {\emph {\bibinfo {title} {Correlated Electron Systems}}}\ (\bibinfo  {publisher} {WORLD SCIENTIFIC},\ \bibinfo {year} {1993})\BibitemShut {NoStop}%
\bibitem [{Gon(1995)}]{Gonzalez1995}%
  \BibitemOpen
  \enquote {\bibinfo {title} {Correspondence from discrete to continuum models},}\ in\ \href {\doibase 10.1007/978-3-540-47678-8_5} {\emph {\bibinfo {booktitle} {Quantum Electron Liquids and High-TcSuperconductivity}}}\ (\bibinfo  {publisher} {Springer Berlin Heidelberg},\ \bibinfo {address} {Berlin, Heidelberg},\ \bibinfo {year} {1995})\ pp.\ \bibinfo {pages} {109--124}\BibitemShut {NoStop}%
\bibitem [{\citenamefont {Son}(2002)}]{Son:2002zn}%
  \BibitemOpen
  \bibfield  {author} {\bibinfo {author} {\bibfnamefont {D.~T.}\ \bibnamefont {Son}},\ }\href@noop {} {\  (\bibinfo {year} {2002})},\ \Eprint {http://arxiv.org/abs/hep-ph/0204199} {arXiv:hep-ph/0204199} \BibitemShut {NoStop}%
\bibitem [{198(1986)}]{1986iv}%
  \BibitemOpen
  in\ \href {\doibase https://doi.org/10.1016/B978-0-08-057069-3.50003-6} {\emph {\bibinfo {booktitle} {Theory of Elasticity (Third Edition)}}},\ \bibinfo {editor} {edited by\ \bibinfo {editor} {\bibfnamefont {E.}~\bibnamefont {LIFSHITZ}}, \bibinfo {editor} {\bibfnamefont {A.}~\bibnamefont {KOSEVICH}}, \ and\ \bibinfo {editor} {\bibfnamefont {L.}~\bibnamefont {PITAEVSKII}}}\ (\bibinfo  {publisher} {Butterworth-Heinemann},\ \bibinfo {address} {Oxford},\ \bibinfo {year} {1986})\ \bibinfo {edition} {third edition}\ ed.,\ p.~\bibinfo {pages} {iv}\BibitemShut {NoStop}%
\bibitem [{\citenamefont {Thorne}\ and\ \citenamefont {Blandford}(2017)}]{2017mcp..book.....T}%
  \BibitemOpen
  \bibfield  {author} {\bibinfo {author} {\bibfnamefont {K.}~\bibnamefont {Thorne}}\ and\ \bibinfo {author} {\bibfnamefont {R.}~\bibnamefont {Blandford}},\ }\href {https://books.google.es/books?id=U1S6BQAAQBAJ} {\emph {\bibinfo {title} {Modern Classical Physics: Optics, Fluids, Plasmas, Elasticity, Relativity, and Statistical Physics}}}\ (\bibinfo  {publisher} {Princeton University Press},\ \bibinfo {year} {2017})\BibitemShut {NoStop}%
\bibitem [{\citenamefont {Coleman}\ and\ \citenamefont {Weinberg}(1973)}]{PhysRevD.7.1888}%
  \BibitemOpen
  \bibfield  {author} {\bibinfo {author} {\bibfnamefont {S.}~\bibnamefont {Coleman}}\ and\ \bibinfo {author} {\bibfnamefont {E.}~\bibnamefont {Weinberg}},\ }\href {\doibase 10.1103/PhysRevD.7.1888} {\bibfield  {journal} {\bibinfo  {journal} {Phys. Rev. D}\ }\textbf {\bibinfo {volume} {7}},\ \bibinfo {pages} {1888} (\bibinfo {year} {1973})}\BibitemShut {NoStop}%
\bibitem [{\citenamefont {Jackiw}(1974)}]{PhysRevD.9.1686}%
  \BibitemOpen
  \bibfield  {author} {\bibinfo {author} {\bibfnamefont {R.}~\bibnamefont {Jackiw}},\ }\href {\doibase 10.1103/PhysRevD.9.1686} {\bibfield  {journal} {\bibinfo  {journal} {Phys. Rev. D}\ }\textbf {\bibinfo {volume} {9}},\ \bibinfo {pages} {1686} (\bibinfo {year} {1974})}\BibitemShut {NoStop}%
\bibitem [{\citenamefont {Wilson}\ and\ \citenamefont {Fisher}(1972)}]{PhysRevLett.28.240}%
  \BibitemOpen
  \bibfield  {author} {\bibinfo {author} {\bibfnamefont {K.~G.}\ \bibnamefont {Wilson}}\ and\ \bibinfo {author} {\bibfnamefont {M.~E.}\ \bibnamefont {Fisher}},\ }\href {\doibase 10.1103/PhysRevLett.28.240} {\bibfield  {journal} {\bibinfo  {journal} {Phys. Rev. Lett.}\ }\textbf {\bibinfo {volume} {28}},\ \bibinfo {pages} {240} (\bibinfo {year} {1972})}\BibitemShut {NoStop}%
\bibitem [{\citenamefont {Loinaz}\ and\ \citenamefont {Willey}(1998)}]{PhysRevD.58.076003}%
  \BibitemOpen
  \bibfield  {author} {\bibinfo {author} {\bibfnamefont {W.}~\bibnamefont {Loinaz}}\ and\ \bibinfo {author} {\bibfnamefont {R.~S.}\ \bibnamefont {Willey}},\ }\href {\doibase 10.1103/PhysRevD.58.076003} {\bibfield  {journal} {\bibinfo  {journal} {Phys. Rev. D}\ }\textbf {\bibinfo {volume} {58}},\ \bibinfo {pages} {076003} (\bibinfo {year} {1998})}\BibitemShut {NoStop}%
\bibitem [{\citenamefont {Schaich}\ and\ \citenamefont {Loinaz}(2009)}]{PhysRevD.79.056008}%
  \BibitemOpen
  \bibfield  {author} {\bibinfo {author} {\bibfnamefont {D.}~\bibnamefont {Schaich}}\ and\ \bibinfo {author} {\bibfnamefont {W.}~\bibnamefont {Loinaz}},\ }\href {\doibase 10.1103/PhysRevD.79.056008} {\bibfield  {journal} {\bibinfo  {journal} {Phys. Rev. D}\ }\textbf {\bibinfo {volume} {79}},\ \bibinfo {pages} {056008} (\bibinfo {year} {2009})}\BibitemShut {NoStop}%
\bibitem [{\citenamefont {Sugihara}(2004)}]{Takanori_Sugihara_2004}%
  \BibitemOpen
  \bibfield  {author} {\bibinfo {author} {\bibfnamefont {T.}~\bibnamefont {Sugihara}},\ }\href {\doibase 10.1088/1126-6708/2004/05/007} {\bibfield  {journal} {\bibinfo  {journal} {Journal of High Energy Physics}\ }\textbf {\bibinfo {volume} {2004}},\ \bibinfo {pages} {007} (\bibinfo {year} {2004})}\BibitemShut {NoStop}%
\bibitem [{\citenamefont {Asit K.~De}\ \emph {et~al.}(2005)\citenamefont {Asit K.~De}, \citenamefont {Maiti},\ and\ \citenamefont {Sinha}}]{montecarlo}%
  \BibitemOpen
  \bibfield  {author} {\bibinfo {author} {\bibfnamefont {A.~H.}\ \bibnamefont {Asit K.~De}}, \bibinfo {author} {\bibfnamefont {J.}~\bibnamefont {Maiti}}, \ and\ \bibinfo {author} {\bibfnamefont {T.}~\bibnamefont {Sinha}},\ }\href {\doibase 10.1103/physrevd.72.094503} {\bibfield  {journal} {\bibinfo  {journal} {Physical Review D}\ }\textbf {\bibinfo {volume} {72}} (\bibinfo {year} {2005}),\ 10.1103/physrevd.72.094503}\BibitemShut {NoStop}%
\bibitem [{\citenamefont {Milsted}\ \emph {et~al.}(2013{\natexlab{a}})\citenamefont {Milsted}, \citenamefont {Haegeman},\ and\ \citenamefont {Osborne}}]{PhysRevD.88.085030}%
  \BibitemOpen
  \bibfield  {author} {\bibinfo {author} {\bibfnamefont {A.}~\bibnamefont {Milsted}}, \bibinfo {author} {\bibfnamefont {J.}~\bibnamefont {Haegeman}}, \ and\ \bibinfo {author} {\bibfnamefont {T.~J.}\ \bibnamefont {Osborne}},\ }\href {\doibase 10.1103/PhysRevD.88.085030} {\bibfield  {journal} {\bibinfo  {journal} {Phys. Rev. D}\ }\textbf {\bibinfo {volume} {88}},\ \bibinfo {pages} {085030} (\bibinfo {year} {2013}{\natexlab{a}})}\BibitemShut {NoStop}%
\bibitem [{\citenamefont {Kadoh}\ \emph {et~al.}(2019{\natexlab{a}})\citenamefont {Kadoh}, \citenamefont {Kuramashi}, \citenamefont {Nakamura}, \citenamefont {Sakai}, \citenamefont {Takeda},\ and\ \citenamefont {Yoshimura}}]{Kadoh2019}%
  \BibitemOpen
  \bibfield  {author} {\bibinfo {author} {\bibfnamefont {D.}~\bibnamefont {Kadoh}}, \bibinfo {author} {\bibfnamefont {Y.}~\bibnamefont {Kuramashi}}, \bibinfo {author} {\bibfnamefont {Y.}~\bibnamefont {Nakamura}}, \bibinfo {author} {\bibfnamefont {R.}~\bibnamefont {Sakai}}, \bibinfo {author} {\bibfnamefont {S.}~\bibnamefont {Takeda}}, \ and\ \bibinfo {author} {\bibfnamefont {Y.}~\bibnamefont {Yoshimura}},\ }\href {\doibase 10.1007/JHEP05(2019)184} {\bibfield  {journal} {\bibinfo  {journal} {Journal of High Energy Physics}\ }\textbf {\bibinfo {volume} {2019}},\ \bibinfo {pages} {184} (\bibinfo {year} {2019}{\natexlab{a}})}\BibitemShut {NoStop}%
\bibitem [{\citenamefont {Bronzin}\ \emph {et~al.}(2019)\citenamefont {Bronzin}, \citenamefont {De~Palma},\ and\ \citenamefont {Guagnelli}}]{PhysRevD.99.034508}%
  \BibitemOpen
  \bibfield  {author} {\bibinfo {author} {\bibfnamefont {S.}~\bibnamefont {Bronzin}}, \bibinfo {author} {\bibfnamefont {B.}~\bibnamefont {De~Palma}}, \ and\ \bibinfo {author} {\bibfnamefont {M.}~\bibnamefont {Guagnelli}},\ }\href {\doibase 10.1103/PhysRevD.99.034508} {\bibfield  {journal} {\bibinfo  {journal} {Phys. Rev. D}\ }\textbf {\bibinfo {volume} {99}},\ \bibinfo {pages} {034508} (\bibinfo {year} {2019})}\BibitemShut {NoStop}%
\bibitem [{\citenamefont {Delcamp}\ and\ \citenamefont {Tilloy}(2020{\natexlab{a}})}]{PhysRevResearch.2.033278}%
  \BibitemOpen
  \bibfield  {author} {\bibinfo {author} {\bibfnamefont {C.}~\bibnamefont {Delcamp}}\ and\ \bibinfo {author} {\bibfnamefont {A.}~\bibnamefont {Tilloy}},\ }\href {\doibase 10.1103/PhysRevResearch.2.033278} {\bibfield  {journal} {\bibinfo  {journal} {Phys. Rev. Res.}\ }\textbf {\bibinfo {volume} {2}},\ \bibinfo {pages} {033278} (\bibinfo {year} {2020}{\natexlab{a}})}\BibitemShut {NoStop}%
\bibitem [{\citenamefont {Vanhecke}\ \emph {et~al.}(2022)\citenamefont {Vanhecke}, \citenamefont {Verstraete},\ and\ \citenamefont {Van~Acoleyen}}]{PhysRevD.106.L071501}%
  \BibitemOpen
  \bibfield  {author} {\bibinfo {author} {\bibfnamefont {B.}~\bibnamefont {Vanhecke}}, \bibinfo {author} {\bibfnamefont {F.}~\bibnamefont {Verstraete}}, \ and\ \bibinfo {author} {\bibfnamefont {K.}~\bibnamefont {Van~Acoleyen}},\ }\href {\doibase 10.1103/PhysRevD.106.L071501} {\bibfield  {journal} {\bibinfo  {journal} {Phys. Rev. D}\ }\textbf {\bibinfo {volume} {106}},\ \bibinfo {pages} {L071501} (\bibinfo {year} {2022})}\BibitemShut {NoStop}%
\bibitem [{\citenamefont {Chang}(1976)}]{PhysRevD.13.2778}%
  \BibitemOpen
  \bibfield  {author} {\bibinfo {author} {\bibfnamefont {S.-J.}\ \bibnamefont {Chang}},\ }\href {\doibase 10.1103/PhysRevD.13.2778} {\bibfield  {journal} {\bibinfo  {journal} {Phys. Rev. D}\ }\textbf {\bibinfo {volume} {13}},\ \bibinfo {pages} {2778} (\bibinfo {year} {1976})}\BibitemShut {NoStop}%
\bibitem [{\citenamefont {Chiaverini}\ \emph {et~al.}(2004)\citenamefont {Chiaverini} \emph {et~al.}}]{Chiaverini2004}%
  \BibitemOpen
  \bibfield  {author} {\bibinfo {author} {\bibfnamefont {J.}~\bibnamefont {Chiaverini}} \emph {et~al.},\ }\href {\doibase 10.1038/nature03074} {\bibfield  {journal} {\bibinfo  {journal} {Nature}\ }\textbf {\bibinfo {volume} {432}},\ \bibinfo {pages} {602} (\bibinfo {year} {2004})}\BibitemShut {NoStop}%
\bibitem [{\citenamefont {Schindler}\ \emph {et~al.}(2011)\citenamefont {Schindler}, \citenamefont {Barreiro}, \citenamefont {Monz}, \citenamefont {Nebendahl}, \citenamefont {Nigg}, \citenamefont {Chwalla}, \citenamefont {Hennrich},\ and\ \citenamefont {Blatt}}]{Schindler1059}%
  \BibitemOpen
  \bibfield  {author} {\bibinfo {author} {\bibfnamefont {P.}~\bibnamefont {Schindler}}, \bibinfo {author} {\bibfnamefont {J.~T.}\ \bibnamefont {Barreiro}}, \bibinfo {author} {\bibfnamefont {T.}~\bibnamefont {Monz}}, \bibinfo {author} {\bibfnamefont {V.}~\bibnamefont {Nebendahl}}, \bibinfo {author} {\bibfnamefont {D.}~\bibnamefont {Nigg}}, \bibinfo {author} {\bibfnamefont {M.}~\bibnamefont {Chwalla}}, \bibinfo {author} {\bibfnamefont {M.}~\bibnamefont {Hennrich}}, \ and\ \bibinfo {author} {\bibfnamefont {R.}~\bibnamefont {Blatt}},\ }\href {\doibase 10.1126/science.1203329} {\bibfield  {journal} {\bibinfo  {journal} {Science}\ }\textbf {\bibinfo {volume} {332}},\ \bibinfo {pages} {1059} (\bibinfo {year} {2011})}\BibitemShut {NoStop}%
\bibitem [{\citenamefont {Nigg}\ \emph {et~al.}(2014)\citenamefont {Nigg}, \citenamefont {M{\"u}ller}, \citenamefont {Martinez}, \citenamefont {Schindler}, \citenamefont {Hennrich}, \citenamefont {Monz}, \citenamefont {Martin-Delgado},\ and\ \citenamefont {Blatt}}]{Nigg302}%
  \BibitemOpen
  \bibfield  {author} {\bibinfo {author} {\bibfnamefont {D.}~\bibnamefont {Nigg}}, \bibinfo {author} {\bibfnamefont {M.}~\bibnamefont {M{\"u}ller}}, \bibinfo {author} {\bibfnamefont {E.~A.}\ \bibnamefont {Martinez}}, \bibinfo {author} {\bibfnamefont {P.}~\bibnamefont {Schindler}}, \bibinfo {author} {\bibfnamefont {M.}~\bibnamefont {Hennrich}}, \bibinfo {author} {\bibfnamefont {T.}~\bibnamefont {Monz}}, \bibinfo {author} {\bibfnamefont {M.~A.}\ \bibnamefont {Martin-Delgado}}, \ and\ \bibinfo {author} {\bibfnamefont {R.}~\bibnamefont {Blatt}},\ }\href {\doibase 10.1126/science.1253742} {\bibfield  {journal} {\bibinfo  {journal} {Science}\ }\textbf {\bibinfo {volume} {345}},\ \bibinfo {pages} {302} (\bibinfo {year} {2014})}\BibitemShut {NoStop}%
\bibitem [{\citenamefont {Linke}\ \emph {et~al.}(2017)\citenamefont {Linke}, \citenamefont {Gutierrez}, \citenamefont {Landsman}, \citenamefont {Figgatt}, \citenamefont {Debnath}, \citenamefont {Brown},\ and\ \citenamefont {Monroe}}]{Linkee1701074}%
  \BibitemOpen
  \bibfield  {author} {\bibinfo {author} {\bibfnamefont {N.~M.}\ \bibnamefont {Linke}}, \bibinfo {author} {\bibfnamefont {M.}~\bibnamefont {Gutierrez}}, \bibinfo {author} {\bibfnamefont {K.~A.}\ \bibnamefont {Landsman}}, \bibinfo {author} {\bibfnamefont {C.}~\bibnamefont {Figgatt}}, \bibinfo {author} {\bibfnamefont {S.}~\bibnamefont {Debnath}}, \bibinfo {author} {\bibfnamefont {K.~R.}\ \bibnamefont {Brown}}, \ and\ \bibinfo {author} {\bibfnamefont {C.}~\bibnamefont {Monroe}},\ }\href {\doibase 10.1126/sciadv.1701074} {\bibfield  {journal} {\bibinfo  {journal} {Science Advances}\ }\textbf {\bibinfo {volume} {3}} (\bibinfo {year} {2017}),\ 10.1126/sciadv.1701074}\BibitemShut {NoStop}%
\bibitem [{\citenamefont {Negnevitsky}\ \emph {et~al.}(2018)\citenamefont {Negnevitsky}, \citenamefont {Marinelli}, \citenamefont {Mehta}, \citenamefont {Lo}, \citenamefont {Fl{\"u}hmann},\ and\ \citenamefont {Home}}]{Negnevitsky2018}%
  \BibitemOpen
  \bibfield  {author} {\bibinfo {author} {\bibfnamefont {V.}~\bibnamefont {Negnevitsky}}, \bibinfo {author} {\bibfnamefont {M.}~\bibnamefont {Marinelli}}, \bibinfo {author} {\bibfnamefont {K.~K.}\ \bibnamefont {Mehta}}, \bibinfo {author} {\bibfnamefont {H.-Y.}\ \bibnamefont {Lo}}, \bibinfo {author} {\bibfnamefont {C.}~\bibnamefont {Fl{\"u}hmann}}, \ and\ \bibinfo {author} {\bibfnamefont {J.~P.}\ \bibnamefont {Home}},\ }\href {\doibase 10.1038/s41586-018-0668-z} {\bibfield  {journal} {\bibinfo  {journal} {Nature}\ }\textbf {\bibinfo {volume} {563}},\ \bibinfo {pages} {527} (\bibinfo {year} {2018})}\BibitemShut {NoStop}%
\bibitem [{\citenamefont {Fluehmann}\ \emph {et~al.}(2019)\citenamefont {Fluehmann}, \citenamefont {Nguyen}, \citenamefont {Marinelli}, \citenamefont {Negnevitsky}, \citenamefont {Mehta},\ and\ \citenamefont {Home}}]{Fluhmann2019}%
  \BibitemOpen
  \bibfield  {author} {\bibinfo {author} {\bibfnamefont {C.}~\bibnamefont {Fluehmann}}, \bibinfo {author} {\bibfnamefont {T.}~\bibnamefont {Nguyen}}, \bibinfo {author} {\bibfnamefont {M.}~\bibnamefont {Marinelli}}, \bibinfo {author} {\bibfnamefont {V.}~\bibnamefont {Negnevitsky}}, \bibinfo {author} {\bibfnamefont {K.}~\bibnamefont {Mehta}}, \ and\ \bibinfo {author} {\bibfnamefont {J.}~\bibnamefont {Home}},\ }\href {\doibase 10.1038/s41586-019-0960-6} {\bibfield  {journal} {\bibinfo  {journal} {Nature}\ }\textbf {\bibinfo {volume} {566}},\ \bibinfo {pages} {513} (\bibinfo {year} {2019})}\BibitemShut {NoStop}%
\bibitem [{\citenamefont {Stricker}\ \emph {et~al.}(2020)\citenamefont {Stricker} \emph {et~al.}}]{Stricker2020}%
  \BibitemOpen
  \bibfield  {author} {\bibinfo {author} {\bibfnamefont {R.}~\bibnamefont {Stricker}} \emph {et~al.},\ }\href {\doibase 10.1038/s41586-020-2667-0} {\bibfield  {journal} {\bibinfo  {journal} {Nature}\ }\textbf {\bibinfo {volume} {585}},\ \bibinfo {pages} {207} (\bibinfo {year} {2020})}\BibitemShut {NoStop}%
\bibitem [{\citenamefont {de~Neeve}\ \emph {et~al.}(2020)\citenamefont {de~Neeve}, \citenamefont {Nguyen}, \citenamefont {Behrle},\ and\ \citenamefont {Home}}]{https://doi.org/10.48550/arxiv.2010.09681}%
  \BibitemOpen
  \bibfield  {author} {\bibinfo {author} {\bibfnamefont {B.}~\bibnamefont {de~Neeve}}, \bibinfo {author} {\bibfnamefont {T.~L.}\ \bibnamefont {Nguyen}}, \bibinfo {author} {\bibfnamefont {T.}~\bibnamefont {Behrle}}, \ and\ \bibinfo {author} {\bibfnamefont {J.}~\bibnamefont {Home}},\ }\href {\doibase 10.1038/s41567-021-01487-7} {\  (\bibinfo {year} {2020}),\ 10.1038/s41567-021-01487-7}\BibitemShut {NoStop}%
\bibitem [{\citenamefont {Erhard}\ \emph {et~al.}(2021)\citenamefont {Erhard}, \citenamefont {Poulsen~Nautrup}, \citenamefont {Meth}, \citenamefont {Postler}, \citenamefont {Stricker}, \citenamefont {Stadler}, \citenamefont {Negnevitsky}, \citenamefont {Ringbauer}, \citenamefont {Schindler}, \citenamefont {Briegel}, \citenamefont {Blatt}, \citenamefont {Friis},\ and\ \citenamefont {Monz}}]{Erhard2021}%
  \BibitemOpen
  \bibfield  {author} {\bibinfo {author} {\bibfnamefont {A.}~\bibnamefont {Erhard}}, \bibinfo {author} {\bibfnamefont {H.}~\bibnamefont {Poulsen~Nautrup}}, \bibinfo {author} {\bibfnamefont {M.}~\bibnamefont {Meth}}, \bibinfo {author} {\bibfnamefont {L.}~\bibnamefont {Postler}}, \bibinfo {author} {\bibfnamefont {R.}~\bibnamefont {Stricker}}, \bibinfo {author} {\bibfnamefont {M.}~\bibnamefont {Stadler}}, \bibinfo {author} {\bibfnamefont {V.}~\bibnamefont {Negnevitsky}}, \bibinfo {author} {\bibfnamefont {M.}~\bibnamefont {Ringbauer}}, \bibinfo {author} {\bibfnamefont {P.}~\bibnamefont {Schindler}}, \bibinfo {author} {\bibfnamefont {H.~J.}\ \bibnamefont {Briegel}}, \bibinfo {author} {\bibfnamefont {R.}~\bibnamefont {Blatt}}, \bibinfo {author} {\bibfnamefont {N.}~\bibnamefont {Friis}}, \ and\ \bibinfo {author} {\bibfnamefont {T.}~\bibnamefont {Monz}},\ }\href {\doibase 10.1038/s41586-020-03079-6} {\bibfield  {journal} {\bibinfo  {journal} {Nature}\ }\textbf {\bibinfo {volume} {589}},\ \bibinfo {pages} {220}
  (\bibinfo {year} {2021})}\BibitemShut {NoStop}%
\bibitem [{\citenamefont {Egan}\ \emph {et~al.}(2021)\citenamefont {Egan}, \citenamefont {Debroy}, \citenamefont {Noel}, \citenamefont {Risinger}, \citenamefont {Zhu}, \citenamefont {Biswas}, \citenamefont {Newman}, \citenamefont {Li}, \citenamefont {Brown}, \citenamefont {Cetina},\ and\ \citenamefont {Monroe}}]{Egan2021}%
  \BibitemOpen
  \bibfield  {author} {\bibinfo {author} {\bibfnamefont {L.}~\bibnamefont {Egan}}, \bibinfo {author} {\bibfnamefont {D.~M.}\ \bibnamefont {Debroy}}, \bibinfo {author} {\bibfnamefont {C.}~\bibnamefont {Noel}}, \bibinfo {author} {\bibfnamefont {A.}~\bibnamefont {Risinger}}, \bibinfo {author} {\bibfnamefont {D.}~\bibnamefont {Zhu}}, \bibinfo {author} {\bibfnamefont {D.}~\bibnamefont {Biswas}}, \bibinfo {author} {\bibfnamefont {M.}~\bibnamefont {Newman}}, \bibinfo {author} {\bibfnamefont {M.}~\bibnamefont {Li}}, \bibinfo {author} {\bibfnamefont {K.~R.}\ \bibnamefont {Brown}}, \bibinfo {author} {\bibfnamefont {M.}~\bibnamefont {Cetina}}, \ and\ \bibinfo {author} {\bibfnamefont {C.}~\bibnamefont {Monroe}},\ }\href {\doibase 10.1038/s41586-021-03928-y} {\bibfield  {journal} {\bibinfo  {journal} {Nature}\ }\textbf {\bibinfo {volume} {598}},\ \bibinfo {pages} {281} (\bibinfo {year} {2021})}\BibitemShut {NoStop}%
\bibitem [{\citenamefont {Debroy}\ \emph {et~al.}(2021)\citenamefont {Debroy}, \citenamefont {Egan}, \citenamefont {Noel}, \citenamefont {Risinger}, \citenamefont {Zhu}, \citenamefont {Biswas}, \citenamefont {Cetina}, \citenamefont {Monroe},\ and\ \citenamefont {Brown}}]{PhysRevLett.127.240501}%
  \BibitemOpen
  \bibfield  {author} {\bibinfo {author} {\bibfnamefont {D.~M.}\ \bibnamefont {Debroy}}, \bibinfo {author} {\bibfnamefont {L.}~\bibnamefont {Egan}}, \bibinfo {author} {\bibfnamefont {C.}~\bibnamefont {Noel}}, \bibinfo {author} {\bibfnamefont {A.}~\bibnamefont {Risinger}}, \bibinfo {author} {\bibfnamefont {D.}~\bibnamefont {Zhu}}, \bibinfo {author} {\bibfnamefont {D.}~\bibnamefont {Biswas}}, \bibinfo {author} {\bibfnamefont {M.}~\bibnamefont {Cetina}}, \bibinfo {author} {\bibfnamefont {C.}~\bibnamefont {Monroe}}, \ and\ \bibinfo {author} {\bibfnamefont {K.~R.}\ \bibnamefont {Brown}},\ }\href {\doibase 10.1103/PhysRevLett.127.240501} {\bibfield  {journal} {\bibinfo  {journal} {Phys. Rev. Lett.}\ }\textbf {\bibinfo {volume} {127}},\ \bibinfo {pages} {240501} (\bibinfo {year} {2021})}\BibitemShut {NoStop}%
\bibitem [{\citenamefont {Ryan-Anderson}\ \emph {et~al.}(2021)\citenamefont {Ryan-Anderson}, \citenamefont {Bohnet}, \citenamefont {Lee}, \citenamefont {Gresh}, \citenamefont {Hankin}, \citenamefont {Gaebler}, \citenamefont {Francois}, \citenamefont {Chernoguzov}, \citenamefont {Lucchetti}, \citenamefont {Brown}, \citenamefont {Gatterman}, \citenamefont {Halit}, \citenamefont {Gilmore}, \citenamefont {Gerber}, \citenamefont {Neyenhuis}, \citenamefont {Hayes},\ and\ \citenamefont {Stutz}}]{PhysRevX.11.041058}%
  \BibitemOpen
  \bibfield  {author} {\bibinfo {author} {\bibfnamefont {C.}~\bibnamefont {Ryan-Anderson}}, \bibinfo {author} {\bibfnamefont {J.~G.}\ \bibnamefont {Bohnet}}, \bibinfo {author} {\bibfnamefont {K.}~\bibnamefont {Lee}}, \bibinfo {author} {\bibfnamefont {D.}~\bibnamefont {Gresh}}, \bibinfo {author} {\bibfnamefont {A.}~\bibnamefont {Hankin}}, \bibinfo {author} {\bibfnamefont {J.~P.}\ \bibnamefont {Gaebler}}, \bibinfo {author} {\bibfnamefont {D.}~\bibnamefont {Francois}}, \bibinfo {author} {\bibfnamefont {A.}~\bibnamefont {Chernoguzov}}, \bibinfo {author} {\bibfnamefont {D.}~\bibnamefont {Lucchetti}}, \bibinfo {author} {\bibfnamefont {N.~C.}\ \bibnamefont {Brown}}, \bibinfo {author} {\bibfnamefont {T.~M.}\ \bibnamefont {Gatterman}}, \bibinfo {author} {\bibfnamefont {S.~K.}\ \bibnamefont {Halit}}, \bibinfo {author} {\bibfnamefont {K.}~\bibnamefont {Gilmore}}, \bibinfo {author} {\bibfnamefont {J.~A.}\ \bibnamefont {Gerber}}, \bibinfo {author} {\bibfnamefont {B.}~\bibnamefont {Neyenhuis}}, \bibinfo {author}
  {\bibfnamefont {D.}~\bibnamefont {Hayes}}, \ and\ \bibinfo {author} {\bibfnamefont {R.~P.}\ \bibnamefont {Stutz}},\ }\href {\doibase 10.1103/PhysRevX.11.041058} {\bibfield  {journal} {\bibinfo  {journal} {Phys. Rev. X}\ }\textbf {\bibinfo {volume} {11}},\ \bibinfo {pages} {041058} (\bibinfo {year} {2021})}\BibitemShut {NoStop}%
\bibitem [{\citenamefont {Hilder}\ \emph {et~al.}(2022)\citenamefont {Hilder}, \citenamefont {Pijn}, \citenamefont {Onishchenko}, \citenamefont {Stahl}, \citenamefont {Orth}, \citenamefont {Lekitsch}, \citenamefont {Rodriguez-Blanco}, \citenamefont {M\"uller}, \citenamefont {Schmidt-Kaler},\ and\ \citenamefont {Poschinger}}]{PhysRevX.12.011032}%
  \BibitemOpen
  \bibfield  {author} {\bibinfo {author} {\bibfnamefont {J.}~\bibnamefont {Hilder}}, \bibinfo {author} {\bibfnamefont {D.}~\bibnamefont {Pijn}}, \bibinfo {author} {\bibfnamefont {O.}~\bibnamefont {Onishchenko}}, \bibinfo {author} {\bibfnamefont {A.}~\bibnamefont {Stahl}}, \bibinfo {author} {\bibfnamefont {M.}~\bibnamefont {Orth}}, \bibinfo {author} {\bibfnamefont {B.}~\bibnamefont {Lekitsch}}, \bibinfo {author} {\bibfnamefont {A.}~\bibnamefont {Rodriguez-Blanco}}, \bibinfo {author} {\bibfnamefont {M.}~\bibnamefont {M\"uller}}, \bibinfo {author} {\bibfnamefont {F.}~\bibnamefont {Schmidt-Kaler}}, \ and\ \bibinfo {author} {\bibfnamefont {U.~G.}\ \bibnamefont {Poschinger}},\ }\href {\doibase 10.1103/PhysRevX.12.011032} {\bibfield  {journal} {\bibinfo  {journal} {Phys. Rev. X}\ }\textbf {\bibinfo {volume} {12}},\ \bibinfo {pages} {011032} (\bibinfo {year} {2022})}\BibitemShut {NoStop}%
\bibitem [{\citenamefont {Bermudez}\ \emph {et~al.}(2017{\natexlab{c}})\citenamefont {Bermudez}, \citenamefont {Xu}, \citenamefont {Nigmatullin}, \citenamefont {O'Gorman}, \citenamefont {Negnevitsky}, \citenamefont {Schindler}, \citenamefont {Monz}, \citenamefont {Poschinger}, \citenamefont {Hempel}, \citenamefont {Home}, \citenamefont {Schmidt-Kaler}, \citenamefont {Biercuk}, \citenamefont {Blatt}, \citenamefont {Benjamin},\ and\ \citenamefont {M\"uller}}]{PhysRevX.7.041061}%
  \BibitemOpen
  \bibfield  {author} {\bibinfo {author} {\bibfnamefont {A.}~\bibnamefont {Bermudez}}, \bibinfo {author} {\bibfnamefont {X.}~\bibnamefont {Xu}}, \bibinfo {author} {\bibfnamefont {R.}~\bibnamefont {Nigmatullin}}, \bibinfo {author} {\bibfnamefont {J.}~\bibnamefont {O'Gorman}}, \bibinfo {author} {\bibfnamefont {V.}~\bibnamefont {Negnevitsky}}, \bibinfo {author} {\bibfnamefont {P.}~\bibnamefont {Schindler}}, \bibinfo {author} {\bibfnamefont {T.}~\bibnamefont {Monz}}, \bibinfo {author} {\bibfnamefont {U.~G.}\ \bibnamefont {Poschinger}}, \bibinfo {author} {\bibfnamefont {C.}~\bibnamefont {Hempel}}, \bibinfo {author} {\bibfnamefont {J.}~\bibnamefont {Home}}, \bibinfo {author} {\bibfnamefont {F.}~\bibnamefont {Schmidt-Kaler}}, \bibinfo {author} {\bibfnamefont {M.}~\bibnamefont {Biercuk}}, \bibinfo {author} {\bibfnamefont {R.}~\bibnamefont {Blatt}}, \bibinfo {author} {\bibfnamefont {S.}~\bibnamefont {Benjamin}}, \ and\ \bibinfo {author} {\bibfnamefont {M.}~\bibnamefont {M\"uller}},\ }\href
  {http://dx.doi.org/10.1103/PhysRevX.7.041061} {\bibfield  {journal} {\bibinfo  {journal} {Phys. Rev. X}\ }\textbf {\bibinfo {volume} {7}},\ \bibinfo {pages} {041061} (\bibinfo {year} {2017}{\natexlab{c}})}\BibitemShut {NoStop}%
\bibitem [{\citenamefont {Gulde}\ \emph {et~al.}(2003)\citenamefont {Gulde}, \citenamefont {Riebe}, \citenamefont {Lancaster}, \citenamefont {Becher}, \citenamefont {Eschner}, \citenamefont {H{\"a}ffner}, \citenamefont {Schmidt-Kaler}, \citenamefont {Chuang},\ and\ \citenamefont {Blatt}}]{Gulde2003}%
  \BibitemOpen
  \bibfield  {author} {\bibinfo {author} {\bibfnamefont {S.}~\bibnamefont {Gulde}}, \bibinfo {author} {\bibfnamefont {M.}~\bibnamefont {Riebe}}, \bibinfo {author} {\bibfnamefont {G.~P.~T.}\ \bibnamefont {Lancaster}}, \bibinfo {author} {\bibfnamefont {C.}~\bibnamefont {Becher}}, \bibinfo {author} {\bibfnamefont {J.}~\bibnamefont {Eschner}}, \bibinfo {author} {\bibfnamefont {H.}~\bibnamefont {H{\"a}ffner}}, \bibinfo {author} {\bibfnamefont {F.}~\bibnamefont {Schmidt-Kaler}}, \bibinfo {author} {\bibfnamefont {I.~L.}\ \bibnamefont {Chuang}}, \ and\ \bibinfo {author} {\bibfnamefont {R.}~\bibnamefont {Blatt}},\ }\href {\doibase 10.1038/nature01336} {\bibfield  {journal} {\bibinfo  {journal} {Nature}\ }\textbf {\bibinfo {volume} {421}},\ \bibinfo {pages} {48} (\bibinfo {year} {2003})}\BibitemShut {NoStop}%
\bibitem [{\citenamefont {Barrett}\ \emph {et~al.}(2004)\citenamefont {Barrett}, \citenamefont {Chiaverini}, \citenamefont {Schaetz}, \citenamefont {Britton}, \citenamefont {Itano}, \citenamefont {Jost}, \citenamefont {Knill}, \citenamefont {Langer}, \citenamefont {Leibfried}, \citenamefont {Ozeri},\ and\ \citenamefont {Wineland}}]{Barrett2004}%
  \BibitemOpen
  \bibfield  {author} {\bibinfo {author} {\bibfnamefont {M.~D.}\ \bibnamefont {Barrett}}, \bibinfo {author} {\bibfnamefont {J.}~\bibnamefont {Chiaverini}}, \bibinfo {author} {\bibfnamefont {T.}~\bibnamefont {Schaetz}}, \bibinfo {author} {\bibfnamefont {J.}~\bibnamefont {Britton}}, \bibinfo {author} {\bibfnamefont {W.~M.}\ \bibnamefont {Itano}}, \bibinfo {author} {\bibfnamefont {J.~D.}\ \bibnamefont {Jost}}, \bibinfo {author} {\bibfnamefont {E.}~\bibnamefont {Knill}}, \bibinfo {author} {\bibfnamefont {C.}~\bibnamefont {Langer}}, \bibinfo {author} {\bibfnamefont {D.}~\bibnamefont {Leibfried}}, \bibinfo {author} {\bibfnamefont {R.}~\bibnamefont {Ozeri}}, \ and\ \bibinfo {author} {\bibfnamefont {D.~J.}\ \bibnamefont {Wineland}},\ }\href {\doibase 10.1038/nature02608} {\bibfield  {journal} {\bibinfo  {journal} {Nature}\ }\textbf {\bibinfo {volume} {429}},\ \bibinfo {pages} {737} (\bibinfo {year} {2004})}\BibitemShut {NoStop}%
\bibitem [{\citenamefont {Riebe}\ \emph {et~al.}(2004)\citenamefont {Riebe}, \citenamefont {H{\"a}ffner}, \citenamefont {Roos}, \citenamefont {H{\"a}nsel}, \citenamefont {Benhelm}, \citenamefont {Lancaster}, \citenamefont {K{\"o}rber}, \citenamefont {Becher}, \citenamefont {Schmidt-Kaler}, \citenamefont {James},\ and\ \citenamefont {Blatt}}]{Riebe2004}%
  \BibitemOpen
  \bibfield  {author} {\bibinfo {author} {\bibfnamefont {M.}~\bibnamefont {Riebe}}, \bibinfo {author} {\bibfnamefont {H.}~\bibnamefont {H{\"a}ffner}}, \bibinfo {author} {\bibfnamefont {C.~F.}\ \bibnamefont {Roos}}, \bibinfo {author} {\bibfnamefont {W.}~\bibnamefont {H{\"a}nsel}}, \bibinfo {author} {\bibfnamefont {J.}~\bibnamefont {Benhelm}}, \bibinfo {author} {\bibfnamefont {G.~P.~T.}\ \bibnamefont {Lancaster}}, \bibinfo {author} {\bibfnamefont {T.~W.}\ \bibnamefont {K{\"o}rber}}, \bibinfo {author} {\bibfnamefont {C.}~\bibnamefont {Becher}}, \bibinfo {author} {\bibfnamefont {F.}~\bibnamefont {Schmidt-Kaler}}, \bibinfo {author} {\bibfnamefont {D.~F.~V.}\ \bibnamefont {James}}, \ and\ \bibinfo {author} {\bibfnamefont {R.}~\bibnamefont {Blatt}},\ }\href {\doibase 10.1038/nature02570} {\bibfield  {journal} {\bibinfo  {journal} {Nature}\ }\textbf {\bibinfo {volume} {429}},\ \bibinfo {pages} {734} (\bibinfo {year} {2004})}\BibitemShut {NoStop}%
\bibitem [{\citenamefont {Chiaverini}\ \emph {et~al.}(2005)\citenamefont {Chiaverini}, \citenamefont {Britton}, \citenamefont {Leibfried}, \citenamefont {Knill}, \citenamefont {Barrett}, \citenamefont {Blakestad}, \citenamefont {Itano}, \citenamefont {Jost}, \citenamefont {Langer}, \citenamefont {Ozeri}, \citenamefont {Schaetz},\ and\ \citenamefont {Wineland}}]{doi:10.1126/science.1110335}%
  \BibitemOpen
  \bibfield  {author} {\bibinfo {author} {\bibfnamefont {J.}~\bibnamefont {Chiaverini}}, \bibinfo {author} {\bibfnamefont {J.}~\bibnamefont {Britton}}, \bibinfo {author} {\bibfnamefont {D.}~\bibnamefont {Leibfried}}, \bibinfo {author} {\bibfnamefont {E.}~\bibnamefont {Knill}}, \bibinfo {author} {\bibfnamefont {M.~D.}\ \bibnamefont {Barrett}}, \bibinfo {author} {\bibfnamefont {R.~B.}\ \bibnamefont {Blakestad}}, \bibinfo {author} {\bibfnamefont {W.~M.}\ \bibnamefont {Itano}}, \bibinfo {author} {\bibfnamefont {J.~D.}\ \bibnamefont {Jost}}, \bibinfo {author} {\bibfnamefont {C.}~\bibnamefont {Langer}}, \bibinfo {author} {\bibfnamefont {R.}~\bibnamefont {Ozeri}}, \bibinfo {author} {\bibfnamefont {T.}~\bibnamefont {Schaetz}}, \ and\ \bibinfo {author} {\bibfnamefont {D.~J.}\ \bibnamefont {Wineland}},\ }\href {\doibase 10.1126/science.1110335} {\bibfield  {journal} {\bibinfo  {journal} {Science}\ }\textbf {\bibinfo {volume} {308}},\ \bibinfo {pages} {997} (\bibinfo {year} {2005})}\BibitemShut {NoStop}%
\bibitem [{\citenamefont {Monz}\ \emph {et~al.}(2016)\citenamefont {Monz}, \citenamefont {Nigg}, \citenamefont {Martinez}, \citenamefont {Brandl}, \citenamefont {Schindler}, \citenamefont {Rines}, \citenamefont {Wang}, \citenamefont {Chuang},\ and\ \citenamefont {Blatt}}]{doi:10.1126/science.aad9480}%
  \BibitemOpen
  \bibfield  {author} {\bibinfo {author} {\bibfnamefont {T.}~\bibnamefont {Monz}}, \bibinfo {author} {\bibfnamefont {D.}~\bibnamefont {Nigg}}, \bibinfo {author} {\bibfnamefont {E.~A.}\ \bibnamefont {Martinez}}, \bibinfo {author} {\bibfnamefont {M.~F.}\ \bibnamefont {Brandl}}, \bibinfo {author} {\bibfnamefont {P.}~\bibnamefont {Schindler}}, \bibinfo {author} {\bibfnamefont {R.}~\bibnamefont {Rines}}, \bibinfo {author} {\bibfnamefont {S.~X.}\ \bibnamefont {Wang}}, \bibinfo {author} {\bibfnamefont {I.~L.}\ \bibnamefont {Chuang}}, \ and\ \bibinfo {author} {\bibfnamefont {R.}~\bibnamefont {Blatt}},\ }\href {\doibase 10.1126/science.aad9480} {\bibfield  {journal} {\bibinfo  {journal} {Science}\ }\textbf {\bibinfo {volume} {351}},\ \bibinfo {pages} {1068} (\bibinfo {year} {2016})}\BibitemShut {NoStop}%
\bibitem [{\citenamefont {Figgatt}\ \emph {et~al.}(2017)\citenamefont {Figgatt}, \citenamefont {Maslov}, \citenamefont {Landsman}, \citenamefont {Linke}, \citenamefont {Debnath},\ and\ \citenamefont {Monroe}}]{Figgatt2017}%
  \BibitemOpen
  \bibfield  {author} {\bibinfo {author} {\bibfnamefont {C.}~\bibnamefont {Figgatt}}, \bibinfo {author} {\bibfnamefont {D.}~\bibnamefont {Maslov}}, \bibinfo {author} {\bibfnamefont {K.~A.}\ \bibnamefont {Landsman}}, \bibinfo {author} {\bibfnamefont {N.~M.}\ \bibnamefont {Linke}}, \bibinfo {author} {\bibfnamefont {S.}~\bibnamefont {Debnath}}, \ and\ \bibinfo {author} {\bibfnamefont {C.}~\bibnamefont {Monroe}},\ }\href {\doibase 10.1038/s41467-017-01904-7} {\bibfield  {journal} {\bibinfo  {journal} {Nature Communications}\ }\textbf {\bibinfo {volume} {8}},\ \bibinfo {pages} {1918} (\bibinfo {year} {2017})}\BibitemShut {NoStop}%
\bibitem [{\citenamefont {Wan}\ \emph {et~al.}(2019)\citenamefont {Wan}, \citenamefont {Kienzler}, \citenamefont {Erickson}, \citenamefont {Mayer}, \citenamefont {Tan}, \citenamefont {Wu}, \citenamefont {Vasconcelos}, \citenamefont {Glancy}, \citenamefont {Knill}, \citenamefont {Wineland}, \citenamefont {Wilson},\ and\ \citenamefont {Leibfried}}]{doi:10.1126/science.aaw9415}%
  \BibitemOpen
  \bibfield  {author} {\bibinfo {author} {\bibfnamefont {Y.}~\bibnamefont {Wan}}, \bibinfo {author} {\bibfnamefont {D.}~\bibnamefont {Kienzler}}, \bibinfo {author} {\bibfnamefont {S.~D.}\ \bibnamefont {Erickson}}, \bibinfo {author} {\bibfnamefont {K.~H.}\ \bibnamefont {Mayer}}, \bibinfo {author} {\bibfnamefont {T.~R.}\ \bibnamefont {Tan}}, \bibinfo {author} {\bibfnamefont {J.~J.}\ \bibnamefont {Wu}}, \bibinfo {author} {\bibfnamefont {H.~M.}\ \bibnamefont {Vasconcelos}}, \bibinfo {author} {\bibfnamefont {S.}~\bibnamefont {Glancy}}, \bibinfo {author} {\bibfnamefont {E.}~\bibnamefont {Knill}}, \bibinfo {author} {\bibfnamefont {D.~J.}\ \bibnamefont {Wineland}}, \bibinfo {author} {\bibfnamefont {A.~C.}\ \bibnamefont {Wilson}}, \ and\ \bibinfo {author} {\bibfnamefont {D.}~\bibnamefont {Leibfried}},\ }\href {\doibase 10.1126/science.aaw9415} {\bibfield  {journal} {\bibinfo  {journal} {Science}\ }\textbf {\bibinfo {volume} {364}},\ \bibinfo {pages} {875} (\bibinfo {year} {2019})}\BibitemShut {NoStop}%
\bibitem [{\citenamefont {Harty}\ \emph {et~al.}(2014)\citenamefont {Harty}, \citenamefont {Ballance}, \citenamefont {Guidoni}, \citenamefont {Janacek}, \citenamefont {Linke}, \citenamefont {Stacey},\ and\ \citenamefont {Lucas}}]{fidelities1}%
  \BibitemOpen
  \bibfield  {author} {\bibinfo {author} {\bibfnamefont {D.~T.~C.}\ \bibnamefont {Harty}, \bibfnamefont {T.~P. et al.~Allcock}}, \bibinfo {author} {\bibfnamefont {C.~J.}\ \bibnamefont {Ballance}}, \bibinfo {author} {\bibfnamefont {L.}~\bibnamefont {Guidoni}}, \bibinfo {author} {\bibfnamefont {H.~A.}\ \bibnamefont {Janacek}}, \bibinfo {author} {\bibfnamefont {N.~M.}\ \bibnamefont {Linke}}, \bibinfo {author} {\bibfnamefont {D.~N.}\ \bibnamefont {Stacey}}, \ and\ \bibinfo {author} {\bibfnamefont {D.~M.}\ \bibnamefont {Lucas}},\ }\href {\doibase 10.1103/PhysRevLett.113.220501} {\bibfield  {journal} {\bibinfo  {journal} {Phys. Rev. Lett.}\ }\textbf {\bibinfo {volume} {113}},\ \bibinfo {pages} {220501} (\bibinfo {year} {2014})}\BibitemShut {NoStop}%
\bibitem [{\citenamefont {Ballance}\ \emph {et~al.}(2016)\citenamefont {Ballance}, \citenamefont {Harty}, \citenamefont {Linke}, \citenamefont {Sepiol},\ and\ \citenamefont {Lucas}}]{PhysRevLett.117.060504}%
  \BibitemOpen
  \bibfield  {author} {\bibinfo {author} {\bibfnamefont {C.~J.}\ \bibnamefont {Ballance}}, \bibinfo {author} {\bibfnamefont {T.~P.}\ \bibnamefont {Harty}}, \bibinfo {author} {\bibfnamefont {N.~M.}\ \bibnamefont {Linke}}, \bibinfo {author} {\bibfnamefont {M.~A.}\ \bibnamefont {Sepiol}}, \ and\ \bibinfo {author} {\bibfnamefont {D.~M.}\ \bibnamefont {Lucas}},\ }\href {\doibase 10.1103/PhysRevLett.117.060504} {\bibfield  {journal} {\bibinfo  {journal} {Phys. Rev. Lett.}\ }\textbf {\bibinfo {volume} {117}},\ \bibinfo {pages} {060504} (\bibinfo {year} {2016})}\BibitemShut {NoStop}%
\bibitem [{\citenamefont {Gaebler}\ \emph {et~al.}(2016)\citenamefont {Gaebler}, \citenamefont {Tan}, \citenamefont {Lin}, \citenamefont {Wan}, \citenamefont {Bowler}, \citenamefont {Keith}, \citenamefont {Glancy}, \citenamefont {Coakley}, \citenamefont {Knill}, \citenamefont {Leibfried},\ and\ \citenamefont {Wineland}}]{PhysRevLett.117.060505}%
  \BibitemOpen
  \bibfield  {author} {\bibinfo {author} {\bibfnamefont {J.~P.}\ \bibnamefont {Gaebler}}, \bibinfo {author} {\bibfnamefont {T.~R.}\ \bibnamefont {Tan}}, \bibinfo {author} {\bibfnamefont {Y.}~\bibnamefont {Lin}}, \bibinfo {author} {\bibfnamefont {Y.}~\bibnamefont {Wan}}, \bibinfo {author} {\bibfnamefont {R.}~\bibnamefont {Bowler}}, \bibinfo {author} {\bibfnamefont {A.~C.}\ \bibnamefont {Keith}}, \bibinfo {author} {\bibfnamefont {S.}~\bibnamefont {Glancy}}, \bibinfo {author} {\bibfnamefont {K.}~\bibnamefont {Coakley}}, \bibinfo {author} {\bibfnamefont {E.}~\bibnamefont {Knill}}, \bibinfo {author} {\bibfnamefont {D.}~\bibnamefont {Leibfried}}, \ and\ \bibinfo {author} {\bibfnamefont {D.~J.}\ \bibnamefont {Wineland}},\ }\href {\doibase 10.1103/PhysRevLett.117.060505} {\bibfield  {journal} {\bibinfo  {journal} {Phys. Rev. Lett.}\ }\textbf {\bibinfo {volume} {117}},\ \bibinfo {pages} {060505} (\bibinfo {year} {2016})}\BibitemShut {NoStop}%
\bibitem [{\citenamefont {Harty}\ \emph {et~al.}(2016)\citenamefont {Harty}, \citenamefont {Sepiol}, \citenamefont {Allcock}, \citenamefont {Ballance}, \citenamefont {Tarlton},\ and\ \citenamefont {Lucas}}]{PhysRevLett.117.140501}%
  \BibitemOpen
  \bibfield  {author} {\bibinfo {author} {\bibfnamefont {T.~P.}\ \bibnamefont {Harty}}, \bibinfo {author} {\bibfnamefont {M.~A.}\ \bibnamefont {Sepiol}}, \bibinfo {author} {\bibfnamefont {D.~T.~C.}\ \bibnamefont {Allcock}}, \bibinfo {author} {\bibfnamefont {C.~J.}\ \bibnamefont {Ballance}}, \bibinfo {author} {\bibfnamefont {J.~E.}\ \bibnamefont {Tarlton}}, \ and\ \bibinfo {author} {\bibfnamefont {D.~M.}\ \bibnamefont {Lucas}},\ }\href {\doibase 10.1103/PhysRevLett.117.140501} {\bibfield  {journal} {\bibinfo  {journal} {Phys. Rev. Lett.}\ }\textbf {\bibinfo {volume} {117}},\ \bibinfo {pages} {140501} (\bibinfo {year} {2016})}\BibitemShut {NoStop}%
\bibitem [{\citenamefont {Erhard}\ \emph {et~al.}(2019)\citenamefont {Erhard}, \citenamefont {Wallman}, \citenamefont {Postler}, \citenamefont {Meth}, \citenamefont {Stricker}, \citenamefont {Martinez}, \citenamefont {Schindler}, \citenamefont {Monz}, \citenamefont {Emerson},\ and\ \citenamefont {Blatt}}]{Erhard2019}%
  \BibitemOpen
  \bibfield  {author} {\bibinfo {author} {\bibfnamefont {A.}~\bibnamefont {Erhard}}, \bibinfo {author} {\bibfnamefont {J.~J.}\ \bibnamefont {Wallman}}, \bibinfo {author} {\bibfnamefont {L.}~\bibnamefont {Postler}}, \bibinfo {author} {\bibfnamefont {M.}~\bibnamefont {Meth}}, \bibinfo {author} {\bibfnamefont {R.}~\bibnamefont {Stricker}}, \bibinfo {author} {\bibfnamefont {E.~A.}\ \bibnamefont {Martinez}}, \bibinfo {author} {\bibfnamefont {P.}~\bibnamefont {Schindler}}, \bibinfo {author} {\bibfnamefont {T.}~\bibnamefont {Monz}}, \bibinfo {author} {\bibfnamefont {J.}~\bibnamefont {Emerson}}, \ and\ \bibinfo {author} {\bibfnamefont {R.}~\bibnamefont {Blatt}},\ }\href {\doibase 10.1038/s41467-019-13068-7} {\bibfield  {journal} {\bibinfo  {journal} {Nature Communications}\ }\textbf {\bibinfo {volume} {10}},\ \bibinfo {pages} {5347} (\bibinfo {year} {2019})}\BibitemShut {NoStop}%
\bibitem [{\citenamefont {Zarantonello}\ \emph {et~al.}(2019)\citenamefont {Zarantonello}, \citenamefont {Hahn}, \citenamefont {Morgner}, \citenamefont {Schulte}, \citenamefont {Bautista-Salvador}, \citenamefont {Werner}, \citenamefont {Hammerer},\ and\ \citenamefont {Ospelkaus}}]{PhysRevLett.123.260503}%
  \BibitemOpen
  \bibfield  {author} {\bibinfo {author} {\bibfnamefont {G.}~\bibnamefont {Zarantonello}}, \bibinfo {author} {\bibfnamefont {H.}~\bibnamefont {Hahn}}, \bibinfo {author} {\bibfnamefont {J.}~\bibnamefont {Morgner}}, \bibinfo {author} {\bibfnamefont {M.}~\bibnamefont {Schulte}}, \bibinfo {author} {\bibfnamefont {A.}~\bibnamefont {Bautista-Salvador}}, \bibinfo {author} {\bibfnamefont {R.~F.}\ \bibnamefont {Werner}}, \bibinfo {author} {\bibfnamefont {K.}~\bibnamefont {Hammerer}}, \ and\ \bibinfo {author} {\bibfnamefont {C.}~\bibnamefont {Ospelkaus}},\ }\href {\doibase 10.1103/PhysRevLett.123.260503} {\bibfield  {journal} {\bibinfo  {journal} {Phys. Rev. Lett.}\ }\textbf {\bibinfo {volume} {123}},\ \bibinfo {pages} {260503} (\bibinfo {year} {2019})}\BibitemShut {NoStop}%
\bibitem [{\citenamefont {Kielpinski}\ \emph {et~al.}(2002)\citenamefont {Kielpinski}, \citenamefont {Monroe},\ and\ \citenamefont {Wineland}}]{Kielpinski2002}%
  \BibitemOpen
  \bibfield  {author} {\bibinfo {author} {\bibfnamefont {D.}~\bibnamefont {Kielpinski}}, \bibinfo {author} {\bibfnamefont {C.}~\bibnamefont {Monroe}}, \ and\ \bibinfo {author} {\bibfnamefont {D.~J.}\ \bibnamefont {Wineland}},\ }\href {\doibase 10.1038/nature00784} {\bibfield  {journal} {\bibinfo  {journal} {Nature}\ }\textbf {\bibinfo {volume} {417}},\ \bibinfo {pages} {709} (\bibinfo {year} {2002})}\BibitemShut {NoStop}%
\bibitem [{\citenamefont {Kaushal}\ \emph {et~al.}(2020)\citenamefont {Kaushal}, \citenamefont {Lekitsch}, \citenamefont {Stahl}, \citenamefont {Hilder}, \citenamefont {Pijn}, \citenamefont {Schmiegelow}, \citenamefont {Bermudez}, \citenamefont {Müller}, \citenamefont {Schmidt-Kaler},\ and\ \citenamefont {Poschinger}}]{doi:10.1116/1.5126186}%
  \BibitemOpen
  \bibfield  {author} {\bibinfo {author} {\bibfnamefont {V.}~\bibnamefont {Kaushal}}, \bibinfo {author} {\bibfnamefont {B.}~\bibnamefont {Lekitsch}}, \bibinfo {author} {\bibfnamefont {A.}~\bibnamefont {Stahl}}, \bibinfo {author} {\bibfnamefont {J.}~\bibnamefont {Hilder}}, \bibinfo {author} {\bibfnamefont {D.}~\bibnamefont {Pijn}}, \bibinfo {author} {\bibfnamefont {C.}~\bibnamefont {Schmiegelow}}, \bibinfo {author} {\bibfnamefont {A.}~\bibnamefont {Bermudez}}, \bibinfo {author} {\bibfnamefont {M.}~\bibnamefont {Müller}}, \bibinfo {author} {\bibfnamefont {F.}~\bibnamefont {Schmidt-Kaler}}, \ and\ \bibinfo {author} {\bibfnamefont {U.}~\bibnamefont {Poschinger}},\ }\href {\doibase 10.1116/1.5126186} {\bibfield  {journal} {\bibinfo  {journal} {AVS Quantum Science}\ }\textbf {\bibinfo {volume} {2}},\ \bibinfo {pages} {014101} (\bibinfo {year} {2020})}\BibitemShut {NoStop}%
\bibitem [{\citenamefont {Home}\ \emph {et~al.}(2009)\citenamefont {Home}, \citenamefont {Hanneke}, \citenamefont {Jost}, \citenamefont {Amini}, \citenamefont {Leibfried},\ and\ \citenamefont {Wineland}}]{doi:10.1126/science.1177077}%
  \BibitemOpen
  \bibfield  {author} {\bibinfo {author} {\bibfnamefont {J.~P.}\ \bibnamefont {Home}}, \bibinfo {author} {\bibfnamefont {D.}~\bibnamefont {Hanneke}}, \bibinfo {author} {\bibfnamefont {J.~D.}\ \bibnamefont {Jost}}, \bibinfo {author} {\bibfnamefont {J.~M.}\ \bibnamefont {Amini}}, \bibinfo {author} {\bibfnamefont {D.}~\bibnamefont {Leibfried}}, \ and\ \bibinfo {author} {\bibfnamefont {D.~J.}\ \bibnamefont {Wineland}},\ }\href {\doibase 10.1126/science.1177077} {\bibfield  {journal} {\bibinfo  {journal} {Science}\ }\textbf {\bibinfo {volume} {325}},\ \bibinfo {pages} {1227} (\bibinfo {year} {2009})}\BibitemShut {NoStop}%
\bibitem [{\citenamefont {Debnath}\ \emph {et~al.}(2016)\citenamefont {Debnath}, \citenamefont {Linke}, \citenamefont {Figgatt}, \citenamefont {Landsman}, \citenamefont {Wright},\ and\ \citenamefont {Monroe}}]{Debnath2016}%
  \BibitemOpen
  \bibfield  {author} {\bibinfo {author} {\bibfnamefont {S.}~\bibnamefont {Debnath}}, \bibinfo {author} {\bibfnamefont {N.~M.}\ \bibnamefont {Linke}}, \bibinfo {author} {\bibfnamefont {C.}~\bibnamefont {Figgatt}}, \bibinfo {author} {\bibfnamefont {K.~A.}\ \bibnamefont {Landsman}}, \bibinfo {author} {\bibfnamefont {K.}~\bibnamefont {Wright}}, \ and\ \bibinfo {author} {\bibfnamefont {C.}~\bibnamefont {Monroe}},\ }\href {\doibase 10.1038/nature18648} {\bibfield  {journal} {\bibinfo  {journal} {Nature}\ }\textbf {\bibinfo {volume} {536}},\ \bibinfo {pages} {63} (\bibinfo {year} {2016})}\BibitemShut {NoStop}%
\bibitem [{\citenamefont {Figgatt}\ \emph {et~al.}(2019)\citenamefont {Figgatt}, \citenamefont {Ostrander}, \citenamefont {Linke}, \citenamefont {Landsman}, \citenamefont {Zhu}, \citenamefont {Maslov},\ and\ \citenamefont {Monroe}}]{Figgatt2019}%
  \BibitemOpen
  \bibfield  {author} {\bibinfo {author} {\bibfnamefont {C.}~\bibnamefont {Figgatt}}, \bibinfo {author} {\bibfnamefont {A.}~\bibnamefont {Ostrander}}, \bibinfo {author} {\bibfnamefont {N.~M.}\ \bibnamefont {Linke}}, \bibinfo {author} {\bibfnamefont {K.~A.}\ \bibnamefont {Landsman}}, \bibinfo {author} {\bibfnamefont {D.}~\bibnamefont {Zhu}}, \bibinfo {author} {\bibfnamefont {D.}~\bibnamefont {Maslov}}, \ and\ \bibinfo {author} {\bibfnamefont {C.}~\bibnamefont {Monroe}},\ }\href {\doibase 10.1038/s41586-019-1427-5} {\bibfield  {journal} {\bibinfo  {journal} {Nature}\ }\textbf {\bibinfo {volume} {572}},\ \bibinfo {pages} {368} (\bibinfo {year} {2019})}\BibitemShut {NoStop}%
\bibitem [{\citenamefont {Pogorelov}\ \emph {et~al.}(2021)\citenamefont {Pogorelov}, \citenamefont {Feldker}, \citenamefont {Marciniak}, \citenamefont {Postler}, \citenamefont {Jacob}, \citenamefont {Krieglsteiner}, \citenamefont {Podlesnic}, \citenamefont {Meth}, \citenamefont {Negnevitsky}, \citenamefont {Stadler}, \citenamefont {H\"ofer}, \citenamefont {W\"achter}, \citenamefont {Lakhmanskiy}, \citenamefont {Blatt}, \citenamefont {Schindler},\ and\ \citenamefont {Monz}}]{PRXQuantum.2.020343}%
  \BibitemOpen
  \bibfield  {author} {\bibinfo {author} {\bibfnamefont {I.}~\bibnamefont {Pogorelov}}, \bibinfo {author} {\bibfnamefont {T.}~\bibnamefont {Feldker}}, \bibinfo {author} {\bibfnamefont {C.~D.}\ \bibnamefont {Marciniak}}, \bibinfo {author} {\bibfnamefont {L.}~\bibnamefont {Postler}}, \bibinfo {author} {\bibfnamefont {G.}~\bibnamefont {Jacob}}, \bibinfo {author} {\bibfnamefont {O.}~\bibnamefont {Krieglsteiner}}, \bibinfo {author} {\bibfnamefont {V.}~\bibnamefont {Podlesnic}}, \bibinfo {author} {\bibfnamefont {M.}~\bibnamefont {Meth}}, \bibinfo {author} {\bibfnamefont {V.}~\bibnamefont {Negnevitsky}}, \bibinfo {author} {\bibfnamefont {M.}~\bibnamefont {Stadler}}, \bibinfo {author} {\bibfnamefont {B.}~\bibnamefont {H\"ofer}}, \bibinfo {author} {\bibfnamefont {C.}~\bibnamefont {W\"achter}}, \bibinfo {author} {\bibfnamefont {K.}~\bibnamefont {Lakhmanskiy}}, \bibinfo {author} {\bibfnamefont {R.}~\bibnamefont {Blatt}}, \bibinfo {author} {\bibfnamefont {P.}~\bibnamefont {Schindler}}, \ and\ \bibinfo {author}
  {\bibfnamefont {T.}~\bibnamefont {Monz}},\ }\href {\doibase 10.1103/PRXQuantum.2.020343} {\bibfield  {journal} {\bibinfo  {journal} {PRX Quantum}\ }\textbf {\bibinfo {volume} {2}},\ \bibinfo {pages} {020343} (\bibinfo {year} {2021})}\BibitemShut {NoStop}%
\bibitem [{\citenamefont {Hempel}\ \emph {et~al.}(2018)\citenamefont {Hempel}, \citenamefont {Maier}, \citenamefont {Romero}, \citenamefont {McClean}, \citenamefont {Monz}, \citenamefont {Shen}, \citenamefont {Jurcevic}, \citenamefont {Lanyon}, \citenamefont {Love}, \citenamefont {Babbush}, \citenamefont {Aspuru-Guzik}, \citenamefont {Blatt},\ and\ \citenamefont {Roos}}]{PhysRevX.8.031022}%
  \BibitemOpen
  \bibfield  {author} {\bibinfo {author} {\bibfnamefont {C.}~\bibnamefont {Hempel}}, \bibinfo {author} {\bibfnamefont {C.}~\bibnamefont {Maier}}, \bibinfo {author} {\bibfnamefont {J.}~\bibnamefont {Romero}}, \bibinfo {author} {\bibfnamefont {J.}~\bibnamefont {McClean}}, \bibinfo {author} {\bibfnamefont {T.}~\bibnamefont {Monz}}, \bibinfo {author} {\bibfnamefont {H.}~\bibnamefont {Shen}}, \bibinfo {author} {\bibfnamefont {P.}~\bibnamefont {Jurcevic}}, \bibinfo {author} {\bibfnamefont {B.~P.}\ \bibnamefont {Lanyon}}, \bibinfo {author} {\bibfnamefont {P.}~\bibnamefont {Love}}, \bibinfo {author} {\bibfnamefont {R.}~\bibnamefont {Babbush}}, \bibinfo {author} {\bibfnamefont {A.}~\bibnamefont {Aspuru-Guzik}}, \bibinfo {author} {\bibfnamefont {R.}~\bibnamefont {Blatt}}, \ and\ \bibinfo {author} {\bibfnamefont {C.~F.}\ \bibnamefont {Roos}},\ }\href {\doibase 10.1103/PhysRevX.8.031022} {\bibfield  {journal} {\bibinfo  {journal} {Phys. Rev. X}\ }\textbf {\bibinfo {volume} {8}},\ \bibinfo {pages} {031022} (\bibinfo
  {year} {2018})}\BibitemShut {NoStop}%
\bibitem [{\citenamefont {Nam}\ \emph {et~al.}(2020)\citenamefont {Nam}, \citenamefont {Chen}, \citenamefont {Pisenti}, \citenamefont {Wright}, \citenamefont {Delaney}, \citenamefont {Maslov}, \citenamefont {Brown}, \citenamefont {Allen}, \citenamefont {Amini}, \citenamefont {Apisdorf}, \citenamefont {Beck}, \citenamefont {Blinov}, \citenamefont {Chaplin}, \citenamefont {Chmielewski}, \citenamefont {Collins}, \citenamefont {Debnath}, \citenamefont {Hudek}, \citenamefont {Ducore}, \citenamefont {Keesan}, \citenamefont {Kreikemeier}, \citenamefont {Mizrahi}, \citenamefont {Solomon}, \citenamefont {Williams}, \citenamefont {Wong-Campos}, \citenamefont {Moehring}, \citenamefont {Monroe},\ and\ \citenamefont {Kim}}]{Nam2020}%
  \BibitemOpen
  \bibfield  {author} {\bibinfo {author} {\bibfnamefont {Y.}~\bibnamefont {Nam}}, \bibinfo {author} {\bibfnamefont {J.-S.}\ \bibnamefont {Chen}}, \bibinfo {author} {\bibfnamefont {N.~C.}\ \bibnamefont {Pisenti}}, \bibinfo {author} {\bibfnamefont {K.}~\bibnamefont {Wright}}, \bibinfo {author} {\bibfnamefont {C.}~\bibnamefont {Delaney}}, \bibinfo {author} {\bibfnamefont {D.}~\bibnamefont {Maslov}}, \bibinfo {author} {\bibfnamefont {K.~R.}\ \bibnamefont {Brown}}, \bibinfo {author} {\bibfnamefont {S.}~\bibnamefont {Allen}}, \bibinfo {author} {\bibfnamefont {J.~M.}\ \bibnamefont {Amini}}, \bibinfo {author} {\bibfnamefont {J.}~\bibnamefont {Apisdorf}}, \bibinfo {author} {\bibfnamefont {K.~M.}\ \bibnamefont {Beck}}, \bibinfo {author} {\bibfnamefont {A.}~\bibnamefont {Blinov}}, \bibinfo {author} {\bibfnamefont {V.}~\bibnamefont {Chaplin}}, \bibinfo {author} {\bibfnamefont {M.}~\bibnamefont {Chmielewski}}, \bibinfo {author} {\bibfnamefont {C.}~\bibnamefont {Collins}}, \bibinfo {author} {\bibfnamefont {S.}~\bibnamefont
  {Debnath}}, \bibinfo {author} {\bibfnamefont {K.~M.}\ \bibnamefont {Hudek}}, \bibinfo {author} {\bibfnamefont {A.~M.}\ \bibnamefont {Ducore}}, \bibinfo {author} {\bibfnamefont {M.}~\bibnamefont {Keesan}}, \bibinfo {author} {\bibfnamefont {S.~M.}\ \bibnamefont {Kreikemeier}}, \bibinfo {author} {\bibfnamefont {J.}~\bibnamefont {Mizrahi}}, \bibinfo {author} {\bibfnamefont {P.}~\bibnamefont {Solomon}}, \bibinfo {author} {\bibfnamefont {M.}~\bibnamefont {Williams}}, \bibinfo {author} {\bibfnamefont {J.~D.}\ \bibnamefont {Wong-Campos}}, \bibinfo {author} {\bibfnamefont {D.}~\bibnamefont {Moehring}}, \bibinfo {author} {\bibfnamefont {C.}~\bibnamefont {Monroe}}, \ and\ \bibinfo {author} {\bibfnamefont {J.}~\bibnamefont {Kim}},\ }\href {\doibase 10.1038/s41534-020-0259-3} {\bibfield  {journal} {\bibinfo  {journal} {npj Quantum Information}\ }\textbf {\bibinfo {volume} {6}},\ \bibinfo {pages} {33} (\bibinfo {year} {2020})}\BibitemShut {NoStop}%
\bibitem [{\citenamefont {Kokail}\ \emph {et~al.}(2019)\citenamefont {Kokail}, \citenamefont {Maier}, \citenamefont {van Bijnen}, \citenamefont {Brydges}, \citenamefont {Joshi}, \citenamefont {Jurcevic}, \citenamefont {Muschik}, \citenamefont {Silvi}, \citenamefont {Blatt}, \citenamefont {Roos},\ and\ \citenamefont {Zoller}}]{Kokail2019}%
  \BibitemOpen
  \bibfield  {author} {\bibinfo {author} {\bibfnamefont {C.}~\bibnamefont {Kokail}}, \bibinfo {author} {\bibfnamefont {C.}~\bibnamefont {Maier}}, \bibinfo {author} {\bibfnamefont {R.}~\bibnamefont {van Bijnen}}, \bibinfo {author} {\bibfnamefont {T.}~\bibnamefont {Brydges}}, \bibinfo {author} {\bibfnamefont {M.~K.}\ \bibnamefont {Joshi}}, \bibinfo {author} {\bibfnamefont {P.}~\bibnamefont {Jurcevic}}, \bibinfo {author} {\bibfnamefont {C.~A.}\ \bibnamefont {Muschik}}, \bibinfo {author} {\bibfnamefont {P.}~\bibnamefont {Silvi}}, \bibinfo {author} {\bibfnamefont {R.}~\bibnamefont {Blatt}}, \bibinfo {author} {\bibfnamefont {C.~F.}\ \bibnamefont {Roos}}, \ and\ \bibinfo {author} {\bibfnamefont {P.}~\bibnamefont {Zoller}},\ }\href {\doibase 10.1038/s41586-019-1177-4} {\bibfield  {journal} {\bibinfo  {journal} {Nature}\ }\textbf {\bibinfo {volume} {569}},\ \bibinfo {pages} {355} (\bibinfo {year} {2019})}\BibitemShut {NoStop}%
\bibitem [{\citenamefont {Wiese}(2013)}]{https://doi.org/10.1002/andp.201300104}%
  \BibitemOpen
  \bibfield  {author} {\bibinfo {author} {\bibfnamefont {U.-J.}\ \bibnamefont {Wiese}},\ }\href {\doibase https://doi.org/10.1002/andp.201300104} {\bibfield  {journal} {\bibinfo  {journal} {Annalen der Physik}\ }\textbf {\bibinfo {volume} {525}},\ \bibinfo {pages} {777} (\bibinfo {year} {2013})}\BibitemShut {NoStop}%
\bibitem [{\citenamefont {Zohar}\ \emph {et~al.}(2015)\citenamefont {Zohar}, \citenamefont {Cirac},\ and\ \citenamefont {Reznik}}]{Zohar_2016}%
  \BibitemOpen
  \bibfield  {author} {\bibinfo {author} {\bibfnamefont {E.}~\bibnamefont {Zohar}}, \bibinfo {author} {\bibfnamefont {J.~I.}\ \bibnamefont {Cirac}}, \ and\ \bibinfo {author} {\bibfnamefont {B.}~\bibnamefont {Reznik}},\ }\href {\doibase 10.1088/0034-4885/79/1/014401} {\bibfield  {journal} {\bibinfo  {journal} {Reports on Progress in Physics}\ }\textbf {\bibinfo {volume} {79}},\ \bibinfo {pages} {014401} (\bibinfo {year} {2015})}\BibitemShut {NoStop}%
\bibitem [{\citenamefont {Dalmonte}\ and\ \citenamefont {Montangero}(2016)}]{doi:10.1080/00107514.2016.1151199}%
  \BibitemOpen
  \bibfield  {author} {\bibinfo {author} {\bibfnamefont {M.}~\bibnamefont {Dalmonte}}\ and\ \bibinfo {author} {\bibfnamefont {S.}~\bibnamefont {Montangero}},\ }\href {\doibase 10.1080/00107514.2016.1151199} {\bibfield  {journal} {\bibinfo  {journal} {Contemporary Physics}\ }\textbf {\bibinfo {volume} {57}},\ \bibinfo {pages} {388} (\bibinfo {year} {2016})}\BibitemShut {NoStop}%
\bibitem [{\citenamefont {Bañuls}\ \emph {et~al.}(2020)\citenamefont {Bañuls}, \citenamefont {Blatt}, \citenamefont {Catani}, \citenamefont {Celi}, \citenamefont {Cirac}, \citenamefont {Dalmonte}, \citenamefont {Fallani}, \citenamefont {Jansen}, \citenamefont {Lewenstein}, \citenamefont {Montangero}, \citenamefont {Muschik}, \citenamefont {Reznik}, \citenamefont {Rico~Ortega}, \citenamefont {Tagliacozzo}, \citenamefont {Acoleyen}, \citenamefont {Verstraete}, \citenamefont {Wiese}, \citenamefont {Wingate}, \citenamefont {Zakrzewski},\ and\ \citenamefont {Zoller}}]{Bannuls2020}%
  \BibitemOpen
  \bibfield  {author} {\bibinfo {author} {\bibfnamefont {M.-C.}\ \bibnamefont {Bañuls}}, \bibinfo {author} {\bibfnamefont {R.}~\bibnamefont {Blatt}}, \bibinfo {author} {\bibfnamefont {J.}~\bibnamefont {Catani}}, \bibinfo {author} {\bibfnamefont {A.}~\bibnamefont {Celi}}, \bibinfo {author} {\bibfnamefont {J.}~\bibnamefont {Cirac}}, \bibinfo {author} {\bibfnamefont {M.}~\bibnamefont {Dalmonte}}, \bibinfo {author} {\bibfnamefont {L.}~\bibnamefont {Fallani}}, \bibinfo {author} {\bibfnamefont {K.}~\bibnamefont {Jansen}}, \bibinfo {author} {\bibfnamefont {M.}~\bibnamefont {Lewenstein}}, \bibinfo {author} {\bibfnamefont {S.}~\bibnamefont {Montangero}}, \bibinfo {author} {\bibfnamefont {C.}~\bibnamefont {Muschik}}, \bibinfo {author} {\bibfnamefont {B.}~\bibnamefont {Reznik}}, \bibinfo {author} {\bibfnamefont {E.}~\bibnamefont {Rico~Ortega}}, \bibinfo {author} {\bibfnamefont {L.}~\bibnamefont {Tagliacozzo}}, \bibinfo {author} {\bibfnamefont {K.}~\bibnamefont {Acoleyen}}, \bibinfo {author} {\bibfnamefont
  {F.}~\bibnamefont {Verstraete}}, \bibinfo {author} {\bibfnamefont {U.-J.}\ \bibnamefont {Wiese}}, \bibinfo {author} {\bibfnamefont {M.}~\bibnamefont {Wingate}}, \bibinfo {author} {\bibfnamefont {J.}~\bibnamefont {Zakrzewski}}, \ and\ \bibinfo {author} {\bibfnamefont {P.}~\bibnamefont {Zoller}},\ }\href {\doibase 10.1140/epjd/e2020-100571-8} {\bibfield  {journal} {\bibinfo  {journal} {The European Physical Journal D}\ }\textbf {\bibinfo {volume} {74}} (\bibinfo {year} {2020}),\ 10.1140/epjd/e2020-100571-8}\BibitemShut {NoStop}%
\bibitem [{\citenamefont {Ba{\~{n}}uls}\ and\ \citenamefont {Cichy}(2020)}]{Carmen_Ba_uls_2020}%
  \BibitemOpen
  \bibfield  {author} {\bibinfo {author} {\bibfnamefont {M.~C.}\ \bibnamefont {Ba{\~{n}}uls}}\ and\ \bibinfo {author} {\bibfnamefont {K.}~\bibnamefont {Cichy}},\ }\href {\doibase 10.1088/1361-6633/ab6311} {\bibfield  {journal} {\bibinfo  {journal} {Reports on Progress in Physics}\ }\textbf {\bibinfo {volume} {83}},\ \bibinfo {pages} {024401} (\bibinfo {year} {2020})}\BibitemShut {NoStop}%
\bibitem [{\citenamefont {Aidelsburger}\ \emph {et~al.}(2022)\citenamefont {Aidelsburger}, \citenamefont {Barbiero}, \citenamefont {Bermudez}, \citenamefont {Chanda}, \citenamefont {Dauphin}, \citenamefont {González-Cuadra}, \citenamefont {Grzybowski}, \citenamefont {Hands}, \citenamefont {Jendrzejewski}, \citenamefont {Jünemann}, \citenamefont {Juzeliūnas}, \citenamefont {Kasper}, \citenamefont {Piga}, \citenamefont {Ran}, \citenamefont {Rizzi}, \citenamefont {Sierra}, \citenamefont {Tagliacozzo}, \citenamefont {Tirrito}, \citenamefont {Zache}, \citenamefont {Zakrzewski}, \citenamefont {Zohar},\ and\ \citenamefont {Lewenstein}}]{doi:10.1098/rsta.2021.0064}%
  \BibitemOpen
  \bibfield  {author} {\bibinfo {author} {\bibfnamefont {M.}~\bibnamefont {Aidelsburger}}, \bibinfo {author} {\bibfnamefont {L.}~\bibnamefont {Barbiero}}, \bibinfo {author} {\bibfnamefont {A.}~\bibnamefont {Bermudez}}, \bibinfo {author} {\bibfnamefont {T.}~\bibnamefont {Chanda}}, \bibinfo {author} {\bibfnamefont {A.}~\bibnamefont {Dauphin}}, \bibinfo {author} {\bibfnamefont {D.}~\bibnamefont {González-Cuadra}}, \bibinfo {author} {\bibfnamefont {P.~R.}\ \bibnamefont {Grzybowski}}, \bibinfo {author} {\bibfnamefont {S.}~\bibnamefont {Hands}}, \bibinfo {author} {\bibfnamefont {F.}~\bibnamefont {Jendrzejewski}}, \bibinfo {author} {\bibfnamefont {J.}~\bibnamefont {Jünemann}}, \bibinfo {author} {\bibfnamefont {G.}~\bibnamefont {Juzeliūnas}}, \bibinfo {author} {\bibfnamefont {V.}~\bibnamefont {Kasper}}, \bibinfo {author} {\bibfnamefont {A.}~\bibnamefont {Piga}}, \bibinfo {author} {\bibfnamefont {S.-J.}\ \bibnamefont {Ran}}, \bibinfo {author} {\bibfnamefont {M.}~\bibnamefont {Rizzi}}, \bibinfo {author}
  {\bibfnamefont {G.}~\bibnamefont {Sierra}}, \bibinfo {author} {\bibfnamefont {L.}~\bibnamefont {Tagliacozzo}}, \bibinfo {author} {\bibfnamefont {E.}~\bibnamefont {Tirrito}}, \bibinfo {author} {\bibfnamefont {T.~V.}\ \bibnamefont {Zache}}, \bibinfo {author} {\bibfnamefont {J.}~\bibnamefont {Zakrzewski}}, \bibinfo {author} {\bibfnamefont {E.}~\bibnamefont {Zohar}}, \ and\ \bibinfo {author} {\bibfnamefont {M.}~\bibnamefont {Lewenstein}},\ }\href {\doibase 10.1098/rsta.2021.0064} {\bibfield  {journal} {\bibinfo  {journal} {Philosophical Transactions of the Royal Society A: Mathematical, Physical and Engineering Sciences}\ }\textbf {\bibinfo {volume} {380}},\ \bibinfo {pages} {20210064} (\bibinfo {year} {2022})}\BibitemShut {NoStop}%
\bibitem [{\citenamefont {Klco}\ \emph {et~al.}(2022)\citenamefont {Klco}, \citenamefont {Roggero},\ and\ \citenamefont {Savage}}]{Klco_2022}%
  \BibitemOpen
  \bibfield  {author} {\bibinfo {author} {\bibfnamefont {N.}~\bibnamefont {Klco}}, \bibinfo {author} {\bibfnamefont {A.}~\bibnamefont {Roggero}}, \ and\ \bibinfo {author} {\bibfnamefont {M.~J.}\ \bibnamefont {Savage}},\ }\href {\doibase 10.1088/1361-6633/ac58a4} {\bibfield  {journal} {\bibinfo  {journal} {Reports on Progress in Physics}\ }\textbf {\bibinfo {volume} {85}},\ \bibinfo {pages} {064301} (\bibinfo {year} {2022})}\BibitemShut {NoStop}%
\bibitem [{\citenamefont {Bauer}\ \emph {et~al.}(2022)\citenamefont {Bauer}, \citenamefont {Davoudi}, \citenamefont {Balantekin}, \citenamefont {Bhattacharya}, \citenamefont {Carena}, \citenamefont {de~Jong}, \citenamefont {Draper}, \citenamefont {El-Khadra}, \citenamefont {Gemelke}, \citenamefont {Hanada}, \citenamefont {Kharzeev}, \citenamefont {Lamm}, \citenamefont {Li}, \citenamefont {Liu}, \citenamefont {Lukin}, \citenamefont {Meurice}, \citenamefont {Monroe}, \citenamefont {Nachman}, \citenamefont {Pagano}, \citenamefont {Preskill}, \citenamefont {Rinaldi}, \citenamefont {Roggero}, \citenamefont {Santiago}, \citenamefont {Savage}, \citenamefont {Siddiqi}, \citenamefont {Siopsis}, \citenamefont {Van~Zanten}, \citenamefont {Wiebe}, \citenamefont {Yamauchi}, \citenamefont {Yeter-Aydeniz},\ and\ \citenamefont {Zorzetti}}]{https://doi.org/10.48550/arxiv.2204.03381}%
  \BibitemOpen
  \bibfield  {author} {\bibinfo {author} {\bibfnamefont {C.~W.}\ \bibnamefont {Bauer}}, \bibinfo {author} {\bibfnamefont {Z.}~\bibnamefont {Davoudi}}, \bibinfo {author} {\bibfnamefont {A.~B.}\ \bibnamefont {Balantekin}}, \bibinfo {author} {\bibfnamefont {T.}~\bibnamefont {Bhattacharya}}, \bibinfo {author} {\bibfnamefont {M.}~\bibnamefont {Carena}}, \bibinfo {author} {\bibfnamefont {W.~A.}\ \bibnamefont {de~Jong}}, \bibinfo {author} {\bibfnamefont {P.}~\bibnamefont {Draper}}, \bibinfo {author} {\bibfnamefont {A.}~\bibnamefont {El-Khadra}}, \bibinfo {author} {\bibfnamefont {N.}~\bibnamefont {Gemelke}}, \bibinfo {author} {\bibfnamefont {M.}~\bibnamefont {Hanada}}, \bibinfo {author} {\bibfnamefont {D.}~\bibnamefont {Kharzeev}}, \bibinfo {author} {\bibfnamefont {H.}~\bibnamefont {Lamm}}, \bibinfo {author} {\bibfnamefont {Y.-Y.}\ \bibnamefont {Li}}, \bibinfo {author} {\bibfnamefont {J.}~\bibnamefont {Liu}}, \bibinfo {author} {\bibfnamefont {M.}~\bibnamefont {Lukin}}, \bibinfo {author} {\bibfnamefont
  {Y.}~\bibnamefont {Meurice}}, \bibinfo {author} {\bibfnamefont {C.}~\bibnamefont {Monroe}}, \bibinfo {author} {\bibfnamefont {B.}~\bibnamefont {Nachman}}, \bibinfo {author} {\bibfnamefont {G.}~\bibnamefont {Pagano}}, \bibinfo {author} {\bibfnamefont {J.}~\bibnamefont {Preskill}}, \bibinfo {author} {\bibfnamefont {E.}~\bibnamefont {Rinaldi}}, \bibinfo {author} {\bibfnamefont {A.}~\bibnamefont {Roggero}}, \bibinfo {author} {\bibfnamefont {D.~I.}\ \bibnamefont {Santiago}}, \bibinfo {author} {\bibfnamefont {M.~J.}\ \bibnamefont {Savage}}, \bibinfo {author} {\bibfnamefont {I.}~\bibnamefont {Siddiqi}}, \bibinfo {author} {\bibfnamefont {G.}~\bibnamefont {Siopsis}}, \bibinfo {author} {\bibfnamefont {D.}~\bibnamefont {Van~Zanten}}, \bibinfo {author} {\bibfnamefont {N.}~\bibnamefont {Wiebe}}, \bibinfo {author} {\bibfnamefont {Y.}~\bibnamefont {Yamauchi}}, \bibinfo {author} {\bibfnamefont {K.}~\bibnamefont {Yeter-Aydeniz}}, \ and\ \bibinfo {author} {\bibfnamefont {S.}~\bibnamefont {Zorzetti}},\ }\href {\doibase
  10.1103/PRXQuantum.4.027001} {\  (\bibinfo {year} {2022}),\ 10.1103/PRXQuantum.4.027001}\BibitemShut {NoStop}%
\bibitem [{\citenamefont {Leibfried}\ \emph {et~al.}(2003{\natexlab{b}})\citenamefont {Leibfried}, \citenamefont {DeMarco}, \citenamefont {Meyer}, \citenamefont {Lucas}, \citenamefont {Barrett}, \citenamefont {Britton}, \citenamefont {Itano}, \citenamefont {Jelenkovi{\'{c}}}, \citenamefont {Langer}, \citenamefont {Rosenband},\ and\ \citenamefont {Wineland}}]{Leibfried2003}%
  \BibitemOpen
  \bibfield  {author} {\bibinfo {author} {\bibfnamefont {D.}~\bibnamefont {Leibfried}}, \bibinfo {author} {\bibfnamefont {B.}~\bibnamefont {DeMarco}}, \bibinfo {author} {\bibfnamefont {V.}~\bibnamefont {Meyer}}, \bibinfo {author} {\bibfnamefont {D.}~\bibnamefont {Lucas}}, \bibinfo {author} {\bibfnamefont {M.}~\bibnamefont {Barrett}}, \bibinfo {author} {\bibfnamefont {J.}~\bibnamefont {Britton}}, \bibinfo {author} {\bibfnamefont {W.~M.}\ \bibnamefont {Itano}}, \bibinfo {author} {\bibfnamefont {B.}~\bibnamefont {Jelenkovi{\'{c}}}}, \bibinfo {author} {\bibfnamefont {C.}~\bibnamefont {Langer}}, \bibinfo {author} {\bibfnamefont {T.}~\bibnamefont {Rosenband}}, \ and\ \bibinfo {author} {\bibfnamefont {D.~J.}\ \bibnamefont {Wineland}},\ }\href {\doibase 10.1038/nature01492} {\bibfield  {journal} {\bibinfo  {journal} {Nature}\ }\textbf {\bibinfo {volume} {422}},\ \bibinfo {pages} {412} (\bibinfo {year} {2003}{\natexlab{b}})}\BibitemShut {NoStop}%
\bibitem [{\citenamefont {Wineland}\ \emph {et~al.}(2003)\citenamefont {Wineland}, \citenamefont {Barrett}, \citenamefont {Britton}, \citenamefont {Chiaverini}, \citenamefont {DeMarco}, \citenamefont {Itano}, \citenamefont {Jelenković}, \citenamefont {Langer}, \citenamefont {Leibfried}, \citenamefont {Meyer}, \citenamefont {Rosenband},\ and\ \citenamefont {Schätz}}]{doi:10.1098/rsta.2003.1205}%
  \BibitemOpen
  \bibfield  {author} {\bibinfo {author} {\bibfnamefont {D.~J.}\ \bibnamefont {Wineland}}, \bibinfo {author} {\bibfnamefont {M.}~\bibnamefont {Barrett}}, \bibinfo {author} {\bibfnamefont {J.}~\bibnamefont {Britton}}, \bibinfo {author} {\bibfnamefont {J.}~\bibnamefont {Chiaverini}}, \bibinfo {author} {\bibfnamefont {B.}~\bibnamefont {DeMarco}}, \bibinfo {author} {\bibfnamefont {W.~M.}\ \bibnamefont {Itano}}, \bibinfo {author} {\bibfnamefont {B.}~\bibnamefont {Jelenković}}, \bibinfo {author} {\bibfnamefont {C.}~\bibnamefont {Langer}}, \bibinfo {author} {\bibfnamefont {D.}~\bibnamefont {Leibfried}}, \bibinfo {author} {\bibfnamefont {V.}~\bibnamefont {Meyer}}, \bibinfo {author} {\bibfnamefont {T.}~\bibnamefont {Rosenband}}, \ and\ \bibinfo {author} {\bibfnamefont {T.}~\bibnamefont {Schätz}},\ }\href {\doibase 10.1098/rsta.2003.1205} {\bibfield  {journal} {\bibinfo  {journal} {Philosophical Transactions of the Royal Society of London. Series A: Mathematical, Physical and Engineering Sciences}\ }\textbf {\bibinfo
  {volume} {361}},\ \bibinfo {pages} {1349} (\bibinfo {year} {2003})}\BibitemShut {NoStop}%
\bibitem [{\citenamefont {Trotzky}\ \emph {et~al.}(2012)\citenamefont {Trotzky}, \citenamefont {Chen}, \citenamefont {Flesch}, \citenamefont {McCulloch}, \citenamefont {Schollw{\"o}ck}, \citenamefont {Eisert},\ and\ \citenamefont {Bloch}}]{Trotzky2012}%
  \BibitemOpen
  \bibfield  {author} {\bibinfo {author} {\bibfnamefont {S.}~\bibnamefont {Trotzky}}, \bibinfo {author} {\bibfnamefont {Y.-A.}\ \bibnamefont {Chen}}, \bibinfo {author} {\bibfnamefont {A.}~\bibnamefont {Flesch}}, \bibinfo {author} {\bibfnamefont {I.~P.}\ \bibnamefont {McCulloch}}, \bibinfo {author} {\bibfnamefont {U.}~\bibnamefont {Schollw{\"o}ck}}, \bibinfo {author} {\bibfnamefont {J.}~\bibnamefont {Eisert}}, \ and\ \bibinfo {author} {\bibfnamefont {I.}~\bibnamefont {Bloch}},\ }\href {\doibase 10.1038/nphys2232} {\bibfield  {journal} {\bibinfo  {journal} {Nature Physics}\ }\textbf {\bibinfo {volume} {8}},\ \bibinfo {pages} {325} (\bibinfo {year} {2012})}\BibitemShut {NoStop}%
\bibitem [{\citenamefont {Flannigan}\ \emph {et~al.}(2022)\citenamefont {Flannigan}, \citenamefont {Pearson}, \citenamefont {Low}, \citenamefont {Buyskikh}, \citenamefont {Bloch}, \citenamefont {Zoller}, \citenamefont {Troyer},\ and\ \citenamefont {Daley}}]{https://doi.org/10.48550/arxiv.2204.13644}%
  \BibitemOpen
  \bibfield  {author} {\bibinfo {author} {\bibfnamefont {S.}~\bibnamefont {Flannigan}}, \bibinfo {author} {\bibfnamefont {N.}~\bibnamefont {Pearson}}, \bibinfo {author} {\bibfnamefont {G.~H.}\ \bibnamefont {Low}}, \bibinfo {author} {\bibfnamefont {A.}~\bibnamefont {Buyskikh}}, \bibinfo {author} {\bibfnamefont {I.}~\bibnamefont {Bloch}}, \bibinfo {author} {\bibfnamefont {P.}~\bibnamefont {Zoller}}, \bibinfo {author} {\bibfnamefont {M.}~\bibnamefont {Troyer}}, \ and\ \bibinfo {author} {\bibfnamefont {A.~J.}\ \bibnamefont {Daley}},\ }\href {\doibase 10.1088/2058-9565/ac88f5} {\  (\bibinfo {year} {2022}),\ 10.1088/2058-9565/ac88f5}\BibitemShut {NoStop}%
\bibitem [{\citenamefont {Trivedi}\ \emph {et~al.}(2022)\citenamefont {Trivedi}, \citenamefont {Rubio},\ and\ \citenamefont {Cirac}}]{https://doi.org/10.48550/arxiv.2212.04924}%
  \BibitemOpen
  \bibfield  {author} {\bibinfo {author} {\bibfnamefont {R.}~\bibnamefont {Trivedi}}, \bibinfo {author} {\bibfnamefont {A.~F.}\ \bibnamefont {Rubio}}, \ and\ \bibinfo {author} {\bibfnamefont {J.~I.}\ \bibnamefont {Cirac}},\ }\href {\doibase 10.48550/ARXIV.2212.04924} {\  (\bibinfo {year} {2022}),\ 10.48550/ARXIV.2212.04924}\BibitemShut {NoStop}%
\bibitem [{\citenamefont {Fradkin}(2021)}]{fradkin}%
  \BibitemOpen
  \bibfield  {author} {\bibinfo {author} {\bibfnamefont {E.}~\bibnamefont {Fradkin}},\ }\href {https://books.google.es/books?id=quEIEAAAQBAJ} {\emph {\bibinfo {title} {Quantum Field Theory: An Integrated Approach}}}\ (\bibinfo  {publisher} {Princeton University Press},\ \bibinfo {year} {2021})\BibitemShut {NoStop}%
\bibitem [{\citenamefont {Peskin}\ and\ \citenamefont {Schroeder}(1995)}]{peskin}%
  \BibitemOpen
  \bibfield  {author} {\bibinfo {author} {\bibfnamefont {M.}~\bibnamefont {Peskin}}\ and\ \bibinfo {author} {\bibfnamefont {D.}~\bibnamefont {Schroeder}},\ }\href {https://books.google.ch/books?id=EVeNNcslvX0C} {\emph {\bibinfo {title} {An Introduction To Quantum Field Theory}}},\ Frontiers in Physics\ (\bibinfo  {publisher} {Avalon Publishing},\ \bibinfo {year} {1995})\BibitemShut {NoStop}%
\bibitem [{\citenamefont {Jordan}\ and\ \citenamefont {Wigner}(1928)}]{Jordan1928}%
  \BibitemOpen
  \bibfield  {author} {\bibinfo {author} {\bibfnamefont {P.}~\bibnamefont {Jordan}}\ and\ \bibinfo {author} {\bibfnamefont {E.}~\bibnamefont {Wigner}},\ }\href {\doibase 10.1007/BF01331938} {\bibfield  {journal} {\bibinfo  {journal} {Zeitschrift f{\"u}r Physik}\ }\textbf {\bibinfo {volume} {47}},\ \bibinfo {pages} {631} (\bibinfo {year} {1928})}\BibitemShut {NoStop}%
\bibitem [{\citenamefont {Kranzl}\ \emph {et~al.}(2022)\citenamefont {Kranzl}, \citenamefont {Joshi}, \citenamefont {Maier}, \citenamefont {Brydges}, \citenamefont {Franke}, \citenamefont {Blatt},\ and\ \citenamefont {Roos}}]{insbruck}%
  \BibitemOpen
  \bibfield  {author} {\bibinfo {author} {\bibfnamefont {F.}~\bibnamefont {Kranzl}}, \bibinfo {author} {\bibfnamefont {M.~K.}\ \bibnamefont {Joshi}}, \bibinfo {author} {\bibfnamefont {C.}~\bibnamefont {Maier}}, \bibinfo {author} {\bibfnamefont {T.}~\bibnamefont {Brydges}}, \bibinfo {author} {\bibfnamefont {J.}~\bibnamefont {Franke}}, \bibinfo {author} {\bibfnamefont {R.}~\bibnamefont {Blatt}}, \ and\ \bibinfo {author} {\bibfnamefont {C.~F.}\ \bibnamefont {Roos}},\ }\href {\doibase 10.1103/PhysRevA.105.052426} {\bibfield  {journal} {\bibinfo  {journal} {Phys. Rev. A}\ }\textbf {\bibinfo {volume} {105}},\ \bibinfo {pages} {052426} (\bibinfo {year} {2022})}\BibitemShut {NoStop}%
\bibitem [{\citenamefont {Ob{\v{s} }il}\ \emph {et~al.}(2019)\citenamefont {Ob{\v{s} }il}, \citenamefont {Le{\v{s}}und{\'{a}}k}, \citenamefont {Pham}, \citenamefont {Lakhmanskiy}, \citenamefont {Podhora}, \citenamefont {Oral}, \citenamefont {{\v{C}}{\'{\i}}p},\ and\ \citenamefont {Slodi{\v{c}}ka}}]{Ob_il_2019}%
  \BibitemOpen
  \bibfield  {author} {\bibinfo {author} {\bibfnamefont {P.}~\bibnamefont {Ob{\v{s} }il}}, \bibinfo {author} {\bibfnamefont {A.}~\bibnamefont {Le{\v{s}}und{\'{a}}k}}, \bibinfo {author} {\bibfnamefont {T.}~\bibnamefont {Pham}}, \bibinfo {author} {\bibfnamefont {K.}~\bibnamefont {Lakhmanskiy}}, \bibinfo {author} {\bibfnamefont {L.}~\bibnamefont {Podhora}}, \bibinfo {author} {\bibfnamefont {M.}~\bibnamefont {Oral}}, \bibinfo {author} {\bibfnamefont {O.}~\bibnamefont {{\v{C}}{\'{\i}}p}}, \ and\ \bibinfo {author} {\bibfnamefont {L.}~\bibnamefont {Slodi{\v{c}}ka}},\ }\href {\doibase 10.1063/1.5104346} {\bibfield  {journal} {\bibinfo  {journal} {Review of Scientific Instruments}\ }\textbf {\bibinfo {volume} {90}},\ \bibinfo {pages} {083201} (\bibinfo {year} {2019})}\BibitemShut {NoStop}%
\bibitem [{\citenamefont {Hankin}\ \emph {et~al.}(2019)\citenamefont {Hankin}, \citenamefont {Clements}, \citenamefont {Huang}, \citenamefont {Brewer}, \citenamefont {Chen}, \citenamefont {Chou}, \citenamefont {Hume},\ and\ \citenamefont {Leibrandt}}]{PhysRevA.100.033419}%
  \BibitemOpen
  \bibfield  {author} {\bibinfo {author} {\bibfnamefont {A.~M.}\ \bibnamefont {Hankin}}, \bibinfo {author} {\bibfnamefont {E.~R.}\ \bibnamefont {Clements}}, \bibinfo {author} {\bibfnamefont {Y.}~\bibnamefont {Huang}}, \bibinfo {author} {\bibfnamefont {S.~M.}\ \bibnamefont {Brewer}}, \bibinfo {author} {\bibfnamefont {J.-S.}\ \bibnamefont {Chen}}, \bibinfo {author} {\bibfnamefont {C.~W.}\ \bibnamefont {Chou}}, \bibinfo {author} {\bibfnamefont {D.~B.}\ \bibnamefont {Hume}}, \ and\ \bibinfo {author} {\bibfnamefont {D.~R.}\ \bibnamefont {Leibrandt}},\ }\href {\doibase 10.1103/PhysRevA.100.033419} {\bibfield  {journal} {\bibinfo  {journal} {Phys. Rev. A}\ }\textbf {\bibinfo {volume} {100}},\ \bibinfo {pages} {033419} (\bibinfo {year} {2019})}\BibitemShut {NoStop}%
\bibitem [{\citenamefont {Leibfried}\ \emph {et~al.}(2001)\citenamefont {Leibfried}, \citenamefont {Roos}, \citenamefont {Barton}, \citenamefont {Rohde}, \citenamefont {Gulde}, \citenamefont {Mundt}, \citenamefont {Reymond}, \citenamefont {Lederbauer}, \citenamefont {Schmidt-Kaler}, \citenamefont {Eschner},\ and\ \citenamefont {Blatt}}]{doi:10.1063/1.1354345}%
  \BibitemOpen
  \bibfield  {author} {\bibinfo {author} {\bibfnamefont {D.}~\bibnamefont {Leibfried}}, \bibinfo {author} {\bibfnamefont {C.}~\bibnamefont {Roos}}, \bibinfo {author} {\bibfnamefont {P.}~\bibnamefont {Barton}}, \bibinfo {author} {\bibfnamefont {H.}~\bibnamefont {Rohde}}, \bibinfo {author} {\bibfnamefont {S.}~\bibnamefont {Gulde}}, \bibinfo {author} {\bibfnamefont {A.~B.}\ \bibnamefont {Mundt}}, \bibinfo {author} {\bibfnamefont {G.}~\bibnamefont {Reymond}}, \bibinfo {author} {\bibfnamefont {M.}~\bibnamefont {Lederbauer}}, \bibinfo {author} {\bibfnamefont {F.}~\bibnamefont {Schmidt-Kaler}}, \bibinfo {author} {\bibfnamefont {J.}~\bibnamefont {Eschner}}, \ and\ \bibinfo {author} {\bibfnamefont {R.}~\bibnamefont {Blatt}},\ }\href {\doibase 10.1063/1.1354345} {\bibfield  {journal} {\bibinfo  {journal} {AIP Conference Proceedings}\ }\textbf {\bibinfo {volume} {551}},\ \bibinfo {pages} {130} (\bibinfo {year} {2001})}\BibitemShut {NoStop}%
\bibitem [{\citenamefont {Joshi}\ \emph {et~al.}(2020)\citenamefont {Joshi}, \citenamefont {Fabre}, \citenamefont {Maier}, \citenamefont {Brydges}, \citenamefont {Kiesenhofer}, \citenamefont {Hainzer}, \citenamefont {Blatt},\ and\ \citenamefont {Roos}}]{polarization}%
  \BibitemOpen
  \bibfield  {author} {\bibinfo {author} {\bibfnamefont {M.~K.}\ \bibnamefont {Joshi}}, \bibinfo {author} {\bibfnamefont {A.}~\bibnamefont {Fabre}}, \bibinfo {author} {\bibfnamefont {C.}~\bibnamefont {Maier}}, \bibinfo {author} {\bibfnamefont {T.}~\bibnamefont {Brydges}}, \bibinfo {author} {\bibfnamefont {D.}~\bibnamefont {Kiesenhofer}}, \bibinfo {author} {\bibfnamefont {H.}~\bibnamefont {Hainzer}}, \bibinfo {author} {\bibfnamefont {R.}~\bibnamefont {Blatt}}, \ and\ \bibinfo {author} {\bibfnamefont {C.~F.}\ \bibnamefont {Roos}},\ }\href {\doibase 10.1088/1367-2630/abb912} {\bibfield  {journal} {\bibinfo  {journal} {New Journal of Physics}\ }\textbf {\bibinfo {volume} {22}},\ \bibinfo {pages} {103013} (\bibinfo {year} {2020})}\BibitemShut {NoStop}%
\bibitem [{\citenamefont {C.}\ \emph {et~al.}(2015)\citenamefont {C.}, \citenamefont {W.C.}, \citenamefont {E.E.}, \citenamefont {R.}, \citenamefont {D.}, \citenamefont {S.}, \citenamefont {A.}, \citenamefont {P.}, \citenamefont {C.},\ and\ \citenamefont {J.}}]{Monroe_varenna}%
  \BibitemOpen
  \bibfield  {author} {\bibinfo {author} {\bibfnamefont {M.}~\bibnamefont {C.}}, \bibinfo {author} {\bibfnamefont {C.}~\bibnamefont {W.C.}}, \bibinfo {author} {\bibfnamefont {E.}~\bibnamefont {E.E.}}, \bibinfo {author} {\bibfnamefont {I.}~\bibnamefont {R.}}, \bibinfo {author} {\bibfnamefont {K.}~\bibnamefont {D.}}, \bibinfo {author} {\bibfnamefont {K.}~\bibnamefont {S.}}, \bibinfo {author} {\bibfnamefont {L.}~\bibnamefont {A.}}, \bibinfo {author} {\bibfnamefont {R.}~\bibnamefont {P.}}, \bibinfo {author} {\bibfnamefont {S.}~\bibnamefont {C.}}, \ and\ \bibinfo {author} {\bibfnamefont {S.}~\bibnamefont {J.}},\ }\href {\doibase 10.3254/978-1-61499-526-5-169} {\bibfield  {journal} {\bibinfo  {journal} {Proceedings of the International School of Physics Enrico Fermi;}\ }\textbf {\bibinfo {volume} {189}},\ \bibinfo {pages} {169–187} (\bibinfo {year} {2015})}\BibitemShut {NoStop}%
\bibitem [{\citenamefont {Benhelm}(2008)}]{Benhelm_thesis}%
  \BibitemOpen
  \bibfield  {author} {\bibinfo {author} {\bibfnamefont {J.}~\bibnamefont {Benhelm}},\ }\href {https://www.quantumoptics.at/en/publications/ph-d-theses.html} {\bibfield  {journal} {\bibinfo  {journal} {PhD thesis}\ } (\bibinfo {year} {2008})}\BibitemShut {NoStop}%
\bibitem [{\citenamefont {Cirac}\ \emph {et~al.}(1992)\citenamefont {Cirac}, \citenamefont {Blatt}, \citenamefont {Zoller},\ and\ \citenamefont {Phillips}}]{PhysRevA.46.2668}%
  \BibitemOpen
  \bibfield  {author} {\bibinfo {author} {\bibfnamefont {J.~I.}\ \bibnamefont {Cirac}}, \bibinfo {author} {\bibfnamefont {R.}~\bibnamefont {Blatt}}, \bibinfo {author} {\bibfnamefont {P.}~\bibnamefont {Zoller}}, \ and\ \bibinfo {author} {\bibfnamefont {W.~D.}\ \bibnamefont {Phillips}},\ }\href {\doibase 10.1103/PhysRevA.46.2668} {\bibfield  {journal} {\bibinfo  {journal} {Phys. Rev. A}\ }\textbf {\bibinfo {volume} {46}},\ \bibinfo {pages} {2668} (\bibinfo {year} {1992})}\BibitemShut {NoStop}%
\bibitem [{\citenamefont {Laine}\ and\ \citenamefont {Vuorinen}(2016)}]{Laine:2016hma}%
  \BibitemOpen
  \bibfield  {author} {\bibinfo {author} {\bibfnamefont {M.}~\bibnamefont {Laine}}\ and\ \bibinfo {author} {\bibfnamefont {A.}~\bibnamefont {Vuorinen}},\ }\href {\doibase 10.1007/978-3-319-31933-9} {\emph {\bibinfo {title} {{Basics of Thermal Field Theory}}}},\ Vol.\ \bibinfo {volume} {925}\ (\bibinfo  {publisher} {Springer},\ \bibinfo {year} {2016})\ \Eprint {http://arxiv.org/abs/1701.01554} {arXiv:1701.01554 [hep-ph]} \BibitemShut {NoStop}%
\bibitem [{\citenamefont {Kapusta}\ and\ \citenamefont {Gale}(2006)}]{kapusta_gale_2006}%
  \BibitemOpen
  \bibfield  {author} {\bibinfo {author} {\bibfnamefont {J.~I.}\ \bibnamefont {Kapusta}}\ and\ \bibinfo {author} {\bibfnamefont {C.}~\bibnamefont {Gale}},\ }\href {\doibase 10.1017/CBO9780511535130} {\emph {\bibinfo {title} {Finite-Temperature Field Theory: Principles and Applications}}},\ \bibinfo {edition} {2nd}\ ed.,\ Cambridge Monographs on Mathematical Physics\ (\bibinfo  {publisher} {Cambridge University Press},\ \bibinfo {year} {2006})\BibitemShut {NoStop}%
\bibitem [{\citenamefont {Gong}\ \emph {et~al.}(2010)\citenamefont {Gong}, \citenamefont {Lin},\ and\ \citenamefont {Duan}}]{PhysRevLett.105.265703}%
  \BibitemOpen
  \bibfield  {author} {\bibinfo {author} {\bibfnamefont {Z.-X.}\ \bibnamefont {Gong}}, \bibinfo {author} {\bibfnamefont {G.-D.}\ \bibnamefont {Lin}}, \ and\ \bibinfo {author} {\bibfnamefont {L.-M.}\ \bibnamefont {Duan}},\ }\href {\doibase 10.1103/PhysRevLett.105.265703} {\bibfield  {journal} {\bibinfo  {journal} {Phys. Rev. Lett.}\ }\textbf {\bibinfo {volume} {105}},\ \bibinfo {pages} {265703} (\bibinfo {year} {2010})}\BibitemShut {NoStop}%
\bibitem [{\citenamefont {Li}\ \emph {et~al.}(2019)\citenamefont {Li}, \citenamefont {Yan}, \citenamefont {Chen}, \citenamefont {Liu}, \citenamefont {Zhou}, \citenamefont {Zhang}, \citenamefont {Yang},\ and\ \citenamefont {Feng}}]{PhysRevA.99.063402}%
  \BibitemOpen
  \bibfield  {author} {\bibinfo {author} {\bibfnamefont {J.}~\bibnamefont {Li}}, \bibinfo {author} {\bibfnamefont {L.~L.}\ \bibnamefont {Yan}}, \bibinfo {author} {\bibfnamefont {L.}~\bibnamefont {Chen}}, \bibinfo {author} {\bibfnamefont {Z.~C.}\ \bibnamefont {Liu}}, \bibinfo {author} {\bibfnamefont {F.}~\bibnamefont {Zhou}}, \bibinfo {author} {\bibfnamefont {J.~Q.}\ \bibnamefont {Zhang}}, \bibinfo {author} {\bibfnamefont {W.~L.}\ \bibnamefont {Yang}}, \ and\ \bibinfo {author} {\bibfnamefont {M.}~\bibnamefont {Feng}},\ }\href {\doibase 10.1103/PhysRevA.99.063402} {\bibfield  {journal} {\bibinfo  {journal} {Phys. Rev. A}\ }\textbf {\bibinfo {volume} {99}},\ \bibinfo {pages} {063402} (\bibinfo {year} {2019})}\BibitemShut {NoStop}%
\bibitem [{\citenamefont {Liu}\ \emph {et~al.}(2020)\citenamefont {Liu}, \citenamefont {Chen}, \citenamefont {Li}, \citenamefont {Zhang}, \citenamefont {Li}, \citenamefont {Zhou}, \citenamefont {Su}, \citenamefont {Yan},\ and\ \citenamefont {Feng}}]{PhysRevA.102.033116}%
  \BibitemOpen
  \bibfield  {author} {\bibinfo {author} {\bibfnamefont {Z.}~\bibnamefont {Liu}}, \bibinfo {author} {\bibfnamefont {L.}~\bibnamefont {Chen}}, \bibinfo {author} {\bibfnamefont {J.}~\bibnamefont {Li}}, \bibinfo {author} {\bibfnamefont {H.}~\bibnamefont {Zhang}}, \bibinfo {author} {\bibfnamefont {C.}~\bibnamefont {Li}}, \bibinfo {author} {\bibfnamefont {F.}~\bibnamefont {Zhou}}, \bibinfo {author} {\bibfnamefont {S.}~\bibnamefont {Su}}, \bibinfo {author} {\bibfnamefont {L.}~\bibnamefont {Yan}}, \ and\ \bibinfo {author} {\bibfnamefont {M.}~\bibnamefont {Feng}},\ }\href {\doibase 10.1103/PhysRevA.102.033116} {\bibfield  {journal} {\bibinfo  {journal} {Phys. Rev. A}\ }\textbf {\bibinfo {volume} {102}},\ \bibinfo {pages} {033116} (\bibinfo {year} {2020})}\BibitemShut {NoStop}%
\bibitem [{\citenamefont {Kiethe}\ \emph {et~al.}(2021)\citenamefont {Kiethe}, \citenamefont {Timm}, \citenamefont {Landa}, \citenamefont {Kalincev}, \citenamefont {Morigi},\ and\ \citenamefont {Mehlst\"aubler}}]{PhysRevB.103.104106}%
  \BibitemOpen
  \bibfield  {author} {\bibinfo {author} {\bibfnamefont {J.}~\bibnamefont {Kiethe}}, \bibinfo {author} {\bibfnamefont {L.}~\bibnamefont {Timm}}, \bibinfo {author} {\bibfnamefont {H.}~\bibnamefont {Landa}}, \bibinfo {author} {\bibfnamefont {D.}~\bibnamefont {Kalincev}}, \bibinfo {author} {\bibfnamefont {G.}~\bibnamefont {Morigi}}, \ and\ \bibinfo {author} {\bibfnamefont {T.~E.}\ \bibnamefont {Mehlst\"aubler}},\ }\href {\doibase 10.1103/PhysRevB.103.104106} {\bibfield  {journal} {\bibinfo  {journal} {Phys. Rev. B}\ }\textbf {\bibinfo {volume} {103}},\ \bibinfo {pages} {104106} (\bibinfo {year} {2021})}\BibitemShut {NoStop}%
\bibitem [{\citenamefont {Ryder}(1996)}]{ryder}%
  \BibitemOpen
  \bibfield  {author} {\bibinfo {author} {\bibfnamefont {L.~H.}\ \bibnamefont {Ryder}},\ }\href {\doibase 10.1017/CBO9780511813900} {\emph {\bibinfo {title} {Quantum Field Theory}}},\ \bibinfo {edition} {2nd}\ ed.\ (\bibinfo  {publisher} {Cambridge University Press},\ \bibinfo {year} {1996})\ Chap.~\bibinfo {chapter} {6}\BibitemShut {NoStop}%
\bibitem [{\citenamefont {Dyson}(1949)}]{PhysRev.75.1736}%
  \BibitemOpen
  \bibfield  {author} {\bibinfo {author} {\bibfnamefont {F.~J.}\ \bibnamefont {Dyson}},\ }\href {\doibase 10.1103/PhysRev.75.1736} {\bibfield  {journal} {\bibinfo  {journal} {Phys. Rev.}\ }\textbf {\bibinfo {volume} {75}},\ \bibinfo {pages} {1736} (\bibinfo {year} {1949})}\BibitemShut {NoStop}%
\bibitem [{\citenamefont {Schwinger}(1951{\natexlab{a}})}]{sch1}%
  \BibitemOpen
  \bibfield  {author} {\bibinfo {author} {\bibfnamefont {J.}~\bibnamefont {Schwinger}},\ }\href {\doibase 10.1073/pnas.37.7.452} {\bibfield  {journal} {\bibinfo  {journal} {Proceedings of the National Academy of Sciences}\ }\textbf {\bibinfo {volume} {37}},\ \bibinfo {pages} {452} (\bibinfo {year} {1951}{\natexlab{a}})}\BibitemShut {NoStop}%
\bibitem [{\citenamefont {Schwinger}(1951{\natexlab{b}})}]{sch2}%
  \BibitemOpen
  \bibfield  {author} {\bibinfo {author} {\bibfnamefont {J.}~\bibnamefont {Schwinger}},\ }\href {\doibase 10.1073/pnas.37.7.455} {\bibfield  {journal} {\bibinfo  {journal} {Proceedings of the National Academy of Sciences}\ }\textbf {\bibinfo {volume} {37}},\ \bibinfo {pages} {455} (\bibinfo {year} {1951}{\natexlab{b}})}\BibitemShut {NoStop}%
\bibitem [{\citenamefont {Coleman}(1985)}]{coleman}%
  \BibitemOpen
  \bibfield  {author} {\bibinfo {author} {\bibfnamefont {S.}~\bibnamefont {Coleman}},\ }\href {\doibase 10.1017/CBO9780511565045} {\emph {\bibinfo {title} {Aspects of Symmetry: Selected Erice Lectures}}}\ (\bibinfo  {publisher} {Cambridge University Press},\ \bibinfo {year} {1985})\BibitemShut {NoStop}%
\bibitem [{\citenamefont {Schnitzer}(1974)}]{PhysRevD.10.2042}%
  \BibitemOpen
  \bibfield  {author} {\bibinfo {author} {\bibfnamefont {H.~J.}\ \bibnamefont {Schnitzer}},\ }\href {\doibase 10.1103/PhysRevD.10.2042} {\bibfield  {journal} {\bibinfo  {journal} {Phys. Rev. D}\ }\textbf {\bibinfo {volume} {10}},\ \bibinfo {pages} {2042} (\bibinfo {year} {1974})}\BibitemShut {NoStop}%
\bibitem [{\citenamefont {Abramowitz}\ and\ \citenamefont {Stegun}(1964)}]{elliptic1}%
  \BibitemOpen
  \bibfield  {author} {\bibinfo {author} {\bibfnamefont {M.}~\bibnamefont {Abramowitz}}\ and\ \bibinfo {author} {\bibfnamefont {I.~A.}\ \bibnamefont {Stegun}},\ }\href@noop {} {\emph {\bibinfo {title} {Handbook of mathematical functions with formulas, graphs, and mathematical tables}}},\ Vol.~\bibinfo {volume} {55}\ (\bibinfo  {publisher} {US Government printing office},\ \bibinfo {year} {1964})\BibitemShut {NoStop}%
\bibitem [{\citenamefont {Matsubara}(1955)}]{10.1143/PTP.14.351}%
  \BibitemOpen
  \bibfield  {author} {\bibinfo {author} {\bibfnamefont {T.}~\bibnamefont {Matsubara}},\ }\href {\doibase 10.1143/PTP.14.351} {\bibfield  {journal} {\bibinfo  {journal} {Progress of Theoretical Physics}\ }\textbf {\bibinfo {volume} {14}},\ \bibinfo {pages} {351} (\bibinfo {year} {1955})}\BibitemShut {NoStop}%
\bibitem [{\citenamefont {Kubo}(1957)}]{doi:10.1143/JPSJ.12.570}%
  \BibitemOpen
  \bibfield  {author} {\bibinfo {author} {\bibfnamefont {R.}~\bibnamefont {Kubo}},\ }\href {\doibase 10.1143/JPSJ.12.570} {\bibfield  {journal} {\bibinfo  {journal} {Journal of the Physical Society of Japan}\ }\textbf {\bibinfo {volume} {12}},\ \bibinfo {pages} {570} (\bibinfo {year} {1957})}\BibitemShut {NoStop}%
\bibitem [{\citenamefont {Martin}\ and\ \citenamefont {Schwinger}(1959)}]{PhysRev.115.1342}%
  \BibitemOpen
  \bibfield  {author} {\bibinfo {author} {\bibfnamefont {P.~C.}\ \bibnamefont {Martin}}\ and\ \bibinfo {author} {\bibfnamefont {J.}~\bibnamefont {Schwinger}},\ }\href {\doibase 10.1103/PhysRev.115.1342} {\bibfield  {journal} {\bibinfo  {journal} {Phys. Rev.}\ }\textbf {\bibinfo {volume} {115}},\ \bibinfo {pages} {1342} (\bibinfo {year} {1959})}\BibitemShut {NoStop}%
\bibitem [{\citenamefont {Yang}(2011)}]{thermal}%
  \BibitemOpen
  \bibfield  {author} {\bibinfo {author} {\bibfnamefont {Y.}~\bibnamefont {Yang}},\ }\emph {\bibinfo {title} {An Introduction to Thermal Field Theory}},\ \href@noop {} {Master's thesis},\ \bibinfo  {school} {Imperial College London} (\bibinfo {year} {2011})\BibitemShut {NoStop}%
\bibitem [{\citenamefont {Delcamp}\ and\ \citenamefont {Tilloy}(2020{\natexlab{b}})}]{Delcamp_2020}%
  \BibitemOpen
  \bibfield  {author} {\bibinfo {author} {\bibfnamefont {C.}~\bibnamefont {Delcamp}}\ and\ \bibinfo {author} {\bibfnamefont {A.}~\bibnamefont {Tilloy}},\ }\href {\doibase 10.1103/physrevresearch.2.033278} {\bibfield  {journal} {\bibinfo  {journal} {Physical Review Research}\ }\textbf {\bibinfo {volume} {2}} (\bibinfo {year} {2020}{\natexlab{b}}),\ 10.1103/physrevresearch.2.033278}\BibitemShut {NoStop}%
\bibitem [{\citenamefont {Milsted}\ \emph {et~al.}(2013{\natexlab{b}})\citenamefont {Milsted}, \citenamefont {Haegeman},\ and\ \citenamefont {Osborne}}]{Milsted_2013}%
  \BibitemOpen
  \bibfield  {author} {\bibinfo {author} {\bibfnamefont {A.}~\bibnamefont {Milsted}}, \bibinfo {author} {\bibfnamefont {J.}~\bibnamefont {Haegeman}}, \ and\ \bibinfo {author} {\bibfnamefont {T.~J.}\ \bibnamefont {Osborne}},\ }\href {\doibase 10.1103/physrevd.88.085030} {\bibfield  {journal} {\bibinfo  {journal} {Physical Review D}\ }\textbf {\bibinfo {volume} {88}} (\bibinfo {year} {2013}{\natexlab{b}}),\ 10.1103/physrevd.88.085030}\BibitemShut {NoStop}%
\bibitem [{\citenamefont {Kadoh}\ \emph {et~al.}(2019{\natexlab{b}})\citenamefont {Kadoh}, \citenamefont {Kuramashi}, \citenamefont {Nakamura}, \citenamefont {Sakai}, \citenamefont {Takeda},\ and\ \citenamefont {Yoshimura}}]{Kadoh_2019}%
  \BibitemOpen
  \bibfield  {author} {\bibinfo {author} {\bibfnamefont {D.}~\bibnamefont {Kadoh}}, \bibinfo {author} {\bibfnamefont {Y.}~\bibnamefont {Kuramashi}}, \bibinfo {author} {\bibfnamefont {Y.}~\bibnamefont {Nakamura}}, \bibinfo {author} {\bibfnamefont {R.}~\bibnamefont {Sakai}}, \bibinfo {author} {\bibfnamefont {S.}~\bibnamefont {Takeda}}, \ and\ \bibinfo {author} {\bibfnamefont {Y.}~\bibnamefont {Yoshimura}},\ }\href {\doibase 10.1007/jhep05(2019)184} {\bibfield  {journal} {\bibinfo  {journal} {Journal of High Energy Physics}\ }\textbf {\bibinfo {volume} {2019}} (\bibinfo {year} {2019}{\natexlab{b}}),\ 10.1007/jhep05(2019)184}\BibitemShut {NoStop}%
\bibitem [{\citenamefont {Serone}\ \emph {et~al.}(2018)\citenamefont {Serone}, \citenamefont {Spada},\ and\ \citenamefont {Villadoro}}]{Serone_2018}%
  \BibitemOpen
  \bibfield  {author} {\bibinfo {author} {\bibfnamefont {M.}~\bibnamefont {Serone}}, \bibinfo {author} {\bibfnamefont {G.}~\bibnamefont {Spada}}, \ and\ \bibinfo {author} {\bibfnamefont {G.}~\bibnamefont {Villadoro}},\ }\href {\doibase 10.1007/jhep08(2018)148} {\bibfield  {journal} {\bibinfo  {journal} {Journal of High Energy Physics}\ }\textbf {\bibinfo {volume} {2018}} (\bibinfo {year} {2018}),\ 10.1007/jhep08(2018)148}\BibitemShut {NoStop}%
\bibitem [{\citenamefont {Elias-Mir{\'{o}}}\ \emph {et~al.}(2017)\citenamefont {Elias-Mir{\'{o}}}, \citenamefont {Rychkov},\ and\ \citenamefont {Vitale}}]{Elias_Mir__2017}%
  \BibitemOpen
  \bibfield  {author} {\bibinfo {author} {\bibfnamefont {J.}~\bibnamefont {Elias-Mir{\'{o}}}}, \bibinfo {author} {\bibfnamefont {S.}~\bibnamefont {Rychkov}}, \ and\ \bibinfo {author} {\bibfnamefont {L.~G.}\ \bibnamefont {Vitale}},\ }\href {\doibase 10.1007/jhep10(2017)213} {\bibfield  {journal} {\bibinfo  {journal} {Journal of High Energy Physics}\ }\textbf {\bibinfo {volume} {2017}} (\bibinfo {year} {2017}),\ 10.1007/jhep10(2017)213}\BibitemShut {NoStop}%
\bibitem [{\citenamefont {Pfeuty}(1970)}]{PFEUTY197079}%
  \BibitemOpen
  \bibfield  {author} {\bibinfo {author} {\bibfnamefont {P.}~\bibnamefont {Pfeuty}},\ }\href {\doibase https://doi.org/10.1016/0003-4916(70)90270-8} {\bibfield  {journal} {\bibinfo  {journal} {Annals of Physics}\ }\textbf {\bibinfo {volume} {57}},\ \bibinfo {pages} {79} (\bibinfo {year} {1970})}\BibitemShut {NoStop}%
\bibitem [{\citenamefont {Jackiw}\ and\ \citenamefont {Rebbi}(1976)}]{PhysRevD.13.3398}%
  \BibitemOpen
  \bibfield  {author} {\bibinfo {author} {\bibfnamefont {R.}~\bibnamefont {Jackiw}}\ and\ \bibinfo {author} {\bibfnamefont {C.}~\bibnamefont {Rebbi}},\ }\href {\doibase 10.1103/PhysRevD.13.3398} {\bibfield  {journal} {\bibinfo  {journal} {Phys. Rev. D}\ }\textbf {\bibinfo {volume} {13}},\ \bibinfo {pages} {3398} (\bibinfo {year} {1976})}\BibitemShut {NoStop}%
\end{thebibliography}%

\end{document}